\documentclass[10pt,letterpaper,english,aps]{revtex4}
\usepackage{times}
\usepackage[T1]{fontenc}
\usepackage[latin1]{inputenc}
\usepackage{float}
\usepackage{amsmath}
\usepackage[dvips]{graphicx}
\usepackage{amssymb}

\makeatletter
\usepackage{babel}
\makeatother
\begin{document}

\title{Casimir Force in Non-Planar Geometric Configurations%
\footnote{This paper is based on my Ph.D. dissertation in physics at Virginia
Polytechnic Institute and State University, April, 2004. The first
two chapters (I and II) are former introduction to the Casimir effect
and the general electromagnetic field quantization scheme. My actual
contributions to the physics start from chapter (III) where the reflection
dynamics is formulated. %
} %
\footnote{This is the third revision to the initial version (v1) which have
been submitted to the arXiv.org in April, 2004. As with the second
revision (v2), this third revision (v3) fixes small typographical
errors found in the previous version. %
}}

\author{Sung Nae Cho}

\email{sucho2@vt.edu}

\address{Department of Physics, Virginia Polytechnic Institute and State University,
Blacksburg, Virginia  24061, USA}

\date{Friday, August 06, 2004 }

\begin{abstract}
\noindent The Casimir force for charge-neutral, perfect conductors
of non-planar geometric configurations have been investigated. The
configurations are: (1) the plate-hemisphere, (2) the hemisphere-hemisphere
and (3) the spherical shell. The resulting Casimir forces for these
physical arrangements have been found to be attractive. The repulsive
Casimir force found by Boyer for a spherical shell is a special case
requiring stringent material property of the sphere, as well as the
specific boundary conditions for the wave modes inside and outside
of the sphere. The necessary criteria in detecting Boyer's repulsive
Casimir force for a sphere are discussed at the end of this investigation. 
\end{abstract}


\maketitle
\tableofcontents{}

\section{Introduction}

The introduction is divided into three parts: (1) \emph{physics},
(2) \emph{applications}, and (3) \emph{developments}. A brief outline
of the physics behind the Casimir effect is discussed in item (1).
In item (2), major impact of Casimir effect on technology and science
is outlined. In item (3), the introduction of this investigation is
concluded with a brief review of the past developments, followed by
a brief outline of the organization of this investigation and its
contributions to the physics.

\subsection{Physics}

When two electrically neutral, conducting plates are placed parallel
to each other, our understanding from classical electrodynamics tells
us that nothing should happen to these plates. The plates are assumed
to be that made of perfect conductors for simplicity. In 1948, H.
B. G. Casimir and D. Polder faced a similar problem in studying forces
between polarizable neutral molecules in colloidal solutions. Colloidal
solutions are viscous materials, such as paint, that contain micron-sized
particles in a liquid matrix. It had been thought that forces between
such polarizable, neutral molecules were governed by the van der Waals
interaction. The van der Waals interaction is also referred to as
the {}``Lennard-Jones interaction.'' It is a long range electrostatic
interaction that acts to attract two nearby polarizable molecules.
Casimir and Polder found to their surprise that there existed an attractive
force which could not be ascribed to the van der Waals theory. Their
experimental result could not be correctly explained unless the retardation
effect was included in van der Waals' theory. This retarded van der
Waals interaction or Lienard-Wiechert dipole-dipole interaction \cite{key-Jackson-EM}
is now known as the Casimir-Polder interaction \cite{key-Casimir-Polder}.
Casimir, following this first work, elaborated on the Casimir-Polder
interaction in predicting the existence of an attractive force between
two electrically neutral, parallel plates of perfect conductors separated
by a small gap \cite{key-Casimir}. This alternative derivation of
the Casimir force is in terms of the difference between the zero-point
energy in vacuum and the zero-point energy in the presence of boundaries.
This force has been confirmed by experiments and the phenomenon is
what is now known as the {}``Casimir Effect.'' The force responsible
for the attraction of two uncharged conducting plates is accordingly
termed the {}``Casimir Force.'' It was shown later that the Casimir
force could be both attractive or repulsive depending on the geometry
and the material property of the conductors \cite{key-Boyer,key-Maclay,key-Kenneth-Klich-Mann-Revzen}. 

\begin{figure}[b]
\begin{center}\includegraphics[%
  scale=0.7]{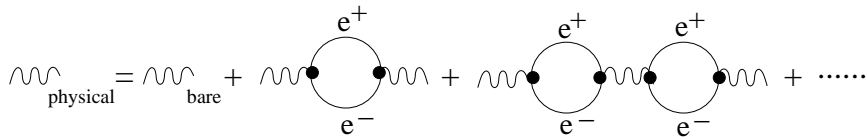}\end{center}

\caption{The vacuum polarization of a photon. \label{cap:vacuum-polarization}}
\end{figure}
 The Casimir effect is regarded as macroscopic manifestation of the
retarded van der Waals interaction between uncharged polarizable molecules
(or atoms). Microscopically, the Casimir effect is due to interactions
between induced multipole moments, where the dipole term is the most
dominant contributor if it is non-vanishing. Therefore, the dipole
interaction is exclusively referred to, unless otherwise explicitly
stated, throughout this investigation. The induced dipole moments
can be qualitatively explained by the concept of {}``vacuum polarization''
in quantum electrodynamics (QED). The idea is that a photon, whether
real or virtual, has a charged particle content. Namely, the internal
loop, illustrated in Figure \ref{cap:vacuum-polarization}, can be
$e^{+}e^{-},$ $\mu^{+}\mu^{-},$ $\tau^{+}\tau^{-},$ $\pi^{+}\pi^{-}$
or $q\bar{q}$ pairs, etc. Its correctness have been born out from
the precision measurements of Lamb shift \cite{key-Lundeen-Pipkin,key-Dubler-et-al}
and photoproduction \cite{key-photo-pro-M-Ya-Amusia,key-photo-pro-F-J-Gilman,key-photo-pro-Lautrup-Peterman-Rafael,key-photo-pro-Krawczyk-Zembrzuski-Staszel}
experiments over a vast range of energies. For the almost zero energy
photons considered in the Casimir effect, these pairs last for a time
interval $\triangle t$ consistent with that given by the Heisenberg
uncertainty principle $\triangle E\cdot\triangle t=h,$ where $\triangle E$
is the energy imbalance and $h$ is the Planck constant. These virtual
charged particles can induce the requisite polarizability on the boundary
of the dielectric (or conducting) plates which explained the Casimir
effect. However, the dipole strength is left as a free parameter in
the calculations because it cannot be readily calculated. Its value
can be determined from experiments. 

Once this idea is taken for granted, one can then move forward to
calculate the effective, temperature averaged, energy due to the dipole-dipole
interactions with the time retardation effect folded in. The energy
between the dielectric (or conducting) media is obtained from the
allowed modes of electromagnetic waves determined by the Maxwell equations
together with the boundary conditions. The Casimir force is then obtained
by taking the negative gradient of the energy in space. This approach,
as opposed to full atomistic treatment of the dielectrics (or conductors),
is justified as long as the most significant photon wavelengths determining
the interaction are large when compared with the spacing of the lattice
points in the media. The effect of all the multiple dipole scattering
by atoms in the dielectric (or conducting) media simply enforces the
macroscopic reflection laws of electromagnetic waves. For instance,
in the case of the two parallel plates, the most significant wavelengths
are those of the order of the plate gap distance. When this wavelength
is large compared with the interatomic distances, the macroscopic
electromagnetic theory can be used with impunity. The geometric configuration
can introduce significant complications, which is the subject matter
this study is going to address.

In order to handle the dipole-dipole interaction Hamiltonian in this
case, the classical electromagnetic fields have to be quantized into
the photon representation first. The photon with non-zero occupation
number have energies in units of $\hbar\omega;$ where $\hbar$ is
the Planck constant divided by $2\pi,$ and $\omega,$ the angular
frequency. The lowest energy state of the electromagnetic fields has
energy $\hbar\omega/2.$ They are called the vacuum or the zero point
energy state, and they play a major role in the Casimir effect. Throughout
this investigation, the terminology {}``photon'' is used to represent
the entity with energy $\hbar\omega,$ or the entity with energy $\hbar\omega/2$
unless explicitly stated otherwise.

\subsection{Applications}

In order to appreciate the importance of the Casimir effect from industry's
point of view, we first examine the theoretical value for the attractive
force between two uncharged conducting parallel plates separated by
a gap of distance $d:$ $F_{C}=-240^{-1}\pi^{2}d^{-4}\hbar c,$ where
$c$ is the speed of light in vacuum and $d$ is the plate gap distance.
To get a sense of the magnitude of this force, two mirrors of an area
of $\sim1\, cm^{2}$ separated by a distance of $\sim1\,\mu m$ would
experience an attractive Casimir force of roughly $\sim10^{-7}\, N,$
which is about the weight of a water droplet of half a millimeter
in diameter. Naturally, the scale of size plays a crucial role in
the Casimir effect. At a gap separation in the ranges of $\sim10\, nm,$
which is roughly about a hundred times the typical size of an atom,
the equivalent Casimir force would be in the range of $1$ atmospheric
pressure. The Casimir force have been verified by Steven Lamoreaux
\cite{key-Lamoreaux} in 1996 to within an experimental uncertainty
of $5$\%. An independent verification of this force have been done
by U. Mohideen and Anushree Roy \cite{key-Mohideen-Roy} in 1998 to
within an experimental uncertainty of $1$\%. It is however emphasized
that these experiments were not done exactly in the same context of
Casimir's original configuration due to technical difficulties associated
with Casimir's idealized perfectly flat surfaces. In 2002, Casimir's
parallel plate configuration have been examined by G. Bressi, G. Carugno,
R. Onofrio, and G. Ruoso \cite{key-Bressi-Carugno-Onofrio-Ruoso}.
Their force coefficient was measured at the 15\% precision level. 

The importance of Casimir effect is most significant for the miniaturization
of modern electronics. The technology already in use that is affected
by the Casimir effect is that of the microelectromechanical systems
(MEMS). These are devices fabricated on the scale of microns and sub-micron
sizes. The order of the magnitude of Casimir force at such a small
length scale can be enormous. It can cause mechanical malfunctions
if the Casimir force is not properly taken into account in the design,
e.g., mechanical parts of a structure could stick together, etc \cite{key-Roukes-Caltech}.
The Casimir force may someday be put to good use in other fields where
nonlinearity is important. Such potential applications requiring nonlinear
phenomena have been demonstrated \cite{key-Chan-Aksyuk-Kleiman-Bishop-Capasso}.
The technology of MEMS hold many promising applications in science
and engineering. With the MEMS soon to be replaced by the next generation
of its kind, the nanoelectromechanical systems or NEMS, understanding
the phenomenon of the Casimir effect become even more crucial. 

Aside from the technology and engineering applications, the Casimir
effect plays a crucial role in accurate force measurements at nanometer
and micrometer scales \cite{key-Gundlach-Merkowtz}. As an example,
if one wants to measure the gravitational force at a distance of atomic
scale, not only the subtraction of the dominant Coulomb force has
to be done, but also the Casimir force, assuming that there is no
effect due to strong and weak interactions. 

Most recently, a new Casimir-like quantum phenomenon have been predicted
by Feigel \cite{key-Feigel}. The contribution of vacuum fluctuations
to the motion of dielectric liquids in crossed electric and magnetic
fields could generate velocities of $\sim50\, nm/s.$ Unlike the ordinary
Casimir effect where its contribution is solely due to low frequency
vacuum modes, the new Casimir-like phenomenon predicted recently by
Feigel is due to the contribution of high frequency vacuum modes.
If this phenomenon is verified, it could be used in the future as
an investigating tool for vacuum fluctuations. Other possible applications
of this new effect lie in fields of microfluidics or precise positioning
of micro-objects such as cold atoms or molecules. 

Everything that was said above dealt with only one aspect of the Casimir
effect, the attractive Casimir force. In spite of many technical challenges
in precision Casimir force measurements \cite{key-Lamoreaux,key-Mohideen-Roy,key-Bressi-Carugno-Onofrio-Ruoso},
the attractive Casimir force is fairly well established. This aspect
of the theory is not however what drives most of the researches in
the field. The Casimir effect also predicts a repulsive force and
many researchers in the field today are focusing on this phenomenon
yet to be confirmed experimentally. Theoretical calculations suggest
that for certain geometric configurations, two neutral conductors
would exhibit repulsive behavior rather than being attractive. The
classic result that started all this is due to Boyer's work on the
Casimir force calculation for an uncharged spherical conducting shell
\cite{key-Boyer}. For a spherical conductor, the net electromagnetic
radiation pressure, which constitute the Casimir force, has a positive
sign, thus being repulsive. This conclusion seems to violate fundamental
principle of physics for the fields outside of the sphere take on
continuum in allowed modes, where as the fields inside the sphere
can only assume discrete wave modes. However, no one has been able
to experimentally confirm this repulsive Casimir force. 

The phenomenon of Casimir effect is too broad, both in theory and
in engineering applications, to be completely summarized here. I hope
this informal brief survey of the phenomenon could motivate people
interested in this remarkable area of quantum physics.

\subsection{Developments}

Casimir's result of attractive force between two uncharged, parallel
conducting plates is thought to be a remarkable application of QED.
This attractive force have been confirmed experimentally to a great
precision as mentioned earlier \cite{key-Lamoreaux,key-Mohideen-Roy,key-Bressi-Carugno-Onofrio-Ruoso}. 

Casimir's attractive force result between two parallel plates has
been unanimously thought to be obvious. Its origin can also be attributed
to the differences in vacuum-field energies between those inside and
outside of the resonator. However, in 1968, T. H. Boyer, then at Harvard
working on his thesis on Casimir effect for an uncharged spherical
shell, had come to a conclusion that the Casimir force was repulsive
for his configuration, which was contrary to popular belief. His result
is the well known repulsive Casimir force prediction for an uncharged
spherical shell of a perfect conductor \cite{key-Boyer}. 

The surprising result of Boyer's work has motivated many physicists,
both in theory and experiment, to search for its evidence. On the
theoretical side, people have tried different configurations, such
as cylinders, cube, etc., and found many more configurations that
can give a repulsive Casimir force \cite{key-Maclay,key-Emre-Tasci-Erkoc,key-Milton}.
Completely different methodologies were developed in striving to correctly
explain the Casimir effect. For example, the {}``Source Theory''
was employed by Schwinger for the explanation of the Casimir effect
\cite{key-Milton,key-Schwinger-DeRaas-Milton,key-Schwinger-Casimir-Light,key-Milonni}.
In spite of the success in finding many boundary geometries that gave
rise to the repulsive Casimir force, the experimental evidence of
a repulsive Casimir effect is yet to be found. The lack of experimental
evidence of a repulsive Casimir force has triggered further examination
of Boyer's work. 

The physics and the techniques employed in the Casimir force calculations
are well established. The Casimir force calculations involve summing
up of the allowed modes of waves in the given resonator. This turned
out to be one of the difficulties in Casimir force calculations. For
the Casimir's original parallel plate configuration, the calculation
was particularly simple due to the fact that zeroes of the sinusoidal
modes are provided by a simple functional relationship, $kd=n\pi,$
where $k$ is the wave number, $d$ is the plate gap distance and
$n$ is a positive integer. This technique can be easily extended
to other boundary geometries such as sphere, cylinder, cone or a cube,
etc. For a sphere, the functional relation that determines the allowed
wave modes in the resonator is $kr_{o}=\alpha_{s,l},$ where $r_{o}$
is the radius of the sphere; and $\alpha_{s,l},$ the $l$th root
of the spherical Bessel function $j_{s}.$ The same convention is
applied to all other Bessel function solutions. The allowed wave modes
of a cylindrical resonator is determined by a simple functional relation
$ka_{o}=\beta_{s,l},$ where $a_{o}$ is the cylinder radius; and
$\beta_{s,l},$ the zeroes of cylindrical Bessel functions $J_{s}.$ 

One of the major difficulties in the Casimir force calculation for
nontrivial boundaries such as those considered in this investigation
is in defining the functional relation that determines the allowed
modes in a  given resonator. For example, for the hemisphere-hemisphere
boundary configuration, the radiation originating from one hemisphere
would enter the other and run through a complex series of reflections
before escaping the hemispherical cavity. The allowed vacuum-field
modes in the resonator is then governed by a functional relation $k\left\Vert \vec{R'}_{2}-\vec{R'}_{1}\right\Vert =n\pi,$
where $\left\Vert \vec{R'}_{2}-\vec{R'}_{1}\right\Vert $ is the distance
between two successive reflection points $\vec{R'}_{1}$ and $\vec{R'}_{2}$
of the resonator, as is illustrated in Figure \ref{cap:cross-sectional-view-plate-and-hemisphere}
of section (III) on reflection dynamics. As will be shown in the subsequent
sections, the actual functional form for $\left\Vert \vec{R'}_{2}-\vec{R'}_{1}\right\Vert $
is not simple even though the physics behind $\left\Vert \vec{R'}_{2}-\vec{R'}_{1}\right\Vert $
is particularly simple: the application of the law of reflections.
The task of obtaining the functional relation $k\left\Vert \vec{R'}_{2}-\vec{R'}_{1}\right\Vert =n\pi$
for the hemisphere-hemisphere, the plate-hemisphere, and the sphere
configuration formed by bringing in two hemispheres together constitutes
the major part of this investigation. 

This investigation is not about questioning the theoretical origin
of the Casimir effect. Instead, its emphasis is on applying the Casimir
effect as already known to determine the sign of Casimir force for
the realistic experiments. In spite of a number of successes in the
theoretical study of repulsive Casimir force, most of the configurations
are unrealistic. In order to experimentally verify Boyer's repulsive
force for a charge-neutral spherical shell made of perfect conductor,
one should consider the case where the sphere is formed by bringing
in two hemispheres together. When the two hemispheres are closed,
it mimics that of Boyer's sphere. It is, however, shown later in this
investigation that a configuration change from hemisphere-hemisphere
to a sphere induces non-spherically symmetric energy flow that is
not present in Boyer's sphere. Because Boyer's sphere gives a repulsive
Casimir force, once those two closed hemispheres are released, they
must repulse if Boyer's prediction were correct. Although the two
hemisphere configuration have been studied for decades, no one has
yet carried out its analytical calculation successfully. The analytical
solutions on two hemispheres, existing so far, was done by considering
the two hemispheres that were separated by an infinitesimal distance.
In this investigation, the consideration of two hemispheres is not
limited to such infinitesimal separations. 

The three physical arrangements being studied in this investigation
are: (1) the plate-hemisphere, (2) the hemisphere-hemisphere and (3)
the sphere formed by brining in two hemispheres together. Although
there are many other boundary configurations that give repulsive Casimir
force, the configurations under consideration were chosen mainly because
of the following reasons: (1) to be able to check experimentally the
Boyer's repulsive Casimir force result for a spherical shell, (2)
the experimental work involving configurations similar to that of
the plate-hemisphere configuration is underway \cite{key-Chan-Aksyuk-Kleiman-Bishop-Capasso};
and (3) to the best of our knowledge, no detailed analytical study
on these three configurations exists to date. 

The motivation behind to mathematically model the plate-hemisphere
system came from the experiment done by a group at the Bell Laboratory
\cite{key-Chan-Aksyuk-Kleiman-Bishop-Capasso} in which they bring
in an atomic-force-probe to a flopping plate to observe the Casimir
force which can affect the motion of the plate. In our derivations
of the equations of motion, the configuration is that of the {}``plate
displaced on upper side of a bowl (hemisphere).'' The Bell Laboratory
apparatus can be easily mimicked by simply displacing the plate to
the under side of the bowl, which we have not done. The motivation
behind the hemisphere-hemisphere system actually arose from an article
by Kenneth and Nussinov \cite{key-Kenneth-Nussinov}. In their paper,
they speculate on how the edges of the hemispheres may produce effects
such that two arbitrarily close hemispheres cannot mimic Boyer's sphere.
This led to their heuristic conclusion which stated that Boyer's sphere
can never be the same as the two arbitrarily close hemispheres. 

To the best of our knowledge, two of the geometrical configurations
investigated in this work have not yet been investigated by others.
They are the plate-hemisphere and the hemisphere-hemisphere configurations.
This does not mean that these boundary configurations were not known
to the researchers in the field, e.g., \cite{key-Kenneth-Nussinov}.
For the case of the hemisphere-hemisphere configuration, people realized
that it could be the best way to test for the existence of a repulsive
Casimir force for a sphere as predicted by Boyer. The sphere configuration
investigated in this work, which is formed by bringing two hemispheres
together, contains non-spherically symmetric energy flows that are
not present in Boyer's sphere. In that regards, the treatment of the
sphere geometry here is different from that of Boyer. 

The basic layout of this investigation is as follows: (1) \emph{Introduction},
(2) \emph{Theory}, (3) \emph{Calculations}, and (4) \emph{Results}.
The formal introduction of the theory is addressed in chapters (1)
and (2). The original developments resulting from this investigation
are contained in chapters (3) and (4). The brief outline of each chapter
is the following: In chapter (1), a brief introduction to the physics
is addressed; and the application importance and major developments
in this field are discussed. In chapter (2), the formal aspect of
the theory is addressed, which includes the detailed outline of the
Casimir-Polder interaction and brief descriptions of various techniques
that are currently used in Casimir force calculations. In chapter
(3), the actual Casimir force calculations pertaining to the boundary
geometries considered in this investigation are derived. The important
functional relation for $\left\Vert \vec{R'}_{2}-\vec{R'}_{1}\right\Vert $
is developed here. The dynamical aspect of the Casimir effect is also
introduced here. Due to the technical nature of the derivations, many
of the results presented are referred to the detailed derivations
contained in the appendices. In chapter (4), the results are summarized.
Lastly, the appendices have been added in order to accommodate the
tedious and lengthy derivations to keep the text from losing focus
due to mathematical details. 

The goal of this investigation is not to embark so much on the theory
side of the Casimir effect. Instead, its emphasis is on bringing forth
the suggestions that might be useful in detecting the repulsive Casimir
effect originally initiated by Boyer on an uncharged spherical shell.
In concluding this brief outline of the motivation behind this investigation,
it must be added that if by any chance someone already did these work
that have been claimed to be the original developments in this investigation,
I was not aware of their work at the time of this work was being prepared.
And, should that turn out to be the case, I would like to express
my sincere apology for not referencing their work in this investigation.

\section{Casimir Effect}

The Casimir effect is divided into two major categories: (1) the electromagnetic
Casimir effect and (2) the fermionic Casimir effect. As the titles
suggest, the electromagnetic Casimir effect is due to the fluctuations
in a massless Maxwell bosonic fields, whereas the fermionic Casimir
effect is due to the fluctuations in a massless Dirac fermionic fields.
The primary distinction between the two types of Casimir effect is
in the boundary conditions among other differences. The boundary conditions
appropriate to the Dirac equations are the so called {}``bag-model''
boundary conditions, whereas the electromagnetic Casimir effect follows
the boundary conditions of the Maxwell equations. The details of the
fermionic force can be found in references \cite{key-Milton,key-Milonni}. 

In this investigation, only the electromagnetic Casimir effect is
considered. As it is inherently an electromagnetic phenomenon, we
begin with a brief introduction to the Maxwell equations, followed
by the quantization of electromagnetic fields.

\subsection{Quantization of Free Maxwell Field}

There are four Maxwell equations: \begin{align}
\vec{\nabla}\cdot\vec{E}\left(\vec{R},t\right) & =4\pi\rho\left(\vec{R},t\right),\label{eq:Maxwell-E-equations-1}\end{align}
 \begin{align}
\vec{\nabla}\cdot\vec{B}\left(\vec{R},t\right) & =0,\label{eq:Maxwell-E-equations-2}\end{align}
\begin{align}
\vec{\nabla}\times\vec{E}\left(\vec{R},t\right) & =-\frac{1}{c}\frac{\partial\vec{B}\left(\vec{R},t\right)}{\partial t},\label{eq:Maxwell-E-equations-3}\end{align}
 \begin{align}
\vec{\nabla}\times\vec{B}\left(\vec{R},t\right) & =\frac{4\pi}{c}\vec{J}\left(\vec{R},t\right)+\frac{1}{c}\frac{\partial\vec{E}\left(\vec{R},t\right)}{\partial t},\label{eq:Maxwell-B-equations-4}\end{align}
 where the Gaussian system of units have been adopted. The electric
and the magnetic field are defined respectively by $\vec{E}=-\vec{\nabla}\Phi-c^{-1}\partial_{t}\vec{A}$
and $\vec{B}=\vec{\nabla}\times\vec{A},$ where $\Phi$ is the scalar
potential and $\vec{A}$ is the vector potential. Equations (\ref{eq:Maxwell-E-equations-1})
through (\ref{eq:Maxwell-B-equations-4}) are combined to give \begin{align*}
\sum_{l=1}^{3}\left\{ 4\pi\partial_{l}\rho+\frac{4\pi}{c^{2}}\partial_{t}J_{l}-\sum_{m=1}^{3}\partial_{m}^{2}\left[\partial_{l}\Phi+\frac{1}{c}\partial_{t}A_{l}+\epsilon_{ljk}\partial_{j}A_{k}\right]\right.\\
\left.+\frac{4\pi}{c}\epsilon_{lmn}\partial_{m}J_{n}+\frac{1}{c^{2}}\partial_{t}^{2}\left[\partial_{l}\Phi+\frac{1}{c}\partial_{t}A_{l}\right]+\frac{1}{c^{2}}\epsilon_{ljk}\partial_{j}\partial_{t}^{2}A_{k}\right\} \hat{e_{l}} & =0,\end{align*}
 where the Einstein summation convention is assumed for repeated indices.
Because the components along basis direction $\hat{e_{l}}$ are independent
of each other, the above vector algebraic relation becomes three equations:
\begin{align}
4\pi\partial_{l}\rho+\frac{4\pi}{c^{2}}\partial_{t}J_{l}-\sum_{m=1}^{3}\partial_{m}^{2}\left[\partial_{l}\Phi+\frac{1}{c}\partial_{t}A_{l}+\epsilon_{ljk}\partial_{j}A_{k}\right]\nonumber \\
+\frac{4\pi}{c}\epsilon_{lmn}\partial_{m}J_{n}+\frac{1}{c^{2}}\partial_{t}^{2}\left[\partial_{l}\Phi+\frac{1}{c}\partial_{t}A_{l}\right]+\frac{1}{c^{2}}\epsilon_{ljk}\partial_{j}\partial_{t}^{2}A_{k} & =0,\label{eq:Maxwell-A-equations}\end{align}
 where $l=1,2,3.$ 

To understand the full implications of electrodynamics, one has to
solve the above set of coupled differential equations. Unfortunately,
they are in general too complicated to solve exactly. The need to
choose an appropriate gauge to approximately solve the above equations
is not only an option, it is a must. Also, for what is concerned with
the vacuum-fields, that is, the radiation from matter when it is in
its lowest energy state, information about the charge density $\rho$
and the current density $\vec{J}$ must be first prescribed. Unfortunately,
to describe properly the charge and current densities of matter is
a major difficulty in its own. Therefore, the charge density $\rho$
and the current density $\vec{J}$ are set to be zero for the sake
of simplicity and the Coulomb gauge, $\vec{\nabla}\cdot\vec{A}=0,$
is adopted. Under these conditions, equation (\ref{eq:Maxwell-A-equations})
is simplified to \begin{eqnarray*}
\partial_{l}^{2}A_{l}-c^{-2}\partial_{t}^{2}A_{l}=0, &  & l=1,2,3.\end{eqnarray*}
 The steady state monochromatic solution is then of the form \begin{align*}
\vec{A}\left(\vec{R},t\right) & =\alpha\left(t\right)\vec{A}_{0}\left(\vec{R}\right)+\alpha^{*}\left(t\right)\vec{A}_{0}^{*}\left(\vec{R}\right)\\
 & =\alpha\left(0\right)\exp\left(-i\omega t\right)\vec{A}_{0}\left(\vec{R}\right)+\alpha^{*}\left(0\right)\exp\left(i\omega t\right)\vec{A}_{0}^{*}\left(\vec{R}\right),\end{align*}
 where $\vec{A}_{0}\left(\vec{R}\right)$ is the solution to the Helmholtz
equation $\nabla^{2}\vec{A}_{0}\left(\vec{R}\right)+c^{-2}\omega^{2}\vec{A}_{0}\left(\vec{R}\right)=0,$
and $\alpha\left(t\right)$ is the solution of the temporal differential
equation satisfying $\ddot{\alpha}\left(t\right)+\omega^{2}\alpha\left(t\right)=0.$
With the solution $\vec{A}\left(\vec{R},t\right),$ the electric and
the magnetic fields are found to be \begin{align*}
\vec{E}\left(\vec{R},t\right) & =-\frac{1}{c}\left[\dot{\alpha}\left(t\right)\vec{A}_{0}\left(\vec{R}\right)+\dot{\alpha}^{*}\left(t\right)\vec{A}_{0}^{*}\left(\vec{R}\right)\right]\end{align*}
 and \begin{align*}
\vec{B}\left(\vec{R},t\right) & =\alpha\left(t\right)\vec{\nabla}\times\vec{A}_{0}\left(\vec{R}\right)+\alpha^{*}\left(t\right)\vec{\nabla}\times\vec{A}_{0}^{*}\left(\vec{R}\right).\end{align*}
 The electromagnetic field Hamiltonian becomes: \begin{align}
\mathcal{H}_{F} & =\frac{1}{8\pi}\int_{V}\left[\vec{E}\left(\vec{R},t\right)\cdot\vec{E}^{*}\left(\vec{R},t\right)+\vec{B}\left(\vec{R},t\right)\cdot\vec{B}^{*}\left(\vec{R},t\right)\right]dV\nonumber \\
 & =\frac{k^{2}}{2\pi}\left\Vert \alpha\left(t\right)\right\Vert ^{2},\label{eq:Field-Kamiltonian-Classical}\end{align}
 where $k$ is a wave number, and $\vec{A}_{0}\left(\vec{R}\right)$
have been normalized such that $\int_{V}\left\Vert \vec{A}_{0}\left(\vec{R}\right)\right\Vert ^{2}dV=1.$

We can transform $\mathcal{H}_{F}$ into the {}``normal coordinate
representation'' through the introduction of {}``creation'' and
{}``annihilation'' operators, $a^{\dagger}$ and $a.$ The resulting
field Hamiltonian $\mathcal{H}_{F}$ of equation (\ref{eq:Field-Kamiltonian-Classical})
is identical in form to that of the canonically transformed simple
harmonic oscillator, $\mathcal{H}_{SH}\propto p^{2}+q^{2}\rightarrow K_{SH}\propto a^{\dagger}a.$
For the free electromagnetic field Hamiltonian, the canonical transformation
is to follow the sequence $K_{SH}\propto\left\Vert \alpha\left(t\right)\right\Vert ^{2}\rightarrow\mathcal{H}_{SH}\propto E^{2}+B^{2}$
under a properly chosen generating function. The result is that with
the following physical quantities, \begin{eqnarray*}
q\left(t\right)=\frac{i}{c\sqrt{4\pi}}\left[\alpha\left(t\right)-\alpha^{*}\left(t\right)\right], &  & p\left(t\right)=\frac{k}{\sqrt{4\pi}}\left[\alpha\left(t\right)+\alpha^{*}\left(t\right)\right],\end{eqnarray*}
 the free field Hamiltonian of equation (\ref{eq:Field-Kamiltonian-Classical})
becomes \begin{align}
\mathcal{H}_{F} & =\frac{1}{2}\left[p^{2}\left(t\right)+\omega^{2}q^{2}\left(t\right)\right],\label{eq:Field-Hamiltonian-Classical}\end{align}
 which is identical to the Hamiltonian of the simple harmonic oscillator.
Then, through a direct comparison and observation with the usual simple
harmonic oscillator Hamiltonian in quantum mechanics, the following
replacements are made \begin{eqnarray*}
\alpha\left(t\right)\rightarrow\sqrt{\frac{2\pi\hbar c^{2}}{\omega}}a\left(t\right), &  & \alpha^{*}\left(t\right)\rightarrow\sqrt{\frac{2\pi\hbar c^{2}}{\omega}}a^{\dagger}\left(t\right),\end{eqnarray*}
 and, the quantized relations for $\vec{A}\left(\vec{R},t\right),$
$\vec{E}\left(\vec{R},t\right)$ and $\vec{B}\left(\vec{R},t\right)$
are found, \begin{align}
\vec{A}\left(\vec{R},t\right) & =\sqrt{\frac{2\pi\hbar c^{2}}{\omega}}\left[a\left(t\right)\vec{A}_{0}\left(\vec{R}\right)+a^{\dagger}\left(t\right)\vec{A}_{0}^{*}\left(\vec{R}\right)\right],\label{eq:monochromatic-A}\end{align}
 \begin{align}
\vec{E}\left(\vec{R},t\right) & =i\sqrt{2\pi\hbar\omega}\left[a\left(t\right)\vec{A}_{0}\left(\vec{R}\right)-a^{\dagger}\left(t\right)\vec{A}_{0}^{*}\left(\vec{R}\right)\right],\label{eq:monochromatic-E}\end{align}
 \begin{align}
\vec{B}\left(\vec{R},t\right) & =\sqrt{\frac{2\pi\hbar c^{2}}{\omega}}\left[a\left(t\right)\vec{\nabla}\times\vec{A}_{0}\left(\vec{R}\right)+a^{\dagger}\left(t\right)\vec{\nabla}\times\vec{A}_{0}^{*}\left(\vec{R}\right)\right],\label{eq:monochromatic-B}\end{align}
 where it is understood that $\vec{A}\left(\vec{R},t\right),$ $\vec{E}\left(\vec{R},t\right)$
and $\vec{B}\left(\vec{R},t\right)$ are now quantum mechanical operators.
The associated field Hamiltonian operator for the photon is then written
\begin{align}
\mathcal{\hat{H}}_{F,mono} & =\left[a^{\dagger}\left(t\right)a\left(t\right)+\frac{1}{2}\right]\hbar\omega,\label{eq:stationary-state-energy-1D-Maxwell-monochromatic}\end{align}
 where the hat $\left(\wedge\right)$ over $\mathcal{H}_{F,mono}$
now denotes an operator.

The generalization of the above quantization procedure for a monochromatic
field to a multimode field is straightforward. We start by redefining
the monochromatic mode function, $\vec{A}_{0}\left(\vec{R}\right),$
with the multimode counterpart, $\vec{A}_{0,\vec{k'},\lambda'}\left(\vec{R}\right),$
\begin{eqnarray*}
\vec{A}_{0,\vec{k'},\lambda'}\left(\vec{R}\right)=\frac{1}{\sqrt{V}}\exp\left(i\vec{k'}\cdot\vec{R}\right)\hat{\epsilon}_{\vec{k'},\lambda'}, &  & \lambda'=1,2,\end{eqnarray*}
 where $V$ is the quantization volume; $\hat{\epsilon}_{\vec{k'},\lambda'},$
the polarization of the field mode, and the subscripts $\vec{k'}$
and $\lambda'$ denotes the particular modes of the wave. Similarly,
the replacement is done for monochromatic $a\left(t\right):$ \begin{eqnarray*}
a\left(t\right)=a\left(0\right)\exp\left(-i\omega t\right) & \rightarrow & a_{\vec{k'},\lambda'}\left(t\right)=a_{\vec{k'},\lambda'}\left(0\right)\exp\left(-i\omega_{k'}t\right),\end{eqnarray*}
 where the angular frequency $\omega$ in multimode representation
have been replaced by $\omega_{k'}$ to denote particular mode of
wave. The replacements for the Hermitian conjugates, $\vec{A}_{0,\vec{k'},\lambda'}^{*}\left(\vec{R}\right)$
and $a_{\vec{k'},\lambda'}^{\dagger}\left(t\right),$ are straightforward.
With $\vec{A}_{0,\vec{k'},\lambda'}\left(\vec{R}\right),$ $\vec{A}_{0,\vec{k'},\lambda'}^{*}\left(\vec{R}\right)$
and $a_{\vec{k'},\lambda'}\left(t\right),$ $a_{\vec{k'},\lambda'}^{\dagger}\left(t\right),$
the monochromatic vector potential $\vec{A}\left(\vec{R},t\right)$
of equation (\ref{eq:monochromatic-A}) is replaced by the multimode
counterpart, the multimode vector potential $\vec{A}_{\vec{k'},\lambda'}\left(\vec{R},t\right):$
\begin{align*}
\vec{A}_{\vec{k'},\lambda'}\left(\vec{R},t\right) & =\sqrt{\frac{2\pi\hbar c^{2}}{\omega_{k'}V}}\left[a_{\vec{k'},\lambda'}\left(t\right)\vec{A}_{0,\vec{k'},\lambda'}\left(\vec{R}\right)+a_{\vec{k'},\lambda'}^{\dagger}\left(t\right)\vec{A}_{0,\vec{k'},\lambda'}^{*}\left(\vec{R}\right)\right],\end{align*}
 or substituting in the explicit expressions for $\vec{A}_{0,\vec{k'},\lambda'}\left(\vec{R}\right),$
$\vec{A}_{0,\vec{k'},\lambda'}^{*}\left(\vec{R}\right)$ and $a_{\vec{k'},\lambda'}\left(t\right),$
$a_{\vec{k'},\lambda'}^{\dagger}\left(t\right),$\begin{align*}
\vec{A}_{\vec{k'},\lambda'}\left(\vec{R},t\right) & =\sqrt{\frac{2\pi\hbar c^{2}}{\omega_{k'}V}}\left[a_{\vec{k'},\lambda'}\left(0\right)\exp\left(i\left[\vec{k'}\cdot\vec{R}-\omega_{k'}t\right]\right)+a_{\vec{k'},\lambda'}^{\dagger}\left(0\right)\exp\left(-i\left[\vec{k'}\cdot\vec{R}-\omega_{k'}t\right]\right)\right]\hat{\epsilon}_{\vec{k'},\lambda'}.\end{align*}
 The linearity of Maxwell's equations then allows us to write for
the total vector potential in free space as \begin{align*}
\vec{A}_{T}\left(\vec{R},t\right) & =\sum_{\vec{k'},\lambda'}\vec{A}_{\vec{k'},\lambda'}\left(\vec{R},t\right)\end{align*}
 or \begin{align}
\vec{A}_{T}\left(\vec{R},t\right) & =\sum_{\vec{k'},\lambda'}\sqrt{\frac{2\pi\hbar c^{2}}{\omega_{k'}V}}\left[a_{\vec{k'},\lambda'}\left(0\right)\exp\left(i\left[\vec{k'}\cdot\vec{R}-\omega_{k'}t\right]\right)+a_{\vec{k'},\lambda'}^{\dagger}\left(0\right)\exp\left(-i\left[\vec{k'}\cdot\vec{R}-\omega_{k'}t\right]\right)\right]\hat{\epsilon}_{\vec{k'},\lambda'}.\label{eq:multimode-A-total}\end{align}
 Similarly, for the total electric and magnetic fields, we find \begin{align}
\vec{E}_{T}\left(\vec{R},t\right) & =i\sum_{\vec{k'},\lambda'}\sqrt{\frac{2\pi\hbar\omega_{k'}}{V}}\left[a_{\vec{k'},\lambda'}\left(0\right)\exp\left(i\left[\vec{k'}\cdot\vec{R}-\omega_{k'}t\right]\right)-a_{\vec{k'},\lambda'}^{\dagger}\left(0\right)\exp\left(-i\left[\vec{k'}\cdot\vec{R}-\omega_{k'}t\right]\right)\right]\hat{\epsilon}_{\vec{k'},\lambda'},\label{eq:monochromatic-E-total}\end{align}
 \begin{align}
\vec{B}_{T}\left(\vec{R},t\right) & =i\sum_{\vec{k'},\lambda'}\sqrt{\frac{2\pi\hbar c^{2}}{\omega_{k'}V}}\left[a_{\vec{k'},\lambda'}\left(0\right)\exp\left(i\left[\vec{k'}\cdot\vec{R}-\omega_{k'}t\right]\right)+a_{\vec{k'},\lambda'}^{\dagger}\left(0\right)\exp\left(-i\left[\vec{k'}\cdot\vec{R}-\omega_{k'}t\right]\right)\right]\vec{k'}\times\hat{\epsilon}_{\vec{k'},\lambda'}.\label{eq:monochromatic-B-total}\end{align}
 The associated total field Hamiltonian operator for the photon is
\begin{align*}
\mathcal{\hat{H}}_{F} & =\frac{1}{8\pi}\int_{V}\left[\vec{E}_{T}\left(\vec{R},t\right)\cdot\vec{E}_{T}^{*}\left(\vec{R},t\right)+\vec{B}_{T}\left(\vec{R},t\right)\cdot\vec{B}_{T}^{*}\left(\vec{R},t\right)\right]dV,\end{align*}
 or using the fact that \begin{align*}
\int_{V}\vec{A}_{0,\vec{k'},\lambda'}\left(\vec{R}\right)\cdot\vec{A}_{0,\vec{k'},\lambda'}^{*}\left(\vec{R}\right)dV & =\delta_{\vec{k'},\vec{k}}^{3}\delta_{\lambda',\lambda},\end{align*}
 along with the commutation relation \begin{align*}
\left[a_{\vec{k'},\lambda'}\left(t\right),a_{\vec{k'},\lambda'}^{\dagger}\left(t\right)\right] & =\delta_{\vec{k'},\vec{k}}^{3}\delta_{\lambda',\lambda},\end{align*}
 \begin{align*}
\left[a_{\vec{k'},\lambda'}\left(t\right),a_{\vec{k'},\lambda'}\left(t\right)\right]=\left[a_{\vec{k'},\lambda'}^{\dagger}\left(t\right),a_{\vec{k'},\lambda'}^{\dagger}\left(t\right)\right] & =0,\end{align*}
 the total field Hamiltonian operator for the photon in free space
is written as \begin{align}
\mathcal{\hat{H}}_{F} & =\sum_{\vec{k'},\lambda'}\left[a_{\vec{k'},\lambda'}^{\dagger}\left(t\right)a_{\vec{k'},\lambda'}\left(t\right)+\frac{1}{2}\right]\hbar\omega_{k'}\nonumber \\
 & =\sum_{\vec{k'},\lambda'}\left[\hat{N}_{\vec{k'},\lambda'}+\frac{1}{2}\right]\hbar ck',\label{eq:stationary-state-energy-1D-Maxwell-multimode}\end{align}
 where $k'\equiv\left\Vert \vec{k'}\right\Vert ,$ and $\hat{N}_{\vec{k'},\lambda'}=a_{\vec{k'},\lambda'}^{\dagger}\left(t\right)a_{\vec{k'},\lambda'}\left(t\right)$
is the occupation number operator. It is understood that in the free
space limit, i.e., $V\rightarrow\infty,$ the \textbf{3D} Kronecker
delta, $\delta_{\vec{k'},\vec{k}}^{3}\equiv\delta_{k'_{x},k_{x}}\delta_{k'_{y},k_{y}}\delta_{k'_{z},k_{z}},$
is to be replaced by the continuum counterpart, $\delta^{3}\left(\vec{k'}-\vec{k}\right)\equiv\delta\left(k'_{x}-k_{x}\right)\delta\left(k'_{y}-k_{y}\right)\delta\left(k'_{z}-k_{z}\right).$
Similarly, the total linear momentum operator for the photon in free
space is written as \begin{align*}
\vec{p}_{F} & =\frac{1}{4\pi c}\int_{V}\vec{E}_{T}\left(\vec{R},t\right)\times\vec{B}_{T}\left(\vec{R},t\right)dV,\end{align*}
 or \begin{align}
\vec{p}_{F} & =\sum_{\vec{k'},\lambda'}\left[a_{\vec{k'},\lambda'}^{\dagger}\left(t\right)a_{\vec{k'},\lambda'}\left(t\right)+\frac{1}{2}\right]\hbar\vec{k'}\nonumber \\
 & =\sum_{\vec{k'},\lambda'}\left[\hat{N}_{\vec{k'},\lambda'}+\frac{1}{2}\right]\hbar\vec{k'}.\label{eq:stationary-state-momentum-1D-Maxwell-multimode}\end{align}
 The eigenvalue of the electromagnetic field energy is then \begin{align*}
\mathcal{H}_{F} & \equiv\sum_{n_{s}=0}^{\infty}\left\langle n_{s};\vec{k'},\lambda'\left|\mathcal{\hat{H}}_{F}\right|n_{s};\vec{k'},\lambda'\right\rangle \\
 & =\sum_{n_{s}=0}^{\infty}\left\langle n_{s};\vec{k'},\lambda'\left|\sum_{\vec{k'},\lambda'}\left\{ \left[\hat{N}_{\vec{k'},\lambda'}+\frac{1}{2}\right]\hbar ck'\right\} \right|n_{s};\vec{k'},\lambda'\right\rangle \\
 & =2\sum_{n_{s}=0}^{\infty}\sum_{\vec{k'}}\left\{ \left[n_{s}+\frac{1}{2}\right]\hbar ck'\right\} ,\end{align*}
 where $n_{s}=\left\langle n_{s};\vec{k'},\lambda'\left|\hat{N}_{\vec{k'},\lambda'}\right|n_{s};\vec{k'},\lambda'\right\rangle .$
The factor of two here comes from the fact that there are two possible
polarizations, i.e., $\left(\lambda'=1,2\right),$ for electromagnetic
fields. With the following definitions, \begin{eqnarray*}
k'\equiv\left\Vert \vec{k'}\right\Vert =\sqrt{\sum_{i=1}^{3}\left[k'_{i}\left(n_{i},L_{i}\right)\right]^{2}}, &  & \sum_{\vec{k'}}\equiv\sum_{n_{1}=0}^{\infty}\sum_{n_{2}=0}^{\infty}\sum_{n_{3}=0}^{\infty},\end{eqnarray*}
 the quantized electromagnetic field energy is written as \begin{align*}
\mathcal{H}_{F} & =2\hbar c\sum_{n_{s}=0}^{\infty}\left[n_{s}+\frac{1}{2}\right]\sum_{n_{1}=0}^{\infty}\sum_{n_{2}=0}^{\infty}\sum_{n_{3}=0}^{\infty}\sqrt{\sum_{i=1}^{3}\left[k'_{i}\left(n_{i},L_{i}\right)\right]^{2}},\end{align*}
 where $L_{i}$ is the quantization length to be determined from the
boundary conditions, and $n_{i}$ is the wave mode number for the
corresponding $k'_{i}.$ The quantized field energy per quantum state
$\left|n_{s};\vec{k'},\lambda'\right\rangle $ is therefore \begin{align}
\mathcal{H}'_{n_{s},b} & \equiv\mathcal{H}_{F}=\left[n_{s}+\frac{1}{2}\right]\hbar c\Theta_{k'}\sum_{n_{1}=0}^{\infty}\sum_{n_{2}=0}^{\infty}\sum_{n_{3}=0}^{\infty}\sqrt{\sum_{i=1}^{3}\left[k'_{i}\left(n_{i},L_{i}\right)\right]^{2}},\label{eq:stationary-state-energy-bounded}\end{align}
 where the subscript $b$ of $\mathcal{H}'_{n_{s},b}$ denotes the
bounded space; and $\Theta_{k'}\equiv2,$ the number of polarizations. 

When the dimensions of boundaries are such that the difference, $\triangle k'_{i}\left(n_{i},L_{i}\right)=k'_{i}\left(n_{i}+1,L_{i}\right)-k'_{i}\left(n_{i},L_{i}\right),$
is infinitesimally small, we can replace the summation in equation
(\ref{eq:stationary-state-energy-bounded}) by integration, \begin{align*}
\sum_{n_{1}=0}^{\infty}\sum_{n_{2}=0}^{\infty}\sum_{n_{3}=0}^{\infty} & \rightarrow\int_{n_{1}=0}^{\infty}\int_{n_{2}=0}^{\infty}\int_{n_{3}=0}^{\infty}dn_{1}dn_{2}dn_{3}\rightarrow\left[f_{1}\left(L_{1}\right)f_{2}\left(L_{2}\right)f_{3}\left(L_{3}\right)\right]^{-1}\int_{0}^{\infty}\int_{0}^{\infty}\int_{0}^{\infty}dk'_{1}dk'_{2}dk'_{3},\end{align*}
 where in the last step the functional definition for $k'_{i}\equiv k'_{i}\left(n_{i},L_{i}\right)=n_{i}f_{i}\left(L_{i}\right)$
have been used to replace $dn_{i}$ by $dk'_{i}/f_{i}\left(L_{i}\right).$
In free space, the electromagnetic field energy for quantum state
$\left|n_{s};\vec{k'},\lambda'\right\rangle $ is given by \begin{align}
\mathcal{H}'_{n_{s},u}\equiv\mathcal{H}'_{n_{s}} & =\frac{\left[n_{s}+\frac{1}{2}\right]\hbar c\Theta_{k'}}{f_{1}\left(L_{1}\right)f_{2}\left(L_{2}\right)f_{3}\left(L_{3}\right)}\int_{0}^{\infty}\int_{0}^{\infty}\int_{0}^{\infty}\sqrt{\sum_{i=1}^{3}\left[k'_{i}\left(n_{i},L_{i}\right)\right]^{2}}dk'_{1}dk'_{2}dk'_{3},\label{eq:stationary-state-energy-unbounded}\end{align}
 where the subscript $u$ of $\mathcal{H}'_{n_{s},u}$ denotes free
or unbounded space, and the functional $f_{i}\left(L_{i}\right)$
in the denominator is equal to $\zeta_{zero}n_{i}^{-1}L_{i}^{-1}$
for a given $L_{i}.$ Here $\zeta_{zero}$ is the zeroes of the function
representing the transversal component of the electric field.

\subsection{Casimir-Polder Interaction}

The phenomenon referred to as Casimir effect has its root in van der
Waals interaction between neutral particles that are polarizable.
The Casimir force may be regarded as a macroscopic manifestations
of the retarded van der Waals force. The energy associated with an
electric dipole moment $\vec{p}_{d}$ in a given electric field $\vec{E}$
is $\mathcal{H}_{d}=-\vec{p}_{d}\cdot\vec{E}.$ When the involved
dipole moment $\vec{p}_{d}$ is that of the induced rather than that
of the permanent one, the induced dipole interaction energy is reduced
by a factor of two, $\mathcal{H}_{d}=-\vec{p}_{d}\cdot\vec{E}/2.$
The role of an external field here is played by the vacuum-field.
Since the polarizability is linearly proportional to the external
field, the average value leads to a factor of one half in the induced
dipole interaction energy. Here the medium of the dielectric is assumed
to be linear. Throughout this investigation, the dipole moments induced
by vacuum polarization are considered as a free parameters. 

\begin{figure}
\begin{center}\includegraphics[%
  scale=0.7]{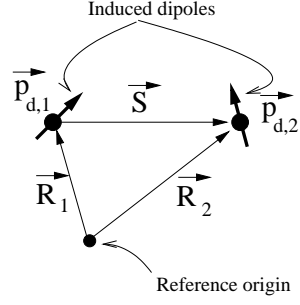}\end{center}

\caption{Two interacting molecules through induced dipole interactions. \label{cap:Two-interacting-dipoles}}
\end{figure}

The interaction energy between two induced dipoles shown in Figure
\ref{cap:Two-interacting-dipoles} are given by \begin{align*}
\mathcal{H}_{int} & =\frac{1}{2}\left\Vert \vec{R}_{2}-\vec{R}_{1}\right\Vert ^{-5}\left\{ \left[\vec{p}_{d,1}\cdot\vec{p}_{d,2}\right]\left\Vert \vec{R}_{2}-\vec{R}_{1}\right\Vert ^{2}-3\left[\vec{p}_{d,1}\cdot\left(\vec{R}_{2}-\vec{R}_{1}\right)\right]\left[\vec{p}_{d,2}\cdot\left(\vec{R}_{2}-\vec{R}_{1}\right)\right]\right\} ,\end{align*}
 where $\vec{R}_{i}$ is the position of $i$th dipole. For an isolated
system, the first order perturbation energy $\left\langle \mathcal{H}_{int}^{\left(1\right)}\right\rangle $
vanishes due to the fact that dipoles are randomly oriented, i.e.,
$\left\langle \vec{p}_{d,i}\right\rangle =0.$ The first non-vanishing
perturbation energy is that of the second order, $U_{eff,static}=\left\langle \mathcal{H}_{int}^{\left(2\right)}\right\rangle =\sum_{m\neq0}\left\langle 0\left|\mathcal{H}_{int}\right|m\right\rangle \left\langle m\left|\mathcal{H}_{int}\right|0\right\rangle \left[E_{0}-E_{m}\right]^{-1},$
which falls off with respect to the separation distance like $U_{eff,static}\propto\left\Vert \vec{R}_{2}-\vec{R}_{1}\right\Vert ^{-6}.$
This is the classical result obtained by F. London for short distance
electrostatic fields. F. London employed quantum mechanical perturbation
approach to reach his result on a static van der Waals interaction
without retardation effect in 1930. 

The electromagnetic interaction can only propagate as fast as the
speed of light in a given medium. This retardation effect due to propagation
time was included by Casimir and Polder in their consideration. It
led to their surprising discovery that the interaction between molecules
falls off like $\left\Vert \vec{R}_{1}-\vec{R}_{2}\right\Vert ^{-7}.$
It became the now well known Casimir-Polder potential \cite{key-Casimir-Polder},
\begin{align*}
U_{eff,retarded} & =-\frac{\hbar c}{4\pi}\left\Vert \vec{R}_{2}-\vec{R}_{1}\right\Vert ^{-7}\left\{ 23\left[\alpha_{E}^{\left(1\right)}\alpha_{E}^{\left(2\right)}+\alpha_{M}^{\left(1\right)}\alpha_{M}^{\left(2\right)}\right]-7\left[\alpha_{E}^{\left(1\right)}\alpha_{M}^{\left(2\right)}+\alpha_{M}^{\left(1\right)}\alpha_{E}^{\left(2\right)}\right]\right\} ,\end{align*}
 where $\alpha_{E}^{\left(i\right)}$ and $\alpha_{M}^{\left(i\right)}$
represents the electric and magnetic polarizability of $i$th particle
(or molecule). 

To understand the Casimir effect, the physics behind the Casimir-Polder
(or retarded van der Waals) interaction is essential. In the expression
of the induced dipole energy $\mathcal{H}_{d}=-\vec{p}_{d}\cdot\vec{E}/2,$
we rewrite $\vec{p}_{d}=\alpha\left(\omega\right)\vec{E}_{\omega}$
for the Fourier component of the dipole moment induced by the Fourier
component $\vec{E}_{\omega}$ of the field. Here $\alpha\left(\omega\right)$
is the polarizability. The induced dipole field energy becomes $\mathcal{H}_{d}=-\alpha\left(\omega\right)\vec{E}_{\omega}^{\dagger}\cdot\vec{E}_{\omega}/2.$
Summing over all possible modes and polarizations, the field energy
due to the induced dipole becomes \begin{align*}
\mathcal{H}_{d,1} & =-\frac{1}{2}\sum_{\vec{k},\lambda}\alpha_{1}\left(\omega_{k}\right)\vec{E}_{1,\vec{k},\lambda}^{\dagger}\left(\vec{R}_{1},t\right)\cdot\vec{E}_{1,\vec{k},\lambda}\left(\vec{R}_{1},t\right),\end{align*}
 where the subscripts $\left(1\right)$ and $\left(1,\vec{k},\lambda\right)$
denote that this is the energy associated with the induced dipole
moment $\vec{p}_{d,1}$ at location $\vec{R}_{1}$ as shown in Figure
\ref{cap:Two-interacting-dipoles}. The total electric field $\vec{E}_{1,\vec{k},\lambda}\left(\vec{R}_{1},t\right)$
in mode $\left(\vec{k},\lambda\right)$ acting on $\vec{p}_{d,1}$
is given by \begin{align*}
\vec{E}_{1,\vec{k},\lambda}\left(\vec{R}_{1},t\right) & =\vec{E}_{o,\vec{k},\lambda}\left(\vec{R}_{1},t\right)+\vec{E}_{2,\vec{k},\lambda}\left(\vec{R}_{1},t\right),\end{align*}
 where $\vec{E}_{o,\vec{k},\lambda}\left(\vec{R}_{1},t\right)$ is
the vacuum-field at location $\vec{R}_{1}$ and $\vec{E}_{2,\vec{k},\lambda}\left(\vec{R}_{1},t\right)$
is the induced dipole field at $\vec{R}_{1}$ due to the neighboring
induced dipole $\vec{p}_{d,2}$ located at $\vec{R}_{2}.$ The effective
Hamiltonian becomes \begin{align*}
\mathcal{H}_{d,1} & =-\frac{1}{2}\sum_{\vec{k},\lambda}\left\{ \alpha_{1}\left(\omega_{k}\right)\left[\vec{E}_{o,\vec{k},\lambda}^{\dagger}\left(\vec{R}_{1},t\right)\cdot\vec{E}_{o,\vec{k},\lambda}\left(\vec{R}_{1},t\right)+\vec{E}_{2,\vec{k},\lambda}^{\dagger}\left(\vec{R}_{1},t\right)\cdot\vec{E}_{2,\vec{k},\lambda}\left(\vec{R}_{1},t\right)\right.\right.\\
 & \left.\left.+\vec{E}_{o,\vec{k},\lambda}^{\dagger}\left(\vec{R}_{1},t\right)\cdot\vec{E}_{2,\vec{k},\lambda}\left(\vec{R}_{1},t\right)+\vec{E}_{2,\vec{k},\lambda}^{\dagger}\left(\vec{R}_{1},t\right)\cdot\vec{E}_{o,\vec{k},\lambda}\left(\vec{R}_{1},t\right)\right]\right\} \\
 & =\mathcal{H}_{o}+\mathcal{H}_{\vec{p}_{d,2}}+\mathcal{H}_{\vec{p}_{d,1},\vec{p}_{d,2}},\end{align*}
 where \begin{align*}
\mathcal{H}_{o} & =-\frac{1}{2}\sum_{\vec{k},\lambda}\alpha_{1}\left(\omega_{k}\right)\vec{E}_{o,\vec{k},\lambda}^{\dagger}\left(\vec{R}_{1},t\right)\cdot\vec{E}_{o,\vec{k},\lambda}\left(\vec{R}_{1},t\right),\end{align*}
 \begin{align*}
\mathcal{H}_{\vec{p}_{d,2}} & =-\frac{1}{2}\sum_{\vec{k},\lambda}\alpha_{1}\left(\omega_{k}\right)\vec{E}_{2,\vec{k},\lambda}^{\dagger}\left(\vec{R}_{1},t\right)\cdot\vec{E}_{2,\vec{k},\lambda}\left(\vec{R}_{1},t\right),\end{align*}
 \begin{align*}
\mathcal{H}_{\vec{p}_{d,1},\vec{p}_{d,2}} & =-\frac{1}{2}\sum_{\vec{k},\lambda}\alpha_{1}\left(\omega_{k}\right)\left[\vec{E}_{o,\vec{k},\lambda}^{\dagger}\left(\vec{R}_{1},t\right)\cdot\vec{E}_{2,\vec{k},\lambda}\left(\vec{R}_{1},t\right)+\vec{E}_{2,\vec{k},\lambda}^{\dagger}\left(\vec{R}_{1},t\right)\cdot\vec{E}_{o,\vec{k},\lambda}\left(\vec{R}_{1},t\right)\right].\end{align*}
 Because only the interaction between the two induced dipoles is relevant
to the Casimir effect, the $\mathcal{H}_{\vec{p}_{d,1},\vec{p}_{d,2}}$
term is considered solely here. In the language of field operators,
the vacuum-field $\vec{E}_{o,\vec{k},\lambda}\left(\vec{R}_{1},t\right)$
is expressed as a sum: \begin{align*}
\vec{E}_{o,\vec{k},\lambda}\left(\vec{R}_{1},t\right) & =\vec{E}_{o,\vec{k},\lambda}^{\left(+\right)}\left(\vec{R}_{1},t\right)+\vec{E}_{o,\vec{k},\lambda}^{\left(-\right)}\left(\vec{R}_{1},t\right),\end{align*}
 where \begin{align*}
\vec{E}_{o,\vec{k},\lambda}^{\left(+\right)}\left(\vec{R}_{1},t\right) & \equiv i\sqrt{\frac{2\pi\hbar\omega_{k}}{V}}a_{\vec{k},\lambda}\left(0\right)\exp\left(-i\omega_{k}t\right)\exp\left(i\vec{k}\cdot\vec{R}_{1}\right)\hat{\epsilon}_{\vec{k},\lambda},\end{align*}
 \begin{alignat*}{1}
\vec{E}_{o,\vec{k},\lambda}^{\left(-\right)}\left(\vec{R}_{1},t\right) & \equiv-i\sqrt{\frac{2\pi\hbar\omega_{k}}{V}}a_{\vec{k},\lambda}^{\dagger}\left(0\right)\exp\left(i\omega_{k}t\right)\exp\left(-i\vec{k}\cdot\vec{R}_{1}\right)\hat{\epsilon}_{\vec{k},\lambda}.\end{alignat*}
 In the above expressions, $a_{\vec{k},\lambda}^{\dagger}$ and $a_{\vec{k},\lambda}$
are the creation and annihilation operators respectively; and $V,$
the quantization volume; $\hat{\epsilon}_{\vec{k},\lambda},$ the
polarization. By convention, $\vec{E}_{o,\vec{k},\lambda}^{\left(+\right)}\left(\vec{R}_{1},t\right)$
is called the positive frequency (annihilation) operator and $\vec{E}_{o,\vec{k},\lambda}^{\left(-\right)}\left(\vec{R}_{1},t\right)$
is called the negative frequency (creation) operator. 

The field operator $\vec{E}_{2,\vec{k},\lambda}\left(\vec{R}_{1},t\right)$
has the same form as the classical field of an induced electric dipole,
\begin{align*}
\vec{E}_{2,\vec{k},\lambda}\left(\vec{R}_{1},t\right) & =\left\{ 3\left[\hat{p}_{d,2}\cdot\hat{S}\right]\hat{S}-\hat{p}_{d,2}\right\} \left[\frac{1}{r^{3}}\left\Vert \vec{p}_{d,2}\left(t-\frac{r}{c}\right)\right\Vert +\frac{1}{cr^{2}}\left\Vert \dot{\vec{p}}_{d,2}\left(t-\frac{r}{c}\right)\right\Vert \right]\\
 & -\frac{1}{c^{2}r}\left\{ \hat{p}_{d,2}-\left[\hat{p}_{d,2}\cdot\hat{S}\right]\hat{S}\right\} \left\Vert \ddot{\vec{p}}_{d,2}\left(t-\frac{r}{c}\right)\right\Vert ,\end{align*}
 where $r=\left\Vert \vec{R}_{2}-\vec{R}_{1}\right\Vert \equiv\left\Vert \vec{S}\right\Vert ,$
$\hat{S}=\left[\vec{R}_{2}-\vec{R}_{1}\right]\left\Vert \vec{R}_{2}-\vec{R}_{1}\right\Vert ^{-1},$
$\hat{p}_{d,2}=\vec{p}_{d,2}\left\Vert \vec{p}_{d,2}\right\Vert ^{-1}$
as shown in Figure \ref{cap:Two-interacting-dipoles}, and $c$ is
the speed of light in vacuum. Because the dipole moment is expressed
as $\vec{p}_{d}=\alpha\left(\omega\right)\vec{E}_{\omega},$ the appropriate
dipole moment in the above expression for $\vec{E}_{2,\vec{k},\lambda}\left(\vec{R}_{1},t\right)$
is to be replaced by \begin{align*}
\vec{p}_{d,2} & =\sum_{\vec{k},\lambda}\alpha_{2}\left(\omega_{k}\right)\left[\vec{E}_{o,\vec{k},\lambda}^{\left(+\right)}\left(\vec{R}_{2},t\right)+\vec{E}_{o,\vec{k},\lambda}^{\left(-\right)}\left(\vec{R}_{2},t\right)\right],\end{align*}
 where $\alpha_{2}\left(\omega_{k}\right)$ is now the polarizability
of the molecule or atom associated with the induced dipole moment
$\vec{p}_{d,2}$ at the location $\vec{R}_{2}.$ With this in place,
$\vec{E}_{2,\vec{k},\lambda}\left(\vec{R}_{1},t\right)$ is now a
quantum mechanical operator.

The interaction Hamiltonian operator $\hat{\mathcal{H}}_{\vec{p}_{d,1},\vec{p}_{d,2}}$
can be written as \begin{align*}
\hat{\mathcal{H}}_{\vec{p}_{d,1},\vec{p}_{d,2}} & =-\frac{1}{2}\sum_{\vec{k},\lambda}\alpha_{1}\left(\omega_{k}\right)\left[\left\langle \vec{E}_{o,\vec{k},\lambda}^{\left(+\right)}\left(\vec{R}_{1},t\right)\cdot\vec{E}_{2,\vec{k},\lambda}\left(\vec{R}_{1},t\right)\right\rangle +\left\langle \vec{E}_{2,\vec{k},\lambda}\left(\vec{R}_{1},t\right)\cdot\vec{E}_{o,\vec{k},\lambda}^{\left(-\right)}\left(\vec{R}_{1},t\right)\right\rangle \right],\end{align*}
 where we have taken into account the fact that $\vec{E}_{o,\vec{k},\lambda}^{\left(+\right)}\left(\vec{R}_{2},t\right)\left|vac\right\rangle =\left\langle vac\right|\vec{E}_{o,\vec{k},\lambda}^{\left(-\right)}\left(\vec{R}_{2},t\right)=0.$
It was shown in \cite{key-Milonni} in great detail that the interaction
energy is given by \begin{align*}
U\left(r\right)\equiv\left\langle \mathcal{H}_{\vec{p}_{d,1},\vec{p}_{d,2}}\right\rangle  & =-\frac{2\pi\hbar}{V}\mathbb{R}_{E}\sum_{\vec{k},\lambda}k^{3}\omega_{k}\alpha_{1}\left(\omega_{k}\right)\alpha_{2}\left(\omega_{k}\right)\exp\left(-ikr\right)\exp\left(i\vec{k}\cdot\vec{r}\right)\\
 & \times\left[\left\{ 1-\left[\hat{\epsilon}_{\vec{k},\lambda}\cdot\hat{S}\right]^{2}\right\} \frac{1}{kr}+\left\{ 3\left[\hat{\epsilon}_{\vec{k},\lambda}\cdot\hat{S}\right]^{2}-1\right\} \left\{ \frac{1}{k^{3}r^{3}}+\frac{i}{k^{2}r^{2}}\right\} \right].\end{align*}
 In the limit of $r\ll c\left|\omega_{mn}\right|^{-1},$ where $\omega_{mn}$
is the transition frequency between the ground state and the first
excited energy state, or the resonance frequency, the above result
becomes \begin{eqnarray*}
U\left(r\right)\cong-\frac{3\hbar\omega_{o}}{4r^{6}}\alpha^{2}, &  & \alpha=\frac{2}{3\hbar\omega_{o}}\left\Vert \left\langle m\left|\vec{p}_{d}\right|0\right\rangle \right\Vert ^{2}.\end{eqnarray*}
 This was also the non-retarded van der Waals potential obtained by
F. London. Here $\omega_{o}$ is the transition frequency, and $\alpha$
is the static ($\omega=0$) polarizability of an atom in the ground
state. Once the retardation effect due to light propagation is taken
into account, the Casimir-Polder potential becomes, \begin{align*}
U\left(r\right) & \cong-\frac{23\hbar c}{4\pi r^{7}}\alpha_{1}\left(\omega\right)\alpha_{2}\left(\omega\right).\end{align*}

What we try to emphasize in this brief derivation is that both retarded
and non-retarded van der Waals interaction may be regarded as a consequence
of the fluctuating vacuum-fields. It arises due to a non-vanishing
correlation of the vacuum-fields over distance of $r=\left\Vert \vec{R}_{2}-\vec{R}_{1}\right\Vert .$
The non-vanishing correlation here is defined by $\left\langle vac\left|\vec{E}_{o,\vec{k},\lambda}^{\left(+\right)}\left(\vec{R}_{1},t\right)\cdot\vec{E}_{o,\vec{k},\lambda}^{\left(-\right)}\left(\vec{R}_{2},t\right)\right|vac\right\rangle \neq0.$
In more physical terms, the vacuum-fields induce fluctuating dipole
moments in polarizable media. The correlated dipole-dipole interaction
is the van der Waals interaction. If the retardation effect is taken
into account, it is called the {}``Casimir-Polder'' interaction. 

In the Casimir-Polder picture, the Casimir force between two neutral
parallel plates of infinite conductivity was found by a simple summation
of the pairwise intermolecular forces. It can be shown that such a
procedure yields for the force between two parallel plates of infinite
conductivity \cite{key-Milonni} \begin{align}
\left\Vert \vec{F}\left(d;L,c\right)_{Casimir-Polder}\right\Vert  & =\frac{207\hbar c}{640\pi^{2}d^{4}}L^{2},\label{eq:Casimir-Polder-Force-Mag}\end{align}
 where it is understood that the sign of the force is attractive.
When this is compared with the force of equation (\ref{eq:Casimir-Force-Euler-McLaurin-Approach})
computed with Casimir's vacuum-field approach, which will be discussed
in the next section, the agreement is within $\sim20$\% \cite{key-Milonni}.
In other words, one can obtain a fairly reasonable estimate of the
Casimir effect by simply adding up the pairwise intermolecular forces.
The recent experimental verification of the Casimir-Polder force can
be found in reference \cite{key-Yale-Group-Casimir-Polder}.

The discrepancy of $\sim20$\% between the two force results of equations
(\ref{eq:Casimir-Polder-Force-Mag}) and (\ref{eq:Casimir-Force-Euler-McLaurin-Approach})
can be attributed to the fact that the force expression of equation
(\ref{eq:Casimir-Polder-Force-Mag}) had been derived under the assumption
that the intermolecular forces were additive in the sense that the
force between two molecules is independent of the presence of a third
molecule \cite{key-Milonni,key-Boer-Hamaker}. The van der Waals forces
are not however simply additive (see section 8.2 of reference \cite{key-Milonni}).
And, the motivation behind the result of equation (\ref{eq:Casimir-Polder-Force-Mag})
is to illustrate the intrinsic connection between Casimir-Polder interaction
and the Casimir effect, but without any rigor put into the derivation. 

It is this discrepancy between the microscopic theories assuming additive
intermolecular forces, and the experimental results reported in the
early 1950s, that motivated Lifshitz in 1956 to develop a macroscopic
theory of the forces between dielectrics \cite{key-Landau-Lifshitz-EM-Book,key-Lifshitz}.
Lifshitz theory assumed that the dielectrics are characterized by
randomly fluctuating sources. From the assumed delta-function correlation
of these sources, the correlation functions for the field were calculated,
and from these in turn the Maxwell stress tensor was determined. The
force per unit area acting on the two dielectrics was then calculated
as the $zz$ component of the stress tensor. In the limiting case
of perfect conductors, the Lifshitz theory correctly reduces to the
Casimir force of equation (\ref{eq:Casimir-Force-Euler-McLaurin-Approach}).

\subsection{Casimir Force Calculation Between Two Neutral Conducting Parallel
Plates}

Although the Casimir force may be regarded as a macroscopic manifestation
of the retarded van der Waals force between two polarizable charge-neutral
molecules (or atoms), it is most often alternatively derived by the
consideration of the vacuum-field energy $\hbar\omega/2$ per mode
of frequency $\omega$ rather than from the summation of the pairwise
intermolecular forces. Three different methods widely used in Casimir
force calculations are presented here. They are: (1) the Euler-Maclaurin
sum approach, (2) the vacuum pressure approach by Milonni, Cook and
Goggin, and lastly, (3) the source theory by Schwinger. The main purpose
here is to exhibit their different calculational techniques.

\subsubsection{Euler-Maclaurin Summation Approach}

\begin{figure}
\begin{center}\includegraphics[%
  scale=0.7]{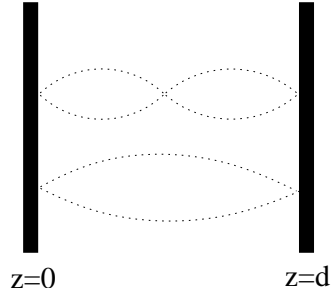}\end{center}

\caption{A cross-sectional view of two infinite parallel conducting plates
separated by a gap distance of $z=d.$ The first two lowest wave modes
are shown. \label{cap:casimir-parallel-plates}}
\end{figure}

For pedagogical reasons and as a brief introduction to the technique,
the Casimir's original configuration (two charge-neutral infinite
parallel conducting plates) shown in Figure \ref{cap:casimir-parallel-plates}
is worked out in detail. 

Since the electromagnetic fields are sinusoidal functions, and the
tangential component of the electric fields vanish at the conducting
surfaces, the functions $f_{i}\left(L_{i}\right)$ have the form $f_{i}\left(L_{i}\right)=\pi L_{i}^{-1}.$
The wave numbers are given by $k'_{i}\left(n_{i},L_{i}\right)=n_{i}f_{i}\left(L_{i}\right)=n_{i}\pi L_{i}^{-1}.$
The vacuum state is given by $\left|n_{s}=0;\vec{k'},\lambda'\right\rangle .$
Then, for $n_{s}=0$ in equation (\ref{eq:stationary-state-energy-bounded}),
the ground state radiation energy is given by \begin{align*}
\mathcal{H}'_{n_{s},b} & =\frac{\hbar c}{2}\Theta_{k'}\sum_{n_{1}=0}^{\infty}\sum_{n_{2}=0}^{\infty}\sum_{n_{3}=0}^{\infty}\sqrt{\sum_{i=1}^{3}\frac{n_{i}^{2}\pi^{2}}{L_{i}^{2}}}.\end{align*}
 For the arrangement shown in Figure \ref{cap:casimir-parallel-plates},
the dimensions are such that $L_{1}\gg L_{3}$ and $L_{2}\gg L_{3},$
where $\left(L_{1},L_{2},L_{3}\right)$ corresponds to $\left(L_{x},L_{y},L_{z}\right).$
The area of the plates are given by $L_{1}\times L_{2}.$ The summation
over $n_{1}$ and $n_{2}$ can be replaced by an integration, \begin{align*}
\mathcal{H}'_{n_{s},b} & =\frac{\hbar c}{2\pi^{2}}L_{1}L_{2}\Theta_{k'}\int_{0}^{\infty}\int_{0}^{\infty}\sum_{n_{3}=0}^{\infty}\sqrt{\left[k'_{x}\right]^{2}+\left[k'_{y}\right]^{2}+\frac{n_{3}^{2}\pi^{2}}{L_{i}^{2}}}dk'_{x}dk'_{y}.\end{align*}
 For simplicity and without any loss of generality, the designation
of $L_{1}=L_{2}=L$ and $L_{3}=d$ yields the result \begin{align*}
\mathcal{H}'_{n_{s},b}\left(d\right) & =\frac{\hbar c}{2\pi^{2}}L^{2}\Theta_{k'}\int_{0}^{\infty}\int_{0}^{\infty}\sum_{n_{3}=0}^{\infty}\sqrt{\left[k'_{x}\right]^{2}+\left[k'_{y}\right]^{2}+\frac{n_{3}^{2}\pi^{2}}{d^{2}}}dk'_{x}dk'_{y}.\end{align*}
 Here $\mathcal{H}'_{n_{s},b}\left(d\right)$ denotes the vacuum electromagnetic
field energy for the cavity when plate gap distance is $d.$ In the
limit the gap distance becomes arbitrarily large, the sum over $n_{3}$
is also replaced by an integral representation to yield \begin{align*}
\mathcal{H}'_{n_{s},b}\left(\infty\right) & =\frac{\hbar c}{2\pi^{2}}L^{2}\Theta_{k'}\lim_{d\rightarrow\infty}\left(\frac{d}{\pi}\int_{0}^{\infty}\int_{0}^{\infty}\int_{0}^{\infty}\sqrt{\left[k'_{x}\right]^{2}+\left[k'_{y}\right]^{2}+\left[k'_{z}\right]^{2}}dk'_{x}dk'_{y}dk'_{z}\right).\end{align*}
 This is the electromagnetic field energy inside an infinitely large
cavity, i.e., free space. 

The work required to bring in the plates from an infinite separation
to a final separation of $d$ is then the potential energy, \begin{align*}
U\left(d\right) & =\mathcal{H}'_{n_{s},b}\left(d\right)-\mathcal{H}'_{n_{s},b}\left(\infty\right)\\
 & =\frac{\hbar c}{2\pi^{2}}L^{2}\Theta_{k'}\left[\int_{0}^{\infty}\int_{0}^{\infty}\sum_{n_{3}=0}^{\infty}\sqrt{\left[k'_{x}\right]^{2}+\left[k'_{y}\right]^{2}+\frac{n_{3}^{2}\pi^{2}}{d^{2}}}dk'_{x}dk'_{y}\right.\\
 & \left.-\lim_{d\rightarrow\infty}\left(\frac{d}{\pi}\int_{0}^{\infty}\int_{0}^{\infty}\int_{0}^{\infty}\sqrt{\left[k'_{x}\right]^{2}+\left[k'_{y}\right]^{2}+\left[k'_{z}\right]^{2}}dk'_{x}dk'_{y}dk'_{z}\right)\right].\end{align*}
 The result is a grossly divergent function. Nonetheless, with a proper
choice of the cutoff function (or regularization function), a finite
value for $U\left(d\right)$ can be obtained. In the polar coordinates
representation $\left(r,\theta\right),$ we define $r^{2}=\left[k'_{x}\right]^{2}+\left[k'_{y}\right]^{2}$
and $dk'_{x}dk'_{y}=rdrd\theta,$ then \begin{align*}
U\left(d\right) & =\frac{\hbar c}{2\pi^{2}}L^{2}\Theta_{k'}\left[\int_{\theta=0}^{\pi/2}\int_{r=0}^{\infty}\sum_{n_{3}=0}^{\infty}\sqrt{r^{2}+\frac{n_{3}^{2}\pi^{2}}{d^{2}}}rdrd\theta\right.\\
 & \left.-\lim_{d\rightarrow\infty}\left(\frac{d}{\pi}\int_{k'_{z}=0}^{\infty}\int_{\theta=0}^{\pi/2}\int_{r=0}^{\infty}\sqrt{r^{2}+\left[k'_{z}\right]^{2}}rdrd\theta dk'_{z}\right)\right],\end{align*}
 where the integration over $\theta$ is done in the range $0\leq\theta\leq\pi/2$
to ensure $k'_{x}\geq0$ and $k'_{y}\geq0.$ For convenience, the
integration over $\theta$ is carried out first, \begin{align*}
U\left(d\right) & =\frac{\hbar c}{4\pi}L^{2}\Theta_{k'}\left[\int_{r=0}^{\infty}\sum_{n_{3}=0}^{\infty}\sqrt{r^{2}+\frac{n_{3}^{2}\pi^{2}}{d^{2}}}rdr-\lim_{d\rightarrow\infty}\left(\frac{d}{\pi}\int_{k'_{z}=0}^{\infty}\int_{r=0}^{\infty}\sqrt{r^{2}+\left[k'_{z}\right]^{2}}rdrdk'_{z}\right)\right].\end{align*}
 As mentioned earlier, $U\left(d\right)$ in current form is grossly
divergent. It is regularized through the use of a regularization function
in the form of $f\left(k'\right)=f\left(\sqrt{r^{2}+\left[k'_{z}\right]^{2}}\right)$
or $f\left(k'\right)=f\left(\sqrt{r^{2}+n_{3}^{2}\pi^{2}d^{-2}}\right)$
with the condition that $f\left(k'\right)=1$ for $k'\ll k'_{cutoff}$
and $f\left(k'\right)=0$ for $k'\gg k'_{cutoff}.$ Mathematically
speaking, this cutoff function $f\left(k'\right)$ is able to regularize
the above divergent function. Physically, introduction of this regularization
takes care of the failure at small distance of the assumption that
plates are perfectly conducting for short wavelengths. It is a good
approximation to assume $k'_{cutoff}\sim1/a_{o},$ where $a_{o}$
is the Bohr radius. In this sense, one is inherently assuming that
Casimir effect is primarily a low-frequency or long wavelength effect.
Hence, with the regularization function substituted in $U\left(d\right)$
above, the potential energy becomes \begin{align*}
U\left(d\right) & =\frac{\hbar c}{4\pi}L^{2}\Theta_{k'}\left[\sum_{n_{3}=0}^{\infty}\int_{r=0}^{\infty}\sqrt{r^{2}+\frac{n_{3}^{2}\pi^{2}}{d^{2}}}f\left(\sqrt{r^{2}+\frac{n_{3}^{2}\pi^{2}}{d^{2}}}\right)rdr\right.\\
 & \left.-\lim_{d\rightarrow\infty}\left(\frac{d}{\pi}\int_{k'_{z}=0}^{\infty}\int_{r=0}^{\infty}\sqrt{r^{2}+\left[k'_{z}\right]^{2}}f\left(\sqrt{r^{2}+\left[k'_{z}\right]^{2}}\right)rdrdk'_{z}\right)\right].\end{align*}
 The summation $\sum_{n_{3}=0}^{\infty}$ and the integral $\int_{r=0}^{\infty}$
in the first term on the right hand side can be interchanged. The
interchange of sums and integrals is justified due to the absolute
convergence in the presence of the regularization function. In terms
of the new definition for the integration variables $x=r^{2}d^{2}\pi^{-2}$
and $\kappa=k'_{z}d\pi^{-1},$ the above expression for $U\left(d\right)$
is rewritten as \begin{align*}
U\left(d\right) & =\frac{\hbar c}{8}\pi^{2}L^{2}\Theta_{k'}\left[\frac{1}{d^{3}}\sum_{n_{3}=0}^{\infty}\int_{x=0}^{\infty}\sqrt{x+n_{3}^{2}}f\left(\frac{\pi}{d}\sqrt{x+n_{3}^{2}}\right)dx\right.\\
 & \left.-\lim_{d\rightarrow\infty}\left(\frac{1}{d^{3}}\int_{\kappa=0}^{\infty}\int_{x=0}^{\infty}\sqrt{x+\kappa^{2}}f\left(\frac{\pi}{d}\sqrt{x+\kappa^{2}}\right)dxd\kappa\right)\right]\\
 & \equiv\frac{\hbar c}{8}\pi^{2}L^{2}\Theta_{k'}\left[\frac{1}{2}F\left(0\right)+\sum_{n_{3}=1}^{\infty}F\left(n_{3}\right)-\int_{\kappa=0}^{\infty}F\left(\kappa\right)d\kappa\right],\end{align*}
 where \begin{align*}
F\left(n_{3}\right) & \equiv\frac{1}{d^{3}}\int_{x=0}^{\infty}\sqrt{x+n_{3}^{2}}f\left(\frac{\pi}{d}\sqrt{x+n_{3}^{2}}\right)dx,\end{align*}
 and \begin{align*}
F\left(\kappa\right) & \equiv\lim_{d\rightarrow\infty}\left(\frac{1}{d^{3}}\int_{x=0}^{\infty}\sqrt{x+\kappa^{2}}f\left(\frac{\pi}{d}\sqrt{x+\kappa^{2}}\right)dx\right).\end{align*}
 Then, according to the Euler-Maclaurin summation formula \cite{key-Euler-Sum-Formula,key-Euler-Sum-Formula-Derivation},
\begin{align*}
\sum_{n_{3}=1}^{\infty}F\left(n_{3}\right)-\int_{\kappa=0}^{\infty}F\left(\kappa\right)d\kappa & =-\frac{1}{2}F\left(0\right)-\frac{1}{12}\frac{dF\left(0\right)}{d\kappa}+\frac{1}{720}\frac{d^{3}F\left(0\right)}{d\kappa^{3}}+\cdots\end{align*}
 for $F\left(\infty\right)\rightarrow0.$ Noting that from $F\left(\kappa\right)=\int_{\kappa^{2}}^{\infty}\sqrt{r}f\left(\frac{\pi}{d}\sqrt{r}\right)dr$
and $dF\left(\kappa\right)/d\kappa=-2\kappa^{2}f\left(\frac{\pi}{d}\kappa\right),$
one can find $dF\left(0\right)/d\kappa=0,$ $d^{3}F\left(0\right)/d\kappa^{3}=-4,$
and all higher order derivatives vanish if one assumes that all derivatives
of the regularization function vanish at $\kappa=0.$ Finally, the
result for the vacuum electromagnetic potential energy $U\left(d\right)$
becomes \begin{align*}
U\left(d;L,c\right) & =-\frac{\hbar c\pi^{2}}{1440d^{3}}L^{2}\Theta_{k'}.\end{align*}
 This result is finite, and it is independent of the regularization
function as it should be. The corresponding Casimir force for the
two infinite parallel conducting plates is given by \begin{align*}
\vec{F}_{z=d}\left(d;L,c\right) & =-\frac{\partial U\left(d;L,c\right)}{\partial d}=-\frac{3\hbar c\pi^{2}}{1440d^{4}}L^{2}\Theta_{k'}\hat{z},\end{align*}
 where this is the force on plate at $z=d$ due to the presence of
plate at $z=0.$ The force on plate at $z=0$ due to the presence
of plate at $z=d$ is, like wise, given by \begin{align*}
\vec{F}_{z=0}\left(d;L,c\right) & =\frac{3\hbar c\pi^{2}}{1440d^{4}}L^{2}\Theta_{k'}\hat{z}.\end{align*}
 This can be easily verified by replacing \begin{align*}
d & \equiv\left\Vert \vec{R}_{d}\left(z=d\right)-\vec{R}_{0}\left(z=0\right)\right\Vert ,\end{align*}
 where $\vec{R}_{d}\left(z=d\right)$ and $\vec{R}_{0}\left(z=0\right)$
are the position vectors for the plates located at $z=d$ and $z=0,$
respectively. Then for the forces $\vec{F}_{z=d}\left(d;L,c\right)$
and $\vec{F}_{z=0}\left(d;L,c\right),$ we have \begin{align*}
\vec{F}_{z=d}\left(d;L,c\right) & =-\frac{\partial U\left(d;L,c\right)}{\partial\vec{R}_{d}\left(z=d\right)}\hat{z}\\
 & =-\frac{\partial U\left(d;L,c\right)}{\partial d}\frac{\partial d}{\partial\vec{R}_{d}\left(z=d\right)}\hat{z}\end{align*}
 and \begin{align*}
\vec{F}_{z=0}\left(d;L,c\right) & =-\frac{\partial U\left(d;L,c\right)}{\partial\vec{R}_{0}\left(z=0\right)}\hat{z}\\
 & =-\frac{\partial U\left(d;L,c\right)}{\partial d}\frac{\partial d}{\partial\vec{R}_{0}\left(z=0\right)}\hat{z}.\end{align*}
 It is easily shown that \begin{align*}
\frac{\partial d}{\partial\vec{R}_{d}\left(z=d\right)} & =-\frac{\partial d}{\partial\vec{R}_{0}\left(z=0\right)},\end{align*}
 where $d\equiv\left\Vert \vec{R}_{d}\left(z=d\right)-\vec{R}_{0}\left(z=0\right)\right\Vert ;$
and, therefore the previous result $\vec{F}_{z=d}\left(d;L,c\right)=-\vec{F}_{z=0}\left(d;L,c\right).$ 

Since the electromagnetic wave has two possible polarizations, we
have $\Theta_{k'}=2.$ The force is therefore written in magnitude
as \begin{align}
\left\Vert \vec{F}_{z=d}\left(d;L,c\right)\right\Vert =\left\Vert \vec{F}_{z=0}\left(d;L,c\right)\right\Vert  & =\frac{\hbar c\pi^{2}}{240d^{4}}L^{2},\label{eq:Casimir-Force-Euler-McLaurin-Approach}\end{align}
 where it is understood that the sign of the force is attractive.
This is the Casimir force between two uncharged parallel conducting
plates \cite{key-Casimir}. 

It is to be noted that the Euler-Maclaurin summation approach discussed
here is just one of the many techniques that can be used in calculating
the Casimir force. One can also employ dimensional regularization
to compute the Casimir force. This technique can be found in section
2.2 of the reference \cite{key-Milton}.

\subsubsection{Vacuum Pressure Approach}

The Casimir force between two perfectly conducting plates can also
be calculated from the radiation pressure exerted by a plane wave
incident normally on one of the plates. Here the radiation pressure
is due to the vacuum electromagnetic fields. The technique discussed
here is due to Milonni, Cook and Goggin \cite{key-Milonni-Cook-Goggin}. 

The Casimir force is regarded as a consequence of the radiation pressure
associated with the zero-point energy of $\hbar\omega/2$ per mode
of the field. The main idea behind this techniques is that since the
zero-point fields have the momentum $p'_{i}=\hbar k'_{i}/2,$ the
pressure exerted by an incident wave normal to the plates is twice
the energy $\mathcal{H}$ per unit volume of the incident field. The
pressure imparted to the plate is twice that of the incident wave
for perfect conductors. If the wave has an angle of incidence $\theta_{inc},$
the radiation pressure is \begin{align*}
P & =FA^{-1}=2\mathcal{H}\cos^{2}\theta_{inc}.\end{align*}
 Two factors of $\cos\theta_{inc}$ appear here because (1) the normal
component of the linear momentum imparted to the plate is proportional
to $\cos\theta_{inc},$ and (2) the element of area $A$ is increased
by $1/\cos\theta_{inc}$ compared with the case of normal incidence.
It can be shown then \begin{align*}
P & =2\mathcal{H}\cos^{2}\theta_{inc}\\
 & =2\times\frac{1}{2}\times\frac{1}{2}\hbar\omega\times V^{-1}\times\cos^{2}\theta_{inc}\\
 & =\frac{\hbar\omega}{2V}\left[k'_{z}\right]^{2}\left\Vert \vec{k'}\right\Vert ^{-2},\end{align*}
 where the factor of half have been inserted because the zero-point
field energy of a mode of energy $\hbar\omega/2$ is divided equally
between waves propagating toward and away from each of the plates.
The $\cos\theta_{inc}$ factor have been rewritten using the fact
that $k'_{z}=\vec{k'}\cdot\hat{e}_{z}=\left\Vert \vec{k'}\right\Vert \cos\theta_{inc},$
where $\hat{e}_{z}$ is the unit vector normal to the plate on the
inside, $\left\Vert \vec{k'}\right\Vert =\omega/c$ and $V$ is the
quantization volume. 

The successive reflections of the radiation off the plates act to
push the plates apart through a pressure $P.$ For large plates where
$k'_{x},$ $k'_{y}$ take on a continuum of values and the component
along the plate gap is $k'_{z}=n\pi/d,$ where $n$ is a positive
integer, the total outward pressure on each plate over all possible
modes can be written as \begin{align*}
P_{out} & =\frac{\hbar c}{2\pi^{2}d}\Theta_{k'}\sum_{n=1}^{\infty}\int_{k'_{y}=0}^{\infty}\int_{k'_{x}=0}^{\infty}\frac{\left[n\pi/d\right]^{2}}{\sqrt{\left[k'_{x}\right]^{2}+\left[k'_{y}\right]^{2}+\left[n\pi/d\right]^{2}}}dk'_{x}dk'_{y},\end{align*}
 where $\Theta_{k'}$ is the number of independent polarizations. 

External to the plates, the allowed field modes take on a continuum
of values. Therefore, by the replacement of $\sum_{n=1}^{\infty}\rightarrow\pi^{-1}d\int_{k'_{z}=0}^{\infty}$
in the above expression, the total inward pressure on each plate over
all possible modes is given by \begin{align*}
P_{in} & =\frac{\hbar c}{2\pi^{3}}\Theta_{k'}\int_{k'_{z}=0}^{\infty}\int_{k'_{y}=0}^{\infty}\int_{k'_{x}=0}^{\infty}\frac{\left[k'_{z}\right]^{2}}{\sqrt{\left[k'_{x}\right]^{2}+\left[k'_{y}\right]^{2}+\left[k'_{z}\right]^{2}}}dk'_{x}dk'_{y}dk'_{z}.\end{align*}

Both $P_{out}$ and $P_{in}$ are infinite, but their difference has
physical meaning. After some algebraic simplifications, the difference
can be written as \begin{align*}
P_{out}-P_{in} & =\frac{\hbar c\pi^{2}}{8d^{4}}\Theta_{k'}\left[\sum_{n=1}^{\infty}n^{2}\int_{x=0}^{\infty}\frac{dx}{\sqrt{x+n^{2}}}-\int_{u=0}^{\infty}\int_{x=0}^{\infty}\frac{u^{2}}{\sqrt{x+u^{2}}}dxdu\right].\end{align*}
 An application of the Euler-Maclaurin summation formula \cite{key-Euler-Sum-Formula,key-Euler-Sum-Formula-Derivation}
leads to the Casimir's result \begin{align*}
P_{out}-P_{in} & =-\frac{\hbar c\pi^{2}}{240d^{4}},\end{align*}
 where $\Theta_{k'}=2$ for two possible polarizations for zero-point
electromagnetic fields.

\subsubsection{The Source Theory Approach}

\begin{figure}[H]
\begin{center}\includegraphics[%
  scale=0.7]{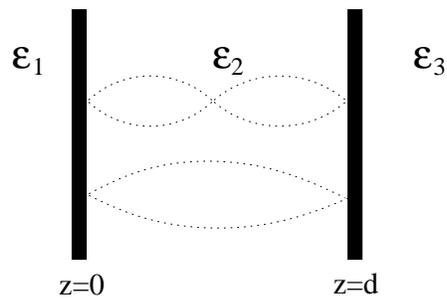}\end{center}

\caption{A cross-sectional view of two infinite parallel conducting plates.
The plates are separated by a gap distance of $z=d.$ Also, the three
regions have different dielectric constants $\varepsilon_{i}\left(\omega\right).$
\label{cap:casimir-plates-source-theory}}
\end{figure}

The Casimir effect can also be explained by the source theory of Schwinger
\cite{key-Milton,key-Schwinger-DeRaas-Milton,key-Milonni}. An induced
dipole $\vec{p}_{d}$ in a field $\vec{E}$ has an energy $\mathcal{H}_{d}=-\vec{p}_{d}\cdot\vec{E}/2.$
The factor of one half comes from the average value of an induced
dipole. When there are $N$ dipoles per unit volume, the associated
polarization is $\vec{P}=N\vec{p}_{d}$ and the expectation value
of the energy in quantum theory is $\left\langle \mathcal{H}_{d}\right\rangle =-\int\left\langle \vec{p}_{d}\cdot\vec{E}/2\right\rangle d^{3}\vec{R}.$
Here the polarizability in $\vec{p}_{d}$ is left as a free parameter
which needs to be determined from the experiment. The expectation
value of the energy is then \begin{align*}
\left\langle \mathcal{H}_{d}\right\rangle  & =-\frac{1}{2}\int\left\langle \vec{p}_{d}\cdot\vec{E}^{\left(+\right)}+\vec{E}^{\left(-\right)}\cdot\vec{p}_{d}\right\rangle d^{3}\vec{R},\end{align*}
 where $\vec{E}^{\left(\pm\right)}\left(\vec{R},t\right)=\vec{E}_{v}^{\left(\pm\right)}\left(\vec{R},t\right)+\vec{E}_{s}^{\left(\pm\right)}\left(\vec{R},t\right).$
Here $\vec{E}_{v}^{\left(\pm\right)}$ is the vacuum-field and $\vec{E}_{s}^{\left(\pm\right)}$
is the field due to other sources. Since $\vec{E}_{v}^{\left(+\right)}\left(\vec{R},t\right)\left|vac\right\rangle =\left\langle vac\right|\vec{E}_{v}^{\left(-\right)}\left(\vec{R},t\right)=0,$
the above expectation value of the energy can be written as \begin{align}
\left\langle \mathcal{H}_{d}\right\rangle  & =-\frac{1}{2}\int\left\langle \vec{p}_{d}\cdot\vec{E}^{\left(+\right)}\right\rangle d^{3}\vec{R}+c.c.,\label{eq:Source-Theory-Energy-1}\end{align}
 where $c.c.$ denotes complex conjugation. From the fact that electric
field operator can be written as an expansion in the mode functions
$\vec{A}_{\alpha}\left(\vec{R}\right),$ \begin{align*}
\vec{E}^{\left(+\right)} & =i\sum_{\alpha}\sqrt{2\pi\hbar\omega_{\alpha}}\left[a_{\alpha}\left(t\right)\vec{A}_{\alpha}\left(\vec{R}\right)-a_{\alpha}^{\dagger}\left(t\right)\vec{A}_{\alpha}^{*}\left(\vec{R}\right)\right],\end{align*}
 the Heisenberg equation of motion for $\dot{a}_{\alpha}\left(t\right)$
and $a_{\alpha,s}\left(t\right)$ are obtained as \begin{align*}
\dot{a}_{\alpha}\left(t\right) & =-i\omega_{\alpha}a_{\alpha}\left(t\right)+\sqrt{\frac{2\pi\omega_{\alpha}}{\hbar}}\int\vec{A}_{\alpha}^{*}\left(\vec{R}\right)\cdot\vec{p}_{d}\left(\vec{R},t\right)d^{3}\vec{R},\end{align*}
 \begin{align*}
a_{\alpha,s}\left(t\right) & =\sqrt{\frac{2\pi\omega_{\alpha}}{\hbar}}\int_{0}^{t}\exp\left(i\omega_{\alpha}\left[t'-t\right]\right)dt'\int\vec{A}_{\alpha}^{*}\left(\vec{R}\right)\cdot\vec{p}_{d}\left(\vec{R},t'\right)d^{3}\vec{R},\end{align*}
 where $a_{\alpha,s}\left(t\right)$ is the source contribution part
of $a_{\alpha}\left(t\right).$ The {}``positive frequency'' or
the photon annihilation part of $\vec{E}_{s}^{\left(+\right)}\left(\vec{R},t\right)$
can then be written as \begin{align*}
\vec{E}_{s}^{\left(+\right)}\left(\vec{R},t\right) & =2\pi i\sum_{\alpha}\omega_{\alpha}\vec{A}_{\alpha}\left(\vec{R}\right)\int_{0}^{t}\exp\left(i\omega_{\alpha}\left[t'-t\right]\right)dt'\int\vec{A}_{\alpha}^{*}\left(\vec{R}'\right)\cdot\vec{p}_{d}\left(\vec{R}',t'\right)d^{3}\vec{R}'\\
 & =2\pi i\int\int_{0}^{t}\sum_{\alpha}\omega_{\alpha}\vec{A}_{\alpha}\left(\vec{R}\right)\vec{A}_{\alpha}^{*}\left(\vec{R}'\right)\exp\left(i\omega_{\alpha}\left[t'-t\right]\right)\cdot\vec{p}_{d}\left(\vec{R}',t'\right)dt'd^{3}\vec{R}'\\
 & \equiv8\pi\int\int_{0}^{t}\overleftrightarrow{G^{\left(+\right)}}\left(\vec{R},\vec{R}';t,t'\right)\cdot\vec{p}_{d}\left(\vec{R}',t'\right)dt'd^{3}\vec{R}',\end{align*}
 where $\overleftrightarrow{G^{\left(+\right)}}\left(\vec{R},\vec{R}';t,t'\right)$
is a dyadic Green function \begin{align}
\overleftrightarrow{G^{\left(+\right)}}\left(\vec{R},\vec{R}';t,t'\right) & =\frac{i}{4}\sum_{\alpha}\omega_{\alpha}\vec{A}_{\alpha}\left(\vec{R}\right)\vec{A}_{\alpha}^{*}\left(\vec{R}'\right)\exp\left(i\omega_{\alpha}\left[t'-t\right]\right).\label{eq:Source-Theory-Energy-GREEN}\end{align}
 Equations (\ref{eq:Source-Theory-Energy-1}) and (\ref{eq:Source-Theory-Energy-GREEN})
lead to the result \begin{align*}
\left\langle \mathcal{H}_{d}\right\rangle  & =-8\pi\mathbb{R}_{E}\int_{\vec{R}}\int_{\vec{R}'}\int_{0}^{t}\overleftrightarrow{G_{ij}^{\left(+\right)}}\left(\vec{R},\vec{R}';t,t'\right)\left\langle \vec{p}_{d,j}\left(\vec{R},t\right)\cdot\vec{p}_{d,i}\left(\vec{R}',t'\right)\right\rangle dt'd^{3}\vec{R}'d^{3}\vec{R},\end{align*}
 where the summation over repeated indices is understood, and $\mathbb{R}_{E}$
denotes the real part. The above result is the energy of the induced
dipoles in a medium due to the source fields produced by the dipoles.
It can be further shown that for the infinitesimal variations in energy,
\begin{align*}
\left\langle \delta\mathcal{H}_{d}\right\rangle  & =-4\mathbb{R}_{E}\int_{\vec{R}}\int_{\vec{R}'}\int_{0}^{t}\int_{0}^{\infty}\Gamma_{ij}\left(\vec{R},\vec{R}',\omega\right)\left\langle \vec{p}_{d,j}\left(\vec{R},t\right)\cdot\vec{p}_{d,i}\left(\vec{R}',t'\right)\right\rangle \\
 & \times\exp\left(i\omega\left[t'-t\right]\right)d\omega dt'd^{3}\vec{R}'d^{3}\vec{R},\end{align*}
 where $\Gamma_{ij}\left(\vec{R},\vec{R}',\omega\right)$ is related
to $\overleftrightarrow{G_{ij}^{\left(+\right)}}\left(\vec{R},\vec{R}';t,t'\right)$
through the relation \begin{align*}
\overleftrightarrow{G^{\left(+\right)}}\left(\vec{R},\vec{R}';t,t'\right) & =\frac{1}{2\pi}\int_{0}^{\infty}\Gamma_{ij}\left(\vec{R},\vec{R}',\omega\right)\exp\left(i\omega\left[t'-t\right]\right)d\omega.\end{align*}
 The force per unit area can then be shown to be \begin{align}
F\left(d\right) & =\frac{i\hbar}{8\pi^{3}}\int_{0}^{\infty}\int_{\vec{k}_{\perp}}\left[\varepsilon_{2}-\varepsilon_{3}\right]\Gamma_{jj}\left(d,d,\vec{k}_{\perp},\omega\right)d^{2}\vec{k}_{\perp}d\omega,\label{eq:Force-Source-Theory}\end{align}
 where the factor $\left[\varepsilon_{2}-\varepsilon_{3}\right]\Gamma_{jj}\left(d,d,\vec{k}_{\perp},\omega\right)$
is given by \begin{align*}
\left[\varepsilon_{2}-\varepsilon_{3}\right]\Gamma_{jj}\left(d,d,\vec{k}_{\perp},\omega\right) & =2\left[K_{3}-K_{2}\right]+2K_{3}\left\{ \left(\left[\frac{K_{1}+K_{3}}{K_{1}-K_{3}}\right]\left[\frac{K_{2}+K_{3}}{K_{2}-K_{3}}\right]\exp\left(2K_{3}d\right)-1\right)^{-1}\right.\\
 & \left.+\left(\left[\frac{\varepsilon_{3}K_{1}+\varepsilon_{1}K_{3}}{\varepsilon_{3}K_{1}-\varepsilon_{1}K_{3}}\right]\left[\frac{\varepsilon_{3}K_{2}+\varepsilon_{2}K_{3}}{\varepsilon_{3}K_{2}-\varepsilon_{2}K_{3}}\right]\exp\left(2K_{3}d\right)-1\right)^{-1}\right\} .\end{align*}
 Here $K^{2}\equiv k_{\perp}^{2}-c^{-2}\omega^{2}\varepsilon\left(\omega\right)$
and $\varepsilon_{i}$ is the dielectric constant corresponding to
the region $i.$ The plate configuration corresponding to the source
theory description discussed above is illustrated in Figure \ref{cap:casimir-plates-source-theory}. 

The expression of force, equation (\ref{eq:Force-Source-Theory}),
is derived from the source theory of Schwinger, Milton and DeRaad
\cite{key-Milton,key-Schwinger-DeRaas-Milton}. It reproduces the
result of Lifshitz \cite{key-Landau-Lifshitz-EM-Book,key-Lifshitz},
which is a generalization of the Casimir force involving perfectly
conducting parallel plates to that involving dielectric media. The
details of this brief outline of the source theory description can
be found in references \cite{key-Milton,key-Milonni}.

\section{Reflection Dynamics}

Once the idea of physics of vacuum polarization is taken for granted,
one can move forward to calculate the effective, temperature-averaged
energy due to the dipole-dipole interactions with the time retardation
effect folded into the van der Waals interaction. The energy between
the dielectric or conducting media is then obtained from the allowed
modes of electromagnetic waves determined by the Maxwell equations
together with the boundary conditions, granted that the most significant
zero-point electromagnetic field wavelengths determining the interaction
are large when compared with the spacing of the lattice points in
the media. Under such an assumption, the effect of all the multiple
dipole scattering by atoms in the dielectric or conducting media is
to simply enforce the macroscopic reflection laws of electromagnetic
waves; and this allows the macroscopic electromagnetic theory to be
used with impunity in the calculation of Casimir force, granted the
classical electromagnetic fields have been quantized. The Casimir
force is then simply obtained by taking the negative gradient of the
energy in space.

In principle, the atomistic approach utilizing the Casimir-Polder
interaction explains the Casimir effect observed between any system.
Unfortunately, the pairwise summation of the intermolecular forces
for systems containing large number of atoms can become very complicated.
H. B. G. Casimir, realizing the linear relationship between the field
and the polarization, devised an easier approach to the calculation
of the Casimir effect for large systems such as two perfectly conducting
parallel plates. This latter development is the description of the
Euler-Maclaurin summation approach shown previously, in which the
Casimir force can be found by utilizing the field boundary conditions
only. The vacuum pressure approach originated by Milonni, Cook and
Goggin \cite{key-Milonni-Cook-Goggin} is a simple elaboration of
Casimir's latter invention utilizing the boundary conditions. The
source theory description of Schwinger is an alternate explanation
of the Casimir effect which can be inherently traced to the retarded
van der Waals interaction. 

Because all four approaches which were previously mentioned, (1) the
Casimir-Polder interaction, (2) the Euler-Maclaurin summation, (3)
the vacuum pressure and (4) the source theory, stem from the same
physics of vacuum polarization, they are equivalent. The preference
of one over another mainly depends on the geometry of the boundaries
being investigated. For the type of physical arrangements of boundary
configurations that are being considered in this investigation, the
vacuum pressure approach provides the most natural route to the Casimir
force calculation. The three physical arrangements for the boundary
configurations considered in this investigation are: (1) the plate-hemisphere,
(2) the hemisphere-hemisphere and (3) a sphere formed by brining two
hemispheres together. Because the geometric configurations of items
(2) and (3) are special versions of the more general, plate-hemisphere
configuration, the basic reflection dynamics needed for the plate-hemisphere
case is worked out first. The results can then be applied to the hemisphere-hemisphere
and the sphere configurations later. 

\begin{figure}[t]
\begin{center}\includegraphics[%
  scale=0.7]{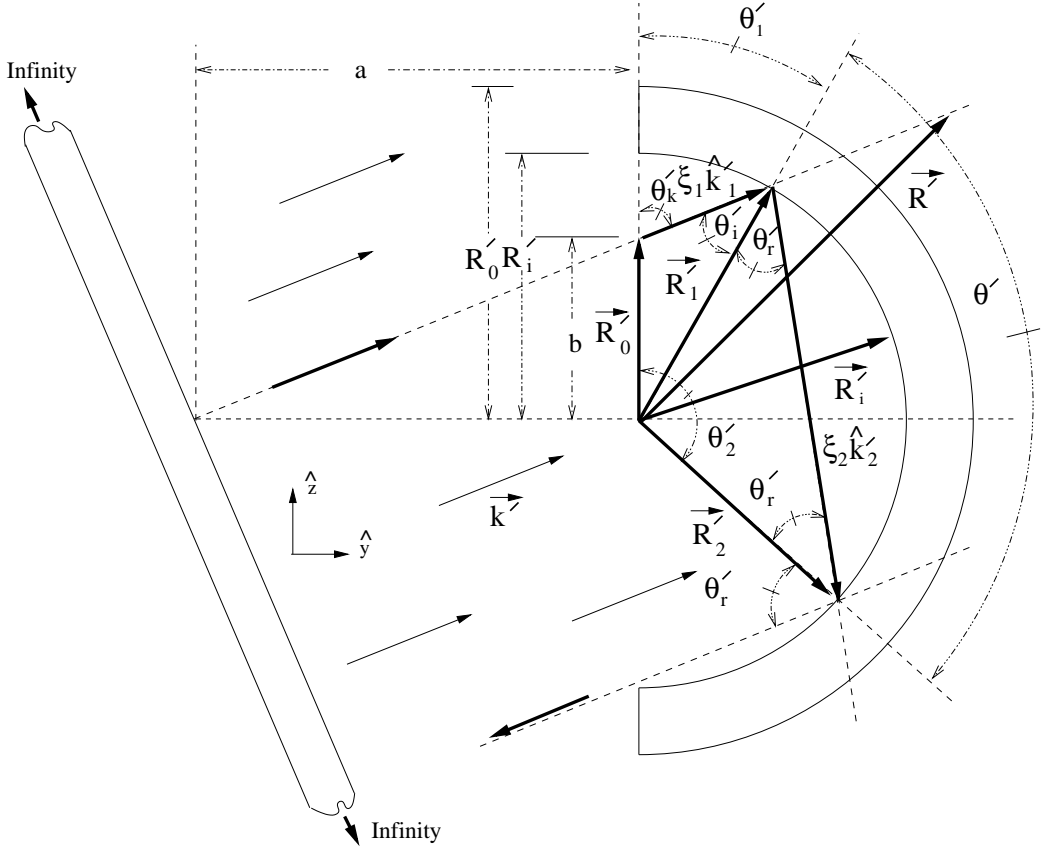}\end{center}

\caption{The plane of incidence view of plate-hemisphere configuration. The
waves that are supported through internal reflections in the hemisphere
cavity must satisfy the relation $\lambda\leq2\left\Vert \vec{R'}_{2}-\vec{R'}_{1}\right\Vert .$
\label{cap:cross-sectional-view-plate-and-hemisphere}}
\end{figure}

The vacuum-fields are subject to the appropriate boundary conditions.
For boundaries made of perfect conductors, the transverse components
of the electric field are zero at the surface. For this simplification,
the skin depth of penetration is considered to be zero. The plate-hemisphere
under consideration is shown in Figure \ref{cap:cross-sectional-view-plate-and-hemisphere}.
The solutions to the vacuum-fields are that of the Cartesian version
of the free Maxwell field vector potential differential equation $\nabla^{2}\vec{A}\left(\vec{R}\right)-c^{-2}\partial_{t}^{2}\vec{A}\left(\vec{R}\right)=0,$
where the Coulomb gauge $\vec{\nabla}\cdot\vec{A}=0$ and the absence
of the source $\Phi\left(\rho,\left\Vert \vec{R}\right\Vert \right)=0$
have been imposed. The electric and the magnetic field component of
the vacuum-field are given by $\vec{E}=-c^{-1}\partial_{t}\vec{A}$
and $\vec{B}=\vec{\nabla}\times\vec{A},$ where $\vec{A}$ is the
free field vector potential. The vanishing of the transversal component
of the electric field at the perfect conductor surface implies that
the solution for $\vec{E}$ is in the form of $\vec{E}\propto\sin\left(2\pi\lambda^{-1}\left\Vert \vec{L}\right\Vert \right),$
where $\lambda$ is the wavelength and $\left\Vert \vec{L}\right\Vert $
is the path length between the boundaries. The wavelength is restricted
by the condition $\lambda\leq2\left\Vert \vec{R'}_{2}-\vec{R'}_{1}\right\Vert \equiv2\xi_{2},$
where $\vec{R'}_{2}$ and $\vec{R'}_{1}$ are two immediate reflection
points in the hemisphere cavity of Figure \ref{cap:cross-sectional-view-plate-and-hemisphere}.
In order to compute the modes allowed inside the hemisphere resonator,
a detailed knowledge of the reflections occurring in the hemisphere
cavity is needed. This is described in the following section.

\subsection{Reflection Points on the Surface of a Resonator}

The wave vector directed along an arbitrary direction in Cartesian
coordinates is written as \begin{eqnarray}
\vec{k'}_{1}\left(k'_{1,x},k'_{1,y},k'_{1,z}\right)=\sum_{i=1}^{3}k'_{1,i}\hat{e_{i}}, &  & k'_{1,i}=\left\{ \begin{array}{cc}
i=1\rightarrow k'_{1,x}, & \hat{e_{1}}=\hat{x},\\
\\i=2\rightarrow k'_{1,y}, & \hat{e_{2}}=\hat{y},\\
\\i=3\rightarrow k'_{1,z}, & \hat{e_{3}}=\hat{z}.\end{array}\right.\label{eq:arbitrary-k-vector-NA}\end{eqnarray}
 Hence, the unit wave vector, $\hat{k'}_{1}=\left\Vert \vec{k'}_{1}\right\Vert ^{-1}\sum_{i=1}^{3}k'_{1,i}\hat{e_{i}}.$
Define the initial position $\vec{R'}_{0}$ for the incident wave
$\vec{k'}_{1},$ \begin{eqnarray}
\vec{R'}_{0}\left(r'_{0,x},r'_{0,y},r'_{0,z}\right)=\sum_{i=1}^{3}r'_{0,i}\hat{e_{i}}, &  & r'_{0,i}=\left\{ \begin{array}{c}
i=1\rightarrow r'_{0,x},\\
\\i=2\rightarrow r'_{0,y},\\
\\i=3\rightarrow r'_{0,z}.\end{array}\right.\label{eq:initial-arbitrary-k-vector-position-NA}\end{eqnarray}
 Here it should be noted that $\vec{R'}_{0}$ really has only components
$r'_{0,x}$ and $r'_{0,z}.$ But nevertheless, one can always set
$r'_{0,y}=0$ whenever needed. Since no particular wave vectors with
specified wave lengths are prescribed initially, it is desirable to
employ a parameterization scheme to represent these wave vectors.
The line segment traced out by this wave vector $\hat{k'}_{1}$ is
formulated in the parametric form \begin{align}
\vec{R'}_{1} & =\xi_{1}\hat{k'}_{1}+\vec{R'}_{0}=\sum_{i=1}^{3}\left[r'_{0,i}+\xi_{1}\left\Vert \vec{k'}_{1}\right\Vert ^{-1}k'_{1,i}\right]\hat{e_{i}},\label{eq:parametrized-line-segment-generalized-NA}\end{align}
 where the variable $\xi_{1}$ is a positive definite parameter. The
restriction $\xi_{1}\geq0$ is necessary because the direction of
the wave propagation is set by $\hat{k'}_{1}.$ Here $\vec{R'}_{1}$
is the first reflection point on the hemisphere. In terms of spherical
coordinate variables, $\vec{R'}_{1}$ takes the form \begin{eqnarray}
\vec{R'}_{1}\left(r'_{i},\theta'_{1},\phi'_{1}\right)=r'_{i}\sum_{i=1}^{3}\Lambda'_{1,i}\hat{e_{i}}, &  & \left\{ \begin{array}{c}
\Lambda'_{1,1}=\sin\theta'_{1}\cos\phi'_{1},\\
\\\Lambda'_{1,2}=\sin\theta'_{1}\sin\phi'_{1},\\
\\\Lambda'_{1,3}=\cos\theta'_{1},\qquad\;\:\end{array}\right.\label{eq:sphere-vector-generalized-NA}\end{eqnarray}
 where $r'_{i}$ is the hemisphere radius, $\theta'_{1}$ and $\phi'_{1}$
are the polar and the azimuthal angle respectively of $\vec{R'}_{1}$
at the first reflection point. Notice that subscript $i$ of $r'_{i}$
denotes {}``inner radius'' not a summation index. 

By combining equations (\ref{eq:parametrized-line-segment-generalized-NA})
and (\ref{eq:sphere-vector-generalized-NA}), we can solve for the
parameter $\xi_{1}.$ It can be shown that \begin{align}
\xi_{1}\equiv\xi_{1,p} & =-\hat{k'}_{1}\cdot\vec{R'}_{0}+\sqrt{\left[\hat{k'}_{1}\cdot\vec{R'}_{0}\right]^{2}+\left[r'_{i}\right]^{2}-\left\Vert \vec{R'}_{0}\right\Vert ^{2}},\label{eq:positive-root-k-NA}\end{align}
 where the positive root for $\xi_{1}$ have been chosen due to the
restriction $\xi_{1}\geq0.$ The detailed proof of equation (\ref{eq:positive-root-k-NA})
is given in Appendix A, where the same equation is designated as equation
(\ref{eq:positive-root-k}). 

Substituting $\xi_{1}$ in equation (\ref{eq:parametrized-line-segment-generalized-NA}),
the first reflection point off the inner hemisphere surface is expressed
as \begin{align}
\vec{R'}_{1}\left(\xi_{1,p};\vec{R'}_{0},\hat{k'}_{1}\right) & =\sum_{i=1}^{3}\left[r'_{0,i}+\xi_{1,p}\left\Vert \vec{k'}_{1}\right\Vert ^{-1}k'_{1,i}\right]\hat{e_{i}},\label{eq:1st-bounce-off-point-NA}\end{align}
 where $\xi_{1,p}$ is from equation (\ref{eq:positive-root-k-NA}). 

The incoming wave vector $\vec{k'}_{i}$ can always be decomposed
into parallel and perpendicular components, $\vec{k'}_{i,\parallel}$
and $\vec{k'}_{i,\perp},$ with respect to the local reflection surface.
It is shown in equation (\ref{eq:K-reflected-in-PARA-n-PERPE-N})
of Appendix A that the reflected wave vector $\vec{k'}_{r}$ has the
form $\vec{k'}_{r}=\alpha_{r,\perp}\left[\hat{n'}\times\vec{k'}_{i}\right]\times\hat{n'}-\alpha_{r,\parallel}\hat{n'}\cdot\vec{k'}_{i}\hat{n'},$
where the quantities $\alpha_{r,\parallel}$ and $\alpha_{r,\perp}$
are the reflection coefficients and $\hat{n'}$ is a unit surface
normal. For the perfect reflecting surfaces, $\alpha_{r,\parallel}=\alpha_{r,\perp}=1.$
In component form, $\vec{k'}_{r}=\sum_{l=1}^{3}\left\{ \alpha_{r,\perp}\left[n'_{n}k'_{i,l}n'_{n}-n'_{l}k'_{i,n}n'_{n}\right]-\alpha_{r,\parallel}n'_{n}k'_{i,n}n'_{l}\right\} \hat{e_{l}},$
where it is understood that $\hat{n'}$ is already normalized and
Einstein summation convention is applied to the index $n.$ The second
reflection point $\vec{R'}_{2}$ is found then by repeating the steps
done for $\vec{R'}_{1}$ and by using the expression $\vec{k'}_{r}\equiv\vec{k'}_{r}/\left\Vert \vec{k'}_{r}\right\Vert ,$
\begin{align*}
\vec{R'}_{2} & =\vec{R'}_{1}+\xi_{2,p}\hat{k'}_{r}=\vec{R'}_{1}+\xi_{2,p}\frac{\alpha_{r,\perp}\left[\hat{n'}\times\vec{k'}_{i}\right]\times\hat{n'}-\alpha_{r,\parallel}\hat{n'}\cdot\vec{k'}_{i}\hat{n'}}{\left\Vert \alpha_{r,\perp}\left[\hat{n'}\times\vec{k'}_{i}\right]\times\hat{n'}-\alpha_{r,\parallel}\hat{n'}\cdot\vec{k'}_{i}\hat{n'}\right\Vert },\end{align*}
 where $\xi_{2,p}$ is the new positive definite parameter for the
second reflection point. 

The incidence plane of reflection is determined solely by the incident
wave $\vec{k'}_{i}$ and the local normal $\vec{n'}_{i}$ of the reflecting
surface. It is important to recognize the fact that the subsequent
successive reflections of this incoming wave will be confined to this
particular incidence plane. This incident plane can be characterized
by a unit normal vector. For the system shown in Figure \ref{cap:cross-sectional-view-plate-and-hemisphere},
$\vec{k'}_{i}=\vec{k'}_{1}$ and $\vec{n'}_{n'_{i},1}=-\xi_{1,p}\hat{k'}_{1}-\vec{R'}_{0}.$
The unit vector which represents the incidence plane is given by $\hat{n'}_{p,1}=-\left\Vert \vec{n'}_{p,1}\right\Vert ^{-1}\sum_{i=1}^{3}\epsilon_{ijk}k'_{1,j}r'_{0,k}\hat{e_{i}},$
where the summations over indices $j$ and $k$ are implicit. If the
plane of incidence is represented by a scalar function $f\left(x',y',z'\right),$
then its unit normal vector $\hat{n'}_{p,1}$ will satisfy the relationship
$\hat{n'}_{p,1}\propto\vec{\nabla'}f_{p,1}\left(x',y',z'\right).$
It is shown from equation (\ref{eq:f-p-i-plane-of-incidence}) of
Appendix A that \begin{eqnarray}
f_{p,1}\left(\nu'_{1},\nu'_{2},\nu'_{3}\right)=-\left\Vert \vec{n'}_{p,1}\right\Vert ^{-1}\sum_{i=1}^{3}\epsilon_{ijk}k'_{1,j}r'_{0,k}\nu'_{i}, &  & i=\left\{ \begin{array}{c}
1\rightarrow\nu'_{1}=x',\\
\\2\rightarrow\nu'_{2}=y',\\
\\3\rightarrow\nu'_{3}=z',\end{array}\right.\label{eq:f-p-i-plane-of-incidence-NA}\end{eqnarray}
 where $-\infty\leq\left\{ \nu'_{1}=x',\nu'_{2}=y',\nu'_{3}=z'\right\} \leq\infty.$ 

\begin{figure}[t]
\begin{center}\includegraphics[%
  scale=0.7]{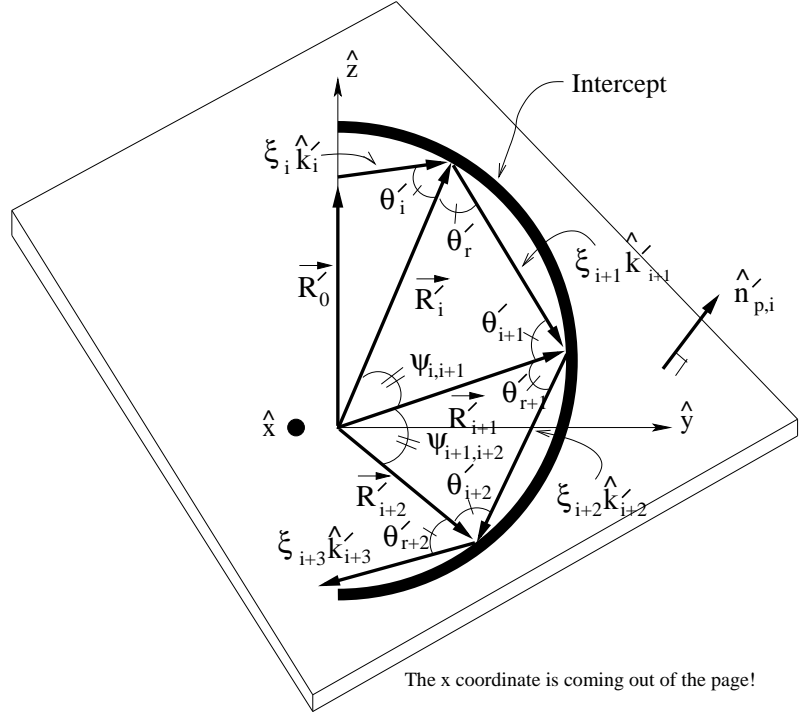}\end{center}

\caption{The thick line shown here represents the intersection between hemisphere
surface and the plane of incidence. The unit vector normal to the
plane of incidence is given by $\hat{n'}_{p,1}=-\left\Vert \vec{n'}_{p,1}\right\Vert ^{-1}\sum_{i=1}^{3}\epsilon_{ijk}k'_{1,j}r'_{0,k}\hat{e_{i}}.$
\label{cap:plane-sphere-intersection}}
\end{figure}

The surface of a sphere or hemisphere is defined through the relation
$f_{hemi}\left(x',y',z'\right)=\left[r'_{i}\right]^{2}-\sum_{i=1}^{3}\left[\nu'_{i}\right]^{2},$
where $r'_{i}$ is the radius of sphere and the subscript $i$ denotes
the inner surface. The intercept of interest is shown in Figure \ref{cap:plane-sphere-intersection}.
The intersection between the hemisphere surface and the incidence
plane $f_{p,1}\left(\nu'_{1},\nu'_{2},\nu'_{3}\right)$ is given by
$f_{hemi}\left(x',y',z'\right)-f_{p,1}\left(x',y',z'\right)=0.$ After
substitution of $f_{p,1}\left(x',y',z'\right)$ and $f_{hemi}\left(x',y',z'\right),$
we have \begin{eqnarray*}
\sum_{i=1}^{3}\left\{ \left[\nu'_{i}\right]^{2}-\left\Vert \vec{n'}_{p,1}\right\Vert ^{-1}\epsilon_{ijk}k'_{1,j}r'_{0,k}\nu'_{i}\right\} -\left[r'_{i}\right]^{2}=0, &  & i=\left\{ \begin{array}{c}
1\rightarrow\nu'_{1}=x',\\
\\2\rightarrow\nu'_{2}=y',\\
\\3\rightarrow\nu'_{3}=z'.\end{array}\right.\end{eqnarray*}
 The term $\left[r'_{i}\right]^{2}$ can be rewritten in the form
$\left[r'_{i}\right]^{2}=\sum_{i=1}^{3}\left[r'_{i,i}\right]^{2},$
where $r'_{i,1}=r'_{i,x'},$ $r'_{i,2}=r'_{i,y'}$ and $r'_{i,3}=r'_{i,z'}.$
Solving for $\nu'_{i},$ it can be shown from equation (\ref{eq:hemi-plane-intercept-root})
of Appendix A that \begin{align}
\nu'_{i} & =\frac{1}{2}\left\Vert \vec{n'}_{p,1}\right\Vert ^{-1}\epsilon_{ijk}k'_{1,j}r'_{0,k}\pm\left\{ \left[\frac{1}{2}\left\Vert \vec{n'}_{p,1}\right\Vert ^{-1}\epsilon_{ijk}k'_{1,j}r'_{0,k}\right]^{2}+\left[r'_{i,i}\right]^{2}\right\} ^{1/2},\label{eq:hemi-plane-intercept-root-NA}\end{align}
 where $i=1,2,3$ and $\epsilon_{ijk}$ is the Levi-Civita symbol.
The result for $\nu'_{i}$ shown above provide a set of discrete reflection
points found by the intercept between the hemisphere and the plane
of incidence. 

Using spherical coordinate representations for the variables $r'_{i,1},$
$r'_{i,2}$ and $r'_{i,3},$ the initial reflection point $\vec{R'}_{1}$
can be expressed in terms of the spherical coordinate variables $\left(r'_{i},\theta'_{1},\phi'_{1}\right)$
(equation (\ref{eq:1st-Reflection-Point}) of Appendix A), \begin{eqnarray}
\vec{R'}_{1}\left(r'_{i},\theta'_{1},\phi'_{1}\right)=\sum_{i=1}^{3}\nu'_{1,i}\left(r'_{i},\theta'_{1},\phi'_{1}\right)\hat{e_{i}}, &  & i=\left\{ \begin{array}{c}
1\rightarrow\nu'_{1,1}=r'_{i}\sin\theta'_{1}\cos\phi'_{1},\\
\\2\rightarrow\nu'_{1,2}=r'_{i}\sin\theta'_{1}\sin\phi'_{1},\\
\\3\rightarrow\nu'_{1,3}=r'_{i}\cos\theta'_{1},\qquad\;\:\end{array}\right.\label{eq:1st-Reflection-Point-NA}\end{eqnarray}
 where $r'_{i}$ is the hemisphere radius, $\phi'_{1}$ and $\theta'_{1},$
the polar and azimuthal angle, respectively. They are defined in equations
(\ref{eq:phi1-prime-0-90-and-180-270-final}), (\ref{eq:phi1-prime-90-180-and-270-360-final}),
(\ref{eq:theta1-prime-0-to-90-degree-final}) and (\ref{eq:theta1-prime-90-to-180-degree-final})
of Appendix A. Similarly, the second reflection point on the inner
hemisphere surface is given by equation (\ref{eq:2nd-Reflection-Point})
of Appendix A: \begin{eqnarray}
\vec{R'}_{2}\left(r'_{i},\theta'_{2},\phi'_{2}\right)=\sum_{i=1}^{3}\nu'_{2,i}\left(r'_{i},\theta'_{2},\phi'_{2}\right)\hat{e_{i}}, &  & i=\left\{ \begin{array}{c}
1\rightarrow\nu'_{2,1}=r'_{i}\sin\theta'_{2}\cos\phi'_{2},\\
\\2\rightarrow\nu'_{2,2}=r'_{i}\sin\theta'_{2}\sin\phi'_{2},\\
\\3\rightarrow\nu'_{2,3}=r'_{i}\cos\theta'_{2},\qquad\;\:\end{array}\right.\label{eq:2nd-Reflection-Point-NA}\end{eqnarray}
 where the spherical angles $\phi'_{2}$ and $\theta'_{2}$ are defined
in equations (\ref{eq:phi2-prime-0-90-and-180-270-final}), (\ref{eq:phi2-prime-90-180-and-270-360-final}),
(\ref{eq:theta2-prime-0-to-90-degree-final}) and (\ref{eq:theta2-prime-90-to-180-degree-final})
of Appendix A. In general, leaving the details to Appendix A, the
$N$th reflection point inside the hemisphere is, from equation (\ref{eq:Nth-Reflection-Point})
of Appendix A, \begin{eqnarray}
\vec{R'}_{N}\left(r'_{i},\theta'_{N},\phi'_{N}\right)=\sum_{i=1}^{3}\nu'_{N,i}\left(r'_{i},\theta'_{N},\phi'_{N}\right)\hat{e_{i}}, &  & i=\left\{ \begin{array}{c}
1\rightarrow\nu'_{N,1}=r'_{i}\sin\theta'_{N}\cos\phi'_{N},\\
\\2\rightarrow\nu'_{N,2}=r'_{i}\sin\theta'_{N}\sin\phi'_{N},\\
\\3\rightarrow\nu'_{N,3}=r'_{i}\cos\theta'_{N},\qquad\;\:\:\end{array}\right.\label{eq:Nth-Reflection-Point-NA}\end{eqnarray}
 where the spherical angles $\theta'_{N}$ and $\phi'_{N}$ are defined
in equations (\ref{eq:thetaN-prime-0-to-90-degree-final}), (\ref{eq:thetaN-prime-90-to-180-degree-final}),
(\ref{eq:phiN-prime-0-90-and-180-270-final}) and (\ref{eq:phiN-prime-90-180-and-270-360-final})
of Appendix A. The details of all the work shown up to this point
can be found in Appendix A. 

\begin{figure}
\begin{center}\includegraphics[%
  scale=0.7]{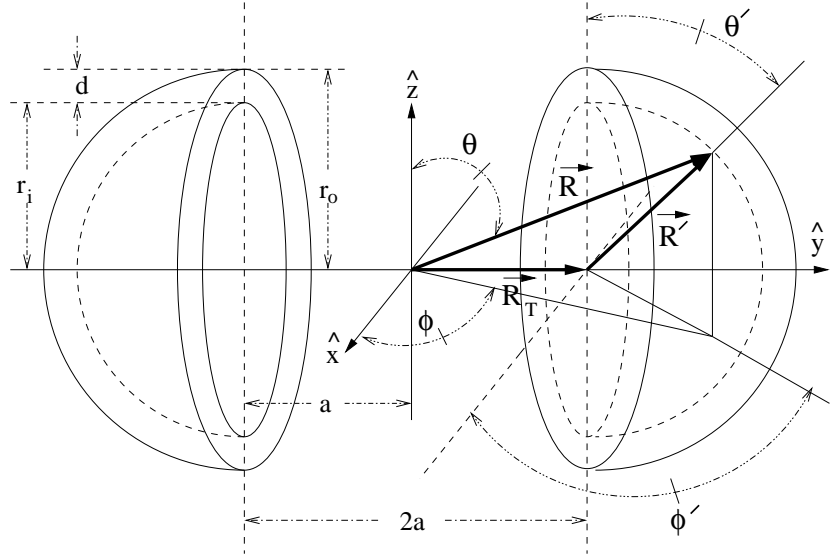}\end{center}

\caption{The surface of the hemisphere-hemisphere configuration can be described
relative to the system origin through $\vec{R},$ or relative to the
hemisphere centers through $\vec{R'}.$ \label{cap:Two-semispheres-without-center-solid-sphere}}
\end{figure}

The previously shown reflection points ($\vec{R'}_{1},$ $\vec{R'}_{2}$
and $\vec{R'}_{N}$) were described relative to the hemisphere center.
In many cases, the preferred choice for the system origin, from which
the variables are defined, depend on the physical arrangements of
the system being considered. For a sphere, the natural choice for
the origin is its center from which the spherical variables $\left(r'_{i},\theta',\phi'\right)$
are prescribed. For more complicated configuration shown in Figure
\ref{cap:Two-semispheres-without-center-solid-sphere}, the preferred
choice for origin really depends on the problem at hand. For this
reason, a set of transformation rules between $\left(r'_{i},\theta',\phi'\right)$
and $\left(r_{i},\theta,\phi\right)$ is sought. Here the primed set
is defined relative to the sphere center and the unprimed set is defined
relative to the origin of the global configuration. In terms of the
Cartesian variables, the two vectors $\vec{R}$ and $\vec{R'}$ describing
an identical point on the hemisphere surface are expressed by \begin{eqnarray}
\vec{R}\left(\nu_{1},\nu_{2},\nu_{3}\right)=\sum_{i=1}^{3}\nu_{i}\hat{e_{i}}, &  & \vec{R'}\left(\nu'_{1},\nu'_{2},\nu'_{3}\right)=\sum_{i=1}^{3}\nu'_{i}\hat{e_{i}},\label{eq:R-and-Rprimed-vector-cartesian-NA}\end{eqnarray}
 where $\left(\nu_{1},\nu_{2},\nu_{3}\right)\rightarrow\left(x,y,z\right),$
$\left(\nu'_{1},\nu'_{2},\nu'_{3}\right)\rightarrow\left(x',y',z'\right)$
and $\left(\hat{e_{1}},\hat{e_{2}},\hat{e_{3}}\right)\rightarrow\left(\hat{x},\hat{y},\hat{z}\right).$
The vectors $\vec{R}$ and $\vec{R'}$ are connected through the relation
$\vec{R}\left(\nu_{1},\nu_{2},\nu_{3}\right)=\sum_{i=1}^{3}\left[\nu_{T,i}+\nu'_{i}\right]\hat{e_{i}}$
with $\vec{R}_{T}\equiv\sum_{i=1}^{3}\nu_{T,i}\hat{e_{i}}$ which
represents the position of hemisphere center relative to the system
origin. As a result, we have $\sum_{i=1}^{3}\left[\nu_{i}-\nu_{T,i}-\nu'_{i}\right]\hat{e_{i}}=0.$
In terms of the spherical coordinate representation for $\left(\nu_{1},\nu_{2},\nu_{3}\right)$
and $\left(\nu'_{1},\nu'_{2},\nu'_{3}\right),$ we can solve for $\theta$
and $\phi.$ As shown from equations (\ref{eq:phi}) and (\ref{eq:theta})
of Appendix B, the result is \begin{align}
\phi\equiv\grave{\phi}\left(r'_{i},\theta',\phi',\nu_{T,1},\nu_{T,2}\right) & =\arctan\left(\frac{\nu_{T,2}+r'_{i}\sin\theta'\sin\phi'}{\nu_{T,1}+r'_{i}\sin\theta'\cos\phi'}\right),\label{eq:phi-NA}\end{align}
 \begin{align}
\theta & \equiv\grave{\theta}\left(r'_{i},\theta',\phi',\vec{R}_{T}\right)\nonumber \\
 & =\arctan\left(\frac{\left\{ \nu_{T,1}+\nu_{T,2}+r'_{i}\sin\theta'\left[\cos\phi'+\sin\phi'\right]\right\} \left[\nu_{T,3}+r'_{i}\cos\theta'\right]^{-1}}{\cos\left(\arctan\left(\frac{\nu_{T,2}+r'_{i}\sin\theta'\sin\phi'}{\nu_{T,1}+r'_{i}\sin\theta'\cos\phi'}\right)\right)+\sin\left(\arctan\left(\frac{\nu_{T,2}+r'_{i}\sin\theta'\sin\phi'}{\nu_{T,1}+r'_{i}\sin\theta'\cos\phi'}\right)\right)}\right),\label{eq:theta-NA}\end{align}
 where the notation $\grave{\phi}$ and $\grave{\theta}$ indicates
that $\phi$ and $\theta$ are explicitly expressed in terms of the
primed variables, respectively. It is to be noticed that for the configuration
shown in Figure \ref{cap:Two-semispheres-without-center-solid-sphere},
the hemisphere center is only shifted along $\hat{y}$ by an amount
of $\nu_{T,2}=a,$ which leads to $\nu_{T,i\neq2}=0.$ Nevertheless,
the derivation have been done for the case where $\nu_{T,i}\neq0,$
$i=1,2,3$ for the purpose of generalization. 

With the magnitude $\left\Vert \vec{R}\right\Vert =\left\{ \sum_{i=1}^{3}\left[\nu_{T,i}+r'_{i}\Lambda'_{i}\right]^{2}\right\} ^{1/2},$
where $\Lambda'_{1}\left(\theta',\phi'\right)=\sin\theta'\cos\phi',$
$\Lambda'_{2}\left(\theta',\phi'\right)=\sin\theta'\sin\phi'$ and
$\Lambda'_{3}\left(\theta'\right)=\cos\theta',$ the vector $\vec{R}\left(r'_{i},\grave{\vec{\Lambda}},\vec{\Lambda}',\vec{R}_{T}\right)$
is given by equation (\ref{eq:Points-on-Hemisphere-R-arbi}) of Appendix
B as \begin{eqnarray}
\vec{R}\left(r'_{i},\grave{\vec{\Lambda}},\vec{\Lambda}',\vec{R}_{T}\right)=\left\{ \sum_{i=1}^{3}\left[\nu_{T,i}+r'_{i}\Lambda'_{i}\right]^{2}\right\} ^{1/2}\sum_{i=1}^{3}\grave{\Lambda}_{i}\hat{e_{i}}, &  & \left\{ \begin{array}{c}
\grave{\Lambda}_{1}\left(\grave{\theta},\grave{\phi}\right)=\sin\grave{\theta}\cos\grave{\phi},\\
\\\grave{\Lambda}_{2}\left(\grave{\theta},\grave{\phi}\right)=\sin\grave{\theta}\sin\grave{\phi},\\
\\\grave{\Lambda}_{3}\left(\grave{\theta}\right)=\cos\grave{\theta}.\;\;\;\;\end{array}\right.\label{eq:Points-on-Hemisphere-R-arbi-NA}\end{eqnarray}
 The details of this section can be found in Appendices A and B.

\subsection{Selected Configurations}

Having found all of the wave reflection points in the hemisphere resonator,
the net momentum imparted on both the inner and outer surfaces by
the incident wave is computed for three configurations: (1) the sphere,
(2) the hemisphere-hemisphere and (3) the plate-hemisphere. The surface
element that is being impinged upon by an incident wave would experience
the net momentum change in an amount proportional to $\triangle\vec{k'}_{inner}\left(;\vec{R'}_{s,1},\vec{R'}_{s,0}\right)$
on the inner side, and $\triangle\vec{k'}_{outer}\left(;\vec{R'}_{s,1}+a\hat{R'}_{s,1}\right)$
on the outer side of the surface. The quantities $\triangle\vec{k'}_{inner}$
and $\triangle\vec{k'}_{outer}$ are due to the contribution from
a single mode of wave traveling in particular direction. The notation
$\left(;\vec{R'}_{s,1},\vec{R'}_{s,0}\right)$ of $\triangle\vec{k'}_{inner}$
denotes that it is defined in terms of the initial reflection point
$\vec{R'}_{s,1}$ on the surface and the initial crossing point $\vec{R'}_{s,0}$
of the hemisphere opening (or the sphere cross-section). The notation
$\left(;\vec{R'}_{s,1}+a\hat{R'}_{s,1}\right)$ of $\triangle\vec{k'}_{outer}$
implies the outer surface reflection point. The total resultant imparted
momentum on the hemisphere or sphere is found by summing over all
modes of wave, over all directions.

\subsubsection{Hollow Spherical Shell}

\begin{figure}
\begin{center}\includegraphics[%
  scale=0.7]{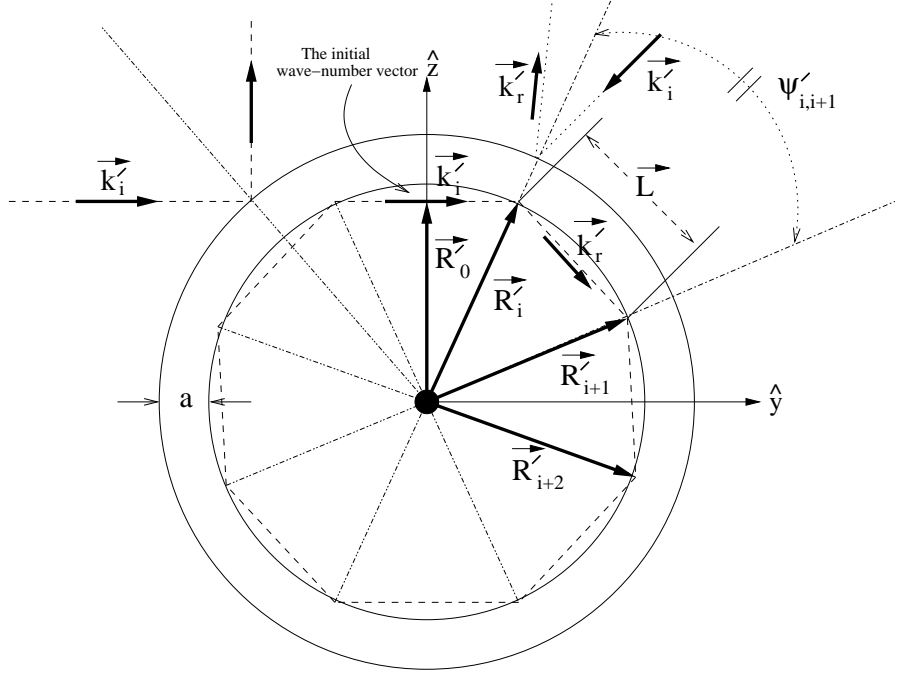}\end{center}

\caption{Inside the cavity, an incident wave $\vec{k'}_{i}$ on first impact
point $\vec{R'}_{i}$ induces a series of reflections that propagate
throughout the entire inner cavity. Similarly, a wave $\vec{k'}_{i}$
incident on the impact point $\vec{R'}_{i}+a\hat{R'}_{i},$ where
$a$ is the thickness of the sphere, induces reflected wave of magnitude
$\left\Vert \vec{k'}_{i}\right\Vert .$ The resultant wave direction
in the external region is along $\vec{R'}_{i}$ and the resultant
wave direction in the resonator is along $-\vec{R'}_{i}$ due to the
fact there is exactly another wave vector traveling in opposite direction
in both regions. In both cases, the reflected and incident waves have
equal magnitude due to the fact that the sphere is assumed to be a
perfect conductor. \label{cap:sphere-reflection-dynamics}}
\end{figure}

A sphere formed by bringing in two hemispheres together is shown in
Figure \ref{cap:sphere-reflection-dynamics}. %
 The resultant change in wave vector direction upon reflection at
the inner surface of the sphere is from the equation (\ref{eq:SPHERE-Delta-K-Inside})
of Appendix C1, \begin{eqnarray}
\triangle\vec{k'}_{inner}\left(;\vec{R'}_{s,1},\vec{R'}_{s,0}\right)=-\frac{4n\pi\cos\theta_{inc}}{\left\Vert \vec{R}_{s,2}\left(r'_{i},\vec{\Lambda}'_{s,2}\right)-\vec{R}_{s,1}\left(r'_{i},\vec{\Lambda}'_{s,1}\right)\right\Vert }\hat{R'}_{s,1}, &  & \left\{ \begin{array}{c}
0\leq\theta_{inc}<\pi/2,\\
\\n=1,2,\cdots,\end{array}\right.\label{eq:SPHERE-Delta-K-Inside-NA}\end{eqnarray}
 where $\theta_{inc}$ is from equation (\ref{eq:angle-of-incidence-exp});
$\vec{R}_{s,1}\left(r'_{i},\vec{\Lambda}'_{s,1}\right)$ and $\vec{R}_{s,2}\left(r'_{i},\vec{\Lambda}'_{s,2}\right)$
follow the generic form shown in the equation (\ref{eq:Nth-Reflection-Point-sphere})
of Appendix C1, \begin{eqnarray}
\vec{R}_{s,N}\left(r'_{i},\vec{\Lambda}'_{s,N}\right)=r'_{i}\sum_{i=1}^{3}\Lambda'_{s,N,i}\hat{e_{i}}, &  & \left\{ \begin{array}{c}
\Lambda'_{s,N,1}\left(\theta'_{s,N},\phi'_{s,N}\right)=\sin\theta'_{s,N}\cos\phi'_{s,N},\\
\\\Lambda'_{s,N,2}\left(\theta'_{s,N},\phi'_{s,N}\right)=\sin\theta'_{s,N}\sin\phi'_{s,N},\\
\\\Lambda'_{s,N,3}\left(\theta'_{s,N}\right)=\cos\theta'_{s,N}.\;\;\;\;\end{array}\right.\label{eq:Nth-Reflection-Point-sphere-NA}\end{eqnarray}
 Here the label $s$ have been attached to denote a sphere and the
obvious index changes in the spherical variables $\theta'_{s,N}$
and $\phi'_{s,N}$ are understood from the set of equations (\ref{eq:thetaN-prime-0-to-90-degree-final}),
(\ref{eq:thetaN-prime-90-to-180-degree-final}), (\ref{eq:phiN-prime-0-90-and-180-270-final})
and (\ref{eq:phiN-prime-90-180-and-270-360-final}). 

Similarly, the resultant change in wave vector direction upon reflection
at the outer surface of the sphere is from equation (\ref{eq:SPHERE-Delta-K-Outside})
of Appendix C1, \begin{eqnarray}
\triangle\vec{k'}_{outer}\left(;\vec{R'}_{s,1}+a\hat{R'}_{s,1}\right)=4\left\Vert \vec{k'}_{i,f}\right\Vert \cos\theta_{inc}\hat{R'}_{s,1}, &  & \left\{ \begin{array}{c}
0\leq\theta_{inc}<\pi/2,\\
\\n=1,2,\cdots.\end{array}\right.\label{eq:SPHERE-Delta-K-Outside-NA}\end{eqnarray}
 The details of this section can be found in Appendix C1.

\subsubsection{Hemisphere-Hemisphere}

For the hemisphere, the changes in wave vector directions after the
reflection at a point $\hat{R'}_{h,1}$ inside the resonator, or after
the reflection at location $\vec{R'}_{h,1}+a\hat{R'}_{h,1}$ outside
the hemisphere, can be found from equations (\ref{eq:SPHERE-Delta-K-Inside-NA})
and (\ref{eq:SPHERE-Delta-K-Outside-NA}) with obvious subscript changes,
\begin{eqnarray}
\triangle\vec{k'}_{inner}\left(;\vec{R'}_{h,1},\vec{R'}_{h,0}\right)=-\frac{4n\pi\cos\theta_{inc}}{\left\Vert \vec{R}_{h,2}\left(r'_{i},\vec{\Lambda}'_{h,2}\right)-\vec{R}_{h,1}\left(r'_{i},\vec{\Lambda}'_{h,1}\right)\right\Vert }\hat{R'}_{h,1}, &  & \left\{ \begin{array}{c}
0\leq\theta_{inc}<\pi/2,\\
\\n=1,2,\cdots;\end{array}\right.\label{eq:HEMISPHERE-Delta-K-Inside-NA}\end{eqnarray}
 \begin{eqnarray}
\triangle\vec{k'}_{outer}\left(;\vec{R'}_{h,1}+a\hat{R'}_{h,1}\right)=4\left\Vert \vec{k'}_{i,f}\right\Vert \cos\theta_{inc}\hat{R'}_{h,1}, &  & \left\{ \begin{array}{c}
0\leq\theta_{inc}<\pi/2,\\
\\n=1,2,\cdots,\end{array}\right.\label{eq:HEMISPHERE-Delta-K-Outside-NA}\end{eqnarray}
 where the reflection location $\vec{R}_{h,N}\left(r'_{i},\grave{\vec{\Lambda}}_{h,N},\vec{\Lambda}'_{h,N},\vec{R}_{T,h}\right)$
follows the generic form as shown in equation (\ref{eq:Points-on-Hemisphere-R})
of Appendix C2, \begin{align}
\vec{R}_{h,N}\left(r'_{i},\grave{\vec{\Lambda}}_{h,N},\vec{\Lambda}'_{h,N},\vec{R}_{T,h}\right) & =\left\{ \sum_{i=1}^{3}\left[\nu_{T,h,i}+r'_{i}\Lambda'_{h,N,i}\right]^{2}\right\} ^{1/2}\sum_{i=1}^{3}\grave{\Lambda}_{h,N,i}\hat{e_{i}}.\label{eq:Points-on-Hemisphere-R-NA}\end{align}
 In the above equation, the subscript $h$ denotes the hemisphere;
and \[
\left\{ \begin{array}{c}
\grave{\Lambda}_{h,N,1}\left(\grave{\theta}_{h,N},\grave{\phi}_{h,N}\right)=\sin\grave{\theta}_{h,N}\cos\grave{\phi}_{h,N},\\
\\\grave{\Lambda}_{h,N,2}\left(\grave{\theta}_{h,N},\grave{\phi}_{h,N}\right)=\sin\grave{\theta}_{h,N}\sin\grave{\phi}_{h,N},\\
\\\grave{\Lambda}_{h,N,3}\left(\grave{\theta}_{h,N}\right)=\cos\grave{\theta}_{h,N}.\;\;\;\;\end{array}\right.\]
 The expressions for $\Lambda'_{h,N,i},$ $i=1,2,3,$ are defined
identically in form. The angular variables in spherical coordinates,
$\grave{\theta}_{h,N}$ and $\grave{\phi}_{h,N},$ can be obtained
from equations (\ref{eq:phi-NA}) and (\ref{eq:theta-NA}), where
the obvious notational changes are understood. The implicit angular
variables, $\theta'_{h,N}$ and $\phi'_{h,N},$ are the sets defined
in Appendix A, equations (\ref{eq:thetaN-prime-0-to-90-degree-final})
and (\ref{eq:thetaN-prime-90-to-180-degree-final}) for $\theta'_{s,N},$
and the sets from equations (\ref{eq:phiN-prime-0-90-and-180-270-final})
and (\ref{eq:phiN-prime-90-180-and-270-360-final}) for $\phi'_{s,N}.$ 

Unlike the sphere situation, the initial wave could eventually escape
the hemisphere resonator after some maximum number of reflections.
It is shown in the Appendix C2 that this maximum number for internal
reflection is given by equation (\ref{eq:N-max-Hemisphere}), \begin{align}
N_{h,max} & =\left[\mathbb{Z}_{h,max}\right]_{G},\label{eq:N-max-Hemisphere-NA}\end{align}
 where the greatest integer function $\left[\mathbb{Z}_{h,max}\right]_{G}$
is defined in equation (\ref{eq:N-max-Hemisphere-Greatest-Integer-Function})
of Appendix C2, \begin{align}
\mathbb{Z}_{h,max} & =\frac{1}{\pi-2\theta_{inc}}\left[\pi-\arccos\left(\frac{1}{2}\left\{ r'_{i}\left\Vert \vec{R'}_{0}\right\Vert ^{-1}+\left[r'_{i}\right]^{-1}\left\Vert \vec{R'}_{0}\right\Vert -\left[r'_{i}\left\Vert \vec{R'}_{0}\right\Vert \right]^{-1}\xi_{1,p}^{2}\right\} \right)\right].\label{eq:N-max-Hemisphere-Greatest-Integer-Function-NA}\end{align}
 Here $\xi_{1,p}$ is given in equation (\ref{eq:positive-root-k-NA})
and $\theta_{inc}$ is from equation (\ref{eq:angle-of-incidence-exp}). 

\begin{figure}
\begin{center}\includegraphics[%
  scale=0.7]{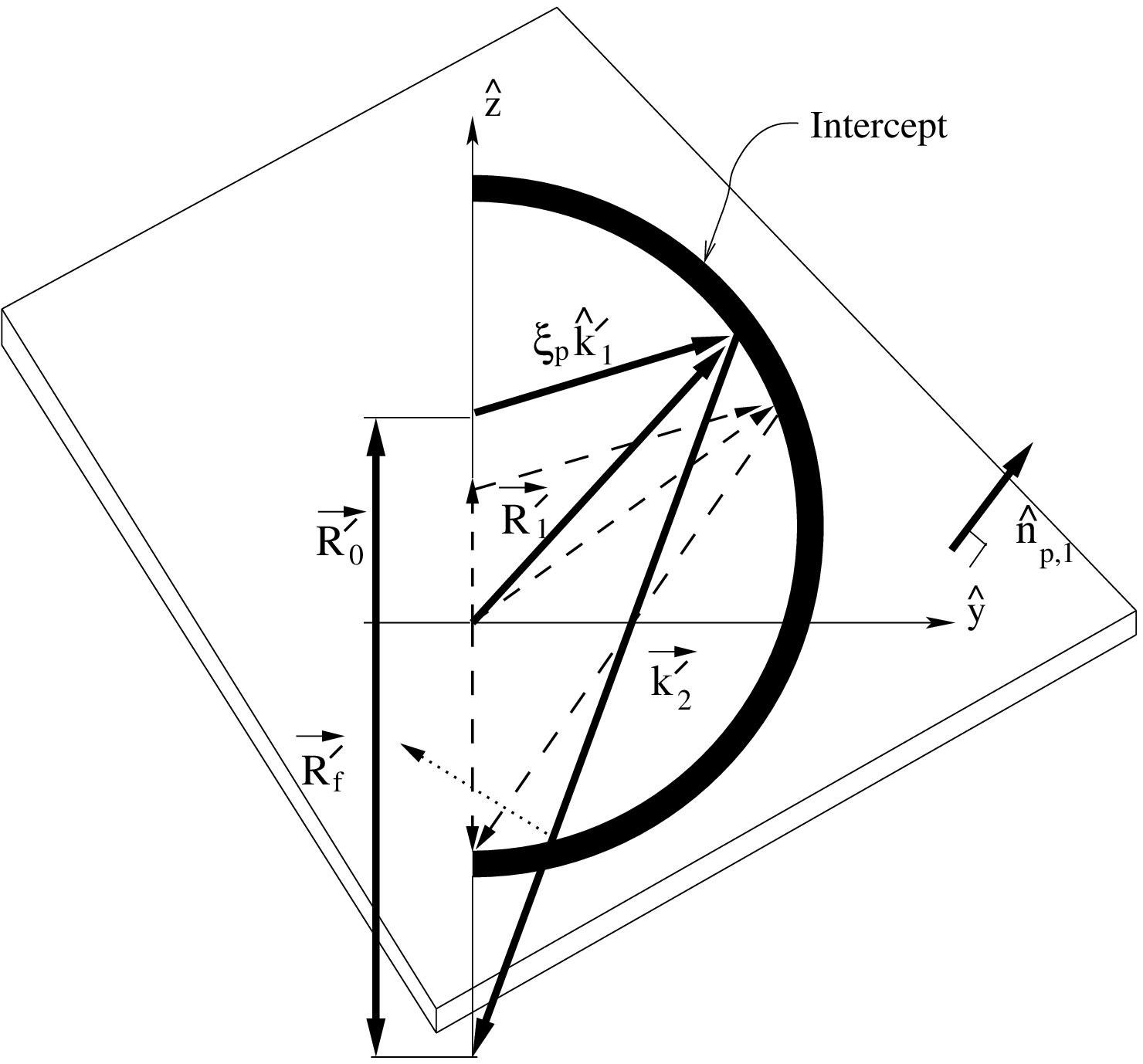}\end{center}

\caption{The dashed line vectors represent the situation where only single
internal reflection occurs. The dark line vectors represent the situation
where multiple internal reflections occur. \label{cap:plane-sphere-intersection-3-Critical-Angle}}
\end{figure}

The above results of $\triangle\vec{k'}_{inner}\left(;\vec{R'}_{h,1},\vec{R'}_{h,0}\right)$
and $\triangle\vec{k'}_{outer}\left(;\vec{R'}_{h,1}+a\hat{R'}_{h,1}\right)$
have been derived based on the fact that there are multiple internal
reflections. For a sphere, the multiple internal reflections are inherent.
However, for a hemisphere, it is not necessarily true that all incoming
waves would result in multiple internal reflections. Naturally, the
criteria for multiple internal reflections are in order. If the initial
direction of the incoming wave vector, $\hat{k'}_{1},$ is given,
the internal reflections can be either single or multiple depending
upon the location of the entry point in the cavity, $\vec{R'}_{0}.$
As shown in Figure \ref{cap:plane-sphere-intersection-3-Critical-Angle},
these are two reflection dynamics where the dashed vectors represent
the single reflection case and the non-dashed vectors represent multiple
reflections case. Because the whole process occurs in the same plane
of incidence, the vector $\vec{R'}_{f}=-\lambda_{0}\vec{R'}_{0}$
where $\lambda_{0}>0.$ The multiple or single internal reflection
criteria can be summarized by the relation found in equation (\ref{eq:Hem-Hemi-Multi-Single-Ref-condi})
of Appendix C2: \begin{align}
\left\Vert \vec{R'}_{f}\right\Vert  & =\frac{1}{2}\left\Vert \vec{R'}_{0}\right\Vert \left[\sum_{n=1}^{3}k'_{1,n}\right]\left\{ \sum_{j=1}^{3}\sum_{l=1}^{3}\left[\left\Vert \vec{k'}_{1}\right\Vert ^{2}-k'_{1,l}\sum_{m=1}^{3}k'_{1,m}\right]r'_{0,l}r'_{0,j}\right\} ^{-1}\sum_{l=1}^{3}\left\{ k'_{1,l}\left[r'_{i}\right]^{2}-\left[r'_{0,l}\right]^{2}\right.\nonumber \\
 & \left.+2\vec{R'}_{0}\cdot\vec{k'}_{1}r'_{0,l}-\left\Vert \vec{R'}_{0}\right\Vert ^{2}k'_{1,l}-2r'_{0,l}\left[\sum_{l=1}^{3}k'_{1,l}\right]^{-1}\sum_{i=1}^{3}\left[\left\Vert \vec{k'}_{1}\right\Vert ^{2}-k'_{1,i}\sum_{m=1}^{3}k'_{1,m}\right]r'_{0,i}\right\} .\label{eq:Hem-Hemi-Multi-Single-Ref-condi-NA}\end{align}
 Finally, because the hemisphere opening has a radius $r'_{i},$ the
following criteria are concluded: \begin{equation}
\left\{ \begin{array}{ccc}
\left\Vert \vec{R'}_{f}\right\Vert <r'_{i}, &  & single\; internal\; reflection,\\
\\\left\Vert \vec{R'}_{f}\right\Vert \geq r'_{i}, &  & multiple\; internal\; reflections,\end{array}\right.\label{eq:Hem-Hemi-Multi-Single-Ref-condi-Interpretation-NA}\end{equation}
 where $\left\Vert \vec{R'}_{f}\right\Vert $ is defined in equation
(\ref{eq:Hem-Hemi-Multi-Single-Ref-condi-NA}). The details of this
section can be found in Appendix C2.

\subsubsection{Plate-Hemisphere}

\begin{figure}
\begin{center}\includegraphics[%
  scale=0.7]{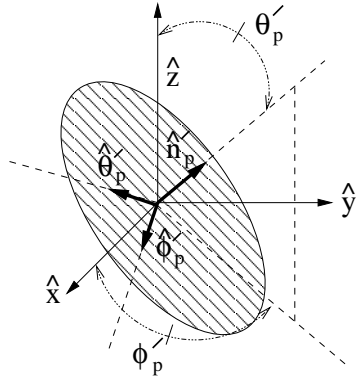}\end{center}

\caption{The orientation of a disk is given through the surface unit normal
$\hat{n'}_{p}.$ The disk is spanned by the two unit vectors $\hat{\theta'}_{p}$
and $\hat{\phi'}_{p}.$ \label{cap:circular-plane}}
\end{figure}

A surface is represented by a unit vector $\hat{n'}_{p},$ which is
normal to the surface locally. For the circular plate shown in Figure
\ref{cap:circular-plane}, its orthonormal triad $\left(\hat{n'}_{p},\hat{\theta'}_{p},\hat{\phi'}_{p}\right)$
has the form \begin{eqnarray*}
\hat{n'}_{p}=\sum_{i=1}^{3}\Lambda'_{p,i}\hat{e}_{i}, &  & \hat{\theta'}_{p}=\sum_{i=1}^{3}\frac{\partial\Lambda'_{p,i}}{\partial\theta'_{p}}\hat{e}_{i},\end{eqnarray*}
 \begin{align*}
\hat{\phi'}_{p} & =\sum_{i=1}^{3}\frac{1}{\sin\theta'_{p}}\frac{\partial\Lambda'_{p,i}}{\partial\phi'_{p}}\hat{e}_{i},\end{align*}
 where $\Lambda'_{p,1}\left(\theta'_{p},\phi'_{p}\right)=\sin\theta'_{p}\cos\phi'_{p},$
$\Lambda'_{p,2}\left(\theta'_{p},\phi'_{p}\right)=\sin\theta'_{p}\sin\phi'_{p}$
and $\Lambda'_{p,3}\left(\theta'_{p}\right)=\cos\theta'_{p}.$ 

\begin{figure}
\begin{center}\includegraphics[%
  scale=0.7]{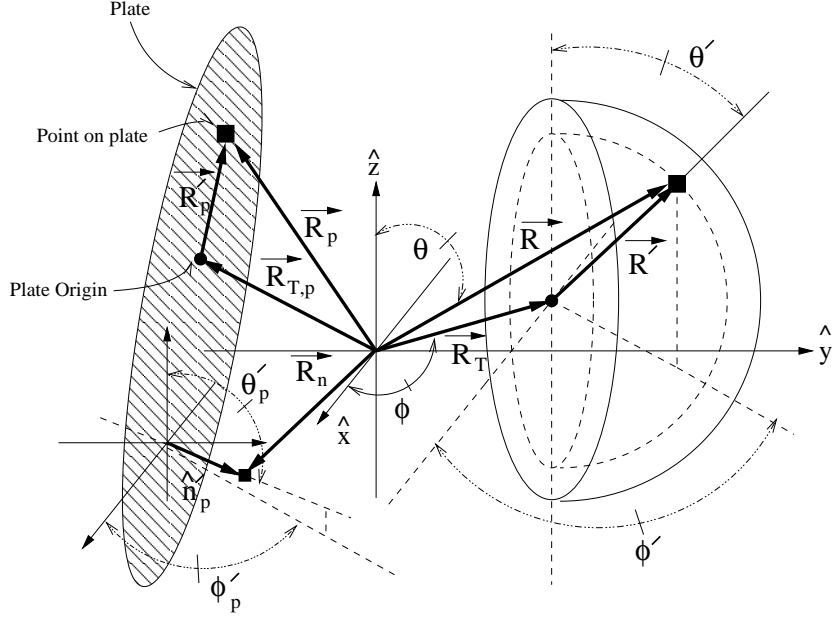}\end{center}

\caption{The plate-hemisphere configuration. \label{cap:plate-and-hemisphere-complex}}
\end{figure}

For the plate-hemisphere configuration shown in Figure \ref{cap:plate-and-hemisphere-complex},
it can be shown that the element $\vec{R}_{p}$ on the plane and its
velocity $d\vec{R}_{p}/dt$ are given by (see equation (\ref{eq:velocity-points-on-circular-plane-NA})
and (\ref{eq:Points-on-Plate-R}) in Appendix C3: \begin{align}
\vec{R}_{p}\left(\grave{\vec{\Lambda}}_{p},\vec{\Lambda}'_{p},\vec{R}_{T,p}\right) & =\left\{ \sum_{i=1}^{3}\left[\nu_{T,p,i}+\nu'_{p,\theta'_{p}}\frac{\partial\Lambda'_{p,i}}{\partial\theta'_{p}}+\frac{\nu'_{p,\phi'_{p}}}{\sin\theta'_{p}}\frac{\partial\Lambda'_{p,i}}{\partial\phi'_{p}}\right]^{2}\right\} ^{1/2}\sum_{i=1}^{3}\grave{\Lambda}_{p,i}\hat{e_{i}},\label{eq:Points-on-Plate-R-NA}\end{align}
\begin{align}
\dot{\vec{R}}_{p}\equiv\frac{d\vec{R}_{p}}{dt} & =\left\{ \sum_{i=1}^{3}\left[\nu_{T,p,i}+\nu'_{p,\theta'_{p}}\frac{\partial\Lambda'_{p,i}}{\partial\theta'_{p}}+\frac{\nu'_{p,\phi'_{p}}}{\sin\theta'_{p}}\frac{\partial\Lambda'_{p,i}}{\partial\phi'_{p}}\right]^{2}\right\} ^{-1/2}\sum_{j=1}^{3}\sum_{k=1}^{3}\left(\left[\nu_{T,p,k}+\nu'_{p,\theta'_{p}}\frac{\partial\Lambda'_{p,k}}{\partial\theta'_{p}}\right.\right.\nonumber \\
 & \left.+\frac{\nu'_{p,\phi'_{p}}}{\sin\theta'_{p}}\frac{\partial\Lambda'_{p,k}}{\partial\phi'_{p}}\right]\left[\dot{\nu}_{T,p,k}+\left\{ \nu'_{p,\theta'_{p}}\frac{\partial^{2}\Lambda'_{p,k}}{\partial\left[\theta'_{p}\right]^{2}}+\frac{\nu'_{p,\phi'_{p}}}{\sin\theta'_{p}}\left(\frac{\partial^{2}\Lambda'_{p,k}}{\partial\theta'_{p}\partial\phi'_{p}}-\cot\theta'_{p}\frac{\partial\Lambda'_{p,k}}{\partial\phi'_{p}}\right)\right\} \dot{\theta'}_{p}\right.\nonumber \\
 & \left.+\left\{ \nu'_{p,\theta'_{p}}\frac{\partial^{2}\Lambda'_{p,k}}{\partial\phi'_{p}\partial\theta'_{p}}+\frac{\nu'_{p,\phi'_{p}}}{\sin\theta'_{p}}\frac{\partial^{2}\Lambda'_{p,k}}{\partial\left[\phi'_{p}\right]^{2}}\right\} \dot{\phi'}_{p}+\dot{\nu'}_{p,\theta'_{p}}\frac{\partial\Lambda'_{p,k}}{\partial\theta'_{p}}+\frac{\dot{\nu'}_{p,\phi'_{p}}}{\sin\theta'_{p}}\frac{\partial\Lambda'_{p,k}}{\partial\phi'_{p}}\right]\grave{\Lambda}_{p,j}\nonumber \\
 & \left.+\sum_{i=1}^{3}\left[\nu_{T,p,i}+\nu'_{p,\theta'_{p}}\frac{\partial\Lambda'_{p,i}}{\partial\theta'_{p}}+\frac{\nu'_{p,\phi'_{p}}}{\sin\theta'_{p}}\frac{\partial\Lambda'_{p,i}}{\partial\phi'_{p}}\right]^{2}\left[\frac{\partial\grave{\Lambda}_{p,j}}{\partial\grave{\theta}_{p}}\frac{\partial\grave{\theta}_{p}}{\partial\phi'_{p}}\dot{\theta}'_{p}+\frac{\partial\grave{\Lambda}_{p,j}}{\partial\grave{\phi}_{p}}\frac{\partial\grave{\phi}_{p}}{\partial\phi'_{p}}\dot{\phi}'_{p}\right]\right)\hat{e_{j}},\label{eq:velocity-points-on-circular-plane-NA}\end{align}
 where $\left(\grave{\Lambda}_{p,1},\grave{\Lambda}_{p,2},\grave{\Lambda}_{p,3}\right)$
is defined in equation (\ref{eq:Capital-Lambda-Plate-Def}) and the
angles $\grave{\phi}_{p}$ and $\grave{\theta}_{p}$ are defined in
equations (\ref{eq:phi-Plate}) and (\ref{eq:theta-Plate}) of Appendix
C3. The subscript $p$ of $\grave{\phi}_{p}$ and $\grave{\theta}_{p}$
indicates that these are spherical variables for the points on the
plate of Figure \ref{cap:plate-and-hemisphere-complex}, not that
of the hemisphere. It is also understood that $\Lambda'_{p,3}$ and
$\grave{\Lambda}_{p,3}$ are independent of $\phi'_{p}$ and $\grave{\phi}_{p},$
respectively. Therefore, their differentiation with respect to $\phi'_{p}$
and $\grave{\phi}_{p}$ respectively vanishes. The quantities $\dot{\theta'}_{p}$
and $\dot{\phi'}_{p}$ are the angular frequencies, and $\dot{\nu}_{T,p,i}$
is the translation speed of the plate relative to the system origin.
The quantities $\dot{\nu'}_{p,\theta'_{p}}$ and $\dot{\nu'}_{p,\phi'_{p}}$
are the lattice vibrations along the directions $\hat{\theta'}_{p}$
and $\hat{\phi'}_{p}$ respectively. For the static plate without
lattice vibrations, $\dot{\nu'}_{p,\theta'_{p}}$ and $\dot{\nu'}_{p,\phi'_{p}}$
vanishes. 

\begin{figure}[t]
\begin{center}\includegraphics[%
  scale=0.7]{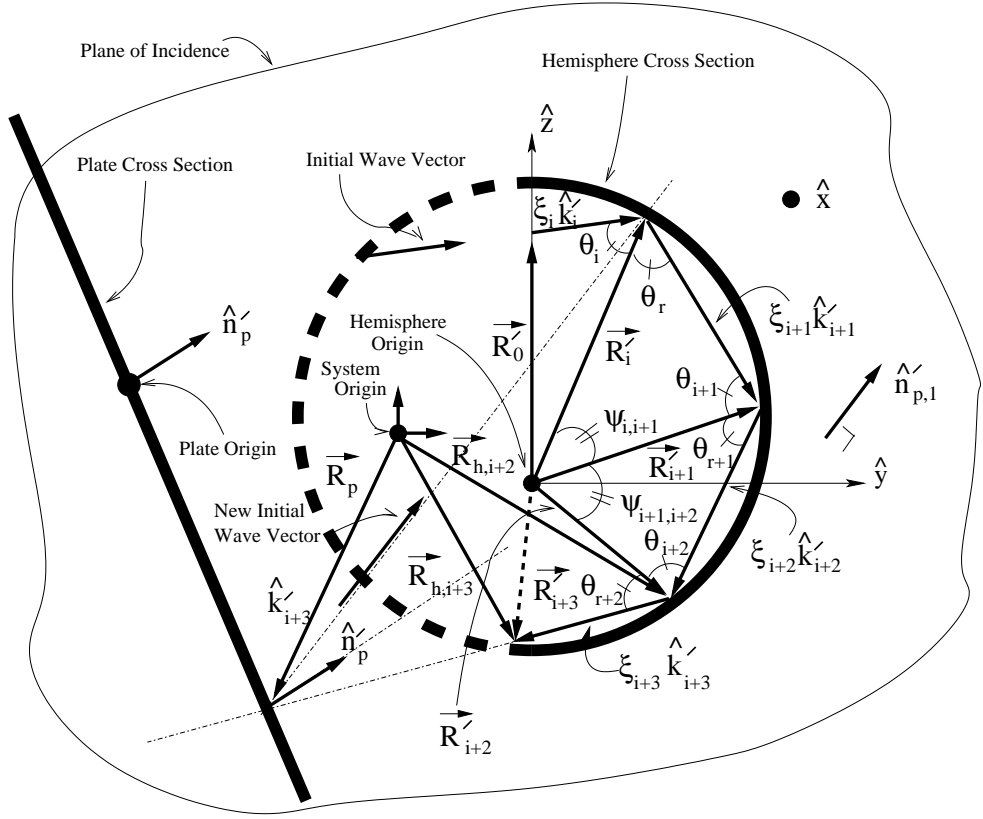}\end{center}

\caption{The intersection between oscillating plate, hemisphere and the plane
of incidence whose normal is $\hat{n'}_{p,1}=-\left\Vert \vec{n'}_{p,1}\right\Vert ^{-1}\sum_{i=1}^{3}\epsilon_{ijk}k'_{1,j}r'_{0,k}\hat{e_{i}}.$
\label{cap:plate-hemisphere-plane-of-incidence-intersect-Complex}}
\end{figure}

In the cross-sectional view of the plate-hemisphere system shown in
Figure \ref{cap:plate-hemisphere-plane-of-incidence-intersect-Complex},
the initial wave vector $\vec{k'}_{i}$ traveling toward the hemisphere
would go through a series of reflections according to the law of reflection
and finally exit the cavity. It would then continue toward the plate,
and depending on the orientation of plate at the time of impact, the
wave-vector, now reflecting off the plate, would either escape to
infinity or re-enter the hemisphere. The process repeats successively.
In order to determine whether the wave that just escaped from the
hemisphere cavity can reflect back from the plate and re-enter the
hemisphere or escape to infinity, the exact location of reflection
on the plate must be known. This reflection point on the plate is
found to be, from equation (\ref{eq:Points-on-Plate-R-Final}) of
Appendix C3, \begin{align}
\vec{R}_{p} & =\left\{ \sum_{s=1}^{3}\left[\frac{\partial\Lambda'_{p,s}}{\partial\phi'_{p}}-\frac{\sum_{i=1}^{3}\frac{\partial\Lambda'_{p,i}}{\partial\phi'_{p}}\left[\Lambda'_{p,i}+\left\Vert \vec{n'}_{p,1}\right\Vert ^{-1}\epsilon_{ijk}k'_{1,j}r'_{0,k}\right]}{\sum_{l=1}^{3}\frac{\partial\Lambda'_{p,l}}{\partial\theta'_{p}}\left[\Lambda'_{p,l}+\left\Vert \vec{n'}_{p,1}\right\Vert ^{-1}\epsilon_{lmn}k'_{1,m}r'_{0,n}\right]}\frac{\partial\Lambda'_{p,s}}{\partial\theta'_{p}}\right]^{2}\right\} ^{1/2}\nonumber \\
 & \times\left[C_{\beta}^{-1}C_{\gamma}^{-1}A_{\gamma}A_{\beta}+\gamma_{o}C_{\beta}^{-1}C_{\gamma}^{-1}B_{\gamma}A_{\beta}+C_{\beta}^{-1}B_{\zeta}B_{\beta}\right]\sum_{i=1}^{3}\grave{\Lambda}_{p,i}\hat{e_{i}},\label{eq:Points-on-Plate-R-Final-NA}\end{align}
 where the translation parameter $\nu_{T,p,j}=0$ and the terms $\left(A_{\zeta},B_{\zeta},C_{\zeta}\right),$
$\left(A_{\gamma},B_{\gamma},C_{\gamma}\right),$ $\left(A_{\beta},B_{\beta},C_{\beta}\right)$
and $\gamma_{o}$ are defined in equations (\ref{eq:Plate-Hemi-exiting-K-scale-ABC-in-Zetai-Def}),
(\ref{eq:Plate-Hemi-exiting-K-scale-ABC-in-GAMMA-Def}), (\ref{eq:Plate-Hemi-exiting-K-scale-ABC-in-BETA-Def})
and (\ref{eq:Plate-Hemi-exiting-K-scale-GAMMA}) of Appendix C3. It
is to be noticed that for a situation where the translation parameter
$\nu_{T,p,j}=0,$ $\grave{\Lambda}$ becomes identical to $\Lambda'$
in form. Results for $\grave{\Lambda}$ can be obtained from $\Lambda'$
by a simple replacement of primed variables with the unprimed ones. 

Leaving the details to the relevant Appendix, the criterion whether
the wave reflecting off the plate at location $\vec{R}_{p}$ can re-enter
the hemisphere cavity or simply escape to infinity is found from the
result shown in equation (\ref{eq:Plate-Hemi-K-Reenter-Criteria})
of Appendix C3, \begin{align}
\xi_{\kappa,i} & =\left(\nu_{T,h,i}+r'_{0,i}-\left\{ \sum_{s=1}^{3}\left[\frac{\partial\Lambda'_{p,s}}{\partial\phi'_{p}}-\frac{\sum_{i=1}^{3}\frac{\partial\Lambda'_{p,i}}{\partial\phi'_{p}}\left[\Lambda'_{p,i}+\left\Vert \vec{n'}_{p,1}\right\Vert ^{-1}\epsilon_{ijk}k'_{1,j}r'_{0,k}\right]}{\sum_{l=1}^{3}\frac{\partial\Lambda'_{p,l}}{\partial\theta'_{p}}\left[\Lambda'_{p,l}+\left\Vert \vec{n'}_{p,1}\right\Vert ^{-1}\epsilon_{lmn}k'_{1,m}r'_{0,n}\right]}\frac{\partial\Lambda'_{p,s}}{\partial\theta'_{p}}\right]^{2}\right\} ^{1/2}\right.\nonumber \\
 & \left.\times\left[C_{\beta}^{-1}C_{\gamma}^{-1}A_{\gamma}A_{\beta}+\gamma_{o}C_{\beta}^{-1}C_{\gamma}^{-1}B_{\gamma}A_{\beta}+C_{\beta}^{-1}B_{\zeta}B_{\beta}\right]\grave{\Lambda}_{p,i}\right)\left(\sum_{k=1}^{3}\left\{ \alpha_{r,\perp}\left[k_{N_{h,max}+1,i}n'_{p,k}n'_{p,k}\right.\right.\right.\nonumber \\
 & \left.\left.\left.-n'_{p,i}k_{N_{h,max}+1,k}n'_{p,k}\right]-\alpha_{r,\parallel}n'_{p,k}k_{N_{h,max}+1,k}n'_{p,i}\right\} \right)^{-1},\label{eq:Plate-Hemi-K-Reenter-Criteria-NA}\end{align}
 where $i=1,2,3$ and $\xi_{\kappa,i}$ is the component of the scale
vector $\vec{\xi}_{\kappa}=\sum_{i=1}^{3}\xi_{\kappa,i}\hat{e_{i}}$
explained in the Appendix C3. 

In the above re-entry criteria, it should be noticed that $\vec{R}_{0}\leq r'_{i}.$
This implies $r'_{0,i}\leq r'_{i},$ where $r'_{i}$ is the radius
of hemisphere. It is then concluded that all waves re-entering the
hemisphere cavity would satisfy the condition $\xi_{\kappa,1}=\xi_{\kappa,2}=\xi_{\kappa,3}.$
On the other hand, those waves that escapes to infinity cannot have
all three $\xi_{\kappa,i}$ equal to a single constant. The re-entry
condition $\xi_{\kappa,1}=\xi_{\kappa,2}=\xi_{\kappa,3}$ is just
another way of stating the existence of a parametric line along the
vector $\vec{k}_{r,N_{h,max}+1}$ that happens to pierce through the
hemisphere opening. In case such a line does not exist, the initial
wave direction has to be rotated into a new direction such that there
is a parametric line that pierces through the hemisphere opening.
That is why all three $\xi_{\kappa,i}$ values cannot be equal to
a single constant. The re-entry criteria are summarized here for bookkeeping
purpose: \begin{equation}
\left\{ \begin{array}{c}
\xi_{\kappa,1}=\xi_{\kappa,2}=\xi_{\kappa,3}:\quad wave\; reenters\; hemisphere,\\
\\ELSE:\quad wave\; escapes\; to\; infinity,\end{array}\right.\label{eq:Plate-Hemi-K-Reenter-Criteria-Interpretation-NA}\end{equation}
 where $ELSE$ is the case where $\xi_{\kappa,1}=\xi_{\kappa,2}=\xi_{\kappa,3}$
cannot be satisfied. The details of this section can be found in Appendix
C3.

\subsection{Dynamical Casimir Force}

\begin{figure}
\begin{center}\includegraphics[%
  scale=0.7]{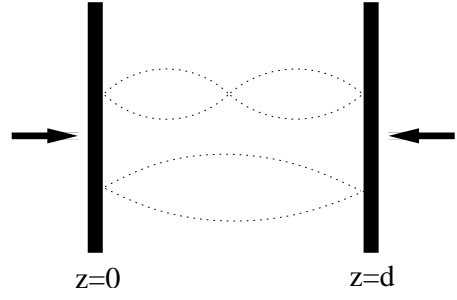}\end{center}

\caption{Because there are more vacuum-field modes in the external regions,
the two charge-neutral conducting plates are accelerated inward till
the two finally stick.\label{cap:casimir-plates-outlook}}
\end{figure}

The phenomenon of Casimir effect is inherently a dynamical effect
due to the fact that it involves radiation, rather than static fields.
One of our original objectives in studying the Casimir effect was
to investigate the physical implications of vacuum-fields on movable
boundaries. Consider the two parallel plates configuration of charge-neutral,
perfect conductors shown in Figure \ref{cap:casimir-plates-outlook}.
Because there are more wave modes in the outer region of the parallel
plate resonator, two loosely restrained (or unfixed in position) plates
will accelerate inward until they finally meet. The energy conservation
would require that the energy initially confined in the resonator
when the two plates were separated be transformed into the heat energy
that acts to raise the temperatures of the two plates. 

Davies in 1975 \cite{key-Davies}, followed by Unruh in 1976 \cite{key-Unruh},
have asked the similar question and came to a conclusion that when
an observer is moving with a constant acceleration in vacuum, the
observer perceives himself to be immersed in a thermal bath at the
temperature $T=\hbar\ddot{R}/\left[2\pi ck'\right],$ where $\ddot{R}$
is the acceleration of the observer and $k',$ the wave number. The
details of the Unruh-Davies effect can also be found in the reference
\cite{key-Milonni}. The other work that dealt with the concept of
dynamical Casimir effect is due to Schwinger in his proposals \cite{key-Milton,key-Schwinger-Casimir-Light}
to explain the phenomenon of sonoluminescense. Sonoluminescense is
a phenomenon in which when a small air bubble filled with noble gas
is under a strong acoustic-field pressure, the bubble will emit an
intense flash of light in the optical range. 

Although the name {}``dynamical Casimir effect'' have been introduced
by Schwinger, the motivation and derivation behind the dynamical Casimir
force in this investigation did not stem from that of Schwinger's
work. Therefore, the dynamical Casimir force here should not have
any resemblance to Schwinger's work to the best of our knowledge.
We have only found out of Schwinger's proposals on sonoluminescense
after the work on dynamical Casimir force have already begun. The
terminology {}``dynamical Casimir force'' seemed to be appealing
enough for its usage at the beginning of this work. After discovering
Schwinger's work on sonoluminescense, we have found that Schwinger
had already introduced the terminology {}``dynamical Casimir effect''
in his papers. Our development to the dynamical Casimir force formalism
is briefly presented in the following sections. The details of the
derivations pertaining to the dynamical Casimir force can be found
in Appendix D.

\subsubsection{Formalism of Zero-Point Energy and its Force}

For massless fields, the energy-momentum relation is $\mathcal{H}'_{n_{s}}\equiv E_{Total}=pc,$
where $p$ is the momentum, $c$ the speed of light, and $\mathcal{H}'_{n_{s}}$
is the quantized field energy for the harmonic fields of equation
(\ref{eq:stationary-state-energy-bounded}) for the bounded space,
or equation (\ref{eq:stationary-state-energy-unbounded}) for the
free space. For the bounded space, the quantized field energy $\mathcal{H}'_{n_{s}}\equiv\mathcal{H}'_{n_{s},b}$
of equation (\ref{eq:stationary-state-energy-bounded}) is a function
of the wave number $k'_{i}\left(n_{i}\right),$ which in turn is a
function of the wave mode value $n_{i}$ and the boundary functional
$f_{i}\left(L_{i}\right),$ where $L_{i}$ is the gap distance in
the direction of $\vec{L}_{i}=\left[\vec{R'}_{2}\cdot\hat{e}_{i}-\vec{R'}_{1}\cdot\hat{e}_{i}\right]\hat{e}_{i}.$
Here $\vec{R'}_{1}$ and $\vec{R'}_{2}$ are the position vectors
for the involved boundaries. As an illustration with the two plate
configuration shown in Figure \ref{cap:casimir-plates-outlook}, $\vec{R'}_{1}$
may represent the plate positioned at $z=0$ and $\vec{R'}_{2}$ may
correspond to the plate at the position $z=d.$ When the position
of these boundaries are changing in time, the quantized field energy
$\mathcal{H}'_{n_{s}}\equiv\mathcal{H}'_{n_{s},b}$ will be modified
accordingly because the wave number functional $k'_{i}\left(n_{i}\right)$
is varying in time, \begin{align*}
\frac{dk'_{i}}{dt} & =\frac{\partial k'_{i}}{\partial n_{i}}\frac{dn_{i}}{dt}f_{i}\left(L_{i}\right)+n_{i}\frac{\partial f_{i}}{\partial L_{i}}\frac{dL_{i}}{dt}=f_{i}\left(L_{i}\right)\frac{\partial k'_{i}}{\partial n_{i}}\dot{n}_{i}+n_{i}\frac{\partial f_{i}}{\partial L_{i}}\dot{L}_{i}.\end{align*}
 Here the term proportional to $\dot{n}_{i}$ refers to the case where
the boundaries remain fixed throughout all times but the number of
wave modes in the resonator are being driven by some active external
influence. The term proportional to $\dot{L}_{i}$ represents the
changes in the number of wave modes due to the moving boundaries. 

For an isolated system, there are no external influences, hence $\dot{n}_{i}=0.$
Then, the dynamical force arising from the fact that the time variation
of the boundaries is given by equation (\ref{eq:dynamical-force-L-dot-ONLY-3D})
of Appendix D1, \begin{align}
\vec{\mathcal{F}'}_{\alpha} & =\sum_{i=1}^{3}\left\{ n_{i}\frac{\partial f_{i}}{\partial L_{i}}\left[C_{\alpha,5}\frac{\partial^{2}\mathcal{H}'_{n_{s}}}{\partial\left[k'_{i}\right]^{2}}+\left(1-\delta_{i\alpha}\right)\left(C_{\alpha,6}-C_{\alpha,7}\left[n_{s}+\frac{1}{2}\right]k'_{i}\right)\left[n_{s}+\frac{1}{2}\right]\right]\dot{L}_{i}\right.\nonumber \\
 & \left.+\sum_{j=1}^{3}\left(1-\delta_{ij}\right)C_{\alpha,5}n_{j}\frac{\partial f_{j}}{\partial L_{j}}\frac{\partial^{2}\mathcal{H}'_{n_{s}}}{\partial k'_{j}\partial k'_{i}}\dot{L}_{j}\right\} \hat{e_{\alpha}},\label{eq:dynamical-force-L-dot-ONLY-3D-NA}\end{align}
 where $C_{\alpha,1},$ $C_{\alpha,2},$ $C_{\alpha,3},$ $C_{\alpha,4},$
$C_{\alpha,5},$ $C_{\alpha,6}$ and $C_{\alpha,7}$ are defined in
equations (\ref{eq:dynamical-force-C1-C2-C3-DEF}), (\ref{eq:dynamical-force-C4-DEF}),
(\ref{eq:dynamical-force-C5-DEF}), (\ref{eq:dynamical-force-C6-DEF})
and (\ref{eq:dynamical-force-C7-DEF}) of Appendix D1. 

The force shown in the above expression vanishes for the one dimensional
case. This is an expected result. To understand why the force vanishes,
we have to refer to the starting point equation (\ref{eq:d-Hamilton-minus-d-momentum-mag-combined})
in the Appendix D1. The summation there obviously runs only once to
arrive at the expression, $\partial\mathcal{H}'_{n_{s}}/\partial k'_{i}=\left[n_{s}+\frac{1}{2}\right]\hbar c.$
This is a classic situation where the problem has been over specified.
For the \textbf{3D} case, equation (\ref{eq:d-Hamilton-minus-d-momentum-mag-combined})
is a combination of two constraints, $\sum_{i=1}^{3}\left[p'_{i}\right]^{2}$
and $\mathcal{H}'_{n_{s}}.$ For the one dimensional case, there is
only one constraint, $\mathcal{H}'_{n_{s}}.$ Therefore, equation
(\ref{eq:d-Hamilton-minus-d-momentum-mag-combined}) becomes an over
specification. In order to avoid the problem caused by over specifications
in this formulation, the one dimensional force expression can be obtained
directly by differentiating equation (\ref{eq:Hamilton-equal-pc})
instead of using the above formulation for the three dimensional case.
The \textbf{1D} dynamical force expression for an isolated, non-driven
systems then becomes (see equation (\ref{eq:dynamical-force-L-dot-ONLY-1D})
of Appendix D1) \begin{align}
\vec{\mathcal{F}'} & =\frac{n}{c}\frac{\partial f}{\partial L}\frac{\partial\mathcal{H}'_{n_{s}}}{\partial k'}\dot{L}\hat{e},\label{eq:dynamical-force-L-dot-ONLY-1D-NA}\end{align}
 where $\vec{\mathcal{F}'}$ is an one dimensional force. Here the
subscript $\alpha$ of $\vec{\mathcal{F}'}_{\alpha}$ have been dropped
for simplicity, since it is a one dimensional force. The details of
this section can be found in Appendix D1.

\subsubsection{Equations of Motion for the Driven Parallel Plates}

\begin{figure}
\begin{center}\includegraphics[%
  scale=0.7]{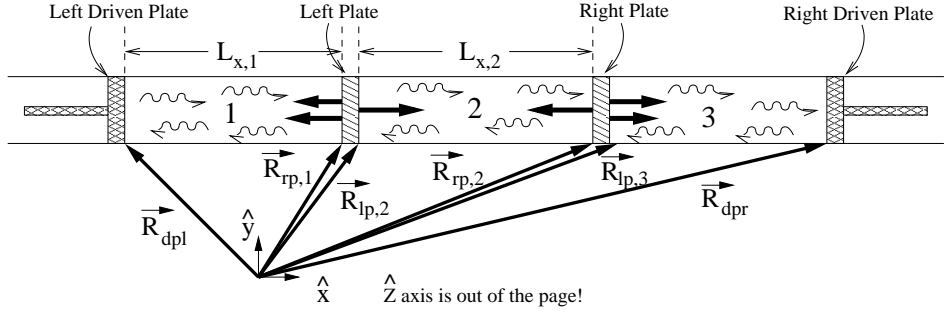}\end{center}

\caption{A one dimensional driven parallel plates configuration. \label{cap:driven-parallel-plates-force}}
\end{figure}

The Unruh-Davies effect states that heating up of an accelerating
conductor plate is proportional to its acceleration through the relation
$T=\hbar\ddot{R}/\left[2\pi ck'\right],$ where $\ddot{R}$ is the
plate acceleration. A one dimensional system of two parallel plates,
shown in Figure \ref{cap:driven-parallel-plates-force}, can be used
as a simple model to demonstrate the complicated sonoluminescense
phenomenon for a bubble subject to a strong acoustic field. 

The dynamical force for the \textbf{1D}, linear coupled system can
be expressed with equation (\ref{eq:dynamical-force-L-dot-ONLY-1D-NA}),
\begin{eqnarray}
\ddot{R}_{1}-\eta_{1}\dot{R}_{1}-\eta_{2}\dot{R}_{2}=\xi_{rp}, &  & \ddot{R}_{2}-\eta_{3}\dot{R}_{2}-\eta_{4}\dot{R}_{1}=\xi_{lp},\label{eq:Eq-of-Mot-Plate-System-Static-NA}\end{eqnarray}
 where the quantities $\eta_{1},$ $\eta_{2},$ $\eta_{3},$ $\eta_{4},$
$\xi_{rp},$ $\xi_{lp},$ $R_{1},$ $R_{2}$ are defined in equation
(\ref{eq:eta-1-2-3-4-ci-lp-rp-DEF}) of Appendix D2. Here $R_{1}$
represents the center of mass position for the {}``Right Plate''
and $R_{2}$ represents the center of mass position for the {}``Left
Plate'' as illustrated in Figure \ref{cap:driven-parallel-plates-force}.
With a slight modification, equation (\ref{eq:Eq-of-Mot-Plate-System-Static-NA})
for this linear coupled system can be written in the matrix form,
(see equations (\ref{eq:Eq-of-Mot-Plate-Static-Coeffi-new-variables}),
(\ref{eq:Eq-of-Mot-Plate-System-Static-Matrix-R1-R2}) and (\ref{eq:Eq-of-Mot-Plate-System-Static-Matrix})
of Appendix D2): \begin{eqnarray*}
R_{1}=\int_{t_{0}}^{t}R_{3}dt', &  & R_{2}=\int_{t_{0}}^{t}R_{4}dt',\end{eqnarray*}
 and \begin{align}
\underbrace{\left[\begin{array}{c}
\dot{R}_{3}\\
\dot{R}_{4}\end{array}\right]}_{\dot{\vec{R}}_{\eta}} & =\underbrace{\left[\begin{array}{cc}
\eta_{1} & \eta_{2}\\
\eta_{4} & \eta_{3}\end{array}\right]}_{\widetilde{M}_{\eta}}\underbrace{\left[\begin{array}{c}
R_{3}\\
R_{4}\end{array}\right]}_{\vec{R}_{\eta}}+\underbrace{\left[\begin{array}{c}
\xi_{rp}\\
\xi_{lp}\end{array}\right]}_{\vec{\xi}},\label{eq:Eq-of-Mot-Plate-System-Static-Matrix-NA}\end{align}
 where \[
\left\{ \begin{array}{c}
\begin{array}{ccc}
\dot{R}_{1}=R_{3}, &  & \dot{R}_{2}=R_{4},\end{array}\\
\\\dot{R}_{3}=\ddot{R}_{1}=\xi_{rp}+\eta_{1}\dot{R}_{1}+\eta_{2}\dot{R}_{2}=\xi_{rp}+\eta_{1}R_{3}+\eta_{2}R_{4},\\
\\\dot{R}_{4}=\ddot{R}_{2}=\xi_{lp}+\eta_{3}\dot{R}_{2}+\eta_{4}\dot{R}_{1}=\xi_{lp}+\eta_{3}R_{4}+\eta_{4}R_{3}.\end{array}\right.\]

The matrix equation has the solutions given by equations (\ref{eq:Eq-of-Mot-R1-Speed})
and (\ref{eq:Eq-of-Mot-R2-Speed}) of Appendix D2: \begin{align}
\dot{R}_{rp,cm,\alpha}\left(t\right) & =\left[\frac{\lambda_{4}\left(;t_{0}\right)-\eta_{1}\left(;t_{0}\right)}{\lambda_{3}\left(;t_{0}\right)-\eta_{1}\left(;t_{0}\right)}-1\right]^{-1}\frac{\psi_{11}\left(t,t_{0}\right)\dot{R}_{rp,cm,\alpha}\left(t_{0}\right)+\psi_{12}\left(t,t_{0}\right)\dot{R}_{lp,cm,\alpha}\left(t_{0}\right)}{\exp\left(\left[\lambda_{3}\left(;t_{0}\right)+\lambda_{4}\left(;t_{0}\right)\right]t_{0}\right)}\nonumber \\
 & +\psi_{11}\left(t,t_{0}\right)\int_{t_{0}}^{t}\frac{\psi_{22}\left(t',t_{0}\right)\xi_{rp}\left(t'\right)-\psi_{12}\left(t',t_{0}\right)\xi_{lp}\left(t'\right)}{\psi_{11}\left(t',t_{0}\right)\psi_{22}\left(t',t_{0}\right)-\psi_{12}\left(t',t_{0}\right)\psi_{21}\left(t',t_{0}\right)}dt'+\psi_{12}\left(t,t_{0}\right)\nonumber \\
 & \times\int_{t_{0}}^{t}\frac{\psi_{11}\left(t',t_{0}\right)\xi_{lp}\left(t'\right)-\psi_{21}\left(t',t_{0}\right)\xi_{rp}\left(t'\right)}{\psi_{11}\left(t',t_{0}\right)\psi_{22}\left(t',t_{0}\right)-\psi_{12}\left(t',t_{0}\right)\psi_{21}\left(t',t_{0}\right)}dt',\label{eq:Eq-of-Mot-R1-Speed-NA}\end{align}
 \begin{align}
\dot{R}_{lp,cm,\alpha}\left(t\right) & =\left[\frac{\lambda_{4}\left(;t_{0}\right)-\eta_{1}\left(;t_{0}\right)}{\lambda_{3}\left(;t_{0}\right)-\eta_{1}\left(;t_{0}\right)}-1\right]^{-1}\frac{\psi_{21}\left(t,t_{0}\right)\dot{R}_{rp,cm,\alpha}+\psi_{22}\left(t,t_{0}\right)\dot{R}_{lp,cm,\alpha}\left(t_{0}\right)}{\exp\left(\left[\lambda_{3}\left(;t_{0}\right)+\lambda_{4}\left(;t_{0}\right)\right]t_{0}\right)}\nonumber \\
 & +\psi_{21}\left(t,t_{0}\right)\int_{t_{0}}^{t}\frac{\psi_{22}\left(t',t_{0}\right)\xi_{rp}\left(t'\right)-\psi_{12}\left(t',t_{0}\right)\xi_{lp}\left(t'\right)}{\psi_{11}\left(t',t_{0}\right)\psi_{22}\left(t',t_{0}\right)-\psi_{12}\left(t',t_{0}\right)\psi_{21}\left(t',t_{0}\right)}dt'+\psi_{22}\left(t,t_{0}\right)\nonumber \\
 & \times\int_{t_{0}}^{t}\frac{\psi_{11}\left(t',t_{0}\right)\xi_{lp}\left(t'\right)-\psi_{21}\left(t',t_{0}\right)\xi_{rp}\left(t'\right)}{\psi_{11}\left(t',t_{0}\right)\psi_{22}\left(t',t_{0}\right)-\psi_{12}\left(t',t_{0}\right)\psi_{21}\left(t',t_{0}\right)}dt',\label{eq:Eq-of-Mot-R2-Speed-NA}\end{align}
 where the terms $\lambda_{3}$ and $\lambda_{4}$ are defined in
equation (\ref{eq:Eq-of-Mot-Plate-System-Static-Matrix-Eigenvalues});
and $\psi_{11}\left(t,t_{0}\right),$ $\psi_{12}\left(t,t_{0}\right),$
$\psi_{21}\left(t,t_{0}\right)$ and $\psi_{22}\left(t,t_{0}\right)$
are defined in equations (\ref{eq:Principal-Matrix-Psi-11}) through
(\ref{eq:Principal-Matrix-Psi-22}) in Appendix D2. The quantities
$\dot{R}_{rp,cm,\alpha}$ and $\dot{R}_{lp,cm,\alpha}$ are the speed
of the center of mass of {}``Right Plate'' and the speed of the
center of mass of the {}``Left Plate,'' respectively, and $\alpha$
defines the particular basis direction. 

The corresponding positions $R_{rp,cm,\alpha}\left(t\right)$ and
$R_{lp,cm,\alpha}\left(t\right)$ are found by integrating equations
(\ref{eq:Eq-of-Mot-R1-Speed-NA}) and (\ref{eq:Eq-of-Mot-R2-Speed-NA})
with respect to time, \begin{align}
R_{rp,cm,\alpha}\left(t\right) & =\left[\frac{\lambda_{4}\left(;t_{0}\right)-\eta_{1}\left(;t_{0}\right)}{\lambda_{3}\left(;t_{0}\right)-\eta_{1}\left(;t_{0}\right)}-1\right]^{-1}\int_{t_{0}}^{t}\left[\frac{\psi_{11}\left(\tau,t_{0}\right)\dot{R}_{rp,cm,\alpha}\left(t_{0}\right)+\psi_{12}\left(\tau,t_{0}\right)\dot{R}_{lp,cm,\alpha}\left(t_{0}\right)}{\exp\left(\left[\lambda_{3}\left(;t_{0}\right)+\lambda_{4}\left(;t_{0}\right)\right]t_{0}\right)}\right.\nonumber \\
 & +\psi_{11}\left(\tau,t_{0}\right)\int_{t_{0}}^{\tau}\frac{\psi_{22}\left(t',t_{0}\right)\xi_{rp}\left(t'\right)-\psi_{12}\left(t',t_{0}\right)\xi_{lp}\left(t'\right)}{\psi_{11}\left(t',t_{0}\right)\psi_{22}\left(t',t_{0}\right)-\psi_{12}\left(t',t_{0}\right)\psi_{21}\left(t',t_{0}\right)}dt'+\psi_{12}\left(\tau,t_{0}\right)\nonumber \\
 & \left.\times\int_{t_{0}}^{\tau}\frac{\psi_{11}\left(t',t_{0}\right)\xi_{lp}\left(t'\right)-\psi_{21}\left(t',t_{0}\right)\xi_{rp}\left(t'\right)}{\psi_{11}\left(t',t_{0}\right)\psi_{22}\left(t',t_{0}\right)-\psi_{12}\left(t',t_{0}\right)\psi_{21}\left(t',t_{0}\right)}dt'\right]d\tau+R_{rp,cm,\alpha}\left(t_{0}\right),\label{eq:Eq-of-Mot-R1-Position-NA}\end{align}
 \begin{align}
R_{lp,cm,\alpha}\left(t\right) & =\left[\frac{\lambda_{4}\left(;t_{0}\right)-\eta_{1}\left(;t_{0}\right)}{\lambda_{3}\left(;t_{0}\right)-\eta_{1}\left(;t_{0}\right)}-1\right]^{-1}\int_{t_{0}}^{t}\left[\frac{\psi_{21}\left(\tau,t_{0}\right)\dot{R}_{rp,cm,\alpha}\left(t_{0}\right)+\psi_{22}\left(\tau,t_{0}\right)\dot{R}_{lp,cm,\alpha}\left(t_{0}\right)}{\exp\left(\left[\lambda_{3}\left(;t_{0}\right)+\lambda_{4}\left(;t_{0}\right)\right]t_{0}\right)}\right.\nonumber \\
 & +\psi_{21}\left(\tau,t_{0}\right)\int_{t_{0}}^{\tau}\frac{\psi_{22}\left(t',t_{0}\right)\xi_{rp}\left(t'\right)-\psi_{12}\left(t',t_{0}\right)\xi_{lp}\left(t'\right)}{\psi_{11}\left(t',t_{0}\right)\psi_{22}\left(t',t_{0}\right)-\psi_{12}\left(t',t_{0}\right)\psi_{21}\left(t',t_{0}\right)}dt'+\psi_{22}\left(\tau,t_{0}\right)\nonumber \\
 & \left.\times\int_{t_{0}}^{\tau}\frac{\psi_{11}\left(t',t_{0}\right)\xi_{lp}\left(t'\right)-\psi_{21}\left(t',t_{0}\right)\xi_{rp}\left(t'\right)}{\psi_{11}\left(t',t_{0}\right)\psi_{22}\left(t',t_{0}\right)-\psi_{12}\left(t',t_{0}\right)\psi_{21}\left(t',t_{0}\right)}dt'\right]d\tau+R_{lp,cm,\alpha}\left(t_{0}\right).\label{eq:Eq-of-Mot-R2-Position-NA}\end{align}
 The remaining integrations are straightforward and the explicit forms
will not be shown here. 

One may argue that for the static case, $\dot{R}_{rp,cm,\alpha}\left(t_{0}\right)$
and $\dot{R}_{lp,cm,\alpha}\left(t_{0}\right)$ must be zero because
the conductors seem to be fixed in position. This argument is flawed,
for any wall totally fixed in position upon impact would require an
infinite amount of energy. One has to consider the conservation of
momentum simultaneously. The wall has to have moved by the amount
$\bigtriangleup R_{wall}=\dot{R}_{wall}\triangle t,$ where $\triangle t$
is the total duration of impact, and $\dot{R}_{wall}$ is calculated
from the momentum conservation and it is non-zero. The same argument
can be applied to the apparatus shown in Figure \ref{cap:driven-parallel-plates-force}.
For that system \begin{eqnarray*}
\left\Vert \vec{p}_{virtual-photon}\right\Vert =\frac{1}{c}\mathcal{H}'_{n_{s},\Re}\left(t_{0}\right), &  & \left\{ \begin{array}{c}
\dot{R}_{rp,cm,\alpha}\left(t_{0}\right)=\left\Vert \dot{\vec{R}}_{lp,3}\left(t_{0}\right)+\dot{\vec{R}}_{rp,2}\left(t_{0}\right)\right\Vert ,\\
\\\dot{R}_{lp,cm,\alpha}\left(t_{0}\right)=\left\Vert \dot{\vec{R}}_{rp,1}\left(t_{0}\right)+\dot{\vec{R}}_{lp,2}\left(t_{0}\right)\right\Vert .\end{array}\right.\end{eqnarray*}
 For simplicity, assuming that the impact is always only in the normal
direction, \begin{eqnarray*}
\dot{R}_{rp,cm,\alpha}\left(t_{0}\right)=\frac{2}{m_{rp}c}\left\Vert \mathcal{H}'_{n_{s},3}\left(t_{0}\right)-\mathcal{H}'_{n_{s},2}\left(t_{0}\right)\right\Vert , &  & \dot{R}_{lp,cm,\alpha}\left(t_{0}\right)=\frac{2}{m_{lp}c}\left\Vert \mathcal{H}'_{n_{s},1}\left(t_{0}\right)-\mathcal{H}'_{n_{s},2}\left(t_{0}\right)\right\Vert ,\end{eqnarray*}
 where the differences under the magnitude symbol imply field energies
from different regions counteract the other. The details of this section
can be found in Appendix D2.

\section{Results and Outlook}

\begin{figure}
\begin{center}\includegraphics[%
  scale=0.7]{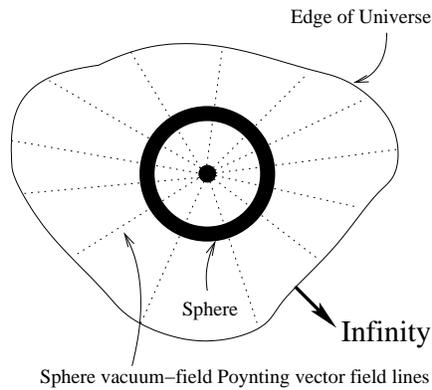}\end{center}

\caption{Boyer's configuration is such that a sphere is the only matter in
the entire universe. His universe extends to the infinity, hence there
are no boundaries. The sense of vacuum-field energy flow is along
the radial vector $\hat{r},$ which is defined with respect to the
sphere center. \label{cap:boyers-sphere-in-universe}}
\end{figure}

The results for the sign of Casimir force on non-planar geometric
configurations considered in this investigation will eventually be
compared with the classic repulsive result obtained by Boyer decades
earlier. For this reason, it is worth reviewing Boyer's original configuration
as shown in Figure \ref{cap:boyers-sphere-in-universe}.

T. H. Boyer in 1968 obtained a repulsive Casimir force result for
his charge-neutral, hollow spherical shell of a perfect conductor
\cite{key-Boyer}. For simplicity, his sphere is the only object in
the entire universe and, therefore, no external boundaries such as
laboratory walls, etc., were defined in his problem. Furthermore,
the zero-point energy flow is always perpendicular to his sphere.
Such restriction constitutes a very stringent condition for the material
property that a sphere has to meet. For example, if one were to look
at Boyer's sphere, he would not see the whole sphere; but instead,
he would see a small spot on the surface of a sphere that happens
to be in his line of sight. This happens because the sphere in Boyer's
configuration can only radiate in a direction normal to the surface.
One could equivalently argue that Boyer's sphere only responds to
the approaching radiation at normal angles of incidence with respect
to the surface of the sphere. When the Casimir force is computed for
such restricted radiation energy flow, the result is repulsive. In
Boyer's picture, this may be attributed to the fact that closer to
the origin of a sphere, the spherically symmetric radiation energy
flow becomes more dense due to the inverse length dependence, and
this density decreases as it gets further away from the sphere center.
This argument, however, seems to be flawed because it inherently implies
existence of the preferred origin for the vacuum fields. As an illustration,
Boyer's sphere is shown in Figure \ref{cap:boyers-sphere-in-universe}.
For the rest of this investigation, {}``Boyer's sphere'' would be
strictly referred to as the sphere made of material with such a property
that it only radiates or responds to vacuum-field radiations at normal
angle of incidence with respect to its surface.

\begin{figure}
\begin{center}\includegraphics[%
  scale=0.7]{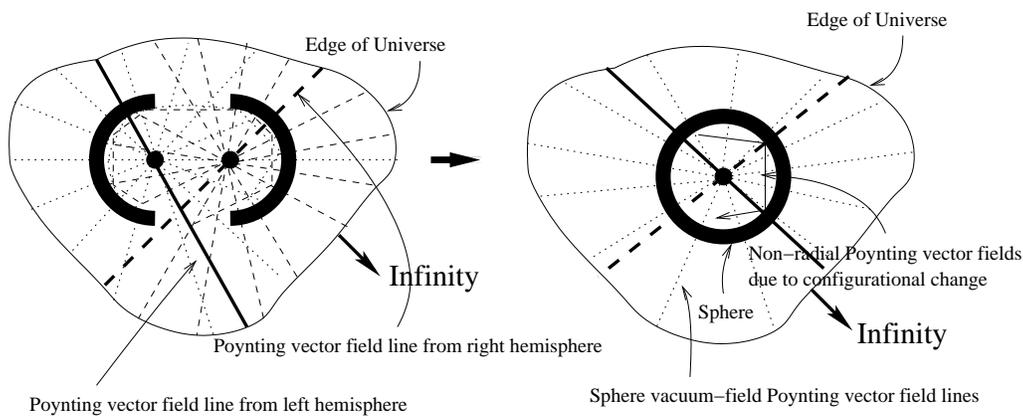}\end{center}

\caption{Manufactured sphere, in which two hemispheres are brought together,
results in small non-spherically symmetric vacuum-field radiation
inside the cavity due to the configuration change. For the hemispheres
made of Boyer's material, these fields in the resonator will eventually
get absorbed by the conductor resulting in heating of the hemispheres.
\label{cap:boyers-sphere-in-universe-hemi-to-sphere}}
\end{figure}

The formation of a sphere by bringing together two nearby hemispheres
satisfying the material property of Boyer's sphere is illustrated
in Figure \ref{cap:boyers-sphere-in-universe-hemi-to-sphere}. Since
Boyer's material property only allow radiation in the normal direction
to its surface, the radiation associated with each hemisphere would
necessarily go through the corresponding hemisphere centers. For clarity,
let us define the unit radial basis vector associated with the left
and right hemispheres by $\hat{r}_{L}$ and $\hat{r}_{R},$ respectively.
If the hemispheres are made of normal conductors the radiation from
one hemisphere entering the other hemisphere cavity would go through
a complex series of reflections before escaping the cavity. Here,
a conductor with Boyer's stringent material property is not considered
normal. Conductors that are normal also radiate in directions non-normal
to their surface, whereas Boyer's conductor can only radiate normal
to its surface. Due to the fact that Boyer's conducting materials
can only respond to radiation impinging at a normal angle of incidence
with respect to its surface, all of the incoming radiation at oblique
angles of incidence with respect to the local surface normal is absorbed
by the host hemisphere. This suggests that for the hemisphere-hemisphere
arrangement made of Boyer's material shown in Figure \ref{cap:boyers-sphere-in-universe-hemi-to-sphere},
the temperature of the two hemispheres would rise indefinitely over
time. This does not happen with ordinary conductors. This suggests
that Boyer's conducting material, of which his sphere is made, is
completely hypothetical. Precisely because of this material assumption,
Boyer's Casimir force is repulsive. 

\begin{figure}
\begin{center}\includegraphics[%
  scale=0.7]{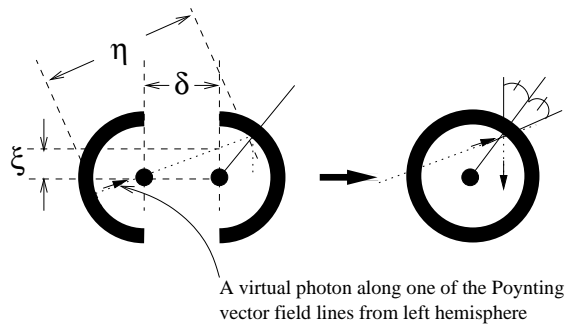}\end{center}

\caption{The process in which a configuration change from hemisphere-hemisphere
to sphere inducing virtual photon in the direction other than $\hat{r}$
is shown. The virtual photon here is referred to as the quanta of
energy associated with the zero-point radiation. \label{cap:boyers-sphere-in-universe-Config-Demo}}
\end{figure}

For the moment, let us relax the stringent Boyer's material property
for the hemispheres to that of ordinary conductors. For the hemispheres
made of ordinary conducting materials, there would result a series
of reflections in one hemisphere cavity due to those radiations entering
the cavity from nearby hemisphere. For simplicity, the ordinary conducting
material referred to here is that of perfect conductors without Boyer's
hypothetical material property requirement. Furthermore, only the
radiation emanating normally with respect to its surface is considered.
The idea is to illustrate that the {}``normally emanated radiation''
from one hemisphere results in elaboration of the effects of {}``obliquely
emanated radiation'' on another hemisphere cavity. Here the obliquely
emanated radiation means those radiation emanating from a surface
not along the local normal of the surface. 

When two such hemispheres are brought together to form a sphere, there
would exist some radiation trapped in the sphere of which the radiation
energy flow lines are not spherically symmetric with respect to the
sphere center. To see how a mere change in configuration invokes such
non-spherically symmetric energy flow, consider the illustration shown
in Figure \ref{cap:boyers-sphere-in-universe-Config-Demo}. For clarity,
only one {}``normally emanated radiation'' energy flow line from
the left hemisphere is shown. When one brings together the two hemispheres
just in time before that quantum of energy escapes the hemisphere
cavity to the right, the trapped energy quantum would continuously
go through series of complex reflections in the cavity obeying the
reflection law. But how fast or how slow one brings in two hemispheres
is irrelevant in invoking such non-spherically symmetric energy flow
because the gap $\delta$ can be chosen arbitrarily. Therefore, there
would always be a stream of energy quanta crossing the hemisphere
opening with $\xi\neq0$ as shown in Figure \ref{cap:boyers-sphere-in-universe-Config-Demo}.
In other words, there is always a time interval $\triangle t$ within
which the hemispheres are separated by an amount $\delta$ before
closure. The quanta of vacuum-field radiation energy created within
that time interval $\triangle t$ would always be satisfying the condition
$\xi\neq0,$ and this results in reflections at oblique angle of incidence
with respect to the local normal of the walls of inner sphere cavity.
Only when the two hemispheres are finally closed, would then $\xi=0$
and the radiation energy produced in the sphere after that moment
would be spherically symmetric and the reflections would be normal
to the surface. However, those trapped quantum of energy that were
produced prior to the closure of the two hemispheres would always
be reflecting from the inner sphere surface at oblique angles of incidence. 

\begin{figure}
\begin{center}\includegraphics[%
  scale=0.7]{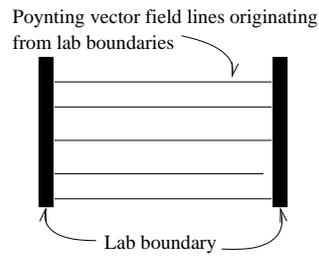}\end{center}

\caption{A realistic laboratory has boundaries, e.g., walls. These boundaries
have effect similar to the field modes between two parallel plates.
In \textbf{3D}, the effects are similar to that of a cubical laboratory,
etc.  \label{cap:reallab}}
\end{figure}

Unlike Boyer's ideal laboratory, realistic laboratories have boundaries
made of ordinary material as illustrated in Figure \ref{cap:reallab}.
One must then take into account, when calculating the Casimir force,
the vacuum-field radiation pressure contributions from the involved
conductors, as well as those contributions from the boundaries such
as laboratory walls, etc. We will examine the physics of placing two
hemispheres inside the laboratory. 

\begin{figure}
\begin{center}\includegraphics[%
  scale=0.7]{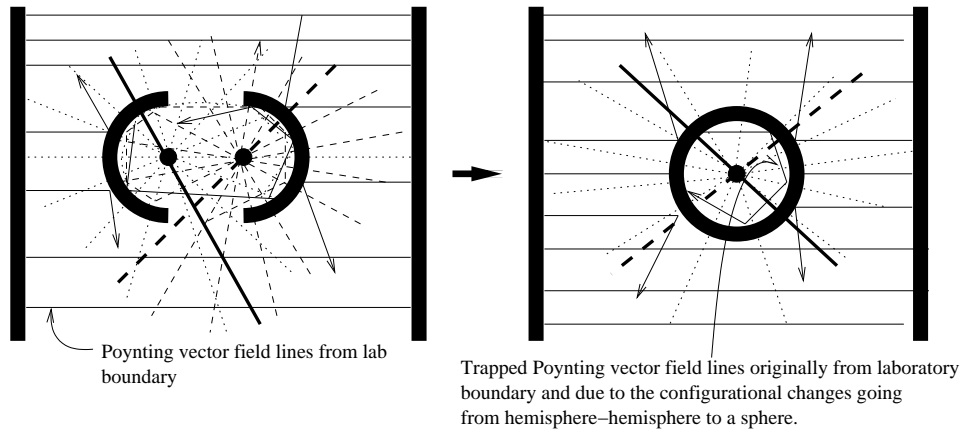}\end{center}

\caption{The schematic of sphere manufacturing process in a realistic laboratory.
\label{cap:boyers-sphere-in-reallab}}
\end{figure}

For simplicity, the boundaries of the laboratory as shown in Figure
\ref{cap:boyers-sphere-in-reallab} are assumed to be simple cubical.
Normally, the dimension of conductors considered in Casimir force
experiment is in the ranges of microns. When this is compared with
the size of the laboratory boundaries such as the walls, the walls
of the laboratory can be treated as a set of infinite parallel plates
and the vacuum-fields inside the the laboratory can be treated as
simple plane waves with impunity. 

The presence of laboratory boundaries induce reflection of energy
flow similar to that between the two parallel plate arrangement. When
the two hemisphere arrangement shown in Figure \ref{cap:boyers-sphere-in-universe-hemi-to-sphere}
is placed in such a laboratory, the result is to elaborate the radiation
pressure contributions from obliquely incident radiations on external
surfaces of the two hemispheres. If the two hemispheres are made of
conducting material satisfying Boyer's material property, the vacuum-field
radiation impinging on hemisphere surfaces at oblique angles of incidence
would cause heating of the hemispheres. It means that Boyer's hemispheres
placed in a realistic laboratory would continue to rise in temperature
as a function of time. However, this does not happen with ordinary
conductors.

If the two hemispheres are made of ordinary perfect conducting materials,
the reflections of radiation at oblique angles of incidence from the
laboratory boundaries would elaborate on the radiation pressure acting
on the external surfaces of two hemispheres at oblique angles of incidence.
Because Boyer's sphere only radiates in the normal direction to its
surface, or only responds to impinging radiation at normal incidence
with respect to the sphere surface, the extra vacuum-field radiation
pressures considered here, i.e., the ones involving oblique angles
of incidence, are missing in his Casimir force calculation for the
sphere.

\subsection{Results}

T. H. Boyer in 1968 have shown that for a charge-neutral, perfect
conductor of hollow spherical shell, the sign of the Casimir force
is positive, which means the force is repulsive. He reached this conclusion
by assuming that all vacuum-field radiation energy flows for his sphere
are spherically symmetric with respect to its center. In other words,
only the wave vectors that are perpendicular to his sphere surface
were included in the Casimir force calculation. In the following sections,
the non-perpendicular wave vector contributions to the Casimir force
that were not accounted for in Boyer's work are considered.

\subsubsection{Hollow Spherical Shell}

\begin{figure}
\begin{center}\includegraphics[%
  scale=0.7]{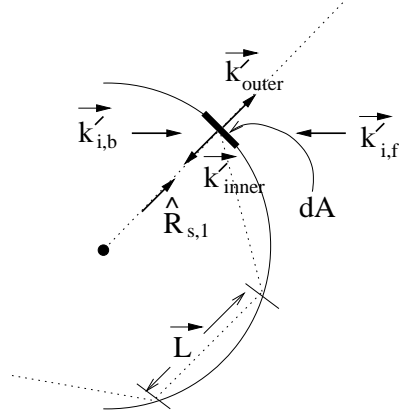}\end{center}

\caption{The vacuum-field wave vectors $\vec{k'}_{i,b}$ and $\vec{k'}_{i,f}$
impart a net momentum of the magnitude $\left\Vert \vec{p}_{net}\right\Vert =\hbar\left\Vert \vec{k'}_{i,b}-\vec{k'}_{i,f}\right\Vert /2$
on differential patch of an area $dA$ on a conducting spherical surface.
\label{cap:effective-momentum}}
\end{figure}

As shown in Figure \ref{cap:effective-momentum}, the vacuum-field
radiation imparts upon a differential patch of an area $dA$ on the
inner wall of the conducting spherical cavity a net momentum of the
amount \begin{eqnarray*}
\triangle\vec{p}_{inner}=-\frac{1}{2}\hbar\triangle\vec{k'}_{inner}\left(;\vec{R'}_{s,1},\vec{R'}_{s,0}\right)=\frac{2n\pi\hbar\cos\theta_{inc}}{\left\Vert \vec{R}_{s,2}\left(r'_{i},\vec{\Lambda}'_{s,2}\right)-\vec{R}_{s,1}\left(r'_{i},\vec{\Lambda}'_{s,1}\right)\right\Vert }\hat{R'}_{s,1}, &  & \left\{ \begin{array}{c}
0\leq\theta_{inc}<\pi/2,\\
\\n=1,2,3,\cdots,\end{array}\right.\end{eqnarray*}
 where $\triangle\vec{k'}_{inner}\left(;\vec{R'}_{s,1},\vec{R'}_{s,0}\right)$
is from equation (\ref{eq:SPHERE-Delta-K-Inside-NA}). The angle of
incidence $\theta_{inc}$ is from equation (\ref{eq:angle-of-incidence-exp});
$\vec{R}_{s,1}\left(r'_{i},\vec{\Lambda}'_{s,1}\right)$ and $\vec{R}_{s,2}\left(r'_{i},\vec{\Lambda}'_{s,2}\right)$
follow the generic form shown in equation (\ref{eq:Nth-Reflection-Point-sphere-NA}). 

Similarly, the vacuum-field radiation imparts upon a differential
patch of an area $dA$ on the outer surface of the conducting spherical
shell a net momentum of the amount \begin{eqnarray*}
\triangle\vec{p}_{outer}=-\frac{1}{2}\hbar\triangle\vec{k'}_{outer}\left(;\vec{R'}_{s,1}+a\hat{R'}_{s,1}\right)=-2\hbar\left\Vert \vec{k'}_{i,f}\right\Vert \cos\theta_{inc}\hat{R'}_{s,1}, &  & \left\{ \begin{array}{c}
0\leq\theta_{inc}<\pi/2,\\
\\n=1,2,3,\cdots,\end{array}\right.\end{eqnarray*}
 where $\triangle\vec{k'}_{outer}\left(;\vec{R'}_{s,1}+a\hat{R'}_{s,1}\right)$
is from equation (\ref{eq:SPHERE-Delta-K-Outside-NA}). 

The net average force per unit time, per initial wave vector direction,
acting on differential element patch of an area $dA$ is given by
\begin{align*}
\vec{\mathcal{F}}_{s,avg} & =\lim_{\triangle t\rightarrow1}\left(\frac{\triangle\vec{p}_{outer}}{\triangle t}+\frac{\triangle\vec{p}_{inner}}{\triangle t}\right)\end{align*}
 or \begin{eqnarray*}
\vec{\mathcal{F}}_{s,avg}=2\hbar\cos\theta_{inc}\left[\frac{n\pi}{\left\Vert \vec{R}_{s,2}\left(r'_{i},\vec{\Lambda}'_{s,2}\right)-\vec{R}_{s,1}\left(r'_{i},\vec{\Lambda}'_{s,1}\right)\right\Vert }-\left\Vert \vec{k'}_{i,f}\right\Vert \right]\hat{R'}_{s,1}, &  & \left\{ \begin{array}{c}
0\leq\theta_{inc}<\pi/2,\\
\\n=1,2,3,\cdots.\end{array}\right.\end{eqnarray*}
 Notice that $\vec{\mathcal{F}}_{s,avg}$ is called a force per initial
wave vector direction because it is computed for $\vec{k'}_{i,b}$
and $\vec{k'}_{i,f}$ along specific initial directions. Here $\vec{k'}_{i,b}$
denotes a particular initial wave vector $\vec{k'}_{i}$ entering
the resonator at $\vec{R}_{s,0}$ as shown in Figure \ref{cap:sphere-reflection-dynamics}.
The subscript $b$ for $\vec{k'}_{i,b}$ denotes the bounded space
inside the resonator. The $\vec{k'}_{i,f}$ denotes a particular initial
wave vector $\vec{k'}_{i}$ impinging upon the surface of the unbounded
region of sphere at point $\vec{R'}_{s,1}+a\hat{R'}_{s,1}$ as shown
in Figure \ref{cap:sphere-reflection-dynamics}. The subscript $f$
for $\vec{k'}_{i,f}$ denotes the free space external to the resonator. 

Because the wave vector $\vec{k'}_{i,f}$ resides in free or unbounded
space, its magnitude $\left\Vert \vec{k'}_{i,f}\right\Vert $ can
take on a continuum of allowed modes, whereas $\left\Vert \vec{k'}_{i,b}\right\Vert $
have been restricted by $\left\Vert \vec{L}\right\Vert =\left\Vert \vec{R}_{s,2}\left(r'_{i},\vec{\Lambda}'_{s,2}\right)-\vec{R}_{s,1}\left(r'_{i},\vec{\Lambda}'_{s,1}\right)\right\Vert $
of equation (\ref{eq:sphere-distance-rela}). The free space limit
is the case where the radius of the sphere becomes very large. Therefore,
by designating $\left\Vert \vec{k'}_{i,f}\right\Vert $ as \begin{align*}
\left\Vert \vec{k'}_{i,f}\right\Vert  & =\lim_{r'_{i}\rightarrow\infty}\frac{n\pi}{\left\Vert \vec{R}_{s,2}\left(r'_{i},\vec{\Lambda}'_{s,2}\right)-\vec{R}_{s,1}\left(r'_{i},\vec{\Lambda}'_{s,1}\right)\right\Vert },\end{align*}
 and summing over all allowed modes, the total average force per unit
time, per initial wave vector direction, per unit area is given by
\begin{align*}
\vec{\mathcal{F}}_{s,avg} & =\left[\sum_{n=1}^{\infty}\frac{n\pi2\hbar\cos\theta_{inc}}{\left\Vert \vec{R}_{s,2}\left(r'_{i},\vec{\Lambda}'_{s,2}\right)-\vec{R}_{s,1}\left(r'_{i},\vec{\Lambda}'_{s,1}\right)\right\Vert }-\lim_{r'_{i}\rightarrow\infty}\sum_{n=1}^{\infty}\frac{n\pi2\hbar\cos\theta_{inc}}{\left\Vert \vec{R}_{s,2}\left(r'_{i},\vec{\Lambda}'_{s,2}\right)-\vec{R}_{s,1}\left(r'_{i},\vec{\Lambda}'_{s,1}\right)\right\Vert }\right]\hat{R'}_{s,1}.\end{align*}
 In the limit $r'_{i}\rightarrow\infty,$ the second summation to
the right can be replaced by an integration, $\sum_{n=1}^{\infty}\rightarrow\int_{0}^{\infty}dn.$
Hence, we have \begin{align*}
\vec{\mathcal{F}}_{s,avg} & =\left[\sum_{n=1}^{\infty}\frac{2\hbar n\pi\cos\theta_{inc}}{\left\Vert \vec{R}_{s,2}\left(r'_{i},\vec{\Lambda}'_{s,2}\right)-\vec{R}_{s,1}\left(r'_{i},\vec{\Lambda}'_{s,1}\right)\right\Vert }-\lim_{r'_{i}\rightarrow\infty}\int_{0}^{\infty}\frac{2\hbar n\pi\cos\theta_{inc}}{\left\Vert \vec{R}_{s,2}\left(r'_{i},\vec{\Lambda}'_{s,2}\right)-\vec{R}_{s,1}\left(r'_{i},\vec{\Lambda}'_{s,1}\right)\right\Vert }dn\right]\hat{R'}_{s,1},\end{align*}
 or with the following substitutions, \begin{eqnarray*}
k'_{i,f}\equiv\frac{n\pi}{\left\Vert \vec{R}_{s,2}\left(r'_{i},\vec{\Lambda}'_{s,2}\right)-\vec{R}_{s,1}\left(r'_{i},\vec{\Lambda}'_{s,1}\right)\right\Vert }, &  & dn=\frac{1}{\pi}\left\Vert \vec{R}_{s,2}\left(r'_{i},\vec{\Lambda}'_{s,2}\right)-\vec{R}_{s,1}\left(r'_{i},\vec{\Lambda}'_{s,1}\right)\right\Vert dk'_{i,f},\end{eqnarray*}
 the total average force per unit time, per initial wave vector direction,
per unit area is written as \begin{align}
\vec{\mathcal{F}}_{s,avg} & =2\hbar\cos\theta_{inc}\left[\sum_{n=1}^{\infty}\frac{n\pi}{\left\Vert \vec{R}_{s,2}\left(r'_{i},\vec{\Lambda}'_{s,2}\right)-\vec{R}_{s,1}\left(r'_{i},\vec{\Lambda}'_{s,1}\right)\right\Vert }\right.\nonumber \\
 & \left.-\frac{1}{\pi}\lim_{r'_{i}\rightarrow\infty}\left\Vert \vec{R}_{s,2}\left(r'_{i},\vec{\Lambda}'_{s,2}\right)-\vec{R}_{s,1}\left(r'_{i},\vec{\Lambda}'_{s,1}\right)\right\Vert \int_{0}^{\infty}k'_{i,f}dk'_{i,f}\right]\hat{R'}_{s,1},\label{eq:sphere-ambient-force-average}\end{align}
 where $0\leq\theta_{inc}<\pi/2$ and $n=1,2,3,\cdots.$ The total
average vacuum-field radiation force per unit time acting on the uncharged
conducting spherical shell is therefore\begin{align*}
\vec{F}_{s,total} & =\sum_{\left\{ \vec{k'}_{i,b},\vec{k'}_{i,f},\vec{R'}_{s,0}\right\} }\int_{S}\vec{\mathcal{F}}_{s,avg}\cdot d\vec{S}_{sphere}\end{align*}
 or \begin{align}
\vec{F}_{s,total} & =\sum_{\left\{ \vec{k'}_{i,b},\vec{k'}_{i,f},\vec{R'}_{s,0}\right\} }\int_{S}\left[\sum_{n=1}^{\infty}\frac{2n\pi\hbar\cos\theta_{inc}}{\left\Vert \vec{R}_{s,2}\left(r'_{i},\vec{\Lambda}'_{s,2}\right)-\vec{R}_{s,1}\left(r'_{i},\vec{\Lambda}'_{s,1}\right)\right\Vert }-\frac{2\hbar}{\pi}\cos\theta_{inc}\right.\nonumber \\
 & \left.\times\lim_{r'_{i}\rightarrow\infty}\left\Vert \vec{R}_{s,2}\left(r'_{i},\vec{\Lambda}'_{s,2}\right)-\vec{R}_{s,1}\left(r'_{i},\vec{\Lambda}'_{s,1}\right)\right\Vert \int_{0}^{\infty}k'_{i,f}dk'_{i,f}\right]\hat{R'}_{s,1}\cdot d\vec{S}_{sphere},\label{eq:sphere-ambient-force-average-TOTAL}\end{align}
 where $d\vec{S}_{sphere}$ is a differential surface element of a
sphere and the integration $\int_{S}$ is over the spherical surface.
The term $\vec{R'}_{s,0}$ is the initial crossing point inside the
sphere as defined in equation (\ref{eq:initial-arbitrary-k-vector-position-NA}).
The notation $\sum_{\left\{ \vec{k'}_{i,b},\vec{k'}_{i,f},\vec{R'}_{s,0}\right\} }$
imply the summation over all initial wave vector directions for both
inside $\left(\vec{k'}_{i,b}\right)$ and outside $\left(\vec{k'}_{i,f}\right)$
of the sphere, over all crossing points given by $\vec{R'}_{s,0}.$ 

It is easy to see that $\vec{\mathcal{F}}_{s,avg}$ of equation (\ref{eq:sphere-ambient-force-average})
is an {}``unregularized'' \textbf{1D} Casimir force expression for
the parallel plates (see the vacuum pressure approach by Milonni,
Cook and Goggin \cite{key-Milonni-Cook-Goggin}). It becomes more
apparent with the substitution $\triangle t=d/c.$ An application
of the Euler-Maclaurin summation formula \cite{key-Euler-Sum-Formula,key-Euler-Sum-Formula-Derivation}
leads to the regularized, finite force expression. The force $\vec{\mathcal{F}}_{s,avg}$
is attractive because \begin{align*}
\cos\theta_{inc} & >0\end{align*}
 and \begin{align*}
\sum_{n=1}^{\infty}\frac{n\pi}{\left\Vert \vec{R}_{s,2}\left(r'_{i},\vec{\Lambda}'_{s,2}\right)-\vec{R}_{s,1}\left(r'_{i},\vec{\Lambda}'_{s,1}\right)\right\Vert } & <\frac{1}{\pi}\lim_{r'_{i}\rightarrow\infty}\left\Vert \vec{R}_{s,2}\left(r'_{i},\vec{\Lambda}'_{s,2}\right)-\vec{R}_{s,1}\left(r'_{i},\vec{\Lambda}'_{s,1}\right)\right\Vert \int_{0}^{\infty}k'_{i,f}dk'_{i,f},\end{align*}
 where $\left\Vert \vec{R}_{s,2}\left(r'_{i},\vec{\Lambda}'_{s,2}\right)-\vec{R}_{s,1}\left(r'_{i},\vec{\Lambda}'_{s,1}\right)\right\Vert $
is a constant for a given initial wave $\vec{k'}_{i,b}$ and the initial
crossing point $\vec{R'}_{s,0}$ in the cross-section of a sphere
(or hemisphere). The total average force $\vec{F}_{s,total},$ which
is really the sum of $\vec{\mathcal{F}}_{s,avg}$ over all $\vec{R'}_{s,0}$
and all initial wave directions, is therefore also attractive. For
the sphere configuration of Figure \ref{cap:sphere-reflection-dynamics},
where the energy flow direction is not restricted to the direction
of local surface normal, the Casimir force problem becomes an extension
of infinite set of parallel plates of a unit area.

\subsubsection{Hemisphere-Hemisphere and Plate-Hemisphere}

Similarly, for the hemisphere-hemisphere and plate-hemisphere configurations,
the expression for the total average force per unit time, per initial
wave vector direction, per unit area is identical to that of the hollow
spherical shell with modifications, \begin{align}
\vec{\mathcal{F}}_{h,avg} & =2\hbar\cos\theta_{inc}\left[\sum_{n=1}^{\infty}\frac{n\pi}{\left\Vert \vec{R}_{h,2}\left(r'_{i},\vec{\Lambda}'_{h,2}\right)-\vec{R}_{h,1}\left(r'_{i},\vec{\Lambda}'_{h,1}\right)\right\Vert }\right.\nonumber \\
 & \left.-\frac{1}{\pi}\lim_{r'_{i}\rightarrow\infty}\left\Vert \vec{R}_{h,2}\left(r'_{i},\vec{\Lambda}'_{h,2}\right)-\vec{R}_{h,1}\left(r'_{i},\vec{\Lambda}'_{h,1}\right)\right\Vert \int_{0}^{\infty}k'_{i,f}dk'_{i,f}\right]\hat{R'}_{h,1},\label{eq:hemisphere-hemisphere-ambient-force-average}\end{align}
 where $\theta_{inc}\leq\pi/2$ and $n=1,2,3,\cdots.$ The incidence
angle $\theta_{inc}$ is from equation (\ref{eq:angle-of-incidence-exp});
$\vec{R}_{h,1}\left(r'_{i},\vec{\Lambda}'_{h,1}\right)$ and $\vec{R}_{h,2}\left(r'_{i},\vec{\Lambda}'_{h,2}\right)$
follow the generic form shown in equation (\ref{eq:Points-on-Hemisphere-R-NA}).
This force is attractive for the same reasons as discussed previously
for the hollow spherical shell case. The total radiation force averaged
over unit time, over all possible initial wave vector directions,
acting on the uncharged conducting hemisphere-hemisphere (plate-hemisphere)
surface is given by \begin{align}
\vec{F}_{h,total} & =\sum_{\left\{ \vec{k'}_{i,b},\vec{k'}_{i,f},\vec{R'}_{h,0}\right\} }\int_{S}\left[\sum_{n=1}^{\infty}\frac{2n\pi\hbar\cos\theta_{inc}}{\left\Vert \vec{R}_{h,2}\left(r'_{i},\vec{\Lambda}'_{h,2}\right)-\vec{R}_{h,1}\left(r'_{i},\vec{\Lambda}'_{h,1}\right)\right\Vert }-\frac{2\hbar}{\pi}\cos\theta_{inc}\right.\nonumber \\
 & \left.\times\lim_{r'_{i}\rightarrow\infty}\left\Vert \vec{R}_{h,2}\left(r'_{i},\vec{\Lambda}'_{h,2}\right)-\vec{R}_{h,1}\left(r'_{i},\vec{\Lambda}'_{h,1}\right)\right\Vert \int_{0}^{\infty}k'_{i,f}dk'_{i,f}\right]\hat{R'}_{h,1}\cdot d\vec{S}_{hemisphere},\label{eq:hemisphere-hemisphere-ambient-force-average-TOTAL}\end{align}
 where $d\vec{S}_{hemisphere}$ is now a differential surface element
of a hemisphere and the integration $\int_{S}$ is over the surface
of the hemisphere. The term $\vec{R'}_{h,0}$ is the initial crossing
point of the hemisphere opening as defined in equation (\ref{eq:initial-arbitrary-k-vector-position-NA}).
The notation $\sum_{\left\{ \vec{k'}_{i,b},\vec{k'}_{i,f},\vec{R'}_{h,0}\right\} }$
imply the summation over all initial wave vector directions for both
inside $\left(\vec{k'}_{i,b}\right)$ and outside $\left(\vec{k'}_{i,f}\right)$
of the hemisphere-hemisphere (or the plate-hemisphere) resonator,
over all crossing points given by $\vec{R'}_{h,0}.$ 

It should be remarked that for the plate-hemisphere configuration,
the total average radiation force remains identical to that of the
hemisphere-hemisphere configuration only for the case where the gap
distance between plate and the center of hemisphere is more than the
hemisphere radius $r'_{i}.$ When the plate is placed closer, the
boundary quantization length $\left\Vert \vec{L}\right\Vert $ must
be chosen carefully to be either \begin{align*}
\left\Vert \vec{L}\right\Vert  & =\left\Vert \vec{R}_{h,2}\left(r'_{i},\vec{\Lambda}'_{h,2}\right)-\vec{R}_{h,1}\left(r'_{i},\vec{\Lambda}'_{h,1}\right)\right\Vert \end{align*}
 or \begin{align*}
\left\Vert \vec{L}\right\Vert  & =\left\Vert \vec{R}_{p}\left(r'_{i},\vec{\Lambda}'_{p}\right)-\vec{R}_{h,N_{h,max}}\left(r'_{i},\vec{\Lambda}'_{h,N_{h,max}}\right)\right\Vert .\end{align*}
 They are illustrated in Figure \ref{cap:plate-hemisphere-plane-of-incidence-intersect-Complex}.
The proper one to use is the smaller of the two. Here $\vec{R}_{p}\left(r'_{i},\vec{\Lambda}'_{p}\right)$
is from equation (\ref{eq:Points-on-Plate-R-Final}) of Appendix C3
and $N_{h,max}$ is defined in equation (\ref{eq:N-max-Hemisphere})
of Appendix C2.

\subsection{Interpretation of the Result}

Because only the specification of boundary is needed in Casimir's
vacuum-field approach as opposed to the use of a polarizability parameter
in Casimir-Polder interaction picture, the Casimir force is sometimes
regarded as a configurational force. On the other hand, the Casimir
effect can be thought of as a macroscopic manifestation of the retarded
van der Waals interaction. And the Casimir force can be equivalently
approximated by a summation of the constituent molecular forces employing
Casimir-Polder interaction. This practice inherently relies on the
material properties of the involved conductors through the use of
polarizability parameters. In this respect, the Casimir force can
be regarded as a material dependent force. 

Boyer's material property is such that the atoms in his conducting
sphere are arranged in such manner to respond only to the impinging
radiation at local normal angle of incidence to the sphere surface,
and they also radiate only along the direction of local normal to
its surface. When the Casimir force is calculated for a sphere made
of Boyer's fictitious material, the force is repulsive. Also, in Boyer's
original work, the laboratory boundary did not exist. When Boyer's
sphere is placed in a realistic laboratory, the net Casimir force
acting on his sphere becomes attractive because the majority of the
radiation from the laboratory boundaries acts to apply inward pressure
on the external surface of sphere when the angle of incidence is oblique
with respect to the local normal. If the sphere is made of ordinary
perfect conductors, the impinging radiation at oblique angles of incidence
would be reflected. In such cases the total radiation pressure applied
to the external local-sphere-surface is twice the pressure exerted
by the incident wave, which is the force found in equation (\ref{eq:sphere-ambient-force-average-TOTAL})
of the previous section. However, Boyer's sphere cannot radiate along
the direction that is not normal to the local-sphere-surface. Therefore,
the total pressure applied to Boyer's sphere is half of the force
given in equation (\ref{eq:sphere-ambient-force-average-TOTAL}) of
the previous section. This peculiar incapability of emission of a
Boyer's sphere would lead to the absorption of the energy and would
cause a rise in the temperature for the sphere. Nonetheless, the extra
pressure due to the waves of oblique angle of incidence is large enough
to change the Casimir force for Boyer's sphere from being repulsive
to attractive. The presence of the laboratory boundaries only act
to enhance the attractive aspect of the Casimir force on a sphere.
The fact that Boyer's sphere cannot irradiate along the direction
that is not normal to the local-sphere-surface, whereas ordinary perfect
conductors irradiate in all directions, implies that his sphere is
made of extraordinarily hypothetical material, and this may be the
reason why the repulsive Casimir force have not been experimentally
observed to date. 

In conclusion, (1) the Casimir force is both boundary and material
property dependent. The particular shape of the conductor, e.g. sphere,
only introduces the preferred direction for radiation. For example,
radiations in direction normal to the local surface has bigger magnitude
than those radiating in other directions. This preference for the
direction of radiation is intrinsically connected to the preferred
directions for the lattice vibrations. And, the characteristic of
lattice vibrations is intrinsically connected to the property of material.
(2) Boyer's sphere is made of extraordinary conducting material, which
is why his Casimir force is repulsive. (3) When the radiation pressures
of all angles of incidence are included in the Casimir force calculation,
the force is attractive for a charge-neutral sphere made of ordinary
perfect conductor.

\subsection{Suggestions on the Detection of Repulsive Casimir Force for a Sphere}

\begin{figure}
\begin{center}\includegraphics[%
  scale=0.7]{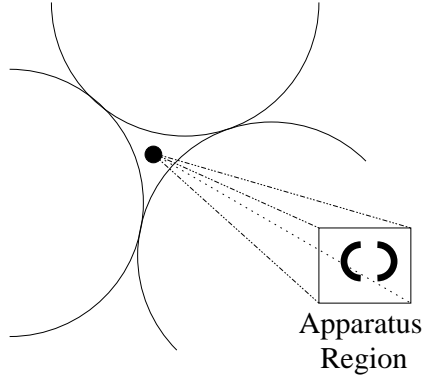}\end{center}

\caption{To deflect away as much possible the vacuum-field radiation emanating
from the laboratory boundaries, the walls, floor and ceiling are constructed
with some optimal curvature to be determined. The apparatus is then
placed within the {}``Apparatus Region.'' \label{cap:casimir-effect-test-lab}}
\end{figure}

The first step in detecting the repulsive Casimir force for a spherical
configuration is to find a conducting material that most closely resembles
the Boyer's material to construct two hemispheres. It has been discussed
previously that even Boyer's sphere can produce attractive Casimir
force when the radiation pressures due to oblique incidence waves
are included in the calculation. Therefore, the geometry of the laboratory
boundaries have to be chosen to deflect away as much as possible the
oblique incident wave as illustrated in Figure \ref{cap:casimir-effect-test-lab}.
Once these conditions are met, the experiment can be conducted in
the region labeled {}``Apparatus Region'' to observe Boyer's repulsive
force. 

\subsection{Outlook}

The Casimir effect has influence in broad range of physics. Here,
we list one such phenomenon known as {}``sonoluminescense,'' and,
finally conclude with the Casimir oscillator.

\subsubsection{Sonoluminescense}

The phenomenon of sonoluminescense remains a poorly understood subject
to date \cite{key-Gaitan-Crum-Church-Roy-Sono-Theory,key-Eberlein}.
When a small air bubble of radius $\sim10^{-3}\, cm$ is injected
into water and subjected to a strong acoustic field of $\sim20\, kHz$
under pressure roughly $\sim1\, atm,$ the bubble emits an intense
flash of light in the optical range, with total energy of roughly
$\sim10^{7}\, eV.$ This emission of light occurs at minimum bubble
radius of roughly $\sim10^{-4}\, cm.$ The flash duration has been
determined to be on the order of $100\, ps$ \cite{key-Gompf-Gunther-Nick-Pecha-Eisen-Sono-Measure,key-Hiller-Putterman-Weninger-Sono-measurement,key-Moran-Sweider-Sono-measure}.
It is to be emphasized that small amounts of noble gases are necessary
in the bubble for sonoluminescense. 

\begin{figure}
\begin{center}\includegraphics[%
  scale=0.7]{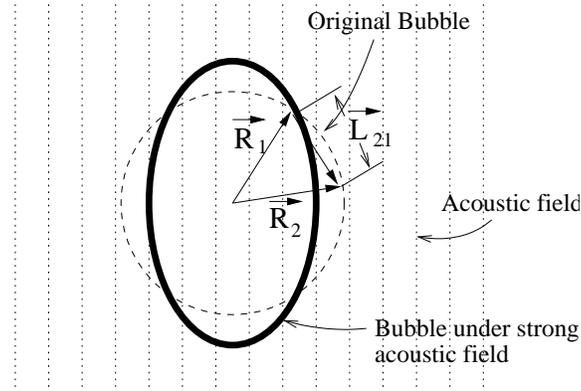}\end{center}

\caption{The original bubble shape shown in dotted lines and the deformed
bubble in solid line under strong acoustic field. \label{cap:sonoluminescense}}
\end{figure}

The bubble in sonoluminescense experiment can be thought of as a deformed
sphere under strong acoustic pressure. The dynamical Casimir effect
arises due to the deformation of the shape; therefore, introducing
a modification to $\vec{L}_{21}=\left\Vert \vec{R}_{2}-\vec{R}_{1}\right\Vert $
from that of the original bubble shape. Here $\vec{L}_{21}$ is the
path length for the reflecting wave in the original bubble shape.
In general $\vec{L}_{21}\equiv\vec{L}_{21}\left(t\right)=\left\Vert \vec{R}_{2}\left(r_{i}\left(t\right),\theta\left(t\right),\phi\left(t\right)\right)-\vec{R}_{1}\left(r_{i}\left(t\right),\theta\left(t\right),\phi\left(t\right)\right)\right\Vert .$
From the relations found in this work for the reflection points $\vec{R}_{1}\left(r_{i}\left(t\right),\theta\left(t\right),\phi\left(t\right)\right)$
and $\vec{R}_{N}\left(r_{i}\left(t\right),\theta\left(t\right),\phi\left(t\right)\right),$
together with the dynamical Casimir force expression of equation (\ref{eq:dynamical-force-L-dot-ONLY-3D-NA}),
the amount of initial radiation energy converted into heat energy
during the deformation process can be found. The bubble deformation
process shown in Figure \ref{cap:sonoluminescense} is a three dimensional
heat generation problem. Current investigation seeks to determine
if the temperature can be raised sufficiently to cause deuterium-tritium
(D-T) fusion to occur, which could provide an alternative approach
to achieve energy generation by this D-T reaction (threshold $\sim17\, KeV$)
\cite{key-A-Prosperetti}. Its theoretical treatment is similar to
that discussed on the \textbf{1D} problem shown in Figure \ref{cap:casimir-plates-outlook}. 

\subsubsection{Casimir Oscillator}

If one can create a laboratory as shown in Figure \ref{cap:casimir-effect-test-lab},
and place in the laboratory hemispheres made of Boyer's material,
then the hemisphere-hemisphere system will execute an oscillatory
motion. When two such hemispheres are separated, the allowed wave
modes in the hemisphere-hemisphere confinement would no longer follow
Boyer's spherical Bessel function restriction. Instead it will be
strictly constrained by the functional relation of $\left\Vert \vec{R}_{2}-\vec{R}_{1}\right\Vert ,$
where $\vec{R}_{1}$ and $\vec{R}_{2}$ are two neighboring reflection
points. Only when the two hemispheres are closed, would the allowed
wave modes obey Boyer's spherical Bessel function restriction. 

\begin{figure}
\begin{center}\includegraphics[%
  scale=0.7]{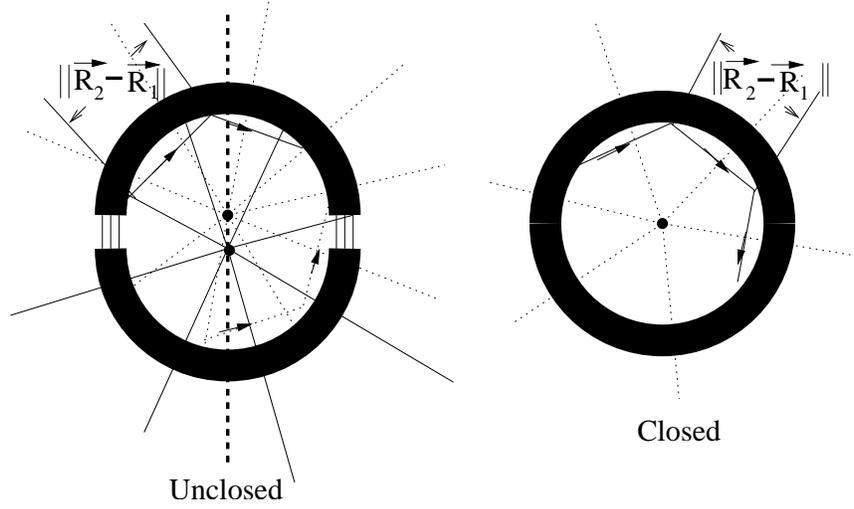}\end{center}

\caption{The vacuum-field radiation energy flows are shown for closed and
unclosed hemispheres. For the hemispheres made of Boyer's material,
the non-radial wave would be absorbed by the hemispheres. \label{cap:cho-casimir-sphere}}
\end{figure}

Assuming that hemispheres are made of Boyer's material and the laboratory
environment is that shown in Figure \ref{cap:casimir-effect-test-lab},
the two closed hemispheres would be repulsing because Boyer's Casimir
force is repulsive. Once the two hemispheres are separated, the allowed
wave modes are governed by the internal reflections at oblique angle
of incidence. Since the hemispheres made of Boyer's material are {}``infinitely
unresponsive'' to oblique incidence waves, all these temporary non-spherical
symmetric waves would be absorbed by the Boyer's hemispheres and the
hemispheres would heat up. The two hemispheres would then attract
each other and the oscillation cycle repeats. Such a mechanical system
may have application. 

\begin{acknowledgments}
I am grateful to Professor Luke W. Mo, from whom I have received so
much help and learned so much physics. The continuing support and
encouragement from Professor J. Ficenec and Mrs. C. Thomas are gracefully
acknowledged. Thanks are due to Professor T. Mizutani for fruitful
discussions which have affected certain aspects of this investigation.
Finally, I express my gratitude for the financial support of the Department
of Physics of Virginia Polytechnic Institute and State University.
\end{acknowledgments}
\appendix

\section*{Appendices on Derivation Details }

The appendices contain our original derivations and developments that
are too tedious and lengthy to be included in the main body of the
paper. There are four appendices: (1) Appendix A, (2) Appendix B,
(3) Appendix C and (4) Appendix D. The appendices C and D are further
divided into subparts C.1, C.2, C.3, D.1 and D.2. The title and the
layout of the appendices closely follow the main body of the paper.

\section{Reflection Points on the Surface of a Resonator}

In this appendix, the original derivations and developments pertaining
to the reflection dynamics used in this paper are included. It is
referenced by the text to supply all the details. 
\vspace{0.25in}

For the configuration shown in Figure \ref{cap:cross-sectional-view-plate-and-hemisphere},
the wave vector directed along an arbitrary direction in Cartesian
coordinates is written as \begin{eqnarray}
\vec{k'}_{1}\left(k'_{1,x},k'_{1,y},k'_{1,z}\right)=\sum_{i=1}^{3}k'_{1,i}\hat{e_{i}}, &  & k'_{1,i}=\left\{ \begin{array}{cc}
i=1\rightarrow k'_{1,x}, & \hat{e_{1}}=\hat{x},\\
\\i=2\rightarrow k'_{1,y}, & \hat{e_{2}}=\hat{y}\\
\\i=3\rightarrow k'_{1,z}, & \hat{e_{3}}=\hat{z}.\end{array}\right.\label{eq:arbitrary-k-vector}\end{eqnarray}
 The unit wave vector is given by \begin{align}
\hat{k'}_{1} & =\left\Vert \vec{k'}_{1}\right\Vert ^{-1}\sum_{i=1}^{3}k'_{1,i}\hat{e_{i}}.\label{eq:normalized-arbitrary-k-vector}\end{align}
 The initial crossing position $\vec{R'}_{0}$ of hemisphere opening
for the incident wave $\vec{k'}_{1}$ is defined as \begin{eqnarray}
\vec{R'}_{0}\left(r'_{0,x},r'_{0,y},r'_{0,z}\right)=\sum_{i=1}^{3}r'_{0,i}\hat{e_{i}}, &  & r'_{0,i}=\left\{ \begin{array}{c}
i=1\rightarrow r'_{0,x},\\
\\i=2\rightarrow r'_{0,y},\\
\\i=3\rightarrow r'_{0,z}.\end{array}\right.\label{eq:initial-arbitrary-k-vector-position}\end{eqnarray}
 It should be noticed here that $\vec{R'}_{0}$ has only two components,
$r'_{0,x}$ and $r'_{0,z}.$ But nevertheless, one can always set
$r'_{0,y}=0$ whenever needed. Since no particular wave with certain
wavelength is prescribed initially, it is desirable to employ a parameterization
scheme to represent these wave vectors. The line segment traced out
by the wave vector $\hat{k'}_{1}$ is formulated in the parametric
form \begin{align}
\vec{R'}_{1} & =\xi_{1}\hat{k'}_{1}+\vec{R'}_{0}=\sum_{i=1}^{3}\left[r'_{0,i}+\xi_{1}\left\Vert \vec{k'}_{1}\right\Vert ^{-1}k'_{1,i}\right]\hat{e_{i}},\label{eq:parametrized-line-segment-generalized}\end{align}
 where the real variable $\xi_{1}$ is a positive definite parameter.
The restriction $\xi_{1}\geq0$ is a necessary condition since the
direction of the wave propagation is set by $\hat{k'}_{1}.$ Here
$\vec{R'}_{1}$ denotes the first reflection point on the hemisphere.
In terms of spherical coordinates, $\vec{R'}_{1}$ takes the form
\begin{eqnarray}
\vec{R'}_{1}\left(r'_{i},\theta'_{1},\phi'_{1}\right)=r'_{i}\sum_{i=1}^{3}\Lambda'_{1,i}\hat{e_{i}}, &  & \left\{ \begin{array}{c}
\Lambda'_{1,1}=\sin\theta'_{1}\cos\phi'_{1},\\
\\\Lambda'_{1,2}=\sin\theta'_{1}\sin\phi'_{1},\\
\\\Lambda'_{1,3}=\cos\theta'_{1},\qquad\;\:\end{array}\right.\label{eq:sphere-vector-generalized}\end{eqnarray}
 where $r'_{i}$ is the hemisphere radius, $\theta'_{1}$ and $\phi'_{1}$
are the polar and azimuthal angles respectively of the first reflection
point $\vec{R'}_{1}.$ The subscript $i$ of $r'_{i}$ denotes {}``inner
radius'' and it is not a summation index. Equations (\ref{eq:parametrized-line-segment-generalized})
and (\ref{eq:sphere-vector-generalized}) are combined as \begin{align}
\sum_{i=1}^{3}\left[r'_{0,i}+\xi_{1}\left\Vert \vec{k'}_{1}\right\Vert ^{-1}k'_{1,i}-r'_{i}\Lambda_{1,i}\right]\hat{e_{i}} & =0.\label{eq:R1-of-zhi-eq-r-of-ri-etc}\end{align}
 Because the basis vectors $\hat{e_{i}}$ are independent of each
other, the above relations are only satisfied when each coefficients
of $\hat{e_{i}}$ vanish independently, \begin{eqnarray}
r'_{0,i}+\xi_{1}\left\Vert \vec{k'}_{1}\right\Vert ^{-1}k'_{1,i}-r'_{i}\Lambda_{1,i}=0, &  & i=1,2,3.\label{eq:xyz-root-condition}\end{eqnarray}
 The three terms $\Lambda_{1,i=1},$ $\Lambda_{1,i=2}$ and $\Lambda_{1,i=3}$
satisfy an identity \begin{align}
\sum_{i=1}^{3}\Lambda_{1,i}^{2} & =1.\label{eq:sphere-cos-sin-identity}\end{align}
 From equation (\ref{eq:xyz-root-condition}), $\Lambda_{1,i}^{2}$
is computed for each $i:$ \begin{eqnarray*}
\Lambda_{1,i}^{2}=\left[r'_{i}\right]^{-2}\left\{ \left[r'_{0,i}\right]^{2}+\xi_{1}^{2}\left\Vert \vec{k'}_{1}\right\Vert ^{-2}\left[k'_{1,i}\right]^{2}+2r'_{0,i}\xi_{1}\left\Vert \vec{k'}_{1}\right\Vert ^{-1}k'_{1,i}\right\} , &  & i=1,2,3.\end{eqnarray*}
 Substituting the above result of $\Lambda_{1,i}^{2}$ into equation
(\ref{eq:sphere-cos-sin-identity}) and after rearrangement, one obtains
\begin{align}
\xi_{1}^{2}\sum_{i=1}^{3}\left\Vert \vec{k'}_{1}\right\Vert ^{-2}\left[k'_{1,i}\right]^{2}+2\xi_{1}\left\Vert \vec{k'}_{1}\right\Vert ^{-1}\sum_{i=1}^{3}r'_{0,i}k'_{1,i}+\sum_{i=1}^{3}\left[r'_{0,i}\right]^{2}-\left[r'_{i}\right]^{2} & =0.\label{eq:root-polynomial-scalar-form}\end{align}
 Further simplifying, it becomes \begin{align}
\xi_{1}^{2}+2\hat{k'}_{1}\cdot\vec{R'}_{0}\xi_{1}+\left\Vert \vec{R'}_{0}\right\Vert ^{2}-\left[r'_{i}\right]^{2} & =0.\label{eq:root-polynomial-vector-form}\end{align}
 There are two roots, \begin{align*}
\xi_{1,a} & =-\hat{k'}_{1}\cdot\vec{R'}_{0}-\sqrt{\left[\hat{k'}_{1}\cdot\vec{R'}_{0}\right]^{2}+\left[r'_{i}\right]^{2}-\left\Vert \vec{R'}_{0}\right\Vert ^{2}}\end{align*}
 and \begin{align*}
\xi_{1,b} & =-\hat{k'}_{1}\cdot\vec{R'}_{0}+\sqrt{\left[\hat{k'}_{1}\cdot\vec{R'}_{0}\right]^{2}+\left[r'_{i}\right]^{2}-\left\Vert \vec{R'}_{0}\right\Vert ^{2}}.\end{align*}
 The root to be used should have a positive value. For the wave reflected
within the hemisphere, $\vec{R'}_{0}\leq r'_{i},$ \begin{align*}
\sqrt{\left[\hat{k'}_{1}\cdot\vec{R'}_{0}\right]^{2}+\left[r'_{i}\right]^{2}-\left\Vert \vec{R'}_{0}\right\Vert ^{2}} & \geq\left|\hat{k'}_{1}\cdot\vec{R'}_{0}\right|\geq-\hat{k'}_{1}\cdot\vec{R'}_{0}\end{align*}
 where the equality $\left|\hat{k'}_{1}\cdot\vec{R'}_{0}\right|=-\hat{k'}_{1}\cdot\vec{R'}_{0}$
happens when $\hat{k'}_{1}\cdot\vec{R'}_{0}\leq0.$ Therefore, $\xi_{1,a}\leq0$
and $\xi_{1,b}\geq0;$ the positive root $\xi_{1,b}$ should be selected.
For bookkeeping purposes, $\xi_{1,b}$ is designated as $\xi_{1,p}:$
\begin{align}
\xi_{1,p} & =-\hat{k'}_{1}\cdot\vec{R'}_{0}+\sqrt{\left[\hat{k'}_{1}\cdot\vec{R'}_{0}\right]^{2}+\left[r'_{i}\right]^{2}-\left\Vert \vec{R'}_{0}\right\Vert ^{2}}.\label{eq:positive-root-k}\end{align}
 Using this positive root, the first reflection point of the inner
hemisphere is found to be \begin{align}
\vec{R'}_{1}\left(\xi_{1,p};\vec{R'}_{0},\hat{k'}_{1}\right) & =\sum_{i=1}^{3}\left[r'_{0,i}+\xi_{1,p}\left\Vert \vec{k'}_{1}\right\Vert ^{-1}k'_{1,i}\right]\hat{e_{i}}.\label{eq:1st-bounce-off-point}\end{align}

\begin{figure}
\begin{center}\includegraphics[%
  scale=0.7]{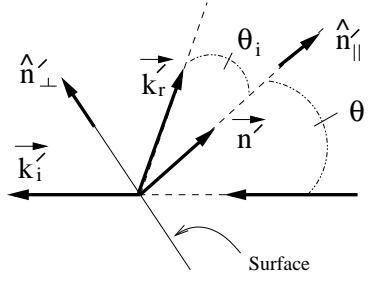}\end{center}

\caption{A simple reflection of incoming wave $\vec{k'}_{i}$ from the surface
defined by a local normal $\vec{n'}.$ \label{cap:Ki-Kr-N-Relation}}
\end{figure}

The incident wave $\vec{k'}_{i},$ shown in Figure \ref{cap:Ki-Kr-N-Relation}
and where $i$ here stands for incident wave, can always be decomposed
into components parallel and perpendicular to the vector $\vec{n'}$
normal to the reflecting surface, \begin{align*}
\vec{k'}_{i} & =\vec{k'}_{i,\parallel}+\vec{k'}_{i,\perp}=\frac{\vec{n'}\cdot\vec{k'}_{i}}{\vec{n'}\cdot\vec{n'}}\vec{n'}+\frac{\left[\vec{n'}\times\vec{k'}_{i}\right]\times\vec{n'}}{\vec{n'}\cdot\vec{n'}}.\end{align*}
 If the local normal $\hat{n'}$ is already normalized to unity, the
above expression reduces to \begin{align}
\vec{k'}_{i} & =\hat{n'}\cdot\vec{k'}_{i}\hat{n'}+\left[\hat{n'}\times\vec{k'}_{i}\right]\times\hat{n'}.\label{eq:K-incident-in-PARA-n-PERPE-N}\end{align}
 Here the angle between $\vec{k'}_{i}$ and $\hat{n'}$ is $\pi-\theta_{i}.$
The action of reflection only modifies $\vec{k'}_{i,\parallel}$ in
the reflected wave. The reflected wave part of $\vec{k'}_{i}$ in
equation (\ref{eq:K-incident-in-PARA-n-PERPE-N}) is \begin{align}
\vec{k'}_{r} & =\alpha_{r,\perp}\vec{k'}_{i,\perp}-\alpha_{r,\parallel}\vec{k'}_{i,\parallel}=\alpha_{r,\perp}\left[\hat{n'}\times\vec{k'}_{i}\right]\times\hat{n'}-\alpha_{r,\parallel}\hat{n'}\cdot\vec{k'}_{i}\hat{n'},\label{eq:K-reflected-in-PARA-n-PERPE-N}\end{align}
 where $\vec{k'}_{i,\parallel}$ have been rotated by $180^{o}$ on
the plane of incidence. The new quantities $\alpha_{r,\parallel}$
and $\alpha_{r,\perp}$ are the reflection coefficients. For a perfect
reflecting surfaces, $\alpha_{r,\parallel}=\alpha_{r,\perp}=1.$ Because
of the frequent usage of the component for $\vec{k'}_{r},$ equation
(\ref{eq:K-reflected-in-PARA-n-PERPE-N}) is also written in component
form. The component of the double cross product $\left[\hat{n'}\times\vec{k'}_{i}\right]\times\hat{n'}$
is computed first, \begin{align*}
\left\{ \left[\hat{n'}\times\vec{k'}_{i}\right]\times\hat{n'}\right\} _{l} & =\epsilon_{lmn}\left[\hat{n'}\times\vec{k'}_{i}\right]_{m}n'_{n}=\epsilon_{lmn}\epsilon_{mqr}n'_{q}k'_{i,r}n'_{n}=\epsilon_{nlm}\epsilon_{qrm}n'_{q}k'_{i,r}n'_{n}\\
 & =\left[\delta_{nq}\delta_{lr}-\delta_{nr}\delta_{lq}\right]n'_{q}k'_{i,r}n'_{n}=\delta_{nq}\delta_{lr}n'_{q}k'_{i,r}n'_{n}-\delta_{nr}\delta_{lq}n'_{q}k'_{i,r}n'_{n}\\
 & =n'_{n}k'_{i,l}n'_{n}-n'_{l}k'_{i,n}n'_{n}\end{align*}
 or \begin{align}
\left[\hat{n'}\times\vec{k'}_{i}\right]\times\hat{n'} & =\sum_{l=1}^{3}\left[n'_{n}k'_{i,l}n'_{n}-n'_{l}k'_{i,n}n'_{n}\right]\hat{e_{l}},\label{eq:n-cross-k-cross-n}\end{align}
 where the summation over the index $n$ is implicit. In component
form, $\vec{k'}_{r}$ is hence expressed as \begin{align}
\vec{k'}_{r} & =\sum_{l=1}^{3}\left\{ \alpha_{r,\perp}\left[n'_{n}k'_{i,l}n'_{n}-n'_{l}k'_{i,n}n'_{n}\right]-\alpha_{r,\parallel}n'_{n}k'_{i,n}n'_{l}\right\} \hat{e_{l}},\label{eq:K-reflected-in-PARA-n-PERPE-N-COMPONENT}\end{align}
 where it is understood $\hat{n'}$ is already normalized. 

The second reflection point $\vec{R'}_{2}$ is found by repeating
the steps done for $\vec{R'}_{1},$ \begin{align*}
\vec{R'}_{2} & =\vec{R'}_{1}+\xi_{2,p}\hat{k'}_{r}=\vec{R'}_{1}+\xi_{2,p}\frac{\alpha_{r,\perp}\left[\hat{n'}\times\vec{k'}_{i}\right]\times\hat{n'}-\alpha_{r,\parallel}\hat{n'}\cdot\vec{k'}_{i}\hat{n'}}{\left\Vert \alpha_{r,\perp}\left[\hat{n'}\times\vec{k'}_{i}\right]\times\hat{n'}-\alpha_{r,\parallel}\hat{n'}\cdot\vec{k'}_{i}\hat{n'}\right\Vert },\end{align*}
 where $\xi_{2,p}$ is the new positive parameter corresponding to
the second reflection point. The procedure can be repeated for any
reflection point. Although this technique is sound, it can be seen
immediately that the technique suffers from the lack of elegance.
For this reason, the scalar field technique will be exclusively used
in studying the reflection dynamics. For a simple plane, the scalar
field function can be inferred rather intuitively. However, for more
complex surfaces, one has to work it out to get the corresponding
scalar field. For the purpose of generalization of the technique to
any arbitrary surfaces, we derive the scalar field functional for
the plane in great detail. 

In simple reflection dynamics, there exists a plane of incidence in
which all reflections occur. The plane of incidence is determined
by the incident wave $\vec{k'}_{i}$ and the local surface normal
$\vec{n'}_{i}.$ For the system shown in Figure \ref{cap:cross-sectional-view-plate-and-hemisphere},
$\vec{k'}_{i}$ and $\vec{n'}_{i}$ are given by \begin{eqnarray*}
\vec{k'}_{i}=\vec{k'}_{1}, &  & \vec{n'}_{n'_{i},1}\equiv-\vec{R'}_{1}\left(\xi_{1,p};\vec{R'}_{0},\hat{k'}_{1}\right)=-\xi_{1,p}\hat{k'}_{1}-\vec{R'}_{0}.\end{eqnarray*}
 The normal to the incidence plane is characterized by the cross product,
\begin{align*}
\vec{n'}_{p,1} & =\vec{k'}_{1}\times\vec{n'}_{n'_{i},1}=\vec{k'}_{1}\times\left[-\xi_{1,p}\hat{k'}_{1}-\vec{R'}_{0}\right]=-\vec{k'}_{1}\times\vec{R'}_{0}=-\sum_{i=1}^{3}\epsilon_{ijk}k'_{1,j}r'_{0,k}\hat{e_{i}},\end{align*}
 where the summations over the indices $j$ and $k$ are implicit.
The normal to the incidence plane is normalized as \begin{align}
\hat{n'}_{p,1} & =-\left\Vert \vec{n'}_{p,1}\right\Vert ^{-1}\sum_{i=1}^{3}\epsilon_{ijk}k'_{1,j}r'_{0,k}\hat{e_{i}}.\label{eq:n-hat-p-i-eq-mag-times-ki-cross-Ro}\end{align}

In order to take advantage of the information given above, the concept
of scalar fields in mathematical sense is in order. A functional $f\left(x',y',z'\right)$
is a scalar field if to each point $\left(x',y',z'\right)$ of a region
in space, there corresponds a number $\lambda.$ The study of a scalar
field is a study of scalar valued functions of three variables. Scalar
fields are connected to its normals, e.g., equation (\ref{eq:n-hat-p-i-eq-mag-times-ki-cross-Ro}),
through the relation \begin{eqnarray}
\hat{n'}_{p,1}\propto\vec{\nabla'}f_{p,1}\left(x',y',z'\right)=\sum_{i=1}^{3}\hat{e_{i}}\frac{\partial}{\partial\nu'_{i}}f_{p,1}\left(x',y',z'\right), &  & i=\left\{ \begin{array}{c}
1\rightarrow\nu'_{1}=x',\\
\\2\rightarrow\nu'_{2}=y',\\
\\3\rightarrow\nu'_{3}=z'.\end{array}\right.\label{eq:n-hat-p-i-prop-beta-del-f-p-i}\end{eqnarray}
 Introducing a constant proportionality factor $\beta_{p,1},$ equation
(\ref{eq:n-hat-p-i-prop-beta-del-f-p-i}) becomes \begin{align}
\hat{n'}_{p,1} & =\beta_{p,1}\sum_{i=1}^{3}\hat{e_{i}}\frac{\partial}{\partial\nu'_{i}}f_{p,1}\left(x',y',z'\right).\label{eq:n-hat-p-i-eq-beta-del-f-p-i}\end{align}
 The proportionality factor $\beta_{p,1}$ is intrinsically connected
to the normalization of $\vec{\nabla'}f_{p,1}.$ Because the vector
$\hat{n'}_{p,1}$ is a unit vector, its magnitude squared is \begin{eqnarray*}
\beta_{p,1}^{2}\sum_{i=1}^{3}\left[\frac{\partial}{\partial\nu'_{i}}f_{p,1}\left(x',y',z'\right)\right]^{2}=1 & \rightarrow & \beta_{p,1}=\pm\left\{ \sum_{i=1}^{3}\left[\frac{\partial}{\partial\nu'_{i}}f_{p,1}\left(x',y',z'\right)\right]^{2}\right\} ^{-1/2}.\end{eqnarray*}
 In equation (\ref{eq:n-hat-p-i-eq-beta-del-f-p-i}), the directions
for vectors $\hat{n'}_{p,1}$ and $\vec{\nabla'}f_{p,1}$ are intrinsically
built in. Therefore, the proportionality factor $\beta_{p,1}$ has
to be a positive quantity, \begin{align}
\beta_{p,1} & =\left\{ \sum_{i=1}^{3}\left[\frac{\partial}{\partial\nu'_{i}}f_{p,1}\left(x',y',z'\right)\right]^{2}\right\} ^{-1/2}.\label{eq:n-hat-p-i-eq-beta-del-f-p-i-The-Beta-Def-Pre}\end{align}
 The exact form of the proportionality coefficient $\beta_{p,1}$
requires the knowledge of $f_{p,1},$ which is yet to be determined.
However, we can use it formally for now until the solution for $f_{p,1}$
is found. 

Substituting the gradient function $\vec{\nabla'}f_{p,1}$ into equation
(\ref{eq:n-hat-p-i-eq-beta-del-f-p-i}), and using equations (\ref{eq:n-hat-p-i-eq-mag-times-ki-cross-Ro})
and (\ref{eq:n-hat-p-i-eq-beta-del-f-p-i}), one arrives at \begin{eqnarray}
\sum_{i=1}^{3}\left[\frac{\partial}{\partial\nu'_{i}}f_{p,1}\left(x',y',z'\right)+\beta_{p,1}^{-1}\left\Vert \vec{n'}_{p,1}\right\Vert ^{-1}\epsilon_{ijk}k'_{1,j}r'_{0,k}\right]\hat{e_{i}}=0, &  & i=\left\{ \begin{array}{c}
1\rightarrow\nu'_{1}=x',\\
\\2\rightarrow\nu'_{2}=y',\\
\\3\rightarrow\nu'_{3}=z'.\end{array}\right.\label{eq:n-hat-p-i-eq-beta-del-f-p-i-poly}\end{eqnarray}
 Because the basis vectors $\hat{e_{i}}$ are linearly independent,
the equation for each component is obtained as \begin{eqnarray}
\frac{\partial}{\partial\nu'_{i}}f_{p,1}\left(\alpha,\beta,\gamma\right)+\beta_{p,1}^{-1}\left\Vert \vec{n'}_{p,1}\right\Vert ^{-1}\epsilon_{ijk}k'_{1,j}r'_{0,k}=0, &  & i=\left\{ \begin{array}{c}
1\rightarrow\nu'_{1}=x'=\alpha,\\
\\2\rightarrow\nu'_{2}=y'=\beta,\\
\\3\rightarrow\nu'_{3}=z'=\gamma.\end{array}\right.\label{eq:n-hat-p-i-eq-beta-del-f-p-i-poly-comp}\end{eqnarray}
 Integrating both sides of equation (\ref{eq:n-hat-p-i-eq-beta-del-f-p-i-poly-comp})
over the variable $\nu'_{i}=\alpha,$ \begin{align*}
\int_{\alpha_{0}}^{\alpha}\frac{\partial}{\partial\alpha'}f_{p,1}\left(\alpha',\beta,\gamma\right)d\alpha' & =-\int_{\alpha_{0}}^{\alpha}\beta_{p,1}^{-1}\left\Vert \vec{n'}_{p,1}\right\Vert ^{-1}\epsilon_{\alpha'jk}k'_{1,j}r'_{0,k}d\alpha',\end{align*}
 where the dummy variable $\alpha'$ is introduced for integration
purpose. The terms $\epsilon_{\alpha'jk}k'_{1,j}r'_{0,k},$ $\beta_{p,1}$
and $\left\Vert \vec{n'}_{p,1}\right\Vert $ are independent of the
dummy variable $\alpha',$ and they can be moved out of the integrand,
\begin{align}
\int_{\alpha_{0}}^{\alpha}\frac{\partial}{\partial\alpha'}f_{p,1}\left(\alpha',\beta,\gamma\right)d\alpha' & =-\beta_{p,1}^{-1}\left\Vert \vec{n'}_{p,1}\right\Vert ^{-1}\epsilon_{\alpha jk}k'_{1,j}r'_{0,k}\int_{\alpha_{0}}^{\alpha}d\alpha'.\label{eq:poly-2-x-inte}\end{align}
 Because the total differential of $f_{p,1}$ is given by \begin{eqnarray}
df_{p,1}=\frac{\partial f_{p,1}}{\partial\alpha'}d\alpha'+\frac{\partial f_{p,1}}{\partial\beta}d\beta+\frac{\partial f_{p,1}}{\partial\gamma}d\gamma, &  & \alpha'\neq\beta\neq\gamma,\label{eq:total-diff-of-f-p-i}\end{eqnarray}
 the term $\left[\partial f_{p,1}/\partial\alpha'\right]d\alpha'$
can be written as \begin{align*}
\frac{\partial f_{p,1}}{\partial\alpha'}d\alpha' & =df_{p,1}-\frac{\partial f_{p,1}}{\partial\beta}d\beta-\frac{\partial f_{p,1}}{\partial\gamma'}d\gamma.\end{align*}
 The integration over the variable $\nu'_{i}=\alpha$ in equation
(\ref{eq:poly-2-x-inte}), with variables $\nu'_{i}\neq\alpha$ fixed,
can be carried out with \begin{eqnarray}
d\beta=d\gamma=0, &  & \frac{\partial}{\partial\alpha'}f_{p,1}\left(\alpha',\beta,\gamma\right)d\alpha'=df_{p,1}\left(\alpha',\beta,\gamma\right)\label{eq:parf-parxp-eq-total-diff-of-f-p-i}\end{eqnarray}
 as \begin{align*}
\int_{\alpha_{0}}^{\alpha}df_{p,1}\left(\alpha',\beta,\gamma\right) & =-\beta_{p,1}^{-1}\left\Vert \vec{n'}_{p,1}\right\Vert ^{-1}\epsilon_{\alpha jk}k'_{1,j}r'_{0,k}\int_{\alpha_{0}}^{\alpha}d\alpha'\end{align*}
 to give \begin{align}
f_{p,1}\left(\alpha,\beta,\gamma\right) & =\beta_{p,1}^{-1}\left\Vert \vec{n'}_{p,1}\right\Vert ^{-1}\epsilon_{\alpha jk}k'_{1,j}r'_{0,k}\left[\alpha_{0}-\alpha\right]+f_{p,1}\left(\alpha_{0},\beta,\gamma\right).\label{eq:f-p-i-of-x-y-z-eq-common-plus-h2-h1}\end{align}
 The two terms $\left[\epsilon_{\alpha jk}k'_{1,j}r'_{0,k}/\left\{ \beta_{p,1}\left\Vert \vec{n'}_{p,1}\right\Vert \right\} \right]\alpha_{0}$
and $f_{p,1}\left(\alpha_{0},\beta,\gamma\right)$ are independent
of $\alpha.$ These terms can only assume values of $\nu'_{i}=\beta$
or $\nu'_{i}=\gamma$. By re-designating $\alpha$ independent terms,
\begin{align}
h_{p,1}\left(\beta,\gamma\right) & =\beta_{p,1}^{-1}\left\Vert \vec{n'}_{p,1}\right\Vert ^{-1}\epsilon_{\alpha jk}k'_{1,j}r'_{0,k}\alpha_{0}+f_{p,1}\left(\alpha_{0},\beta,\gamma\right),\label{eq:h-p-i-of-y-z-eq-h0-eq-h2-h1}\end{align}
 equation (\ref{eq:f-p-i-of-x-y-z-eq-common-plus-h2-h1}) can be rewritten
for bookkeeping purposes as \begin{align}
f_{p,1}\left(\alpha,\beta,\gamma\right) & =h_{p,1}\left(\beta,\gamma\right)-\beta_{p,1}^{-1}\left\Vert \vec{n'}_{p,1}\right\Vert ^{-1}\epsilon_{\alpha jk}k'_{1,j}r'_{0,k}\alpha.\label{eq:f-p-i-of-x-y-z-eq-common-plus-h-p-i-of-y-z}\end{align}
 Substituting the result into equation (\ref{eq:n-hat-p-i-eq-beta-del-f-p-i-poly-comp}),
a differentiation with respect to the variable $\nu'_{i}=\beta$ gives
\begin{align*}
\frac{\partial}{\partial\beta}\left[h_{p,1}\left(\beta,\gamma\right)-\beta_{p,1}^{-1}\left\Vert \vec{n'}_{p,1}\right\Vert ^{-1}\epsilon_{\alpha jk}k'_{1,j}r'_{0,k}\alpha\right]+\beta_{p,1}^{-1}\left\Vert \vec{n'}_{p,1}\right\Vert ^{-1}\epsilon_{\beta jk}k'_{1,j}r'_{0,k} & =0\end{align*}
 or \begin{align*}
\frac{\partial}{\partial\beta}h_{p,1}\left(\beta,\gamma\right) & =-\beta_{p,1}^{-1}\left\Vert \vec{n'}_{p,1}\right\Vert ^{-1}\epsilon_{\beta jk}k'_{1,j}r'_{0,k}.\end{align*}
 The integration of both sides with respect to the variable $\nu'_{i}=\beta$
yields the result \begin{align*}
\int_{\beta_{0}}^{\beta}\frac{\partial}{\partial\beta'}h_{p,1}\left(\beta',\gamma\right)d\beta' & =-\int_{\beta_{0}}^{\beta}\beta_{p,1}^{-1}\left\Vert \vec{n'}_{p,1}\right\Vert ^{-1}\epsilon_{\beta'jk}k'_{1,j}r'_{0,k}d\beta'=-\beta_{p,1}^{-1}\left\Vert \vec{n'}_{p,1}\right\Vert ^{-1}\epsilon_{\beta jk}k'_{1,j}r'_{0,k}\int_{\beta_{0}}^{\beta}d\beta',\end{align*}
 where the dummy variable $\beta'$ is introduced for integration
purpose and the terms $\epsilon_{\beta jk}k'_{1,j}r'_{0,k},$ $\beta_{p,1}$
and $\left\Vert \vec{n'}_{p,1}\right\Vert $ have been taken out of
the integrand because they are independent of $\beta'.$ Following
the same procedure used in equations (\ref{eq:total-diff-of-f-p-i})
through (\ref{eq:parf-parxp-eq-total-diff-of-f-p-i}), the integrand
$\left[\partial h_{p,1}/\partial\beta'\right]d\beta'$ on the left
hand side of the integral is \begin{align*}
\frac{\partial}{\partial\beta'}h_{p,1}\left(\beta',\gamma\right)d\beta' & =dh_{p,1}\left(\beta',\gamma\right).\end{align*}
 Consequently, $h_{p,1}\left(\beta,\gamma\right)$ is given by \begin{align}
h_{p,1}\left(\beta,\gamma\right) & =\beta_{p,1}^{-1}\left\Vert \vec{n'}_{p,1}\right\Vert ^{-1}\epsilon_{\beta jk}k'_{1,j}r'_{0,k}\left[\beta_{0}-\beta\right]+h_{p,1}\left(\beta_{0},\gamma\right).\label{eq:h-p-i-of-y-z-eq-common-plus-g2-g1}\end{align}
 The two terms $\left[\epsilon_{\beta jk}k'_{1,j}r'_{0,k}/\left\{ \beta_{p,1}\left\Vert \vec{n'}_{p,1}\right\Vert \right\} \right]\beta_{0}$
and $h_{p,1}\left(\beta_{0},\gamma\right)$ are independent of $\beta.$
The $\beta$ independent terms can be re-designated as \begin{align}
g_{p,1}\left(\gamma\right) & =\beta_{p,1}^{-1}\left\Vert \vec{n'}_{p,1}\right\Vert ^{-1}\epsilon_{\beta jk}k'_{1,j}r'_{0,k}\beta_{0}+h_{p,1}\left(\beta_{0},\gamma\right).\label{eq:g-p-i-of-z-eq-g0-eq-g2-g1}\end{align}
 For bookkeeping purposes, equation (\ref{eq:h-p-i-of-y-z-eq-common-plus-g2-g1})
is rewritten as \begin{align}
h_{p,1}\left(\beta,\gamma\right) & =g_{p,1}\left(\gamma\right)-\beta_{p,1}^{-1}\left\Vert \vec{n'}_{p,1}\right\Vert ^{-1}\epsilon_{\beta jk}k'_{1,j}r'_{0,k}\beta.\label{eq:h-p-i-of-y-z-eq-common-plus-g-p-i-of-z}\end{align}
 Substitution of $h_{p,1}\left(\beta,\gamma\right)$ into equation
(\ref{eq:f-p-i-of-x-y-z-eq-common-plus-h-p-i-of-y-z}) gives \begin{align}
f_{p,1}\left(\alpha,\beta,\gamma\right) & =g_{p,1}\left(\gamma\right)-\beta_{p,1}^{-1}\left\Vert \vec{n'}_{p,1}\right\Vert ^{-1}\left[\epsilon_{\beta jk}k'_{1,j}r'_{0,k}\beta+\epsilon_{\alpha jk}k'_{1,j}r'_{0,k}\alpha\right].\label{eq:f-p-i-of-x-y-z-eq-common-plus-h-p-i-of-y-z-sub-g}\end{align}
 Substituting $f_{p,1}\left(\alpha,\beta,\gamma\right)$ into equation
(\ref{eq:n-hat-p-i-eq-beta-del-f-p-i-poly-comp}) once more, and performing
the differentiation with respect to the variable $\nu'_{i}=\gamma,$
where $\gamma\neq\alpha\neq\beta,$ we obtain \begin{align*}
\frac{d}{d\gamma}\left[g_{p,1}\left(\gamma\right)-\beta_{p,1}^{-1}\left\Vert \vec{n'}_{p,1}\right\Vert ^{-1}\left\{ \epsilon_{\beta jk}k'_{1,j}r'_{0,k}\beta+\epsilon_{\alpha jk}k'_{1,j}r'_{0,k}\right\} \alpha\right]+\beta_{p,1}^{-1}\left\Vert \vec{n'}_{p,1}\right\Vert ^{-1}\epsilon_{\gamma jk}k'_{1,j}r'_{0,k} & =0\end{align*}
 or \begin{align*}
\frac{d}{d\gamma}g_{p,1}\left(\gamma\right) & =-\beta_{p,1}^{-1}\left\Vert \vec{n'}_{p,1}\right\Vert ^{-1}\epsilon_{\gamma jk}k'_{1,j}r'_{0,k},\end{align*}
 where the differentiation have been changed from $\partial$ to $d$
because $g_{p,1}$ is a function of single variable. The integration
of both sides with respect to the variable $\nu'_{i}=\gamma$ then
gives \begin{align*}
\int_{\gamma_{0}}^{\gamma}\frac{d}{d\gamma'}g_{p,1}\left(\gamma'\right)d\gamma' & =-\int_{\gamma_{0}}^{\gamma}\beta_{p,1}^{-1}\left\Vert \vec{n'}_{p,1}\right\Vert ^{-1}\epsilon_{\gamma'jk}k'_{1,j}r'_{0,k}d\gamma'=-\beta_{p,1}^{-1}\left\Vert \vec{n'}_{p,1}\right\Vert ^{-1}\epsilon_{\gamma jk}k'_{1,j}r'_{0,k}\int_{\gamma_{0}}^{\gamma}d\gamma',\end{align*}
 where the dummy variable $\gamma'$ have been introduced for integration
purpose and the terms $\epsilon_{\gamma jk}k'_{1,j}r'_{0,k},$ $\beta_{p,1}$
and $\left\Vert \vec{n'}_{p,1}\right\Vert $ have been taken out of
the integrand because they are independent of $\gamma'.$ Knowing
$\left[dg_{p,1}/d\gamma'\right]d\gamma'=dg_{p,1},$ the integration
is carried out to yield \begin{align}
g_{p,1}\left(\gamma\right) & =\beta_{p,1}^{-1}\left\Vert \vec{n'}_{p,1}\right\Vert ^{-1}\epsilon_{\gamma jk}k'_{1,j}r'_{0,k}\left[\gamma_{0}-\gamma\right]+g_{p,1}\left(\gamma_{0}\right).\label{eq:g-p-i-of-z-eq-common-plus-b2-b1}\end{align}
 The two terms $\left[\epsilon_{\gamma jk}k'_{1,j}r'_{0,k}/\left\{ \beta_{p,1}\left\Vert \vec{n'}_{p,1}\right\Vert \right\} \right]\gamma_{0}$
and $g_{p,1}\left(\gamma_{0}\right)$ are independent of $\gamma.$
The $\gamma$ independent terms are re-designated as \begin{align}
b_{0} & =\beta_{p,1}^{-1}\left\Vert \vec{n'}_{p,1}\right\Vert ^{-1}\epsilon_{\gamma jk}k'_{1,j}r'_{0,k}\gamma_{0}+g_{p,1}\left(\gamma_{0}\right).\label{eq:b0-eq-b2-b1}\end{align}
 For bookkeeping purposes, equation (\ref{eq:g-p-i-of-z-eq-common-plus-b2-b1})
is rewritten as \begin{align}
g_{p,1}\left(\gamma\right) & =b_{0}-\beta_{p,1}^{-1}\left\Vert \vec{n'}_{p,1}\right\Vert ^{-1}\epsilon_{\gamma jk}k'_{1,j}r'_{0,k}\gamma.\label{eq:g-p-i-of-z-eq-common-plus-b0}\end{align}
 Substituting $g_{p,1}\left(\gamma\right)$ in equation (\ref{eq:f-p-i-of-x-y-z-eq-common-plus-h-p-i-of-y-z-sub-g}),
the result for $f_{p,1}\left(\alpha,\beta,\gamma\right)$ is found
to be \begin{eqnarray}
f_{p,1}\left(\alpha,\beta,\gamma\right)=b_{0}-\beta_{p,1}^{-1}\left\Vert \vec{n'}_{p,1}\right\Vert ^{-1}\sum_{i=1}^{3}\epsilon_{ijk}k'_{1,j}r'_{0,k}\nu'_{i}, &  & i=\left\{ \begin{array}{c}
1\rightarrow\nu'_{1}=\alpha=x',\\
\\2\rightarrow\nu'_{2}=\beta=y',\\
\\3\rightarrow\nu'_{3}=\gamma=z'.\end{array}\right.\label{eq:f-p-i-of-x-y-z-result}\end{eqnarray}
 The cross product expressed in terms of the Levi-Civita symbol is
expanded to give \begin{align}
\epsilon_{x'jk}k'_{1,j}r'_{0,k} & =k'_{1,j=y'}r'_{0,k=z'}-k'_{1,k=z'}r'_{0,j=y'},\label{eq:Levi-Civita-x-prime}\end{align}
 \begin{align}
\epsilon_{y'jk}k'_{1,j}r'_{0,k} & =k'_{1,j=z'}r'_{0,k=x'}-k'_{1,k=x'}r'_{0,k=z'},\label{eq:Levi-Civita-y-prime}\end{align}
 \begin{align}
\epsilon_{z'jk}k'_{1,j}r'_{0,k} & =k'_{1,j=x'}r'_{0,k=y'}-k'_{1,k=y'}r'_{0,j=x'}.\label{eq:Levi-Civita-z-prime}\end{align}
 It is important to understand that the functional $f_{p,1}$ in equation
(\ref{eq:f-p-i-of-x-y-z-result}) is a scalar field description of
an infinite family of parallel planes characterized by the normal
given in equation (\ref{eq:n-hat-p-i-eq-mag-times-ki-cross-Ro}),
\begin{align*}
\hat{n'}_{p,1} & =-\left\Vert \vec{n'}_{p,1}\right\Vert ^{-1}\sum_{i=1}^{3}\epsilon_{ijk}k'_{1,j}r'_{0,k}\hat{e_{i}}.\end{align*}
 Because the normal $\hat{n'}_{p,1}$ is a cross product of the two
vectors $\vec{k'}_{1}$ and $\vec{R'}_{0},$ the surface represented
by the scalar field $f_{p,1}$ is a plane spanned by all the scattered
wave vectors. %
\begin{figure}
\begin{center}\includegraphics[%
  scale=0.7]{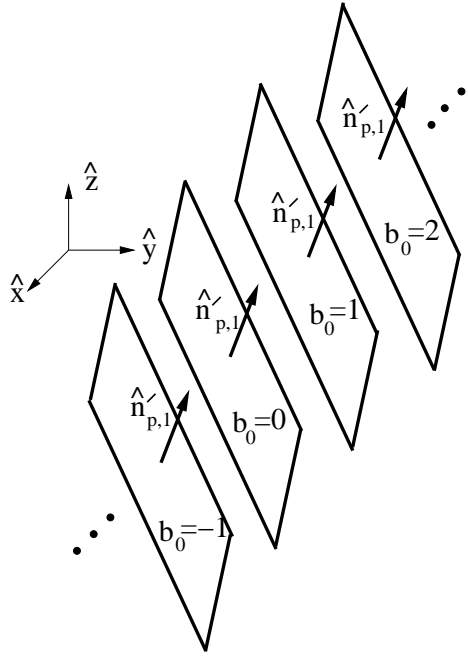}\end{center}

\caption{Parallel planes characterized by a normal $\hat{n'}_{p,1}=-\left\Vert \vec{n'}_{p,1}\right\Vert ^{-1}\sum_{i=1}^{3}\epsilon_{ijk}k'_{1,j}r'_{0,k}\hat{e_{i}}.$
\label{cap:scalar-fields-planes}}
\end{figure}
 The graphical plot of the functional $f_{p,1}$ is illustrated in
Figure \ref{cap:scalar-fields-planes}. The three coefficients, \[
\begin{array}{ccccc}
\epsilon_{\alpha jk}k'_{1,j}r'_{0,k}, &  & \epsilon_{\beta jk}k'_{1,j}r'_{0,k}, &  & \epsilon_{\gamma jk}k'_{1,j}r'_{0,k},\end{array}\]
 of the independent variables $\alpha,$ $\beta$ and $\gamma$ define
the slopes along the respective bases $\hat{\alpha},$ $\hat{\beta}$
and $\hat{\gamma}.$ The integration constant $b_{0}$ provides infinite
set of parallel planes whose common normal is $\hat{n'}_{p,1}.$ For
what is concerned with here, only one of them containing the coordinate
origin is required. It is convenient to choose the plane with $b_{0}=0.$
With the plane of $b_{0}=0$ chosen, the scalar field of equation
(\ref{eq:f-p-i-of-x-y-z-result}) is rewritten for the sake of bookkeeping
purposes: \begin{eqnarray}
f_{p,1}\left(\alpha,\beta,\gamma\right)=-\beta_{p,1}^{-1}\left\Vert \vec{n'}_{p,1}\right\Vert ^{-1}\sum_{i=1}^{3}\epsilon_{ijk}k'_{1,j}r'_{0,k}\nu'_{i}, &  & i=\left\{ \begin{array}{c}
1\rightarrow\nu'_{1}=\alpha=x',\\
\\2\rightarrow\nu'_{2}=\beta=y',\\
\\3\rightarrow\nu'_{3}=\gamma=z'.\end{array}\right.\label{eq:f-p-i-plane-of-incidence-Pre}\end{eqnarray}
 where $-\infty\leq\left\{ \alpha,\beta,\gamma\right\} \leq\infty.$ 

As mentioned before, $f_{p,1}$ of equation (\ref{eq:f-p-i-plane-of-incidence-Pre})
is a scalar field where $\vec{k'}_{1}$ and $\vec{R'}_{0}$ are the
two initially known vectors which span locally the reflection surface.
Other than $\vec{k'}_{1}$ and $\vec{R'}_{0},$ any member vectors
of the spanning set for the incidence plane can also be used to determine
the orientation of a surface. Then it is always true that \begin{align}
f_{p,1}\left(\alpha,\beta,\gamma;\vec{k'}_{1},\vec{R'}_{0}\right) & =f_{p,2}\left(\alpha,\beta,\gamma;\vec{k'}_{2},\vec{R'}_{1}\right)=\cdots=f_{p,N}\left(\alpha,\beta,\gamma;\vec{k'}_{N},\vec{R'}_{N}\right),\label{eq:f-p-i-plane-of-incidence-equality-conditions}\end{align}
 where the integer index $N$ of $\vec{k'}_{N}$ is used to enumerate
the sequence of reflections; and the integer index $N$ of $\vec{R'}_{N}$
is used to enumerate the sequence of reflection points. 

What is still yet undetermined in equation (\ref{eq:f-p-i-plane-of-incidence-Pre})
is the proportionality factor $\beta_{p,1}.$ From equation (\ref{eq:n-hat-p-i-eq-beta-del-f-p-i-The-Beta-Def-Pre}),
the $\beta_{p,1}$ has an algebraic definition \begin{eqnarray*}
\beta_{p,1}=\left\{ \sum_{i=1}^{3}\left[\frac{\partial}{\partial\nu'_{i}}f_{p,1}\left(\alpha,\beta,\gamma\right)\right]^{2}\right\} ^{-1/2}, &  & i=\left\{ \begin{array}{c}
1\rightarrow\nu'_{1}=\alpha,\\
\\2\rightarrow\nu'_{2}=\beta,\\
\\3\rightarrow\nu'_{3}=\gamma.\end{array}\right.\end{eqnarray*}
 The partial derivatives $\partial f_{p,1}/\partial\nu'_{i}$ are
calculated with the solution of $f_{p,1}$ in equation (\ref{eq:f-p-i-plane-of-incidence-Pre}),
\begin{align*}
\frac{\partial}{\partial\nu'_{i}}f_{p,1}\left(\alpha,\beta,\gamma\right) & =-\beta_{p,1}^{-1}\left\Vert \vec{n'}_{p,1}\right\Vert ^{-1}\frac{\partial}{\partial\nu'_{i}}\sum_{i=1}^{3}\epsilon_{ijk}k'_{1,j}r'_{0,k}\nu'_{i}=-\beta_{p,1}^{-1}\left\Vert \vec{n'}_{p,1}\right\Vert ^{-1}\epsilon_{ijk}k'_{1,j}r'_{0,k};\end{align*}
 which reduces to \begin{align}
\beta_{p,1}\left[1-\left\Vert \vec{n'}_{p,1}\right\Vert \left\{ \sum_{i=1}^{3}\left[\epsilon_{ijk}k'_{1,j}r'_{0,k}\right]^{2}\right\} ^{-1/2}\right] & =0.\label{eq:n-hat-p-i-eq-beta-del-f-p-i-The-Beta-Def-Pre-2}\end{align}
 The two possible solutions are either $\beta_{p,1}=0,$ or the term
enclosed in the outermost square bracket must vanish. Because $\beta_{p,1}=0$
is a useless trivial solution, the second solution has to be adopted.
Since $\left\Vert \vec{n'}_{p,1}\right\Vert =\left\{ \sum_{i=1}^{3}\left[\epsilon_{ijk}k'_{1,j}r'_{0,k}\right]^{2}\right\} ^{1/2},$
the equation (\ref{eq:n-hat-p-i-eq-beta-del-f-p-i-The-Beta-Def-Pre-2})
is already satisfied for any value of $\beta_{p,1}.$ Therefore, $\beta_{p,1}$
is an arbitrary quantity, and it is simply chosen to be unity which
makes the gradient function $\vec{\nabla'}f_{p,1}$ automatically
normalized. The scalar field solution $f_{p,1}$ of equation (\ref{eq:f-p-i-plane-of-incidence-Pre})
is now restated with $\beta_{p,1}=1,$ \begin{eqnarray}
f_{p,1}\left(\alpha,\beta,\gamma\right)=-\left\Vert \vec{n'}_{p,1}\right\Vert ^{-1}\sum_{i=1}^{3}\epsilon_{ijk}k'_{1,j}r'_{0,k}\nu'_{i}, &  & i=\left\{ \begin{array}{c}
1\rightarrow\nu'_{1}=\alpha=x',\\
\\2\rightarrow\nu'_{2}=\beta=y',\\
\\3\rightarrow\nu'_{3}=\gamma=z',\end{array}\right.\label{eq:f-p-i-plane-of-incidence}\end{eqnarray}
 where $-\infty\leq\left\{ \nu'_{1}=\alpha=x',\nu'_{2}=\beta=y',\nu'_{3}=\gamma=z'\right\} \leq\infty.$ 

The intercept between the plane of incidence, defined by equation
(\ref{eq:f-p-i-plane-of-incidence}), and the hemisphere is found
through the algebraic relation \begin{align*}
x'^{2}+y'^{2}+z'^{2} & =\left[r'_{i}\right]^{2},\end{align*}
 which can be written as \begin{eqnarray}
\left[r'_{i}\right]^{2}-\sum_{i=1}^{3}\left[\nu'_{i}\right]^{2}=0, &  & i=\left\{ \begin{array}{c}
1\rightarrow\nu'_{1}=x',\\
\\2\rightarrow\nu'_{2}=y',\\
\\3\rightarrow\nu'_{3}=z',\end{array}\right.\label{eq:algebraic-hemisphere-relation}\end{eqnarray}
 where the $r'_{i}$ is the radius of sphere and the index $i$ denotes
the radius of the inner surface. The intercept of interest is shown
in Figure \ref{cap:plane-sphere-intersection}. 

One may be tempted to incorporate the surface vibration into equation
(\ref{eq:algebraic-hemisphere-relation}) through a slight modification
\begin{align*}
\left[r'_{i}\left(t\right)\right]^{2}-\sum_{i=1}^{3}\left[\nu'_{i}\right]^{2} & =0,\end{align*}
 where the vibration have been introduced through the time variations
in radius. Some have employed such a model in describing the {}``Casimir
radiation,'' as well as the phenomenon of sonoluminescense mainly
due to its simplicity from the mathematical point of view \cite{key-Milton,key-Eberlein}.
In general, if one wishes to incorporate the vibration of a surface,
the description of such system could be represented in the form \begin{align*}
\left[r'_{i}\left(\theta',\phi',t\right)\right]^{2}-\sum_{i=1}^{3}\left[\nu'_{i}\left(\theta',\phi'\right)\right]^{2} & =0.\end{align*}
 Since the radius function varies with $\theta',$ $\phi'$ and $t,$
its treatment has to be postponed until the surface function can be
found in later sections. In the present discussion, the hemisphere
is regarded as having no vibration. 

Returning from the above short digression, the surface function of
the sphere is expressed as the null function from equation (\ref{eq:algebraic-hemisphere-relation}),
\begin{eqnarray}
f_{hemi}\left(x',y',z'\right)=\left[r'_{i}\right]^{2}-\sum_{i=1}^{3}\left[\nu'_{i}\right]^{2}=0, &  & i=\left\{ \begin{array}{c}
1\rightarrow\nu'_{1}=x',\\
\\2\rightarrow\nu'_{2}=y',\\
\\3\rightarrow\nu'_{3}=z'.\end{array}\right.\label{eq:hemisphere-surface-function}\end{eqnarray}
 The intersection between the two surfaces, the plane of incidence
defined in equation (\ref{eq:f-p-i-plane-of-incidence}) and the hemisphere
defined in equation (\ref{eq:hemisphere-surface-function}), is found
through the relation \begin{align}
f_{p,1}\left(x',y',z'\right)-f_{hemi}\left(x',y',z'\right) & =0,\label{eq:hemi-plane-of-i-intersect-rela}\end{align}
 which is equivalent to setting the scalar function $f_{p,1}\left(x',y',z'\right)=0.$
Substituting expressions for $f_{p,1}\left(x',y',z'\right)$ and $f_{hemi}\left(x',y',z'\right)$
given in equations (\ref{eq:f-p-i-plane-of-incidence}) and (\ref{eq:hemisphere-surface-function}),
we arrive at \begin{eqnarray}
\sum_{i=1}^{3}\left\{ \left[\nu'_{i}\right]^{2}-\left\Vert \vec{n'}_{p,1}\right\Vert ^{-1}\epsilon_{ijk}k'_{1,j}r'_{0,k}\nu'_{i}\right\} -\left[r'_{i}\right]^{2}=0, &  & i=\left\{ \begin{array}{c}
1\rightarrow\nu'_{1}=\alpha=x',\\
\\2\rightarrow\nu'_{2}=\beta=y',\\
\\3\rightarrow\nu'_{3}=\gamma=z'.\end{array}\right.\label{eq:hemi-plane-of-i-intersect-rela-explicit-pre}\end{eqnarray}
 It is convenient to rewrite $\left[r'_{i}\right]^{2}$ in the form
\begin{eqnarray}
\left[r'_{i}\right]^{2}=\left[r'_{i,x'}\right]^{2}+\left[r'_{i,y'}\right]^{2}+\left[r'_{i,z'}\right]^{2}=\sum_{i=1}^{3}\left[r'_{i,i}\right]^{2}, &  & i=\left\{ \begin{array}{c}
1\rightarrow r'_{i,1}=r'_{i,x'},\\
\\2\rightarrow r'_{i,2}=r'_{i,y'},\\
\\3\rightarrow r'_{i,3}=r'_{i,z'}.\end{array}\right.\label{eq:r-of-i-square-x-y-z-comp}\end{eqnarray}
 Equation (\ref{eq:hemi-plane-of-i-intersect-rela-explicit-pre})
can then be written as \begin{eqnarray}
\sum_{i=1}^{3}\left\{ \left[\nu'_{i}\right]^{2}-\left\Vert \vec{n'}_{p,1}\right\Vert ^{-1}\epsilon_{ijk}k'_{1,j}r'_{0,k}\nu'_{i}-\left[r'_{i,i}\right]^{2}\right\} =0, &  & i=\left\{ \begin{array}{c}
1\rightarrow\nu'_{1}=\alpha=x';r'_{i,1}=r'_{i,x'},\\
\\2\rightarrow\nu'_{2}=\beta=y';r'_{i,2}=r'_{i,y'},\\
\\3\rightarrow\nu'_{3}=\gamma=z';r'_{i,3}=r'_{i,z'}.\end{array}\right.\label{eq:hemi-plane-of-i-intersect-rela-explicit}\end{eqnarray}
 Since the first two terms are already known, we can set each braced
term equal to zero, \begin{eqnarray}
\left[\nu'_{i}\right]^{2}-\left\Vert \vec{n'}_{p,1}\right\Vert ^{-1}\epsilon_{ijk}k'_{1,j}r'_{0,k}\nu'_{i}-\left[r'_{i,i}\right]^{2}=0, &  & i=1,2,3.\label{eq:hemi-plane-of-i-intersect-rela-explicit-set}\end{eqnarray}
 The above relation, equation (\ref{eq:hemi-plane-of-i-intersect-rela-explicit-set}),
is valid in determining the set of discrete reflection points. The
solutions of this quadratic equation are \begin{eqnarray}
\nu'_{i}=\frac{1}{2}\left\Vert \vec{n'}_{p,1}\right\Vert ^{-1}\epsilon_{ijk}k'_{1,j}r'_{0,k}\pm\left\{ \left[\frac{1}{2}\left\Vert \vec{n'}_{p,1}\right\Vert ^{-1}\epsilon_{ijk}k'_{1,j}r'_{0,k}\right]^{2}+\left[r'_{i,i}\right]^{2}\right\} ^{1/2}, &  & i=1,2,3,\label{eq:hemi-plane-intercept-root}\end{eqnarray}
 where the summation over the indices $j$ and $k$ is implicit. The
restriction of $\nu'_{i}$ being real imposes the condition \begin{eqnarray}
\left[\frac{1}{2}\left\Vert \vec{n'}_{p,1}\right\Vert ^{-1}\epsilon_{ijk}k'_{1,j}r'_{0,k}\right]^{2}+\left[r'_{i,i}\right]^{2}\geq0, &  & i=1,2,3.\label{eq:real-valued-criteria-compact}\end{eqnarray}
 In spherical coordinates, the three radial vector components $r'_{i,1},$
$r'_{i,2}$ and $r'_{i,3}$ are \begin{equation}
\begin{array}{c}
\begin{array}{ccccc}
r'_{i,1}=r'_{i}\sin\theta'\cos\phi', &  & r'_{i,2}=r'_{i}\sin\theta'\sin\phi', &  & r'_{i,3}=r'_{i}\cos\theta',\end{array}\end{array}\label{eq:r-of-i-square-x-y-z-comp-in-spherical-variables}\end{equation}
 where $r'_{i,1}=r'_{i,x'},$ $r'_{i,2}=r'_{i,y'}$ and $r'_{i,3}=r'_{i,z'}.$
Here the terms $r'_{i},$ $\theta'$ and $\phi'$ are the usual radial
length, the polar and the azimuthal angle. This guarantees that $\vec{R'}=\sum_{i=1}^{3}r'_{i,i}\hat{e_{i}}$
is on the sphere, and justifies the step taken in equation (\ref{eq:hemi-plane-of-i-intersect-rela-explicit-set})
since we are only interested in the conditions of the discriminants
expressed by equation (\ref{eq:real-valued-criteria-compact}). With
$r'_{i,i}$ redefined in terms of spherical coordinates, the reality
condition of $\nu'_{i}$ in equation (\ref{eq:real-valued-criteria-compact})
becomes \begin{align}
\left[\frac{1}{2}\left\Vert \vec{n'}_{p,1}\right\Vert ^{-1}\epsilon_{1jk}k'_{1,j}r'_{0,k}\right]^{2}+\left[r'_{i}\right]^{2}\sin^{2}\theta'\cos^{2}\phi' & \geq0,\label{eq:hemi-plane-intercept-root-x-real-criteria-x}\end{align}
 \begin{align}
\left[\frac{1}{2}\left\Vert \vec{n'}_{p,1}\right\Vert ^{-1}\epsilon_{2jk}k'_{1,j}r'_{0,k}\right]^{2}+\left[r'_{i}\right]^{2}\sin^{2}\theta'\sin^{2}\phi' & \geq0,\label{eq:hemi-plane-intercept-root-y-real-criteria-y}\end{align}
 \begin{align}
\left[\frac{1}{2}\left\Vert \vec{n'}_{p,1}\right\Vert ^{-1}\epsilon_{3jk}k'_{1,j}r'_{0,k}\right]^{2}+\left[r'_{i}\right]^{2}\cos^{2}\theta' & \geq0.\label{eq:hemi-plane-intercept-root-z-real-criteria-z}\end{align}
 Equations (\ref{eq:hemi-plane-intercept-root-x-real-criteria-x})
through (\ref{eq:hemi-plane-intercept-root-z-real-criteria-z}) provide
allowed range of $\nu'_{i}$ to ensure its value being real. The solution
of equation (\ref{eq:hemi-plane-intercept-root-z-real-criteria-z})
is \begin{align}
\cos^{2}\theta' & \geq-\left[\frac{1}{2r'_{i}}\left\Vert \vec{n'}_{p,1}\right\Vert ^{-1}\epsilon_{3jk}k'_{1,j}r'_{0,k}\right]^{2},\label{eq:hemi-plane-intercept-root-z-real-criteria-z-cos-sq-theta}\end{align}
 which leads to two inequalities, \begin{eqnarray}
\cos\theta'\geq i\frac{1}{2r'_{i}}\left\Vert \vec{n'}_{p,1}\right\Vert ^{-1}\epsilon_{3jk}k'_{1,j}r'_{0,k}, &  & \cos\theta'\leq-i\frac{1}{2r'_{i}}\left\Vert \vec{n'}_{p,1}\right\Vert ^{-1}\epsilon_{3jk}k'_{1,j}r'_{0,k}.\label{eq:cosine-theta-prime-inequalities}\end{eqnarray}
 These two inequalities cannot be satisfied simultaneously by the
two vectors $\vec{R'}_{0}$ and $\vec{k'}_{1}.$ We have to look for
$\sin\theta'$ by combining equations (\ref{eq:hemi-plane-intercept-root-x-real-criteria-x})
and (\ref{eq:hemi-plane-intercept-root-y-real-criteria-y}) to give
\begin{align*}
\frac{1}{4}\left\Vert \vec{n'}_{p,1}\right\Vert ^{-2}\left\{ \left[\epsilon_{1jk}k'_{1,j}r'_{0,k}\right]^{2}+\left[\epsilon_{2mn}k'_{1,m}r'_{0,n}\right]^{2}\right\} +\left[r'_{i}\right]^{2}\sin^{2}\theta' & \geq0,\end{align*}
 which yields \begin{align}
\sin^{2}\theta' & \geq-\frac{1}{4}\left[r'_{i}\right]^{-2}\left\Vert \vec{n'}_{p,1}\right\Vert ^{-2}\left\{ \left[\epsilon_{1jk}k'_{1,j}r'_{0,k}\right]^{2}+\left[\epsilon_{2mn}k'_{1,m}r'_{0,n}\right]^{2}\right\} .\label{eq:hemi-plane-intercept-root-z-real-criteria-z-sin-sq-theta}\end{align}
 The solutions are again two inequalities, \begin{align}
\sin\theta' & \geq\frac{i}{2r'_{i}}\left\Vert \vec{n'}_{p,1}\right\Vert ^{-1}\left\{ \left[\epsilon_{1jk}k'_{1,j}r'_{0,k}\right]^{2}+\left[\epsilon_{2mn}k'_{1,m}r'_{0,n}\right]^{2}\right\} ,\label{eq:sine-theta-prime-inequality-plus}\end{align}
 \begin{align}
\sin\theta' & \leq-\frac{i}{2r'_{i}}\left\Vert \vec{n'}_{p,1}\right\Vert ^{-1}\left\{ \left[\epsilon_{1jk}k'_{1,j}r'_{0,k}\right]^{2}+\left[\epsilon_{2mn}k'_{1,m}r'_{0,n}\right]^{2}\right\} ,\label{eq:sine-theta-prime-inequality-minus}\end{align}
 which cannot be simultaneously satisfied by the vectors $\vec{R'}_{0}$
and $\vec{k'}_{1}.$ We have to combine equations (\ref{eq:hemi-plane-intercept-root-z-real-criteria-z-cos-sq-theta})
and (\ref{eq:hemi-plane-intercept-root-z-real-criteria-z-sin-sq-theta})
to give \begin{align}
\tan^{2}\theta' & \geq\left[\epsilon_{3qr}k'_{1,q}r'_{0,r}\right]^{-2}\left\{ \left[\epsilon_{1jk}k'_{1,j}r'_{0,k}\right]^{2}+\left[\epsilon_{2mn}k'_{1,m}r'_{0,n}\right]^{2}\right\} ,\label{eq:tangent-squared-theta-prime-inequality}\end{align}
 which leads to another two inequalities, \begin{align}
\tan\theta' & \geq\left[\epsilon_{3qr}k'_{1,q}r'_{0,r}\right]^{-1}\left\{ \left[\epsilon_{1jk}k'_{1,j}r'_{0,k}\right]^{2}+\left[\epsilon_{2mn}k'_{1,m}r'_{0,n}\right]^{2}\right\} ^{1/2},\label{eq:tangent-theta-prime-inequality-plus}\end{align}
 \begin{align}
\tan\theta' & \leq-\left[\epsilon_{3qr}k'_{1,q}r'_{0,r}\right]^{-1}\left\{ \left[\epsilon_{1jk}k'_{1,j}r'_{0,k}\right]^{2}+\left[\epsilon_{2mn}k'_{1,m}r'_{0,n}\right]^{2}\right\} ^{1/2}.\label{eq:tangent-theta-prime-inequality-minus}\end{align}
 In the specified range for $\theta',$ $0\leq\theta'\leq\pi,$ the
tangent function has the limits \begin{align}
\lim_{\varepsilon\rightarrow0}\left(0\leq\theta'\leq\frac{1}{2}\pi-\left|\varepsilon\right|\right) & \Rightarrow0\leq\tan\theta'\leq\infty,\label{eq:limit-tangent-theta-prime-positive-domain}\end{align}
 \begin{align}
\lim_{\varepsilon\rightarrow0}\left(\frac{1}{2}\pi+\left|\varepsilon\right|\leq\theta'\leq\pi\right) & \Rightarrow-\infty\leq\tan\theta'\leq0,\label{eq:limit-tangent-theta-prime-negative-domain}\end{align}
 where $\varepsilon$ is infinitesimal quantity introduced for limiting
purposes. Since there is no guarantee that $\epsilon_{3qr}k'_{1,q}r'_{0,r}>0$
in equations (\ref{eq:tangent-theta-prime-inequality-plus}) and (\ref{eq:tangent-theta-prime-inequality-minus}),
one has to consider both cases where $\epsilon_{3qr}k'_{1,q}r'_{0,r}>0$
and $\epsilon_{3qr}k'_{1,q}r'_{0,r}<0.$ Therefore, for the positive
denominator case where $\epsilon_{3qr}k'_{1,q}r'_{0,r}>0,$ we have
\begin{equation}
\left\{ \begin{array}{c}
\begin{array}{ccc}
\epsilon_{3qr}k'_{1,q}r'_{0,r}\geq0, &  & \lim_{\varepsilon\rightarrow0}\left(0\leq\theta'\leq\frac{1}{2}\pi-\left|\varepsilon\right|\right),\end{array}\\
\\0\leq\left(\tan\theta'\geq\left[\epsilon_{3qr}k'_{1,q}r'_{0,r}\right]^{-1}\left\{ \left[\epsilon_{1jk}k'_{1,j}r'_{0,k}\right]^{2}+\left[\epsilon_{2mn}k'_{1,m}r'_{0,n}\right]^{2}\right\} ^{1/2}\right)\leq\infty;\end{array}\right.\label{eq:tan-theta-pos-deno-limit-plus}\end{equation}
 \begin{equation}
\left\{ \begin{array}{c}
\begin{array}{ccc}
\epsilon_{3qr}k'_{1,q}r'_{0,r}\geq0, &  & \lim_{\varepsilon\rightarrow0}\left(\frac{1}{2}\pi+\left|\varepsilon\right|\leq\theta'\leq\pi\right),\end{array}\\
\\-\infty\leq\left(\tan\theta'\leq-\left[\epsilon_{3qr}k'_{1,q}r'_{0,r}\right]^{-1}\left\{ \left[\epsilon_{1jk}k'_{1,j}r'_{0,k}\right]^{2}+\left[\epsilon_{2mn}k'_{1,m}r'_{0,n}\right]^{2}\right\} ^{1/2}\right)\leq0.\end{array}\right.\label{eq:tan-theta-pos-deno-limit-minus}\end{equation}
 For the negative denominator case where $\epsilon_{3qr}k'_{1,q}r'_{0,r}<0,$
we rewrite equations (\ref{eq:tangent-theta-prime-inequality-plus})
and (\ref{eq:tangent-theta-prime-inequality-minus}) in the form \begin{eqnarray*}
\epsilon_{3qr}k'_{1,q}r'_{0,r}\leq0 & \Rightarrow & -\left|\epsilon_{3qr}k'_{1,q}r'_{0,r}\right|\leq0,\end{eqnarray*}
\begin{align}
\tan\theta' & \geq-\left|\epsilon_{3qr}k'_{1,q}r'_{0,r}\right|^{-1}\left\{ \left[\epsilon_{1jk}k'_{1,j}r'_{0,k}\right]^{2}+\left[\epsilon_{2mn}k'_{1,m}r'_{0,n}\right]^{2}\right\} ^{1/2},\label{eq:tangent-theta-prime-inequality-plus-neg-domain}\end{align}
 \begin{align}
\tan\theta' & \leq\left|\epsilon_{3qr}k'_{1,q}r'_{0,r}\right|^{-1}\left\{ \left[\epsilon_{1jk}k'_{1,j}r'_{0,k}\right]^{2}+\left[\epsilon_{2mn}k'_{1,m}r'_{0,n}\right]^{2}\right\} ^{1/2}.\label{eq:tangent-theta-prime-inequality-minus-neg-domain}\end{align}
 The tangent function in the domain $0\leq\theta'\leq\pi$ has a discontinuity
at $\theta'=\pi/2,$ the inequality (\ref{eq:tangent-theta-prime-inequality-plus-neg-domain})
has the limit $0\geq\tan\theta'\geq-\infty,$ and the inequality (\ref{eq:tangent-theta-prime-inequality-minus-neg-domain})
has the limit $\infty\geq\tan\theta'\geq0.$ Therefore, the limits
for a negative denominator case where $\epsilon_{3qr}k'_{1,q}r'_{0,r}<0$,
\begin{equation}
\left\{ \begin{array}{c}
\begin{array}{ccc}
\epsilon_{3qr}k'_{1,q}r'_{0,r}\leq0, &  & \lim_{\varepsilon\rightarrow0}\left(0\leq\theta'\leq\frac{1}{2}\pi-\left|\varepsilon\right|\right),\end{array}\\
\\0\leq\left(\tan\theta'\leq\left|\epsilon_{3qr}k'_{1,q}r'_{0,r}\right|^{-1}\left\{ \left[\epsilon_{1jk}k'_{1,j}r'_{0,k}\right]^{2}+\left[\epsilon_{2mn}k'_{1,m}r'_{0,n}\right]^{2}\right\} ^{1/2}\right)\leq\infty;\end{array}\right.\label{eq:tan-theta-prime-neg-deno-limit-plus}\end{equation}
 \begin{equation}
\left\{ \begin{array}{c}
\begin{array}{ccc}
\epsilon_{3qr}k'_{1,q}r'_{0,r}\leq0, &  & \lim_{\varepsilon\rightarrow0}\left(\frac{1}{2}\pi+\left|\varepsilon\right|\leq\theta'\leq\pi\right),\end{array}\\
\\-\infty\leq\left(\tan\theta'\geq-\left|\epsilon_{3qr}k'_{1,q}r'_{0,r}\right|^{-1}\left\{ \left[\epsilon_{1jk}k'_{1,j}r'_{0,k}\right]^{2}+\left[\epsilon_{2mn}k'_{1,m}r'_{0,n}\right]^{2}\right\} ^{1/2}\right)\leq0.\end{array}\right.\label{eq:tan-theta-prime-neg-deno-limit-minus}\end{equation}
 Comparing equations (\ref{eq:tan-theta-pos-deno-limit-plus}), (\ref{eq:tan-theta-pos-deno-limit-minus}),
(\ref{eq:tan-theta-prime-neg-deno-limit-plus}) and (\ref{eq:tan-theta-prime-neg-deno-limit-minus}),
we see that two of them are identical when rewritten in terms of the
later convention where $\epsilon_{3qr}k'_{1,q}r'_{0,r}\leq0$ is expressed
as $-\left|\epsilon_{3qr}k'_{1,q}r'_{0,r}\right|\leq0.$ The two tangent
function inequality limits are summarized below for bookkeeping purposes:
\begin{equation}
\left\{ \begin{array}{c}
\lim_{\varepsilon\rightarrow0}\left(0\leq\theta'\leq\frac{1}{2}\pi-\left|\varepsilon\right|\right),\\
\\0\leq\left(\tan\theta'\leq\left|\epsilon_{3qr}k'_{1,q}r'_{0,r}\right|^{-1}\left\{ \left[\epsilon_{1jk}k'_{1,j}r'_{0,k}\right]^{2}+\left[\epsilon_{2mn}k'_{1,m}r'_{0,n}\right]^{2}\right\} ^{1/2}\right)\leq\infty;\end{array}\right.\label{eq:tan-theta-prime-neg-deno-limit-plus-0-to-90-degree}\end{equation}
 \begin{equation}
\left\{ \begin{array}{c}
\lim_{\varepsilon\rightarrow0}\left(\frac{1}{2}\pi+\left|\varepsilon\right|\leq\theta'\leq\pi\right),\\
\\-\infty\leq\left(\tan\theta'\geq-\left|\epsilon_{3qr}k'_{1,q}r'_{0,r}\right|^{-1}\left\{ \left[\epsilon_{1jk}k'_{1,j}r'_{0,k}\right]^{2}+\left[\epsilon_{2mn}k'_{1,m}r'_{0,n}\right]^{2}\right\} ^{1/2}\right)\leq0.\end{array}\right.\label{eq:tan-theta-prime-neg-deno-limit-minus-90-to-180-degree}\end{equation}
 The corresponding arguments for inequalities in (\ref{eq:tan-theta-prime-neg-deno-limit-plus-0-to-90-degree})
and (\ref{eq:tan-theta-prime-neg-deno-limit-minus-90-to-180-degree})
are \begin{equation}
\left\{ \begin{array}{c}
\lim_{\varepsilon\rightarrow0}\left(0\leq\theta'\leq\frac{1}{2}\pi-\left|\varepsilon\right|\right),\\
\\\theta'=\arctan\left(\left|\epsilon_{3qr}k'_{1,q}r'_{0,r}\right|^{-1}\left\{ \left[\epsilon_{1jk}k'_{1,j}r'_{0,k}\right]^{2}+\left[\epsilon_{2mn}k'_{1,m}r'_{0,n}\right]^{2}\right\} ^{1/2}\right);\end{array}\right.\label{eq:theta-prime-0-to-90-degree}\end{equation}
 \begin{equation}
\left\{ \begin{array}{c}
\lim_{\varepsilon\rightarrow0}\left(\frac{1}{2}\pi+\left|\varepsilon\right|\leq\theta'\leq\pi\right),\\
\\\theta'=\arctan\left(-\left|\epsilon_{3qr}k'_{1,q}r'_{0,r}\right|^{-1}\left\{ \left[\epsilon_{1jk}k'_{1,j}r'_{0,k}\right]^{2}+\left[\epsilon_{2mn}k'_{1,m}r'_{0,n}\right]^{2}\right\} ^{1/2}\right).\end{array}\right.\label{eq:theta-prime-90-to-180-degree}\end{equation}
 The spherical coordinate representation is incomplete without the
azimuthal angle $\phi'.$ We have to solve for the allowed range for
the azimuthal angle $\phi'$ by combining equations (\ref{eq:hemi-plane-intercept-root-x-real-criteria-x})
and (\ref{eq:hemi-plane-intercept-root-y-real-criteria-y}). From
equation (\ref{eq:hemi-plane-intercept-root-x-real-criteria-x}),
we have \begin{align*}
\cos^{2}\phi' & \geq-\left[\frac{1}{2}\left\Vert \vec{n'}_{p,1}\right\Vert ^{-1}\epsilon_{1jk}k'_{1,j}r'_{0,k}\right]^{2}\left[r'_{i}\sin^{2}\theta'\right]^{-2}\end{align*}
 and from equation (\ref{eq:hemi-plane-intercept-root-y-real-criteria-y}),
\begin{align*}
\sin^{2}\phi' & \geq-\left[\frac{1}{2}\left\Vert \vec{n'}_{p,1}\right\Vert ^{-1}\epsilon_{2jk}k'_{1,j}r'_{0,k}\right]^{2}\left[r'_{i}\sin^{2}\theta'\right]^{-2}.\end{align*}
 They are combined to give \begin{align}
\tan^{2}\phi' & =\frac{\sin^{2}\phi'}{\cos^{2}\phi'}\geq\left[\epsilon_{1jk}k'_{1,j}r'_{0,k}\right]^{-2}\left[\epsilon_{2mn}k'_{1,m}r'_{0,n}\right]^{2}.\label{eq:tangent-squared-phi-prime-inequality}\end{align}
 The two inequalities are derived from the last equation, \begin{eqnarray}
\tan\phi'\geq\left[\epsilon_{1jk}k'_{1,j}r'_{0,k}\right]^{-1}\epsilon_{2mn}k'_{1,m}r'_{0,n}, &  & \tan\phi'\leq-\left[\epsilon_{1jk}k'_{1,j}r'_{0,k}\right]^{-1}\epsilon_{2mn}k'_{1,m}r'_{0,n}.\label{eq:tangent-phi-prime-inequality-plus-minus}\end{eqnarray}
 In the range of $\phi',$ $0\leq\phi'\leq2\pi,$ the tangent function
has the limits \begin{align}
\lim_{\varepsilon\rightarrow0}\left(0\leq\phi'\leq\frac{1}{2}\pi-\left|\varepsilon\right|\right) & \Rightarrow\left[0\leq\tan\phi'\leq\infty\right],\label{eq:limit-tangent-phi-prime-positive-domain-0-to-90}\end{align}
 \begin{align}
\lim_{\varepsilon\rightarrow0}\left(\frac{1}{2}\pi+\left|\varepsilon\right|\leq\phi'\leq\pi-\varepsilon\right) & \Rightarrow\left[-\infty\leq\tan\phi'\leq0\right],\label{eq:limit-tangent-phi-prime-negative-domain-90-to-180}\end{align}
 \begin{align}
\lim_{\varepsilon\rightarrow0}\left(\pi+\left|\varepsilon\right|\leq\phi'\leq\frac{3}{2}\pi-\left|\varepsilon\right|\right) & \Rightarrow\left[0\leq\tan\phi'\leq\infty\right],\label{eq:limit-tangent-phi-prime-positive-domain-180-to-270}\end{align}
 \begin{align}
\lim_{\varepsilon\rightarrow0}\left(\frac{3}{2}\pi+\left|\varepsilon\right|\leq\phi'<2\pi-\left|\varepsilon\right|\right) & \Rightarrow\left[-\infty\leq\tan\phi'<0\right],\label{eq:limit-tangent-phi-prime-negative-domain-270-to-360}\end{align}
 where $\varepsilon$ is an infinitesimal number used in the limiting
process. Because discontinuities occur at $\phi'=\pi/2$ and $\phi'=3\pi/2,$
the inequalities (\ref{eq:tangent-phi-prime-inequality-plus-minus})
has the limits \[
0\leq\left(\tan\phi'\geq\left[\epsilon_{1jk}k'_{1,j}r'_{0,k}\right]^{-1}\epsilon_{2mn}k'_{1,m}r'_{0,n}\right)\leq\infty,\]
 and\[
-\infty\leq\left(\tan\phi'\leq-\left[\epsilon_{1jk}k'_{1,j}r'_{0,k}\right]^{-1}\epsilon_{2mn}k'_{1,m}r'_{0,n}\right)\leq0.\]
 The ranges for inequalities in (\ref{eq:limit-tangent-phi-prime-positive-domain-0-to-90})
through (\ref{eq:limit-tangent-phi-prime-negative-domain-270-to-360})
can now be expressed explicitly as \begin{equation}
\left\{ \begin{array}{c}
\begin{array}{ccc}
\lim_{\varepsilon\rightarrow0}\left(0\leq\phi'\leq\frac{1}{2}\pi-\left|\varepsilon\right|\right), &  & \lim_{\varepsilon\rightarrow0}\left(\pi+\left|\varepsilon\right|\leq\phi'\leq\frac{3}{2}\pi-\left|\varepsilon\right|\right),\end{array}\\
\\0\leq\left(\tan\phi'\geq\left[\epsilon_{1jk}k'_{1,j}r'_{0,k}\right]^{-1}\epsilon_{2mn}k'_{1,m}r'_{0,n}\right)\leq\infty;\end{array}\right.\label{eq:tan-phi-prime-limit-0-90-and-180-270}\end{equation}
 \begin{equation}
\left\{ \begin{array}{c}
\begin{array}{ccc}
\lim_{\varepsilon\rightarrow0}\left(\frac{1}{2}\pi+\left|\varepsilon\right|\leq\phi'\leq\pi-\varepsilon\right), &  & \lim_{\varepsilon\rightarrow0}\left(\frac{3}{2}\pi+\left|\varepsilon\right|\leq\phi'<2\pi-\left|\varepsilon\right|\right),\end{array}\\
\\-\infty\leq\left(\tan\phi'\leq-\left[\epsilon_{1jk}k'_{1,j}r'_{0,k}\right]^{-1}\epsilon_{2mn}k'_{1,m}r'_{0,n}\right)<0.\end{array}\right.\label{eq:tan-phi-prime-limit-90-180-and-270-360}\end{equation}
 The solutions for $\phi'$ are \begin{equation}
\left\{ \begin{array}{c}
\begin{array}{ccc}
\lim_{\varepsilon\rightarrow0}\left(0\leq\phi'\leq\frac{1}{2}\pi-\left|\varepsilon\right|\right), &  & \lim_{\varepsilon\rightarrow0}\left(\pi+\left|\varepsilon\right|\leq\phi'\leq\frac{3}{2}\pi-\left|\varepsilon\right|\right),\end{array}\\
\\\phi'=\arctan\left(\left[\epsilon_{1jk}k'_{1,j}r'_{0,k}\right]^{-1}\epsilon_{2mn}k'_{1,m}r'_{0,n}\right);\end{array}\right.\label{eq:phi-prime-0-90-and-180-270}\end{equation}
 \begin{equation}
\left\{ \begin{array}{c}
\begin{array}{ccc}
\lim_{\varepsilon\rightarrow0}\left(\frac{1}{2}\pi+\left|\varepsilon\right|\leq\phi'\leq\pi-\varepsilon\right), &  & \lim_{\varepsilon\rightarrow0}\left(\frac{3}{2}\pi+\left|\varepsilon\right|\leq\phi'<2\pi-\left|\varepsilon\right|\right),\end{array}\\
\\\phi'=\arctan\left(-\left[\epsilon_{1jk}k'_{1,j}r'_{0,k}\right]^{-1}\epsilon_{2mn}k'_{1,m}r'_{0,n}\right).\end{array}\right.\label{eq:phi-prime-90-180-and-270-360}\end{equation}
 In order to have $\nu'_{i}$ values being real, the allowed range
of $\theta'$ is determined by equations (\ref{eq:theta-prime-0-to-90-degree})
and (\ref{eq:theta-prime-90-to-180-degree}) and the allowed range
of $\phi'$ is determined by equations (\ref{eq:phi-prime-0-90-and-180-270})
and (\ref{eq:phi-prime-90-180-and-270-360}). Having found valid ranges
of $\theta'$ and $\phi'$ in which $\nu'_{i}$ is real, the task
is now shifted in locating reflection points on the inner hemisphere
surface in spherical coordinates. To distinguish one reflection point
from the other, the notation $\nu'_{i}$ is modified to $\nu'_{i}\rightarrow\nu'_{1,i}$
in equation (\ref{eq:hemi-plane-intercept-root}). The first index
$1$ of $\nu'_{1,i}$ denotes the first reflection point. In this
notation, the second reflection point would be $\nu'_{2,i}$ and the
$N$th reflection point, $\nu'_{N,i}.$ Then equation (\ref{eq:r-of-i-square-x-y-z-comp-in-spherical-variables})
is used to rewrite $r'_{i,i}$ in terms of spherical coordinates.
The Cartesian coordinate variables $x',$ $y'$ and $z'$ in equation
(\ref{eq:hemi-plane-intercept-root}) are expressed as \begin{align}
\nu'_{1,1}\equiv x'_{1} & =\frac{1}{2}\left\Vert \vec{n'}_{p,1}\right\Vert ^{-1}\epsilon_{1jk}k'_{1,j}r'_{0,k}\pm\left\{ \left[\frac{1}{2}\left\Vert \vec{n'}_{p,1}\right\Vert ^{-1}\epsilon_{1jk}k'_{1,j}r'_{0,k}\right]^{2}+\left[r'_{i}\right]^{2}\sin^{2}\theta'_{1}\cos^{2}\phi'_{1}\right\} ^{1/2},\label{eq:hemi-plane-intercept-root-x-prime}\end{align}
 \begin{align}
\nu'_{1,2}\equiv y'_{1} & =\frac{1}{2}\left\Vert \vec{n'}_{p,1}\right\Vert ^{-1}\epsilon_{2jk}k'_{1,j}r'_{0,k}\pm\left\{ \left[\frac{1}{2}\left\Vert \vec{n'}_{p,1}\right\Vert ^{-1}\epsilon_{2jk}k'_{1,j}r'_{0,k}\right]^{2}+\left[r'_{i}\right]^{2}\sin^{2}\theta'_{1}\sin^{2}\phi'_{1}\right\} ^{1/2},\label{eq:hemi-plane-intercept-root-y-prime}\end{align}
 \begin{align}
\nu'_{1,3}\equiv z'_{1} & =\frac{1}{2}\left\Vert \vec{n'}_{p,1}\right\Vert ^{-1}\epsilon_{3jk}k'_{1,j}r'_{0,k}\pm\left\{ \left[\frac{1}{2}\left\Vert \vec{n'}_{p,1}\right\Vert ^{-1}\epsilon_{3jk}k'_{1,j}r'_{0,k}\right]^{2}+\left[r'_{i}\right]^{2}\cos^{2}\theta'_{1}\right\} ^{1/2}.\label{eq:hemi-plane-intercept-root-z-prime}\end{align}
 Although the first reflection point on hemisphere is fully described
by $\vec{R'}_{1}$ in equation (\ref{eq:1st-bounce-off-point}), it
is not convenient to use $\vec{R'}_{1}$ in its current form. The
most effective representation of $\vec{R'}_{1}$ is in spherical coordinates.
We set $\theta'=\theta'_{1}$ and $\phi'=\phi'_{1}$ that describe
the same reflection point $\vec{R'}_{1}$ on the hemisphere. The subscript
on the angular variables $\theta'_{1}$ and $\phi'_{1}$ denotes first
reflection point on the hemisphere surface. In terms of Cartesian
variables $x',$ $y'$ and $z',$ the first reflection point on hemisphere
is given by \begin{eqnarray}
\vec{R'}_{1}\left(x'_{1},y'_{1},z'_{1}\right)=\sum_{i=1}^{3}\nu'_{1,i}\hat{e_{i}}, &  & i=\left\{ \begin{array}{c}
1\rightarrow\nu'_{1,1}=x'_{1},\\
\\2\rightarrow\nu'_{1,2}=y'_{1},\\
\\3\rightarrow\nu'_{1,3}=z'_{1}.\end{array}\right.\label{eq:cartesian-R-of-1-in-x-y-z}\end{eqnarray}
 The same point on the hemisphere, defined by equation (\ref{eq:cartesian-R-of-1-in-x-y-z}),
can be expressed in terms of a parametric representation of equation
(\ref{eq:1st-bounce-off-point}), \begin{align}
\vec{R'}_{1}\left(\xi_{1,p};\vec{R'}_{0},\hat{k'}_{1}\right) & =\sum_{i=1}^{3}\left[r'_{0,i}+\xi_{1,p}\left\Vert \vec{k'}_{1}\right\Vert ^{-1}k'_{1,i}\right]\hat{e_{i}}=\sum_{i=1}^{3}\Upsilon_{1,i}\hat{e_{i}},\label{eq:1st-Reflection-Point-in-Upsilon}\end{align}
 where \begin{eqnarray}
\Upsilon_{1,i}=r'_{0,i}+\xi_{1,p}\left\Vert \vec{k'}_{1}\right\Vert ^{-1}k'_{1,i}, &  & i=1,2,3.\label{eq:Upsilon-Def}\end{eqnarray}
 Both representations, $\vec{R'}_{1}\left(x'_{1},y'_{1},z'_{1}\right)$
and $\vec{R'}_{1}\left(\xi_{1,p};\vec{R'}_{0},\hat{k'}_{1}\right),$
describe the same point on the hemisphere. Therefore, we have \begin{eqnarray*}
\vec{R'}_{1}\left(x'_{1},y'_{1},z'_{1}\right)=\vec{R'}_{1}\left(\xi_{1,p};\vec{R'}_{0},\hat{k'}_{1}\right) & \rightarrow & \sum_{i=1}^{3}\left[\nu_{1,i}-\Upsilon_{1,i}\right]\hat{e_{i}}=0.\end{eqnarray*}
 The components of the last equation are \begin{eqnarray}
\nu'_{1,i}-\Upsilon_{1,i}=0, &  & i=1,2,3.\label{eq:x-upsilon-rela-algebraic-condi}\end{eqnarray}
 Substituting expression of $\nu'_{1,i}$ from equations (\ref{eq:hemi-plane-intercept-root-x-prime}),
(\ref{eq:hemi-plane-intercept-root-y-prime}) and (\ref{eq:hemi-plane-intercept-root-z-prime})
into the above equation, we obtain \begin{align}
\frac{1}{2}\left\Vert \vec{n'}_{p,1}\right\Vert ^{-1}\epsilon_{1jk}k'_{1,j}r'_{0,k}\pm\left\{ \left[\frac{1}{2}\left\Vert \vec{n'}_{p,1}\right\Vert ^{-1}\epsilon_{1jk}k'_{1,j}r'_{0,k}\right]^{2}+\left[r'_{i}\right]^{2}\sin^{2}\theta'_{1}\cos^{2}\phi'_{1}\right\} ^{1/2}-\Upsilon_{1,1} & =0,\label{eq:x-upsilon-rela-2}\end{align}
 \begin{align}
\frac{1}{2}\left\Vert \vec{n'}_{p,1}\right\Vert ^{-1}\epsilon_{2jk}k'_{1,j}r'_{0,k}\pm\left\{ \left[\frac{1}{2}\left\Vert \vec{n'}_{p,1}\right\Vert ^{-1}\epsilon_{2jk}k'_{1,j}r'_{0,k}\right]^{2}+\left[r'_{i}\right]^{2}\sin^{2}\theta'_{1}\sin^{2}\phi'_{1}\right\} ^{1/2}-\Upsilon_{1,2} & =0,\label{eq:y-upsilon-rela-2}\end{align}
 \begin{align}
\frac{1}{2}\left\Vert \vec{n'}_{p,1}\right\Vert ^{-1}\epsilon_{3jk}k'_{1,j}r'_{0,k}\pm\left\{ \left[\frac{1}{2}\left\Vert \vec{n'}_{p,1}\right\Vert ^{-1}\epsilon_{3jk}k'_{1,j}r'_{0,k}\right]^{2}+\left[r'_{i}\right]^{2}\cos^{2}\theta'_{1}\right\} ^{1/2}-\Upsilon_{1,3} & =0,\label{eq:z-upsilon-rela-2}\end{align}
 where $\Upsilon_{1,i}$ is defined in equation (\ref{eq:Upsilon-Def}).
To solve for $\theta'_{1},$ equation (\ref{eq:z-upsilon-rela-2})
is first rearranged, \begin{align*}
\pm\left\{ \left[\frac{1}{2}\left\Vert \vec{n'}_{p,1}\right\Vert ^{-1}\epsilon_{3jk}k'_{1,j}r'_{0,k}\right]^{2}+\left[r'_{i}\right]^{2}\cos^{2}\theta'_{1}\right\} ^{1/2} & =\Upsilon_{1,3}-\frac{1}{2}\left\Vert \vec{n'}_{p,1}\right\Vert ^{-1}\epsilon_{3jk}k'_{1,j}r'_{0,k}.\end{align*}
 Square both sides and solve for $\cos^{2}\theta'_{1},$ the result
is \begin{align}
\cos^{2}\theta'_{1} & =\left[r'_{i}\right]^{-2}\left[\Upsilon_{1,3}^{2}-\Upsilon_{1,3}\left\Vert \vec{n'}_{p,1}\right\Vert ^{-1}\epsilon_{3jk}k'_{1,j}r'_{0,k}\right].\label{eq:hemi-plane-cos-squared-theta1}\end{align}
 For reasons discussed earlier, $\theta'_{1}$ information from the
sine function is also needed. Following the earlier procedures, equations
(\ref{eq:x-upsilon-rela-2}) and (\ref{eq:y-upsilon-rela-2}) are
combined to yield the relation, \begin{align*}
\frac{1}{2}\left\Vert \vec{n'}_{p,1}\right\Vert ^{-1}\epsilon_{1jk}k'_{1,j}r'_{0,k}\pm\left\{ \left[\frac{1}{2}\left\Vert \vec{n'}_{p,1}\right\Vert ^{-1}\epsilon_{1jk}k'_{1,j}r'_{0,k}\right]^{2}+\left[r'_{i}\right]^{2}\sin^{2}\theta'_{1}\cos^{2}\phi'_{1}\right\} ^{1/2}-\Upsilon_{1,1}\\
+\frac{1}{2}\left\Vert \vec{n'}_{p,1}\right\Vert ^{-1}\epsilon_{2mn}k'_{1,m}r'_{0,n}\pm\left\{ \left[\frac{1}{2}\left\Vert \vec{n'}_{p,1}\right\Vert ^{-1}\epsilon_{2mn}k'_{1,m}r'_{0,n}\right]^{2}+\left[r'_{i}\right]^{2}\sin^{2}\theta'_{1}\sin^{2}\phi'_{1}\right\} ^{1/2}-\Upsilon_{1,2} & =0,\end{align*}
 The equation is not easy to solve for $\sin^{2}\phi'_{1}.$ Fortunately,
there is another way to extract the sine function which requires the
knowledge of $\phi'_{1}.$ The solution of $\theta'_{1}$ is postponed
until a solution of $\phi'_{1}$ is found. To solve for $\phi'_{1},$
it is desirable to solve for $\cos^{2}\phi'_{1}$ and $\sin^{2}\phi'_{1}$
from equations (\ref{eq:x-upsilon-rela-2}) and (\ref{eq:y-upsilon-rela-2})
first. Rearranging equation (\ref{eq:x-upsilon-rela-2}), \begin{align*}
\pm\left\{ \left[\frac{1}{2}\left\Vert \vec{n'}_{p,1}\right\Vert ^{-1}\epsilon_{1jk}k'_{1,j}r'_{0,k}\right]^{2}+\left[r'_{i}\right]^{2}\sin^{2}\theta'_{1}\cos^{2}\phi'_{1}\right\} ^{1/2} & =\Upsilon_{1,1}-\frac{1}{2}\left\Vert \vec{n'}_{p,1}\right\Vert ^{-1}\epsilon_{1jk}k'_{1,j}r'_{0,k},\end{align*}
 and followed by squaring both sides, then $\cos^{2}\phi'_{1}$ can
be found to be \begin{align}
\cos^{2}\phi'_{1} & =\left[r'_{i}\sin\theta'_{1}\right]^{-2}\left[\Upsilon_{1,1}^{2}-\Upsilon_{1,1}\left\Vert \vec{n'}_{p,1}\right\Vert ^{-1}\epsilon_{1jk}k'_{1,j}r'_{0,k}\right].\label{eq:hemi-plane-cos-squared-phi1}\end{align}
 Similarly, rearranging equation (\ref{eq:y-upsilon-rela-2}), \begin{align*}
\pm\left\{ \left[\frac{1}{2}\left\Vert \vec{n'}_{p,1}\right\Vert ^{-1}\epsilon_{2jk}k'_{1,j}r'_{0,k}\right]^{2}+\left[r'_{i}\right]^{2}\sin^{2}\theta'_{1}\sin^{2}\phi'_{1}\right\} ^{1/2} & =\Upsilon_{1,2}-\frac{1}{2}\left\Vert \vec{n'}_{p,1}\right\Vert ^{-1}\epsilon_{2jk}k'_{1,j}r'_{0,k}\end{align*}
 and squaring both sides, then $\sin^{2}\phi'_{1}$ can be found to
be \begin{align}
\sin^{2}\phi'_{1} & =\left[r'_{i}\sin\theta'_{1}\right]^{-2}\left[\Upsilon_{1,2}^{2}-\Upsilon_{1,2}\left\Vert \vec{n'}_{p,1}\right\Vert ^{-1}\epsilon_{2jk}k'_{1,j}r'_{0,k}\right].\label{eq:hemi-plane-sin-squared-phi1}\end{align}
 The function $\tan^{2}\phi'_{1}$ can be obtained by combining equations
(\ref{eq:hemi-plane-sin-squared-phi1}) and (\ref{eq:hemi-plane-cos-squared-phi1}),
\begin{align}
\tan^{2}\phi'_{1} & =\left[\Upsilon_{1,1}^{2}-\Upsilon_{1,1}\left\Vert \vec{n'}_{p,1}\right\Vert ^{-1}\epsilon_{1jk}k'_{1,j}r'_{0,k}\right]^{-1}\left[\Upsilon_{1,2}^{2}-\Upsilon_{1,2}\left\Vert \vec{n'}_{p,1}\right\Vert ^{-1}\epsilon_{2mn}k'_{1,m}r'_{0,n}\right].\label{eq:tangent-squared-phi1-prime}\end{align}
 Finally, the azimuthal angle $\phi'_{1}$ is found to be \begin{align}
\phi'_{1} & =\arctan\left(\pm\left[\frac{\Upsilon_{1,2}^{2}-\Upsilon_{1,2}\left\Vert \vec{n'}_{p,1}\right\Vert ^{-1}\epsilon_{2mn}k'_{1,m}r'_{0,n}}{\Upsilon_{1,1}^{2}-\Upsilon_{1,1}\left\Vert \vec{n'}_{p,1}\right\Vert ^{-1}\epsilon_{1jk}k'_{1,j}r'_{0,k}}\right]^{1/2}\right).\label{eq:phi1-prime-general}\end{align}
 The restriction of $\phi'_{1}$ being real imposes the condition
\begin{align*}
\frac{\Upsilon_{1,2}^{2}-\Upsilon_{1,2}\left\Vert \vec{n'}_{p,1}\right\Vert ^{-1}\epsilon_{2mn}k'_{1,m}r'_{0,n}}{\Upsilon_{1,1}^{2}-\Upsilon_{1,1}\left\Vert \vec{n'}_{p,1}\right\Vert ^{-1}\epsilon_{1jk}k'_{1,j}r'_{0,k}} & \geq\varsigma\end{align*}
 or \begin{align*}
\Upsilon_{1,2}^{2}-\Upsilon_{1,2}\left\Vert \vec{n'}_{p,1}\right\Vert ^{-1}\epsilon_{2mn}k'_{1,m}r'_{0,n} & \geq\varsigma\Upsilon_{1,1}^{2}-\varsigma\Upsilon_{1,1}\left\Vert \vec{n'}_{p,1}\right\Vert ^{-1}\epsilon_{1jk}k'_{1,j}r'_{0,k},\end{align*}
 where $\varsigma\geq0.$ Following equations (\ref{eq:tangent-squared-phi-prime-inequality})
through (\ref{eq:phi-prime-90-180-and-270-360}), the following results
are obtained: \begin{equation}
\left\{ \begin{array}{c}
\begin{array}{ccc}
\lim_{\varepsilon\rightarrow0}\left(0\leq\phi'_{1}\leq\frac{1}{2}\pi-\left|\varepsilon\right|\right), &  & \lim_{\varepsilon\rightarrow0}\left(\pi+\left|\varepsilon\right|\leq\phi'_{1}\leq\frac{3}{2}\pi-\left|\varepsilon\right|\right),\end{array}\\
\\\phi'_{1}=\arctan\left(\left[\frac{\Upsilon_{1,2}^{2}-\Upsilon_{1,2}\left\Vert \vec{n'}_{p,1}\right\Vert ^{-1}\epsilon_{2mn}k'_{1,m}r'_{0,n}}{\Upsilon_{1,1}^{2}-\Upsilon_{1,1}\left\Vert \vec{n'}_{p,1}\right\Vert ^{-1}\epsilon_{1jk}k'_{1,j}r'_{0,k}}\right]^{1/2}\right);\end{array}\right.\label{eq:phi1-prime-0-90-and-180-270-final}\end{equation}
 \begin{equation}
\left\{ \begin{array}{c}
\begin{array}{ccc}
\lim_{\varepsilon\rightarrow0}\left(\frac{1}{2}\pi+\left|\varepsilon\right|\leq\phi'_{1}\leq\pi-\varepsilon\right), &  & \lim_{\varepsilon\rightarrow0}\left(\frac{3}{2}\pi+\left|\varepsilon\right|\leq\phi'_{1}<2\pi-\left|\varepsilon\right|\right),\end{array}\\
\\\phi'_{1}=\arctan\left(-\left[\frac{\Upsilon_{1,2}^{2}-\Upsilon_{1,2}\left\Vert \vec{n'}_{p,1}\right\Vert ^{-1}\epsilon_{2mn}k'_{1,m}r'_{0,n}}{\Upsilon_{1,1}^{2}-\Upsilon_{1,1}\left\Vert \vec{n'}_{p,1}\right\Vert ^{-1}\epsilon_{1jk}k'_{1,j}r'_{0,k}}\right]^{1/2}\right).\end{array}\right.\label{eq:phi1-prime-90-180-and-270-360-final}\end{equation}
 Having found the solution for $\phi'_{1},$ we can proceed to finalize
the task of solving for the polar angle $\theta'_{1}.$ Combining
the results for $\cos^{2}\phi'_{1}$ and $\sin^{2}\phi'_{1}$ found
in equations (\ref{eq:hemi-plane-cos-squared-phi1}) and (\ref{eq:hemi-plane-sin-squared-phi1}),
$\sin^{2}\theta'_{1}$ can be found to be \begin{align}
\sin^{2}\theta'_{1} & =\left[r'_{i}\right]^{-2}\left[\Upsilon_{1,1}^{2}+\Upsilon_{1,2}^{2}-\left\Vert \vec{n'}_{p,1}\right\Vert ^{-1}\left\{ \Upsilon_{1,1}\epsilon_{1jk}k'_{1,j}r'_{0,k}+\Upsilon_{1,2}\epsilon_{2mn}k'_{1,m}r'_{0,n}\right\} \right].\label{eq:hemi-plane-sin-squared-theta1}\end{align}
 The function $\tan^{2}\theta'_{1}$ is constructed with equations
(\ref{eq:hemi-plane-cos-squared-theta1}) and (\ref{eq:hemi-plane-sin-squared-theta1}),
\begin{align}
\tan^{2}\theta'_{1} & =\frac{\Upsilon_{1,1}^{2}+\Upsilon_{1,2}^{2}-\left\Vert \vec{n'}_{p,1}\right\Vert ^{-1}\left\{ \Upsilon_{1,1}\epsilon_{1jk}k'_{1,j}r'_{0,k}+\Upsilon_{1,2}\epsilon_{2mn}k'_{1,m}r'_{0,n}\right\} }{\Upsilon_{1,3}^{2}-\Upsilon_{1,3}\left\Vert \vec{n'}_{p,1}\right\Vert ^{-1}\epsilon_{3qr}k'_{1,q}r'_{0,r}}.\label{eq:tangent-squared-theta1-prime}\end{align}
 Then $\theta'_{1}$ is given by \begin{align}
\theta'_{1} & =\arctan\left(\pm\left[\frac{\Upsilon_{1,1}^{2}+\Upsilon_{1,2}^{2}-\left\Vert \vec{n'}_{p,1}\right\Vert ^{-1}\left\{ \Upsilon_{1,1}\epsilon_{1jk}k'_{1,j}r'_{0,k}+\Upsilon_{1,2}\epsilon_{2mn}k'_{1,m}r'_{0,n}\right\} }{\Upsilon_{1,3}^{2}-\Upsilon_{1,3}\left\Vert \vec{n'}_{p,1}\right\Vert ^{-1}\epsilon_{3qr}k'_{1,q}r'_{0,r}}\right]^{1/2}\right).\label{eq:theta1-prime-general}\end{align}
 Following equations (\ref{eq:tangent-squared-theta-prime-inequality})
through (\ref{eq:theta-prime-90-to-180-degree}), we arrive at \begin{equation}
\left\{ \begin{array}{c}
\lim_{\varepsilon\rightarrow0}\left(0\leq\theta'_{1}\leq\frac{1}{2}\pi-\left|\varepsilon\right|\right),\\
\\\theta'_{1}=\arctan\left(\left[\frac{\Upsilon_{1,1}^{2}+\Upsilon_{1,2}^{2}-\left\Vert \vec{n'}_{p,1}\right\Vert ^{-1}\left\{ \Upsilon_{1,1}\epsilon_{1jk}k'_{1,j}r'_{0,k}+\Upsilon_{1,2}\epsilon_{2mn}k'_{1,m}r'_{0,n}\right\} }{\Upsilon_{1,3}^{2}-\Upsilon_{1,3}\left\Vert \vec{n'}_{p,1}\right\Vert ^{-1}\epsilon_{3qr}k'_{1,q}r'_{0,r}}\right]^{1/2}\right);\end{array}\right.\label{eq:theta1-prime-0-to-90-degree-final}\end{equation}
 \begin{equation}
\left\{ \begin{array}{c}
\lim_{\varepsilon\rightarrow0}\left(\frac{1}{2}\pi+\left|\varepsilon\right|\leq\theta'_{1}\leq\pi\right),\\
\\\theta'_{1}=\arctan\left(-\left[\frac{\Upsilon_{1,1}^{2}+\Upsilon_{1,2}^{2}-\left\Vert \vec{n'}_{p,1}\right\Vert ^{-1}\left\{ \Upsilon_{1,1}\epsilon_{1jk}k'_{1,j}r'_{0,k}+\Upsilon_{1,2}\epsilon_{2mn}k'_{1,m}r'_{0,n}\right\} }{\Upsilon_{1,3}^{2}-\Upsilon_{1,3}\left\Vert \vec{n'}_{p,1}\right\Vert ^{-1}\epsilon_{3qr}k'_{1,q}r'_{0,r}}\right]^{1/2}\right).\end{array}\right.\label{eq:theta1-prime-90-to-180-degree-final}\end{equation}
 The allowed angular values are all defined now: $\phi'_{1}$ by equations
(\ref{eq:phi1-prime-0-90-and-180-270-final}) and (\ref{eq:phi1-prime-90-180-and-270-360-final});
and $\theta'_{1}$ by equations (\ref{eq:theta1-prime-0-to-90-degree-final})
and (\ref{eq:theta1-prime-90-to-180-degree-final}). The initial reflection
point on the inner hemisphere surface can be calculated by the equation:
\begin{eqnarray}
\vec{R'}_{1}\left(r'_{i},\theta'_{1},\phi'_{1}\right)=\sum_{i=1}^{3}\nu'_{1,i}\left(r'_{i},\theta'_{1},\phi'_{1}\right)\hat{e_{i}}, &  & i=\left\{ \begin{array}{c}
1\rightarrow\nu'_{1,1}=r'_{i}\sin\theta'_{1}\cos\phi'_{1},\\
\\2\rightarrow\nu'_{1,2}=r'_{i}\sin\theta'_{1}\sin\phi'_{1},\\
\\3\rightarrow\nu'_{1,3}=r'_{i}\cos\theta'_{1}.\qquad\;\:\end{array}\right.\label{eq:1st-Reflection-Point}\end{eqnarray}

We still have to determine the maximum wavelength that can fit the
hemispherical cavity. It is determined from the distance between two
immediate reflection points once they are found. We have to find expression
describing the second reflection point $\vec{R'}_{2}.$ In Figure
\ref{cap:plane-sphere-intersection}, the angle $\psi_{1,2}$ satisfies
the relation \begin{align}
\psi_{1,2}+\theta_{2}+\theta_{r} & =\pi.\label{eq:psi-i-i-plus-1-plus-theta-i-plus-theta-r-eq-pi}\end{align}
 Angles $\theta_{2}$ and $\theta_{r}$ are equal due to the law of
reflection, consequently \begin{eqnarray}
\theta_{2}=\theta_{r}=\theta_{1}, &  & \psi_{1,2}=\pi-2\theta_{i}.\label{eq:psi-i-i-plus-1-equal-pi-minus-2-theta-i-pre}\end{eqnarray}
 It is important not to confuse the angle $\theta_{i}$ above with
that of spherical polar angle $\theta_{i}$ which was previously denoted
with an index $i$ to indicate particular reflection point $\vec{R'}_{i}.$
The $\theta_{i}$ in equation (\ref{eq:psi-i-i-plus-1-equal-pi-minus-2-theta-i-pre})
is an angle of incidence, not a polar angle. In order to avoid any
further confusion in notation, equation (\ref{eq:psi-i-i-plus-1-equal-pi-minus-2-theta-i-pre})
is restated with modifications applied to the indexing convention
for angle of incidence, \begin{eqnarray}
\theta_{i+1}=\theta_{r}=\theta_{inc}, &  & \psi_{i,i+1}=\pi-2\theta_{inc}.\label{eq:psi-i-i-plus-1-equal-pi-minus-2-theta-i}\end{eqnarray}
 The relation that connects angle of incidence to known quantities
$\vec{k'}_{1}$ and $\vec{R'}_{1}$ is \begin{align*}
\vec{k'}_{1}\cdot\vec{R'}_{1} & =\sum_{i=1}^{3}k'_{1,i}\nu'_{1,i}=r'_{i}\left\Vert \vec{k'}_{1}\right\Vert \cos\theta_{inc},\end{align*}
 where $\left\Vert \vec{R'}_{1}\right\Vert =r'_{i},$ and the index
$i$ is not summed over. The incident angle $\theta_{inc}$ is given
by \begin{align}
\theta_{inc} & =\arccos\left(\left[r'_{i}\left\Vert \vec{k'}_{1}\right\Vert \right]^{-1}\sum_{i=1}^{3}k'_{1,i}\nu'_{1,i}\right).\label{eq:angle-of-incidence-pre}\end{align}
 Substituting the explicit expression of $\nu'_{1,i}:$ \begin{equation}
\begin{array}{ccccc}
\nu'_{1,1}=x'_{1}=r'_{i}\sin\theta'_{1}\cos\phi'_{1}, &  & \nu'_{1,2}=y'_{1}=r'_{i}\sin\theta'_{1}\sin\phi'_{1}, &  & \nu'_{1,3}=z'_{1}=r'_{i}\cos\theta'_{1},\end{array}\label{eq:x1-y1-z1-primes-in-theta1-phi1-ri-primes}\end{equation}
 into equation (\ref{eq:angle-of-incidence-pre}), the incident angle
is evaluated as \begin{align}
\theta_{inc} & =\arccos\left(\frac{\sin\theta'_{1}\left[k'_{x'_{1}}\cos\phi'_{1}+k'_{y'_{1}}\sin\phi'_{1}\right]+k'_{z'_{1}}\cos\theta'_{1}}{\sqrt{\left[k'_{x'_{1}}\right]^{2}+\left[k'_{y'_{1}}\right]^{2}+\left[k'_{z'_{1}}\right]^{2}}}\right),\label{eq:angle-of-incidence-exp}\end{align}
 where $k'_{1,1}=k'_{x'_{1}},$ $k'_{1,2}=k'_{y'_{1}}$ and $k'_{1,3}=k'_{z'_{1}}.$
The second reflection point $\vec{R'}_{2}$ has the form \begin{eqnarray}
\vec{R'}_{2}\left(\nu'_{2,1},\nu'_{2,2},\nu'_{2,3}\right)=\sum_{i=1}^{3}\nu'_{2,i}\left(\nu'_{1,1},\nu'_{1,2},\nu'_{1,3}\right)\hat{e_{i}}, &  & i=\left\{ \begin{array}{cc}
1\rightarrow\nu'_{2,1}=x'_{2}, & \nu'_{1,1}=x'_{1},\\
\\2\rightarrow\nu'_{2,2}=y'_{2}, & \nu'_{1,2}=y'_{1},\\
\\3\rightarrow\nu'_{2,3}=z'_{2}, & \nu'_{1,3}=z'_{1}.\end{array}\right.\label{eq:R-i-plus-1-of-x-y-z-carte}\end{eqnarray}
 The relation that connects two vectors $\vec{R'}_{1}$ and $\vec{R'}_{2}$
is \begin{align}
\vec{R'}_{1}\cdot\vec{R'}_{2} & =\left[r'_{i}\right]^{2}\cos\psi_{1,2},\label{eq:Ri-Ri-plus-1-dot-product-1-equiv-1-psi-pre}\end{align}
 where $\left\Vert \vec{R'}_{1}\right\Vert =\left\Vert \vec{R'}_{2}\right\Vert =r'_{i}$
for a rigid hemisphere. Equivalently, this expression can be evaluated
using $\psi_{1,2},$ given in equation (\ref{eq:psi-i-i-plus-1-equal-pi-minus-2-theta-i}),
as \begin{align}
\left[r'_{i}\right]^{2}\cos\left(\pi-2\theta_{inc}\right)-\sum_{i=1}^{3}\nu'_{1,i}\nu'_{2,i} & =0.\label{eq:polynomial-1}\end{align}
 Equation (\ref{eq:polynomial-1}) serves as one of the two needed
relations. The other relation can be found from the cross product
of $\vec{R'}_{1}$ and $\vec{R'}_{2},$ \begin{align}
\vec{R'}_{1}\times\vec{R'}_{2} & =\sum_{i=1}^{3}\epsilon_{ijk}\nu'_{1,j}\nu'_{2,k}\hat{e_{i}}.\label{eq:Ri-cross-R-of-i-plus-1}\end{align}
 Since $\vec{R'}_{1}$ and $\vec{R'}_{2}$ span the plane of incidence
whose unit normal is given by equation (\ref{eq:n-hat-p-i-eq-mag-times-ki-cross-Ro}),
\begin{align*}
\hat{n'}_{p,1} & =-\left\Vert \vec{n'}_{p,1}\right\Vert ^{-1}\sum_{i=1}^{3}\epsilon_{ijk}k'_{1,j}r'_{0,k}\hat{e_{i}},\end{align*}
 the cross product of $\vec{R'}_{1}$ and $\vec{R'}_{2}$ can be equivalently
expressed as \begin{align}
\vec{R'}_{1}\times\vec{R'}_{2} & =-\Gamma_{1,2}\left\Vert \vec{n'}_{p,1}\right\Vert ^{-1}\sum_{i=1}^{3}\epsilon_{ijk}k'_{1,j}r'_{0,k}\hat{e_{i}},\label{eq:Ri-cross-R-i-plus-1-equal-gamma}\end{align}
 where $\Gamma_{1,2}$ is a proportionality factor. The factor $\Gamma_{1,2}$
can be found simply by noticing \begin{eqnarray*}
\vec{R'}_{1}\times\vec{R'}_{2}=\Gamma_{1,2}\hat{n'}_{p,1} & \rightarrow & \left\Vert \vec{R'}_{1}\times\vec{R'}_{2}\right\Vert =\Gamma_{1,2}\left\Vert \hat{n'}_{p,1}\right\Vert =\Gamma_{1,2},\end{eqnarray*}
 which leads to \begin{align}
\Gamma_{1,2} & =\left\Vert \vec{R'}_{1}\times\vec{R'}_{2}\right\Vert =\left[r'_{i}\right]^{2}\sin\left(\pi-2\theta_{inc}\right).\label{eq:Ri-cross-R-i-plus-1-Gamma2-Value}\end{align}
 Equations (\ref{eq:Ri-cross-R-of-i-plus-1}) and (\ref{eq:Ri-cross-R-i-plus-1-equal-gamma})
are combined as \begin{align}
\sum_{i=1}^{3}\left[\epsilon_{ijk}\nu'_{1,j}\nu'_{2,k}+\Gamma_{1,2}\left\Vert \vec{n'}_{p,1}\right\Vert ^{-1}\epsilon_{ijk}k'_{1,j}r'_{0,k}\right]\hat{e_{i}} & =0.\label{eq:R1-cross-R-of-i-plus-2}\end{align}
 The individual component equation is given by \begin{eqnarray}
\epsilon_{ijk}\nu'_{1,j}\nu'_{2,k}+\Gamma_{1,2}\left\Vert \vec{n'}_{p,1}\right\Vert ^{-1}\epsilon_{ijk}k'_{1,j}r'_{0,k}=0, &  & i=1,2,3.\label{eq:R1-cross-R-of-i-plus-3-Algebraic}\end{eqnarray}
 Equations (\ref{eq:R1-cross-R-of-i-plus-3-Algebraic}) and (\ref{eq:polynomial-1})
together provide the needed relations to specify the second reflection
point $\vec{R'}_{2}$ in terms of the known quantities, $\vec{R'}_{0},$
$\vec{k'}_{1}$ and $\vec{R'}_{1}.$ It is convenient to expand equations
(\ref{eq:polynomial-1}) and (\ref{eq:R1-cross-R-of-i-plus-3-Algebraic})
as \begin{align}
\overbrace{\left[r'_{i}\right]^{2}\cos\left(\pi-2\theta_{inc}\right)}^{-\left[d\Gamma_{1,2}/d\theta_{inc}\right]/2}-\nu'_{1,1}\nu'_{2,1}-\nu'_{1,2}\nu'_{2,2}-\nu'_{1,3}\nu'_{2,3} & =0,\label{eq:polynomial-1-expanded}\end{align}
 \begin{align}
\nu'_{1,2}\nu'_{2,3}-\nu'_{1,3}\nu'_{2,2}+\Gamma_{1,2}\left\Vert \vec{n'}_{p,1}\right\Vert ^{-1}\epsilon_{1jk}k'_{1,j}r'_{0,k} & =0,\label{eq:R1-cross-R-of-i-plus-3-Algebraic-x}\end{align}
 \begin{align}
\nu'_{1,3}\nu'_{2,1}-\nu'_{1,1}\nu'_{2,3}+\Gamma_{1,2}\left\Vert \vec{n'}_{p,1}\right\Vert ^{-1}\epsilon_{2jk}k'_{1,j}r'_{0,k} & =0,\label{eq:R1-cross-R-of-i-plus-3-Algebraic-y}\end{align}
 \begin{align}
\nu'_{1,1}\nu'_{2,2}-\nu'_{1,2}\nu'_{2,1}+\Gamma_{1,2}\left\Vert \vec{n'}_{p,1}\right\Vert ^{-1}\epsilon_{3jk}k'_{1,j}r'_{0,k} & =0.\label{eq:R1-cross-R-of-i-plus-3-Algebraic-z}\end{align}
 Equations (\ref{eq:polynomial-1-expanded}) and (\ref{eq:R1-cross-R-of-i-plus-3-Algebraic-x})
are added to yield \begin{align}
\nu'_{1,1}\nu'_{2,1}+\left[\nu'_{1,2}+\nu'_{1,3}\right]\nu'_{2,2}+\left[\nu'_{1,3}-\nu'_{1,2}\right]\nu_{2,3} & =\Gamma_{1,2}\left\Vert \vec{n'}_{p,1}\right\Vert ^{-1}\epsilon_{1jk}k'_{1,j}r'_{0,k}-\frac{1}{2}\frac{d\Gamma_{1,2}}{d\theta_{inc}}.\label{eq:R1-cross-R-Algebraic-poly-x-combo}\end{align}
 Equations (\ref{eq:polynomial-1-expanded}) and (\ref{eq:R1-cross-R-of-i-plus-3-Algebraic-y})
are added to give \begin{align}
\left[\nu'_{1,1}-\nu'_{1,3}\right]\nu'_{2,1}+\nu'_{1,2}\nu'_{2,2}+\left[\nu'_{1,3}+\nu'_{1,1}\right]\nu'_{2,3} & =\Gamma_{1,2}\left\Vert \vec{n'}_{p,1}\right\Vert ^{-1}\epsilon_{2jk}k'_{1,j}r'_{0,k}-\frac{1}{2}\frac{d\Gamma_{1,2}}{d\theta_{inc}}.\label{eq:R1-cross-R-Algebraic-poly-y-combo}\end{align}
 Similarly, equations (\ref{eq:polynomial-1-expanded}) and (\ref{eq:R1-cross-R-of-i-plus-3-Algebraic-z})
are combined to give \begin{align}
\left[\nu'_{1,1}+\nu'_{1,2}\right]\nu'_{2,1}+\left[\nu'_{1,2}-\nu'_{1,1}\right]\nu'_{2,2}+\nu'_{1,3}\nu'_{2,3} & =\Gamma_{1,2}\left\Vert \vec{n'}_{p,1}\right\Vert ^{-1}\epsilon_{3jk}k'_{1,j}r'_{0,k}-\frac{1}{2}\frac{d\Gamma_{1,2}}{d\theta_{inc}}.\label{eq:R1-cross-R-Algebraic-poly-z-combo}\end{align}
 Define the quantities \begin{equation}
\left\{ \begin{array}{ccccc}
\alpha_{1}=\nu'_{1,2}+\nu'_{1,3}, &  & \alpha_{2}=\nu'_{1,3}-\nu'_{1,2}, &  & \zeta_{1}=\Gamma_{1,2}\left\Vert \vec{n'}_{p,1}\right\Vert ^{-1}\epsilon_{1jk}k'_{1,j}r'_{0,k}-\frac{1}{2}\frac{d\Gamma_{1,2}}{d\theta_{inc}},\\
\\\alpha_{3}=\nu'_{1,1}-\nu'_{1,3}, &  & \alpha_{4}=\nu'_{1,3}+\nu'_{1,1}, &  & \zeta_{2}=\Gamma_{1,2}\left\Vert \vec{n'}_{p,1}\right\Vert ^{-1}\epsilon_{2jk}k'_{1,j}r'_{0,k}-\frac{1}{2}\frac{d\Gamma_{1,2}}{d\theta_{inc}},\\
\\\alpha_{5}=\nu'_{1,1}+\nu'_{1,2}, &  & \alpha_{6}=\nu'_{1,2}-\nu'_{1,1}, &  & \zeta_{3}=\Gamma_{1,2}\left\Vert \vec{n'}_{p,1}\right\Vert ^{-1}\epsilon_{3jk}k'_{1,j}r'_{0,k}-\frac{1}{2}\frac{d\Gamma_{1,2}}{d\theta_{inc}},\end{array}\right.\label{eq:R1-cross-R-of-i-plus-3-Algebraic-Def}\end{equation}
 where $\Gamma_{1,2}$ is defined in equation (\ref{eq:Ri-cross-R-i-plus-1-Gamma2-Value}).
Equations (\ref{eq:R1-cross-R-Algebraic-poly-x-combo}), (\ref{eq:R1-cross-R-Algebraic-poly-y-combo})
and (\ref{eq:R1-cross-R-Algebraic-poly-z-combo}) form a reduced set
\[
\begin{array}{ccccc}
\nu'_{1,1}\nu'_{2,1}+\alpha_{1}\nu'_{2,2}+\alpha_{2}\nu'_{2,3}=\zeta_{1}, &  & \alpha_{3}\nu'_{2,1}+\nu'_{1,2}\nu'_{2,2}+\alpha_{4}\nu'_{2,3}=\zeta_{2}, &  & \alpha_{5}\nu'_{2,1}+\alpha_{6}\nu'_{2,2}+\nu'_{1,3}\nu'_{2,3}=\zeta_{3}\end{array}.\]
 In matrix form it reads \begin{align}
\underbrace{\left[\begin{array}{ccc}
\nu'_{1,1} & \alpha_{1} & \alpha_{2}\\
\alpha_{3} & \nu'_{1,2} & \alpha_{4}\\
\alpha_{5} & \alpha_{6} & \nu'_{1,3}\end{array}\right]}_{\widetilde{M}_{0}}\cdot\left[\begin{array}{c}
\nu'_{2,1}\\
\nu'_{2,2}\\
\nu'_{2,3}\end{array}\right] & =\left[\begin{array}{c}
\zeta_{1}\\
\zeta_{2}\\
\zeta_{3}\end{array}\right],\label{eq:R1-cross-R-of-i-Matrix-M0}\end{align}
 and its determinant is expressed as \begin{align}
\det\left(\widetilde{M}_{0}\right) & =\left[\nu'_{1,1}+\nu'_{1,2}+\nu'_{1,3}\right]\left\{ \left[\nu'_{1,1}\right]^{2}+\left[\nu'_{1,2}\right]^{2}+\left[\nu'_{1,3}\right]^{2}\right\} =\left[r'_{i}\right]^{2}\left[\nu'_{1,1}+\nu'_{1,2}+\nu'_{1,3}\right].\label{eq:R1-cross-R-of-i-DET-M0}\end{align}
 Three new matrices are then defined here as \[
\begin{array}{ccccc}
\widetilde{M}_{1}=\left[\begin{array}{ccc}
\zeta_{1} & \alpha_{1} & \alpha_{2}\\
\zeta_{2} & \nu'_{1,2} & \alpha_{4}\\
\zeta_{3} & \alpha_{6} & \nu'_{1,3}\end{array}\right], &  & \widetilde{M}_{2}=\left[\begin{array}{ccc}
\nu'_{1,1} & \zeta_{1} & \alpha_{2}\\
\alpha_{3} & \zeta_{2} & \alpha_{4}\\
\alpha_{5} & \zeta_{3} & \nu'_{1,3}\end{array}\right], &  & \widetilde{M}_{3}=\left[\begin{array}{ccc}
\nu'_{1,1} & \alpha_{1} & \zeta_{1}\\
\alpha_{3} & \nu'_{1,2} & \zeta_{2}\\
\alpha_{5} & \alpha_{6} & \zeta_{3}\end{array}\right].\end{array}\]
 The variables $\nu'_{2,1},$ $\nu'_{2,2}$ and $\nu'_{2,3}$ are
solved with the Cramer's Rule as \[
\begin{array}{ccccc}
\nu'_{2,1}=\det\left(\widetilde{M}_{1}\right)/\det\left(\widetilde{M}_{0}\right), &  & \nu'_{2,2}=\det\left(\widetilde{M}_{2}\right)/\det\left(\widetilde{M}_{0}\right), &  & \nu'_{2,3}=\det\left(\widetilde{M}_{3}\right)/\det\left(\widetilde{M}_{0}\right).\end{array}\]
 Explicitly, they are given by \begin{align}
\grave{\nu}'_{2,1}\equiv\nu'_{2,1} & =\left(\nu'_{1,1}\left[\nu'_{1,1}-\nu'_{1,2}+\nu'_{1,3}\right]\zeta_{1}+\left\{ \nu'_{1,1}\left[\nu'_{1,2}-\nu'_{1,3}\right]-\left[\nu'_{1,2}\right]^{2}-\left[\nu'_{1,3}\right]^{2}\right\} \zeta_{2}\right.\nonumber \\
 & \left.+\left\{ \nu'_{1,1}\left[\nu'_{1,2}+\nu'_{1,3}\right]+\left[\nu'_{1,2}\right]^{2}+\left[\nu'_{1,3}\right]^{2}\right\} \zeta_{3}\right)\left[\nu'_{1,1}+\nu'_{1,2}+\nu'_{1,3}\right]^{-1}\left[r'_{i}\right]^{-2},\label{eq:x2-prime}\end{align}
 \begin{align}
\grave{\nu}'_{2,2}\equiv\nu'_{2,2} & =\left(\left\{ \nu'_{1,2}\left[\nu'_{1,1}+\nu'_{1,3}\right]+\left[\nu'_{1,1}\right]^{2}+\left[\nu'_{1,3}\right]^{2}\right\} \zeta_{1}+\left\{ \nu'_{1,2}\left[\nu'_{1,1}-\nu'_{1,3}\right]+\left[\nu'_{1,2}\right]^{2}\right\} \zeta_{2}\right.\nonumber \\
 & \left.+\left\{ \nu'_{1,2}\left[\nu'_{1,3}-\nu'_{1,1}\right]-\left[\nu'_{1,1}\right]^{2}-\left[\nu'_{1,3}\right]^{2}\right\} \zeta_{3}\right)\left[\nu'_{1,1}+\nu'_{1,2}+\nu'_{1,3}\right]^{-1}\left[r'_{i}\right]^{-2},\label{eq:y2-prime}\end{align}
 \begin{align}
\grave{\nu}'_{2,3}\equiv\nu'_{2,3} & =\left(\left\{ \left[\nu'_{1,1}-\nu'_{1,2}\right]\nu'_{1,3}-\left[\nu'_{1,1}\right]^{2}-\left[\nu'_{1,2}\right]^{2}\right\} \zeta_{1}+\left\{ \left[\nu'_{1,1}+\nu'_{1,2}\right]\nu'_{1,3}+\left[\nu'_{1,1}\right]^{2}+\left[\nu'_{1,2}\right]^{2}\right\} \zeta_{2}\right.\nonumber \\
 & \left.+\left\{ \left[\nu'_{1,2}-\nu'_{1,1}\right]\nu'_{1,3}+\left[\nu'_{1,3}\right]^{2}\right\} \zeta_{3}\right)\left[\nu'_{1,1}+\nu'_{1,2}+\nu'_{1,3}\right]^{-1}\left[r'_{i}\right]^{-2},\label{eq:z2-prime}\end{align}
 where \begin{eqnarray}
\nu'_{1,1}+\nu'_{1,2}+\nu'_{1,3}\neq0, &  & \begin{array}{ccc}
\left[\begin{array}{c}
\nu'_{1,1}=x'_{1}\left(r'_{i},\theta_{1},\phi_{1}\right)\\
\\\nu'_{1,2}=y'_{1}\left(r'_{i},\theta_{1},\phi_{1}\right)\\
\\\nu'_{1,3}=z'_{1}\left(r'_{i},\theta_{1},\phi_{1}\right)\end{array}\right], &  & \left[\begin{array}{c}
\nu'_{2,1}\left(\nu'_{1,1},\nu'_{1,2},\nu'_{1,3}\right)=x'_{2}\\
\\\nu'_{2,2}\left(\nu'_{1,1},\nu'_{1,2},\nu'_{1,3}\right)=y'_{2}\\
\\\nu'_{2,3}\left(\nu'_{1,1},\nu'_{1,2},\nu'_{1,3}\right)=z'_{2}\end{array}\right].\end{array}\label{eq:x1-y2-z2-prime-condition}\end{eqnarray}
 In the above set of equations, $\grave{\nu}'_{2,i}$ has been used
to indicate that $\nu'_{2,i}$ is now expressed explicitly in terms
of the Cartesian coordinates $\left(\nu'_{1,1},\nu'_{1,2},\nu'_{1,3}\right)$
instead of spherical coordinates corresponding to the second reflection
point, $\left(r'_{i},\theta_{2},\phi_{2}\right).$ The second reflection
point inside the hemisphere is then from equation (\ref{eq:R-i-plus-1-of-x-y-z-carte}),
\begin{align*}
\vec{R'}_{2}\left(\grave{\nu}'_{2,1},\grave{\nu}'_{2,2},\grave{\nu}'_{2,3}\right) & =\sum_{i=1}^{3}\grave{\nu}'_{2,i}\hat{e_{i}},\end{align*}
 where $\grave{\nu}'_{2,i},$ $i=1,2,3$ are given in equations (\ref{eq:x2-prime})
through (\ref{eq:z2-prime}) with restriction given in equation (\ref{eq:x1-y2-z2-prime-condition}).
In general, all subsequent reflection points $\vec{R'}_{N}$ can be
expressed in generic form \begin{align*}
\vec{R'}_{N}\left(\grave{\nu}'_{N,1},\grave{\nu}'_{N,2},\grave{\nu}'_{N,3}\right) & =\sum_{i=1}^{3}\grave{\nu}'_{N,i}\hat{e_{i}}\end{align*}
 through iterative applications of the result $\grave{\nu}'_{2,i},$
$i=1,2,3.$ This however proves to be very inefficient technique.
A better way is to express $\vec{R'}_{N}$ in terms of spherical coordinates.
Because $\vec{R'}_{2}$ belongs to a spanning set for the plane of
incidence whose unit normal is $\hat{n'}_{p,1}$ defined in equation
(\ref{eq:n-hat-p-i-eq-mag-times-ki-cross-Ro}), the component relations
$\grave{\nu}'_{2,i}$ of equations (\ref{eq:x2-prime}), (\ref{eq:y2-prime})
and (\ref{eq:z2-prime}) satisfy the intercept relation given in equation
(\ref{eq:hemi-plane-intercept-root}), \begin{eqnarray*}
\grave{\nu}'_{2,i}=\frac{1}{2}\left\Vert \vec{n'}_{p,1}\right\Vert ^{-1}\epsilon_{ijk}k'_{1,j}r'_{0,k}\pm\left\{ \left[\frac{1}{2}\left\Vert \vec{n'}_{p,1}\right\Vert ^{-1}\epsilon_{ijk}k'_{1,j}r'_{0,k}\right]^{2}+\left[r'_{i,i}\right]^{2}\right\} ^{1/2}, &  & i=1,2,3,\end{eqnarray*}
 where $r'_{i,1}=r'_{i}\sin\theta'_{2}\cos\phi'_{2},$ $r'_{i,2}=r'_{i}\sin\theta'_{2}\sin\phi'_{2}$
and $r'_{i,3}=r'_{i}\cos\theta'_{2}.$ Here the subscript $2$ of
angular variables $\theta'_{2}$ and $\phi'_{2}$ denote the second
reflection point. In terms of the angular variables, using the above
expression for $\grave{\nu}'_{2,i},$ the $\grave{\nu}'_{2,1},$ $\grave{\nu}'_{2,2}$
and $\grave{\nu}'_{2,3}$ are expressed as \begin{align}
\mp\left\{ \left[\frac{1}{2}\left\Vert \vec{n'}_{p,1}\right\Vert ^{-1}\epsilon_{1jk}k'_{1,j}r'_{0,k}\right]^{2}+\left[r'_{i}\right]^{2}\sin^{2}\theta'_{2}\cos^{2}\phi'_{2}\right\} ^{1/2} & =\frac{1}{2}\left\Vert \vec{n'}_{p,1}\right\Vert ^{-1}\epsilon_{1jk}k'_{1,j}r'_{0,k}-\grave{\nu}'_{2,1},\label{eq:hemi-plane-intercept-root-x2-prime}\end{align}
 \begin{align}
\mp\left\{ \left[\frac{1}{2}\left\Vert \vec{n'}_{p,1}\right\Vert ^{-1}\epsilon_{2jk}k'_{1,j}r'_{0,k}\right]^{2}+\left[r'_{i}\right]^{2}\sin^{2}\theta'_{2}\sin^{2}\phi'_{2}\right\} ^{1/2} & =\frac{1}{2}\left\Vert \vec{n'}_{p,1}\right\Vert ^{-1}\epsilon_{2jk}k'_{1,j}r'_{0,k}-\grave{\nu}'_{2,2},\label{eq:hemi-plane-intercept-root-y2-prime}\end{align}
 \begin{align}
\mp\left\{ \left[\frac{1}{2}\left\Vert \vec{n'}_{p,1}\right\Vert ^{-1}\epsilon_{3jk}k'_{1,j}r'_{0,k}\right]^{2}+\left[r'_{i}\right]^{2}\cos^{2}\theta'_{2}\right\} ^{1/2} & =\frac{1}{2}\left\Vert \vec{n'}_{p,1}\right\Vert ^{-1}\epsilon_{3jk}k'_{1,j}r'_{0,k}-\grave{\nu}'_{2,3}.\label{eq:hemi-plane-intercept-root-z2-prime}\end{align}
 Square both sides of equation (\ref{eq:hemi-plane-intercept-root-x2-prime}),
$\cos^{2}\phi'_{2}$ can be solved as \begin{align}
\cos^{2}\phi'_{2} & =\left[r'_{i}\sin\theta'_{2}\right]^{-2}\left\{ \left[\frac{1}{2}\left\Vert \vec{n'}_{p,1}\right\Vert ^{-1}\epsilon_{1jk}k'_{1,j}r'_{0,k}-\grave{\nu}'_{2,1}\right]^{2}-\left[\frac{1}{2}\left\Vert \vec{n'}_{p,1}\right\Vert ^{-1}\epsilon_{1jk}k'_{1,j}r'_{0,k}\right]^{2}\right\} .\label{eq:cosine-squared-phi2-prime}\end{align}
 Similarly, square both sides of equation (\ref{eq:hemi-plane-intercept-root-y2-prime}),
$\sin^{2}\phi'_{2}$ can be solved as \begin{align}
\sin^{2}\phi'_{2} & =\left[r'_{i}\sin\theta'_{2}\right]^{-2}\left\{ \left[\frac{1}{2}\left\Vert \vec{n'}_{p,1}\right\Vert ^{-1}\epsilon_{2jk}k'_{1,j}r'_{0,k}-\grave{\nu}'_{2,2}\right]^{2}-\left[\frac{1}{2}\left\Vert \vec{n'}_{p,1}\right\Vert ^{-1}\epsilon_{2jk}k'_{1,j}r'_{0,k}\right]^{2}\right\} .\label{eq:sine-squared-phi2-prime}\end{align}
 The last two equations are divided to give \begin{align*}
\tan^{2}\phi'_{2} & =\frac{\left[\frac{1}{2}\left\Vert \vec{n'}_{p,1}\right\Vert ^{-1}\epsilon_{2mn}k'_{1,m}r'_{0,n}-\grave{\nu}'_{2,2}\right]^{2}-\left[\frac{1}{2}\left\Vert \vec{n'}_{p,1}\right\Vert ^{-1}\epsilon_{2mn}k'_{1,m}r'_{0,n}\right]^{2}}{\left[\frac{1}{2}\left\Vert \vec{n'}_{p,1}\right\Vert ^{-1}\epsilon_{1jk}k'_{1,j}r'_{0,k}-\grave{\nu}'_{2,1}\right]^{2}-\left[\frac{1}{2}\left\Vert \vec{n'}_{p,1}\right\Vert ^{-1}\epsilon_{1jk}k'_{1,j}r'_{0,k}\right]^{2}}.\end{align*}
 The azimuthal angle $\phi'_{2}$ is obtained as \begin{align*}
\phi'_{2} & =\arctan\left(\pm\left\{ \frac{\left[\frac{1}{2}\left\Vert \vec{n'}_{p,1}\right\Vert ^{-1}\epsilon_{2mn}k'_{1,m}r'_{0,n}-\grave{\nu}'_{2,2}\right]^{2}-\left[\frac{1}{2}\left\Vert \vec{n'}_{p,1}\right\Vert ^{-1}\epsilon_{2mn}k'_{1,m}r'_{0,n}\right]^{2}}{\left[\frac{1}{2}\left\Vert \vec{n'}_{p,1}\right\Vert ^{-1}\epsilon_{1jk}k'_{1,j}r'_{0,k}-\grave{\nu}'_{2,1}\right]^{2}-\left[\frac{1}{2}\left\Vert \vec{n'}_{p,1}\right\Vert ^{-1}\epsilon_{1jk}k'_{1,j}r'_{0,k}\right]^{2}}\right\} ^{1/2}\right).\end{align*}
 Following the procedures used in equations (\ref{eq:tangent-squared-phi-prime-inequality})
through (\ref{eq:phi-prime-90-180-and-270-360}), the following results
are arrived at \begin{equation}
\left\{ \begin{array}{c}
\begin{array}{ccc}
\lim_{\varepsilon\rightarrow0}\left(0\leq\phi'_{2}\leq\frac{1}{2}\pi-\left|\varepsilon\right|\right), &  & \lim_{\varepsilon\rightarrow0}\left(\pi+\left|\varepsilon\right|\leq\phi'_{2}\leq\frac{3}{2}\pi-\left|\varepsilon\right|\right),\end{array}\\
\\\phi'_{2}=\arctan\left(\left\{ \frac{\left[\frac{1}{2}\left\Vert \vec{n'}_{p,1}\right\Vert ^{-1}\epsilon_{2mn}k'_{1,m}r'_{0,n}-\grave{\nu}'_{2,2}\right]^{2}-\left[\frac{1}{2}\left\Vert \vec{n'}_{p,1}\right\Vert ^{-1}\epsilon_{2mn}k'_{1,m}r'_{0,n}\right]^{2}}{\left[\frac{1}{2}\left\Vert \vec{n'}_{p,1}\right\Vert ^{-1}\epsilon_{1jk}k'_{1,j}r'_{0,k}-\grave{\nu}'_{2,1}\right]^{2}-\left[\frac{1}{2}\left\Vert \vec{n'}_{p,1}\right\Vert ^{-1}\epsilon_{1jk}k'_{1,j}r'_{0,k}\right]^{2}}\right\} ^{1/2}\right);\end{array}\right.\label{eq:phi2-prime-0-90-and-180-270-final}\end{equation}
 \begin{equation}
\left\{ \begin{array}{c}
\begin{array}{ccc}
\lim_{\varepsilon\rightarrow0}\left(\frac{1}{2}\pi+\left|\varepsilon\right|\leq\phi'_{2}\leq\pi-\varepsilon\right), &  & \lim_{\varepsilon\rightarrow0}\left(\frac{3}{2}\pi+\left|\varepsilon\right|\leq\phi'_{2}<2\pi-\left|\varepsilon\right|\right),\end{array}\\
\\\phi'_{2}=\arctan\left(-\left\{ \frac{\left[\frac{1}{2}\left\Vert \vec{n'}_{p,1}\right\Vert ^{-1}\epsilon_{2mn}k'_{1,m}r'_{0,n}-\grave{\nu}'_{2,2}\right]^{2}-\left[\frac{1}{2}\left\Vert \vec{n'}_{p,1}\right\Vert ^{-1}\epsilon_{2mn}k'_{1,m}r'_{0,n}\right]^{2}}{\left[\frac{1}{2}\left\Vert \vec{n'}_{p,1}\right\Vert ^{-1}\epsilon_{1jk}k'_{1,j}r'_{0,k}-\grave{\nu}'_{2,1}\right]^{2}-\left[\frac{1}{2}\left\Vert \vec{n'}_{p,1}\right\Vert ^{-1}\epsilon_{1jk}k'_{1,j}r'_{0,k}\right]^{2}}\right\} ^{1/2}\right).\end{array}\right.\label{eq:phi2-prime-90-180-and-270-360-final}\end{equation}
 The solution for $\phi'_{2}$ forms a generic structure for any subsequent
reflection points on the inner hemisphere surface. The $N$th azimuthal
angle $\phi'_{N}$ is found following a prescribed sequential steps
\begin{equation}
\phi'_{1}\rightarrow\phi'_{2}\rightarrow\phi'_{3}\rightarrow\cdots\rightarrow\phi'_{N-1}\rightarrow\phi'_{N}.\label{eq:sequence-rela-phi123-primes}\end{equation}
 By reversing the direction of sequence, $\phi'_{N}$ can be expressed
in terms of the initial azimuthal angle $\phi'_{1},$ $\phi'_{N}=\phi'_{N}\left(\phi'_{1}\right).$
The polar angle $\theta'_{2}$ of the second reflection point can
be found by squaring equation (\ref{eq:hemi-plane-intercept-root-z2-prime}),
which yields \begin{align}
\cos^{2}\theta'_{2} & =\left[r'_{i}\right]^{-2}\left\{ \left[\frac{1}{2}\left\Vert \vec{n'}_{p,1}\right\Vert ^{-1}\epsilon_{3jk}k'_{1,j}r'_{0,k}-\grave{\nu}'_{2,3}\right]^{2}-\left[\frac{1}{2}\left\Vert \vec{n'}_{p,1}\right\Vert ^{-1}\epsilon_{3jk}k'_{1,j}r'_{0,k}\right]^{2}\right\} .\label{eq:cosine-squared-theta2-prime}\end{align}
 Add together equations (\ref{eq:cosine-squared-phi2-prime}) and
(\ref{eq:sine-squared-phi2-prime}), $\sin\theta'_{2}$ can be solved
as \begin{align}
\sin^{2}\theta'_{2} & =\left[r'_{i}\right]^{-2}\left\{ \left[\frac{1}{2}\left\Vert \vec{n'}_{p,1}\right\Vert ^{-1}\epsilon_{1jk}k'_{1,j}r'_{0,k}-\grave{\nu}'_{2,1}\right]^{2}-\left[\frac{1}{2}\left\Vert \vec{n'}_{p,1}\right\Vert ^{-1}\epsilon_{1jk}k'_{1,j}r'_{0,k}\right]^{2}\right.\nonumber \\
 & \left.+\left[\frac{1}{2}\left\Vert \vec{n'}_{p,1}\right\Vert ^{-1}\epsilon_{2mn}k'_{1,m}r'_{0,n}-\grave{\nu}'_{2,2}\right]^{2}-\left[\frac{1}{2}\left\Vert \vec{n'}_{p,1}\right\Vert ^{-1}\epsilon_{2mn}k'_{1,m}r'_{0,n}\right]^{2}\right\} .\label{eq:sine-squared-theta2-prime}\end{align}
 By dividing equations (\ref{eq:cosine-squared-theta2-prime}) and
(\ref{eq:sine-squared-theta2-prime}), we get \begin{align*}
\tan^{2}\theta'_{2} & =\left\{ \left[\frac{1}{2}\left\Vert \vec{n'}_{p,1}\right\Vert ^{-1}\epsilon_{1jk}k'_{1,j}r'_{0,k}-\grave{\nu}'_{2,1}\right]^{2}-\left[\frac{1}{2}\left\Vert \vec{n'}_{p,1}\right\Vert ^{-1}\epsilon_{1jk}k'_{1,j}r'_{0,k}\right]^{2}\right.\\
 & \left.+\left[\frac{1}{2}\left\Vert \vec{n'}_{p,1}\right\Vert ^{-1}\epsilon_{2mn}k'_{1,m}r'_{0,n}-\grave{\nu}'_{2,2}\right]^{2}-\left[\frac{1}{2}\left\Vert \vec{n'}_{p,1}\right\Vert ^{-1}\epsilon_{2mn}k'_{1,m}r'_{0,n}\right]^{2}\right\} \\
 & \times\left\{ \left[\frac{1}{2}\left\Vert \vec{n'}_{p,1}\right\Vert ^{-1}\epsilon_{3qr}k'_{1,q}r'_{0,r}-\grave{\nu}'_{2,3}\right]^{2}-\left[\frac{1}{2}\left\Vert \vec{n'}_{p,1}\right\Vert ^{-1}\epsilon_{3qr}k'_{1,q}r'_{0,r}\right]^{2}\right\} ^{-1}.\end{align*}
 The polar angle $\theta'_{2}$ is given by \begin{align*}
\theta'_{2} & =\arctan\left(\pm\left\{ \left[\frac{1}{2}\left\Vert \vec{n'}_{p,1}\right\Vert ^{-1}\epsilon_{1jk}k'_{1,j}r'_{0,k}-\grave{\nu}'_{2,1}\right]^{2}-\left[\frac{1}{2}\left\Vert \vec{n'}_{p,1}\right\Vert ^{-1}\epsilon_{1jk}k'_{1,j}r'_{0,k}\right]^{2}\right.\right.\\
 & \left.+\left[\frac{1}{2}\left\Vert \vec{n'}_{p,1}\right\Vert ^{-1}\epsilon_{2mn}k'_{1,m}r'_{0,n}-\grave{\nu}'_{2,2}\right]^{2}-\left[\frac{1}{2}\left\Vert \vec{n'}_{p,1}\right\Vert ^{-1}\epsilon_{2mn}k'_{1,m}r'_{0,n}\right]^{2}\right\} ^{1/2}\\
 & \left.\times\left\{ \left[\frac{1}{2}\left\Vert \vec{n'}_{p,1}\right\Vert ^{-1}\epsilon_{3qr}k'_{1,q}r'_{0,r}-\grave{\nu}'_{2,3}\right]^{2}-\left[\frac{1}{2}\left\Vert \vec{n'}_{p,1}\right\Vert ^{-1}\epsilon_{3qr}k'_{1,q}r'_{0,r}\right]^{2}\right\} ^{-1/2}\right).\end{align*}
 Following the procedures given in equations (\ref{eq:tangent-squared-theta-prime-inequality})
through (\ref{eq:theta-prime-90-to-180-degree}), the following results
are obtained \begin{equation}
\left\{ \begin{array}{c}
\lim_{\varepsilon\rightarrow0}\left(0\leq\theta'_{2}\leq\frac{1}{2}\pi-\left|\varepsilon\right|\right),\\
\\\theta'_{2}=\arctan\left(\left\{ \left[\frac{1}{2}\left\Vert \vec{n'}_{p,1}\right\Vert ^{-1}\epsilon_{1jk}k'_{1,j}r'_{0,k}-\grave{\nu}'_{2,1}\right]^{2}-\left[\frac{1}{2}\left\Vert \vec{n'}_{p,1}\right\Vert ^{-1}\epsilon_{1jk}k'_{1,j}r'_{0,k}\right]^{2}\right.\right.\\
\left.+\left[\frac{1}{2}\left\Vert \vec{n'}_{p,1}\right\Vert ^{-1}\epsilon_{2mn}k'_{1,m}r'_{0,n}-\grave{\nu}'_{2,2}\right]^{2}-\left[\frac{1}{2}\left\Vert \vec{n'}_{p,1}\right\Vert ^{-1}\epsilon_{2mn}k'_{1,m}r'_{0,n}\right]^{2}\right\} ^{1/2}\\
\left.\times\left\{ \left[\frac{1}{2}\left\Vert \vec{n'}_{p,1}\right\Vert ^{-1}\epsilon_{3qr}k'_{1,q}r'_{0,r}-\grave{\nu}'_{2,3}\right]^{2}-\left[\frac{1}{2}\left\Vert \vec{n'}_{p,1}\right\Vert ^{-1}\epsilon_{3qr}k'_{1,q}r'_{0,r}\right]^{2}\right\} ^{-1/2}\right);\end{array}\right.\label{eq:theta2-prime-0-to-90-degree-final}\end{equation}
 \begin{equation}
\left\{ \begin{array}{c}
\lim_{\varepsilon\rightarrow0}\left(\frac{1}{2}\pi+\left|\varepsilon\right|\leq\theta'_{2}\leq\pi\right),\\
\\\theta'_{2}=\arctan\left(-\left\{ \left[\frac{1}{2}\left\Vert \vec{n'}_{p,1}\right\Vert ^{-1}\epsilon_{1jk}k'_{1,j}r'_{0,k}-\grave{\nu}'_{2,1}\right]^{2}-\left[\frac{1}{2}\left\Vert \vec{n'}_{p,1}\right\Vert ^{-1}\epsilon_{1jk}k'_{1,j}r'_{0,k}\right]^{2}\right.\right.\\
\left.+\left[\frac{1}{2}\left\Vert \vec{n'}_{p,1}\right\Vert ^{-1}\epsilon_{2mn}k'_{1,m}r'_{0,n}-\grave{\nu}'_{2,2}\right]^{2}-\left[\frac{1}{2}\left\Vert \vec{n'}_{p,1}\right\Vert ^{-1}\epsilon_{2mn}k'_{1,m}r'_{0,n}\right]^{2}\right\} ^{1/2}\\
\left.\times\left\{ \left[\frac{1}{2}\left\Vert \vec{n'}_{p,1}\right\Vert ^{-1}\epsilon_{3qr}k'_{1,q}r'_{0,r}-\grave{\nu}'_{2,3}\right]^{2}-\left[\frac{1}{2}\left\Vert \vec{n'}_{p,1}\right\Vert ^{-1}\epsilon_{3qr}k'_{1,q}r'_{0,r}\right]^{2}\right\} ^{-1/2}\right).\end{array}\right.\label{eq:theta2-prime-90-to-180-degree-final}\end{equation}
 The above result of $\theta'_{2}$ forms a generic structure for
any subsequent reflection points on the inner hemisphere surface.
The $N$th polar angle $\theta'_{N}$ can be obtained by following
the sequential steps \begin{equation}
\theta'_{1}\rightarrow\theta'_{2}\rightarrow\theta'_{3}\rightarrow\cdots\rightarrow\theta'_{N-1}\rightarrow\theta'_{N}.\label{eq:sequence-rela-theta123-primes}\end{equation}
 Equivalently, reversing the direction of sequence, $\theta'_{N}$
can be expressed as a function of the initial polar angle $\theta'_{1},$
$\theta'_{N}=\theta'_{N}\left(\theta'_{1}\right).$ With angular variable
$\phi'_{2}$ defined in equations (\ref{eq:phi2-prime-0-90-and-180-270-final})
and (\ref{eq:phi2-prime-90-180-and-270-360-final}), and $\theta'_{2}$
defined in equations (\ref{eq:theta2-prime-0-to-90-degree-final})
and (\ref{eq:theta2-prime-90-to-180-degree-final}), the second reflection
point on the inner hemisphere surface is given by \begin{eqnarray}
\vec{R'}_{2}\left(r'_{i},\theta'_{2},\phi'_{2}\right)=\sum_{i=1}^{3}\nu'_{2,i}\left(r'_{i},\theta'_{2},\phi'_{2}\right)\hat{e_{i}}, &  & i=\left\{ \begin{array}{c}
1\rightarrow\nu'_{2,1}=r'_{i}\sin\theta'_{2}\cos\phi'_{2},\\
\\2\rightarrow\nu'_{2,2}=r'_{i}\sin\theta'_{2}\sin\phi'_{2},\\
\\3\rightarrow\nu'_{2,3}=r'_{i}\cos\theta'_{2}.\qquad\;\:\end{array}\right.\label{eq:2nd-Reflection-Point}\end{eqnarray}

\begin{figure}[t]
\begin{center}\includegraphics[%
  scale=0.7]{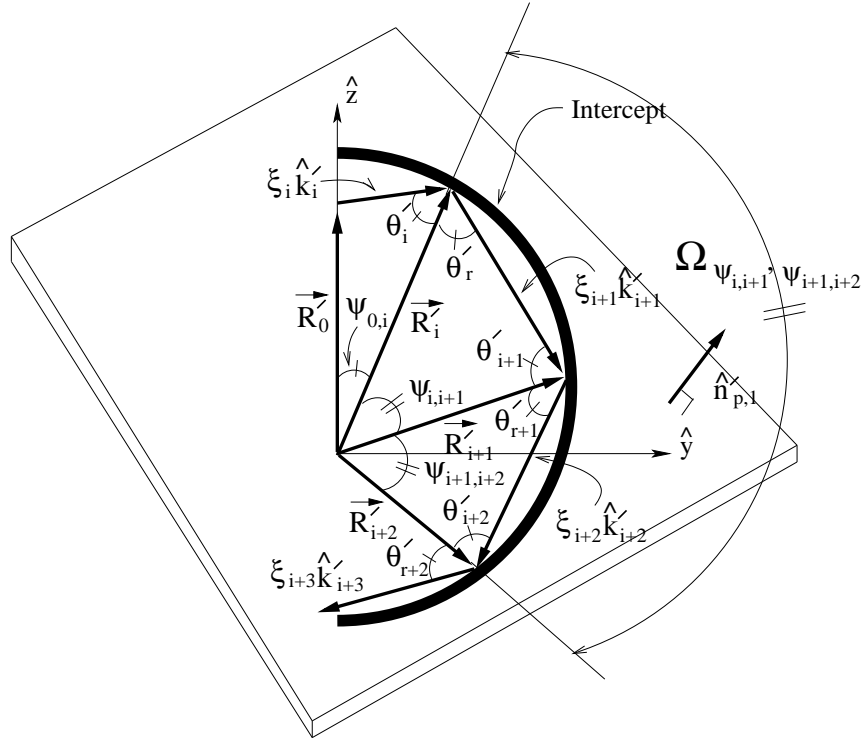}\end{center}

\caption{The two immediate neighboring reflection points $\vec{R'}_{1}$ and
$\vec{R'}_{2}$ are connected through the angle $\psi_{1,2}.$ Similarly,
the two distant neighbor reflection points $\vec{R'}_{i}$ and $\vec{R'}_{i+2}$
are connected through the angle $\Omega_{\psi_{i,i+1},\psi_{i+1,i+2}}.$
\label{cap:plane-sphere-intersection-2}}
\end{figure}

As shown in Figure \ref{cap:plane-sphere-intersection-2}, two reflection
points $\vec{R'}_{1}$ and $\vec{R'}_{2}$ are related through $\psi_{1,2},$
which is the angle measured between the two. Since $\vec{R'}_{j},$
where the index $j=1,2,\cdots,N_{max}$ and $N_{max}$ is the last
count of reflection before a repeat in cycle, belongs to a spanning
set for a plane of incidence whose unit normal is $\hat{n'}_{p,1}$
given in equation (\ref{eq:n-hat-p-i-eq-mag-times-ki-cross-Ro}),
all reflections occur on the same plane of incidence. The task of
determining the $N$th subsequent reflection point $\vec{R'}_{N}$
is therefore particularly simple. The needed connection formulae between
the initial reflection point $\vec{R'}_{1}$ and the $N$th subsequent
reflection point $\vec{R'}_{N}$ is found through both scalar and
vector cross product relations similar to those given in equations
(\ref{eq:Ri-Ri-plus-1-dot-product-1-equiv-1-psi-pre}) and (\ref{eq:Ri-cross-R-of-i-plus-1}).
In order to generalize the previous result for $\vec{R'}_{2}$ to
$\vec{R'}_{N},$ recall the set from equations (\ref{eq:Ri-Ri-plus-1-dot-product-1-equiv-1-psi-pre})
and (\ref{eq:Ri-cross-R-of-i-plus-1}), \begin{eqnarray}
\left.\begin{array}{c}
\vec{R'}_{1}\cdot\vec{R'}_{2}\\
\\\vec{R'}_{1}\times\vec{R'}_{2}\end{array}\right\}  & \Leftrightarrow & \left\{ \begin{array}{c}
\left[r'_{i}\right]^{2}\cos\psi_{1,2}-\sum_{i=1}^{3}\nu'_{1,i}\nu'_{2,i}=0,\\
\\\sum_{i=1}^{3}\left[\epsilon_{ijk}\nu'_{1,j}\nu'_{2,k}+\Gamma_{1,2}\left\Vert \vec{n'}_{p,1}\right\Vert ^{-1}\epsilon_{ijk}k'_{1,j}r'_{0,k}\right]\hat{e_{i}}=0,\end{array}\right.\label{eq:R1-cross-R2-and-R1-dot-R2-set}\end{eqnarray}
 where $\Gamma_{1,2}=\left[r'_{i}\right]^{2}\sin\psi_{1,2}$ and $\psi_{1,2}=\pi-2\theta_{inc}.$
Because $\vec{R'}_{N},$ $\vec{R'}_{1}$ and $\vec{R'}_{2}$ belong
to a same spanning set forming the plane, it is true that \begin{align*}
\vec{R'}_{1}\times\vec{R'}_{N} & \propto\vec{R'}_{1}\times\vec{R'}_{2}.\end{align*}
 Therefore, we can write \begin{align*}
\vec{R'}_{1}\times\vec{R'}_{N} & =\Gamma'_{1,N}\vec{R'}_{1}\times\vec{R'}_{2}=\Gamma'_{1,N}\Gamma_{1,2}\hat{n'}_{p,1}=\Gamma_{1,N}\hat{n'}_{p,1},\end{align*}
 where $\Gamma_{1,N}=\Gamma'_{1,N}\Gamma_{1,2}$ is a proportionality
factor. Comparing the results, \begin{eqnarray*}
\left\{ \vec{R'}_{1}\times\vec{R'}_{N}=\Gamma_{1,N}\hat{n'}_{p,1}\right\}  & \rightleftharpoons & \left\{ \vec{R'}_{1}\times\vec{R'}_{2}=\Gamma_{1,2}\hat{n'}_{p,1}\right\} ,\end{eqnarray*}
 one obtains a set of relations similar to equation (\ref{eq:R1-cross-R2-and-R1-dot-R2-set})
for $\vec{R'}_{N},$ \begin{eqnarray}
\left.\begin{array}{c}
\vec{R'}_{1}\cdot\vec{R'}_{N}\\
\\\vec{R'}_{1}\times\vec{R'}_{N}\end{array}\right\}  & \Leftrightarrow & \left\{ \begin{array}{c}
\left[r'_{i}\right]^{2}\cos\psi_{1,N}-\sum_{i=1}^{3}\nu'_{1,i}\nu'_{N,i}=0,\\
\\\sum_{i=1}^{3}\left[\epsilon_{ijk}\nu'_{1,j}\nu'_{N,k}+\Gamma_{1,N}\left\Vert \vec{n'}_{p,1}\right\Vert ^{-1}\epsilon_{ijk}k'_{1,j}r'_{0,k}\right]\hat{e_{i}}=0.\end{array}\right.\label{eq:R1-cross-RN-and-R1-dot-RN-set}\end{eqnarray}
 In the above expression the index $N$ on $\nu'_{N,i}$ and $\nu'_{N,k}$
denotes components corresponding to $\vec{R'}_{N}$ and $\psi_{1,N}$
is the angle measured between $\vec{R'}_{1}$ and $\vec{R'}_{N}.$
The proportionality factor $\Gamma_{1,N}$ is found to be \begin{eqnarray*}
\vec{R'}_{1}\times\vec{R'}_{N}=\Gamma_{1,N}\hat{n'}_{p,1} & \rightarrow & \left\Vert \vec{R'}_{1}\times\vec{R'}_{N}\right\Vert =\Gamma_{1,N}\left\Vert \hat{n'}_{p,1}\right\Vert =\Gamma_{1,N},\end{eqnarray*}
 which yields \begin{align*}
\Gamma_{1,N} & =\left\Vert \vec{R'}_{1}\times\vec{R'}_{N}\right\Vert =\left[r'_{i}\right]^{2}\sin\psi_{1,N}.\end{align*}
 The angle $\psi_{1,N}$ is contained in $\Omega_{\psi_{1,2},\psi_{N-1,N}}$
as shown in Figure \ref{cap:plane-sphere-intersection-2}, \begin{align}
\Omega_{\vec{R'}_{1},\vec{R'}_{N}}\equiv\Omega_{\psi_{1,2},\psi_{N-1,N}} & =\psi_{1,2}+\psi_{2,3}+\cdots+\psi_{N-2,N-1}+\psi_{N-1,N}.\label{eq:OMEGA-of-psi-psi-pre}\end{align}
 For each $\psi_{i,i+1},$ the sum of inner angles of a triangle gives
\[
\left\{ \begin{array}{c}
\psi_{1,2}+\theta_{2}+\theta_{r}=\pi,\\
\\\psi_{2,3}+\theta_{3}+\theta_{r+1}=\pi,\\
\\\vdots\\
\\\psi_{N-2,N-1}+\theta_{N-1}+\theta_{r+N-2}=\pi,\\
\\\psi_{N-1,N}+\theta_{N}+\theta_{r+N-1}=\pi.\end{array}\right.\]
 The law of reflection gives \begin{align*}
\theta_{2}=\theta_{3}=\cdots=\theta_{N-1}=\theta_{N} & =\theta_{r}=\theta_{r+1}=\cdots=\theta_{r+N-2}=\theta_{r+N-1}=\theta_{inc}.\end{align*}
 Hence, the angles $\psi_{i,i+1}$ are found to be \begin{align}
\psi_{1,2}=\psi_{2,3}=\cdots=\psi_{N-2,N-1}=\psi_{N-1,N} & =\pi-2\theta_{inc}.\label{eq:Psi-12-Psi-23-dot-dot-dot-Psi-N-1-N-equal-Pi-minus-2-Theta-inc}\end{align}
 The angle $\Omega_{\psi_{1,2},\psi_{N-1,N}}$ is expressed as \begin{align*}
\Omega_{\psi_{1,2},\psi_{N-1,N}} & =\psi_{1,2}+\psi_{2,3}+\cdots+\psi_{N-1,N}=\left[\pi-2\theta_{inc}\right]+\left[\pi-2\theta_{inc}\right]+\cdots+\left[\pi-2\theta_{inc}\right]\end{align*}
 or \begin{align}
\psi_{1,N}\equiv\Omega_{\psi_{1,2},\psi_{N-1,N}} & =\left[N-1\right]\left[\pi-2\theta_{inc}\right].\label{eq:OMEGA-of-psi-psi}\end{align}
 Hence for $\Gamma_{1,N},$ we have the result: \begin{align}
\Gamma_{1,N} & =\left[r'_{i}\right]^{2}\sin\left(\left[N-1\right]\left[\pi-2\theta_{inc}\right]\right).\label{eq:R1-cross-RN-GammaN-Value}\end{align}
 The angular variables $\theta'_{N}$ and $\phi'_{N}$ corresponding
to $N$th reflection point $\vec{R'}_{N}$ are given as \begin{equation}
\left\{ \begin{array}{c}
\begin{array}{ccc}
\lim_{\varepsilon\rightarrow0}\left(0\leq\theta'_{N}\leq\frac{1}{2}\pi-\left|\varepsilon\right|\right), &  & N\geq2,\end{array}\\
\\\theta'_{N\geq2}=\arctan\left(\left\{ \left[\frac{1}{2}\left\Vert \vec{n'}_{p,1}\right\Vert ^{-1}\epsilon_{1jk}k'_{1,j}r'_{0,k}-\grave{\nu}'_{N,1}\right]^{2}-\left[\frac{1}{2}\left\Vert \vec{n'}_{p,1}\right\Vert ^{-1}\epsilon_{1jk}k'_{1,j}r'_{0,k}\right]^{2}\right.\right.\\
\left.+\left[\frac{1}{2}\left\Vert \vec{n'}_{p,1}\right\Vert ^{-1}\epsilon_{2mn}k'_{1,m}r'_{0,n}-\grave{\nu}'_{N,2}\right]^{2}-\left[\frac{1}{2}\left\Vert \vec{n'}_{p,1}\right\Vert ^{-1}\epsilon_{2mn}k'_{1,m}r'_{0,n}\right]^{2}\right\} ^{1/2}\\
\left.\times\left\{ \left[\frac{1}{2}\left\Vert \vec{n'}_{p,1}\right\Vert ^{-1}\epsilon_{3qr}k'_{1,q}r'_{0,r}-\grave{\nu}'_{N,3}\right]^{2}-\left[\frac{1}{2}\left\Vert \vec{n'}_{p,1}\right\Vert ^{-1}\epsilon_{3qr}k'_{1,q}r'_{0,r}\right]^{2}\right\} ^{-1/2}\right);\end{array}\right.\label{eq:thetaN-prime-0-to-90-degree-final}\end{equation}
 \begin{equation}
\left\{ \begin{array}{c}
\begin{array}{ccc}
\lim_{\varepsilon\rightarrow0}\left(\frac{1}{2}\pi+\left|\varepsilon\right|\leq\theta'_{N}\leq\pi\right), &  & N\geq2,\end{array}\\
\\\theta'_{N\geq2}=\arctan\left(-\left\{ \left[\frac{1}{2}\left\Vert \vec{n'}_{p,1}\right\Vert ^{-1}\epsilon_{1jk}k'_{1,j}r'_{0,k}-\grave{\nu}'_{N,1}\right]^{2}-\left[\frac{1}{2}\left\Vert \vec{n'}_{p,1}\right\Vert ^{-1}\epsilon_{1jk}k'_{1,j}r'_{0,k}\right]^{2}\right.\right.\\
\left.+\left[\frac{1}{2}\left\Vert \vec{n'}_{p,1}\right\Vert ^{-1}\epsilon_{2mn}k'_{1,m}r'_{0,n}-\grave{\nu}'_{N,2}\right]^{2}-\left[\frac{1}{2}\left\Vert \vec{n'}_{p,1}\right\Vert ^{-1}\epsilon_{2mn}k'_{1,m}r'_{0,n}\right]^{2}\right\} ^{1/2}\\
\left.\times\left\{ \left[\frac{1}{2}\left\Vert \vec{n'}_{p,1}\right\Vert ^{-1}\epsilon_{3qr}k'_{1,q}r'_{0,r}-\grave{\nu}'_{N,3}\right]^{2}-\left[\frac{1}{2}\left\Vert \vec{n'}_{p,1}\right\Vert ^{-1}\epsilon_{3qr}k'_{1,q}r'_{0,r}\right]^{2}\right\} ^{-1/2}\right).\end{array}\right.\label{eq:thetaN-prime-90-to-180-degree-final}\end{equation}
 \begin{equation}
\left\{ \begin{array}{c}
\begin{array}{ccccc}
\lim_{\varepsilon\rightarrow0}\left(0\leq\phi'_{N}\leq\frac{1}{2}\pi-\left|\varepsilon\right|\right), &  & \lim_{\varepsilon\rightarrow0}\left(\pi+\left|\varepsilon\right|\leq\phi'_{N}\leq\frac{3}{2}\pi-\left|\varepsilon\right|\right), &  & N\geq2,\end{array}\\
\\\phi'_{N\geq2}=\arctan\left(\left\{ \frac{\left[\frac{1}{2}\left\Vert \vec{n'}_{p,1}\right\Vert ^{-1}\epsilon_{2mn}k'_{1,m}r'_{0,n}-\grave{\nu}'_{N,2}\right]^{2}-\left[\frac{1}{2}\left\Vert \vec{n'}_{p,1}\right\Vert ^{-1}\epsilon_{2mn}k'_{1,m}r'_{0,n}\right]^{2}}{\left[\frac{1}{2}\left\Vert \vec{n'}_{p,1}\right\Vert ^{-1}\epsilon_{1jk}k'_{1,j}r'_{0,k}-\grave{\nu}'_{N,1}\right]^{2}-\left[\frac{1}{2}\left\Vert \vec{n'}_{p,1}\right\Vert ^{-1}\epsilon_{1jk}k'_{1,j}r'_{0,k}\right]^{2}}\right\} ^{1/2}\right);\end{array}\right.\label{eq:phiN-prime-0-90-and-180-270-final}\end{equation}
 \begin{equation}
\left\{ \begin{array}{c}
\begin{array}{ccccc}
\lim_{\varepsilon\rightarrow0}\left(\frac{1}{2}\pi+\left|\varepsilon\right|\leq\phi'_{N}\leq\pi-\varepsilon\right), &  & \lim_{\varepsilon\rightarrow0}\left(\frac{3}{2}\pi+\left|\varepsilon\right|\leq\phi'_{N}<2\pi-\left|\varepsilon\right|\right), &  & N\geq2,\end{array}\\
\\\phi'_{N\geq2}=\arctan\left(-\left\{ \frac{\left[\frac{1}{2}\left\Vert \vec{n'}_{p,1}\right\Vert ^{-1}\epsilon_{2mn}k'_{1,m}r'_{0,n}-\grave{\nu}'_{N,2}\right]^{2}-\left[\frac{1}{2}\left\Vert \vec{n'}_{p,1}\right\Vert ^{-1}\epsilon_{2mn}k'_{1,m}r'_{0,n}\right]^{2}}{\left[\frac{1}{2}\left\Vert \vec{n'}_{p,1}\right\Vert ^{-1}\epsilon_{1jk}k'_{1,j}r'_{0,k}-\grave{\nu}'_{N,1}\right]^{2}-\left[\frac{1}{2}\left\Vert \vec{n'}_{p,1}\right\Vert ^{-1}\epsilon_{1jk}k'_{1,j}r'_{0,k}\right]^{2}}\right\} ^{1/2}\right),\end{array}\right.\label{eq:phiN-prime-90-180-and-270-360-final}\end{equation}
 where $\left\{ \grave{\nu}'_{N,i}:i=1,2,3\right\} $ are given in
equations (\ref{eq:x2-prime}), (\ref{eq:y2-prime}) and (\ref{eq:z2-prime})
together with the modification $\grave{\nu}'_{2,i}\rightarrow\grave{\nu}'_{N,i},$
$\zeta_{1}\left(\Gamma_{1,2}\right)\rightarrow\zeta_{1}\left(\Gamma_{1,N}\right),$
$\zeta_{2}\left(\Gamma_{1,2}\right)\rightarrow\zeta_{2}\left(\Gamma_{1,N}\right)$
and $\zeta_{3}\left(\Gamma_{1,2}\right)\rightarrow\zeta_{3}\left(\Gamma_{1,N}\right),$
where $\Gamma_{1,N}$ is given in equation (\ref{eq:R1-cross-RN-GammaN-Value}).
With angular variable $\theta'_{N\geq2}$ defined in equations (\ref{eq:thetaN-prime-0-to-90-degree-final})
and (\ref{eq:thetaN-prime-90-to-180-degree-final}) and $\phi'_{N\geq2}$
defined in equations (\ref{eq:phiN-prime-0-90-and-180-270-final})
and (\ref{eq:phiN-prime-90-180-and-270-360-final}), the $N$th reflection
point on the inner hemisphere surface is given by \begin{eqnarray}
\vec{R'}_{N}\left(r'_{i},\theta'_{N},\phi'_{N}\right)=\sum_{i=1}^{3}\nu'_{N,i}\left(r'_{i},\theta'_{N},\phi'_{N}\right)\hat{e_{i}}, &  & i=\left\{ \begin{array}{c}
1\rightarrow\nu'_{N,1}=r'_{i}\sin\theta'_{N}\cos\phi'_{N},\\
\\2\rightarrow\nu'_{N,2}=r'_{i}\sin\theta'_{N}\sin\phi'_{N},\\
\\3\rightarrow\nu'_{N,3}=r'_{i}\cos\theta'_{N},\qquad\;\:\;\end{array}\right.\label{eq:Nth-Reflection-Point}\end{eqnarray}
 where the initial reflection point $\vec{R'}_{1}$ is given in equation
(\ref{eq:1st-Reflection-Point}). 

For a sphere, the maximum number of reflections are given by the equation
\begin{align*}
N_{s,max}\psi_{N-2,N-1} & =2\pi,\end{align*}
 where $\psi_{N-2,N-1}$ is the angle between two neighboring reflection
points $\vec{R'}_{N-1}$ and $\vec{R'}_{N-2};$ the subscript $N_{s,max}$
denotes the maximum number of reflection points for a sphere. The
above result is a statement that the sum of all angles is equal to
$2\pi.$ Application of the rule shown in equation (\ref{eq:Psi-12-Psi-23-dot-dot-dot-Psi-N-1-N-equal-Pi-minus-2-Theta-inc})
for $\psi_{N-2,N-1}$ gives \begin{eqnarray}
N_{max}\left[\pi-2\theta_{inc}\right]=2\pi & \rightarrow & N_{s,max}=\frac{2\pi}{\pi-2\theta_{inc}},\label{eq:N-max-Sphere}\end{eqnarray}
 where $\theta_{inc}$ is given in equation (\ref{eq:angle-of-incidence-exp}).
In explicit form $N_{s,max}$ is given by \begin{align}
N_{s,max} & =2\pi\left[\pi-2\arccos\left(\frac{\sin\theta'_{1}\left[k'_{x'_{1}}\cos\phi'_{1}+k'_{y'_{1}}\sin\phi'_{1}\right]+k'_{z'_{1}}\cos\theta'_{1}}{\sqrt{\left[k'_{x'_{1}}\right]^{2}+\left[k'_{y'_{1}}\right]^{2}+\left[k'_{z'_{1}}\right]^{2}}}\right)\right]^{-1}.\label{eq:N-max-Sphere-exp}\end{align}

\section{Mapping Between Sets $\left(r,\theta,\phi\right)$ and $\left(r',\theta',\phi'\right)$}

In this appendix, the original derivations and developments pertaining
to the mapping between the sets $\left(r,\theta,\phi\right)$ and
$\left(r',\theta',\phi'\right)$ used in this paper are described
in detail. 
\vspace{0.25in}

For a sphere, the natural choice for origin is the sphere center from
which the spherical coordinates $\left(r'_{i},\theta',\phi'\right)$
are prescribed. For more complicated configurations, as shown in Figure
\ref{cap:Two-semispheres-without-center-solid-sphere}, the preferred
choice for the origin depends upon the problem in hand. For this reason,
this section is devoted in deriving a set of transformations between
$\left(r'_{i},\theta',\phi'\right)$ and $\left(r_{i},\theta,\phi\right),$
where the primed set is defined relative to the sphere center, and
the unprimed set is defined relative to the global configuration origin.
In Cartesian coordinates, the two vectors $\vec{R}$ and $\vec{R'}$
describing an identical point on the hemisphere surface are expressed
as \begin{eqnarray}
\vec{R}\left(\nu_{1},\nu_{2},\nu_{3}\right)=\sum_{i=1}^{3}\nu_{i}\hat{e_{i}}, &  & \vec{R'}\left(\nu'_{1},\nu'_{2},\nu'_{3}\right)=\sum_{i=1}^{3}\nu'_{i}\hat{e_{i}},\label{eq:R-and-Rprimed-vector-cartesian}\end{eqnarray}
 where \[
\begin{array}{ccccc}
\left[\begin{array}{c}
\nu_{1}=x\\
\\\nu_{2}=y\\
\\\nu_{3}=z\end{array}\right], &  & \left[\begin{array}{c}
\nu'_{1}=x'\\
\\\nu'_{2}=y'\\
\\\nu'_{3}=z'\end{array}\right], &  & \left[\begin{array}{c}
\hat{e_{1}}=\hat{x}\\
\\\hat{e_{2}}=\hat{y}\\
\\\hat{e_{3}}=\hat{z}\end{array}\right].\end{array}\]
 Here $\vec{R}$ and $\vec{R'}$ are the position vectors of the same
location relative to the system origin and the hemisphere center,
respectively. The two vectors are related through a translation, \begin{align}
\vec{R}\left(\nu_{1},\nu_{2},\nu_{3}\right) & =\vec{R}_{T}\left(\nu_{T,1},\nu_{T,2},\nu_{T,3}\right)+\vec{R'}\left(\nu'_{1},\nu'_{2},\nu'_{3}\right)=\sum_{i=1}^{3}\left[\nu_{T,i}+\nu'_{i}\right]\hat{e_{i}},\label{eq:r-vector-cart-in-prime}\end{align}
 where $\vec{R}_{T}=\sum_{i=1}^{3}\nu_{T,i}\hat{e_{i}}$ is the position
of hemisphere center relative to the system origin. Equation (\ref{eq:r-vector-cart-in-prime})
can be written as \begin{align}
\sum_{i=1}^{3}\left[\nu_{i}-\nu_{T,i}-\nu'_{i}\right]\hat{e_{i}} & =0.\label{eq:r-vector-cart-in-prime-exp}\end{align}
 and the component equations are \begin{eqnarray}
\nu_{i}-\nu_{T,i}-\nu'_{i}=0, &  & i=1,2,3.\label{eq:component-relation}\end{eqnarray}
 It is to be emphasized that in the configuration shown in Figure
\ref{cap:Two-semispheres-without-center-solid-sphere}, the hemisphere
center is only shifted along $\hat{y}$ by the distance $\nu_{T,2}=a,$
therefore $\nu_{T,i\neq2}=0.$ Nevertheless, the derivation is done
for the case where $\nu_{T,i}\neq0,$ $i=1,2,3$ for general purpose.
In explicit forms, they are written as \begin{equation}
\begin{array}{ccccc}
\nu_{1}-\nu_{T,1}-\nu'_{1}=0, &  & \nu_{2}-\nu_{T,2}-\nu'_{2}=0, &  & \nu_{3}-\nu_{T,3}-\nu'_{3}=0.\end{array}\label{eq:x-y-z-relation-to-xp-yp-zp}\end{equation}
 In spherical coordinates, \begin{eqnarray}
\left[\begin{array}{c}
\nu_{1}=r_{i}\sin\theta\cos\phi=r_{i}\Lambda_{1}\left(\theta,\phi\right)\\
\\\nu_{2}=r_{i}\sin\theta\sin\phi=r_{i}\Lambda_{2}\left(\theta,\phi\right)\\
\\\nu_{3}=r_{i}\cos\theta=r_{i}\Lambda_{3}\left(\theta\right)\end{array}\right], &  & \left[\begin{array}{c}
\nu'_{1}=r'_{i}\sin\theta'\cos\phi'=r'_{i}\Lambda'_{1}\left(\theta',\phi'\right)\\
\\\nu'_{2}=r'_{i}\sin\theta'\sin\phi'=r'_{i}\Lambda'_{2}\left(\theta',\phi'\right)\\
\\\nu'_{3}=r'_{i}\cos\theta'=r'_{i}\Lambda'_{3}\left(\theta'\right)\end{array}\right],\label{eq:spherical-variables-v1-v1-v3-and-v1-v2-v3-prime}\end{eqnarray}
 equation (\ref{eq:x-y-z-relation-to-xp-yp-zp}) is written as \begin{align}
r_{i}\sin\theta\cos\phi-\nu_{T,1}-r'_{i}\sin\theta'\cos\phi' & =0,\label{eq:sincos-eq-sincos-p}\end{align}
\begin{align}
r_{i}\sin\theta\sin\phi-\nu_{T,2}-r'_{i}\sin\theta'\sin\phi' & =0,\label{eq:sinsin-eq-a-p-sinsin-p}\end{align}
\begin{align}
r_{i}\cos\theta-\nu_{T,3}-r'_{i}\cos\theta' & =0,\label{eq:cos-eq-cosp}\end{align}
 where the Cartesian variables $\left\{ \nu_{i},\nu'_{i}:i=1,2,3\right\} $
were expressed in terms of the spherical coordinates. The $\cos\phi$
and $\sin\phi$ functions are obtained from equations (\ref{eq:sincos-eq-sincos-p})
and (\ref{eq:sinsin-eq-a-p-sinsin-p}), \begin{eqnarray*}
\cos\phi=\frac{\nu_{T,1}+r'_{i}\sin\theta'\cos\phi'}{r_{i}\sin\theta}, &  & \sin\phi=\frac{\nu_{T,2}+r'_{i}\sin\theta'\sin\phi'}{r_{i}\sin\theta}.\end{eqnarray*}
 The azimuthal angle $\phi$ is given by \begin{align}
\phi\equiv\grave{\phi}\left(r'_{i},\theta',\phi',\nu_{T,1},\nu_{T,2}\right) & =\arctan\left(\frac{\nu_{T,2}+r'_{i}\sin\theta'\sin\phi'}{\nu_{T,1}+r'_{i}\sin\theta'\cos\phi'}\right),\label{eq:phi}\end{align}
 where the notation $\grave{\phi}$ indicates that $\phi$ is explicitly
expressed in terms of primed variables. Combining equations (\ref{eq:sincos-eq-sincos-p})
and (\ref{eq:sinsin-eq-a-p-sinsin-p}), we have \begin{align*}
r_{i}\sin\theta\left[\cos\phi+\sin\phi\right]-\nu_{T,1}-\nu_{T,2}-r'_{i}\sin\theta'\left[\cos\phi'+\sin\phi'\right] & =0\end{align*}
 which leads to \begin{align*}
\sin\theta & =\frac{\nu_{T,1}+\nu_{T,2}+r'_{i}\sin\theta'\left[\cos\phi'+\sin\phi'\right]}{r_{i}\left[\cos\phi+\sin\phi\right]}.\end{align*}
 From equation (\ref{eq:cos-eq-cosp}), we have \begin{align*}
\cos\theta & =r_{i}^{-1}\left[\nu_{T,3}+r'_{i}\cos\theta'\right].\end{align*}
 Combining the above results for $\sin\theta$ and $\cos\theta;$
and, solving for the argument $\theta,$ \begin{align}
\theta & =\arctan\left(\frac{\nu_{T,1}+\nu_{T,2}+r'_{i}\sin\theta'\left[\cos\phi'+\sin\phi'\right]}{\left[\cos\phi+\sin\phi\right]\left[\nu_{T,3}+r'_{i}\cos\theta'\right]}\right),\label{eq:theta-pre}\end{align}
 where $\phi$ is to be substituted in from equation (\ref{eq:phi}).
For convenience, the above result for $\theta$ is rewritten explicitly
in terms of primed variables, \begin{align}
\grave{\theta}\left(r'_{i},\theta',\phi',\vec{R}_{T}\right) & =\arctan\left(\frac{\left\{ \nu_{T,1}+\nu_{T,2}+r'_{i}\sin\theta'\left[\cos\phi'+\sin\phi'\right]\right\} \left[\nu_{T,3}+r'_{i}\cos\theta'\right]^{-1}}{\cos\left(\arctan\left(\frac{\nu_{T,2}+r'_{i}\sin\theta'\sin\phi'}{\nu_{T,1}+r'_{i}\sin\theta'\cos\phi'}\right)\right)+\sin\left(\arctan\left(\frac{\nu_{T,2}+r'_{i}\sin\theta'\sin\phi'}{\nu_{T,1}+r'_{i}\sin\theta'\cos\phi'}\right)\right)}\right).\label{eq:theta}\end{align}
 Here the notation $\grave{\theta}$ indicates that $\theta$ is explicitly
expressed in terms of primed variables. The magnitude of a vector
describing hemisphere relative to system origin is found from equation
(\ref{eq:r-vector-cart-in-prime}), \begin{eqnarray*}
r_{i}\left(r'_{i},\vec{\Lambda}',\vec{R}_{T}\right)\equiv\left\Vert \vec{R}\right\Vert =\left\{ \sum_{i=1}^{3}\left[\nu_{T,i}+r'_{i}\Lambda'_{i}\right]^{2}\right\} ^{1/2}, &  & \left\{ \begin{array}{c}
\Lambda'_{1}\left(\theta',\phi'\right)=\sin\theta'\cos\phi',\\
\\\Lambda'_{2}\left(\theta',\phi'\right)=\sin\theta'\sin\phi',\\
\\\Lambda'_{3}\left(\theta'\right)=\cos\theta'.\;\;\;\;\end{array}\right.\end{eqnarray*}
 In terms of spherical coordinates, the position vector is expressed
as \begin{eqnarray}
\vec{R}\left(r'_{i},\grave{\vec{\Lambda}},\vec{\Lambda}',\vec{R}_{T}\right)=\left\{ \sum_{i=1}^{3}\left[\nu_{T,i}+r'_{i}\Lambda'_{i}\right]^{2}\right\} ^{1/2}\sum_{i=1}^{3}\grave{\Lambda}_{i}\hat{e_{i}}, &  & \left\{ \begin{array}{c}
\grave{\Lambda}_{1}\left(\grave{\theta},\grave{\phi}\right)=\sin\grave{\theta}\cos\grave{\phi},\\
\\\grave{\Lambda}_{2}\left(\grave{\theta},\grave{\phi}\right)=\sin\grave{\theta}\sin\grave{\phi},\\
\\\grave{\Lambda}_{3}\left(\grave{\theta}\right)=\cos\grave{\theta}.\;\;\;\;\end{array}\right.\label{eq:Points-on-Hemisphere-R-arbi}\end{eqnarray}

\section{Selected Configurations}

In this appendix, the original derivations and developments in this
paper pertaining to the selected configurations: (1) the hollow spherical
shell, (2) the hemisphere-hemisphere and (3) the plate-hemisphere,
are described in detail. 
\vspace{0.25in}

\subsection{Hollow Spherical Shell}

For the reflection problem in a sphere as shown in Figure \ref{cap:sphere-reflection-dynamics},
the natural choice for a system origin is that of the sphere center,
$\vec{R'}=0.$ The $N$th reflection point inside the sphere is given
by equation (\ref{eq:Nth-Reflection-Point}) as \begin{eqnarray*}
\vec{R'}_{s,N}\left(r'_{i},\theta'_{s,N},\phi'_{s,N}\right)=\sum_{i=1}^{3}\nu'_{s,N,i}\left(r'_{i},\theta'_{s,N},\phi'_{s,N}\right)\hat{e_{i}}, &  & \left\{ \begin{array}{c}
\nu'_{s,N,1}=r'_{i}\sin\theta'_{s,N}\cos\phi'_{s,N},\\
\\\nu'_{s,N,2}=r'_{i}\sin\theta'_{s,N}\sin\phi'_{s,N},\\
\\\nu'_{s,N,3}=r'_{i}\cos\theta'_{s,N},\qquad\quad\;\:\end{array}\right.\end{eqnarray*}
 where the label $s$ have been attached to denote the sphere. Keeping
in mind the obvious index changes, the angular variable $\theta'_{s,N}$
is defined in equations (\ref{eq:thetaN-prime-0-to-90-degree-final})
and (\ref{eq:thetaN-prime-90-to-180-degree-final}), and $\phi'_{s,N},$
in equations (\ref{eq:phiN-prime-0-90-and-180-270-final}) and (\ref{eq:phiN-prime-90-180-and-270-360-final}).
Staying with the notation of equation (\ref{eq:Points-on-Hemisphere-R-arbi}),
$\vec{R'}_{s,N}$ is rewritten as \begin{eqnarray}
\vec{R}_{s,N}\left(r'_{i},\vec{\Lambda}'_{s,N}\right)=r'_{i}\sum_{i=1}^{3}\Lambda'_{s,N,i}\hat{e_{i}}, &  & \left\{ \begin{array}{c}
\Lambda'_{s,N,1}\left(\theta'_{s,N},\phi'_{s,N}\right)=\sin\theta'_{s,N}\cos\phi'_{s,N},\\
\\\Lambda'_{s,N,2}\left(\theta'_{s,N},\phi'_{s,N}\right)=\sin\theta'_{s,N}\sin\phi'_{s,N},\\
\\\Lambda'_{s,N,3}\left(\theta'_{s,N}\right)=\cos\theta'_{s,N},\;\;\;\;\end{array}\right.\label{eq:Nth-Reflection-Point-sphere}\end{eqnarray}
 where the relations $\nu_{T,s,i}=0,$ and $\sum_{i=1}^{3}\left[\Lambda'_{s,N,i}\right]^{2}=1$
are used. 

The maximum number of internal reflections for a spherical cavity
before a repeat in cycle is given by equation (\ref{eq:N-max-Sphere}),
\begin{align*}
N_{s,max} & =\frac{2\pi}{\pi-2\theta_{inc}},\end{align*}
 where $\theta_{inc}$ is given in equation (\ref{eq:angle-of-incidence-exp}). 

The distance $\left\Vert \vec{L}\right\Vert $ between two immediate
neighboring reflection points on a sphere is \begin{align}
\left\Vert \vec{L}\right\Vert  & =\left\Vert \vec{R}_{s,2}\left(r'_{i},\vec{\Lambda}'_{s,2}\right)-\vec{R}_{s,1}\left(r'_{i},\vec{\Lambda}'_{s,1}\right)\right\Vert .\label{eq:sphere-distance-rela}\end{align}
 It should be noted that \begin{eqnarray}
\left\Vert \vec{R}_{s,2}-\vec{R}_{s,1}\right\Vert =\left\Vert \vec{R}_{s,j}-\vec{R}_{s,j-1}\right\Vert , &  & j=3,\cdots,N_{s,max}.\label{eq:sphere-distance-rela-equiv}\end{eqnarray}
 The only reason that $\vec{R}_{s,1}$ and $\vec{R}_{s,2}$ are used
is for the purpose of convenience. 

To compute the resultant wave vector, $\vec{k'}_{inner},$ acting
at the point $\vec{R'}_{s,1},$ the incident wave is first decomposed
into components parallel and perpendicular to the local normal vector,
$-\hat{R'}_{s,1},$ of the inner surface \begin{align*}
\vec{k'}_{i,+}\equiv\vec{k'}_{i} & =\vec{k'}_{i\parallel}+\vec{k'}_{i\perp}=\left[\vec{k'}_{i}\cdot\hat{R'}_{s,1}\right]\hat{R'}_{s,1}+\left[\hat{R'}_{s,1}\times\vec{k'}_{i}\right]\times\hat{R'}_{s,1},\end{align*}
 where the subscript $\left(+\right)$ of $\vec{k'}_{i,+}$ denotes
the particular contribution where the incident wave $\vec{k'}_{i}$
is approaching $\vec{R'}_{s,1}$ from $\vec{R'}_{s,0}.$ From equation
(\ref{eq:K-reflected-in-PARA-n-PERPE-N}) of Appendix A, the corresponding
reflected wave vector can be expressed in terms of the incident wave
as \begin{align*}
\vec{k'}_{r,+}\equiv\vec{k'}_{r} & =\left[\hat{R'}_{s,1}\times\left\{ \left[\vec{k'}_{i}\cdot\hat{R'}_{s,1}\right]\hat{R'}_{s,1}+\left[\hat{R'}_{s,1}\times\vec{k'}_{i}\right]\times\hat{R'}_{s,1}\right\} \right]\times\hat{R'}_{s,1}\\
 & -\hat{R'}_{s,1}\cdot\left\{ \left[\vec{k'}_{i}\cdot\hat{R'}_{s,1}\right]\hat{R'}_{s,1}+\left[\hat{R'}_{s,1}\times\vec{k'}_{i}\right]\times\hat{R'}_{s,1}\right\} \hat{R'}_{s,1}\\
 & =\left[\hat{R'}_{s,1}\times\left[\vec{k'}_{i}\cdot\hat{R'}_{s,1}\right]\hat{R'}_{s,1}+\hat{R'}_{s,1}\times\left\{ \left[\hat{R'}_{s,1}\times\vec{k'}_{i}\right]\times\hat{R'}_{s,1}\right\} \right]\times\hat{R'}_{s,1}\\
 & -\left[\hat{R'}_{s,1}\cdot\left[\vec{k'}_{i}\cdot\hat{R'}_{s,1}\right]\hat{R'}_{s,1}+\hat{R'}_{s,1}\cdot\left\{ \left[\hat{R'}_{s,1}\times\vec{k'}_{i}\right]\times\hat{R'}_{s,1}\right\} \right]\hat{R'}_{s,1},\end{align*}
 where $\alpha_{r,\perp}=\alpha_{r,\parallel}=1$ and $\hat{n'}\rightarrow-\hat{R'}_{s,1}.$
Because $\hat{R'}_{s,1}\perp\left\{ \left[\hat{R'}_{s,1}\times\vec{k'}_{i}\right]\times\hat{R'}_{s,1}\right\} $
and $\hat{R'}_{s,1}\parallel\left[\vec{k'}_{i}\cdot\hat{R'}_{s,1}\right]\hat{R'}_{s,1},$
the above expression is simplified to \begin{align*}
\vec{k'}_{r,+} & =\left[\hat{R'}_{s,1}\times\left\{ \left[\hat{R'}_{s,1}\times\vec{k'}_{i}\right]\times\hat{R'}_{s,1}\right\} \right]\times\hat{R'}_{s,1}-\left[\vec{k'}_{i}\cdot\hat{R'}_{s,1}\right]\hat{R'}_{s,1}.\end{align*}
 The changes in resultant wave vector $\vec{k'}_{inner}$ at the point
$\vec{R'}_{s,1}$ due to $\vec{k'}_{i,+}$ at location $\vec{R'}_{s,0}$
is given by \begin{align*}
\triangle\vec{k'}_{inner,+}\left(;\vec{R'}_{s,1},\vec{R'}_{s,0}\right) & =\vec{k'}_{r,+}-\vec{k'}_{i,+}\\
 & =\left[\hat{R'}_{s,1}\times\left\{ \left[\hat{R'}_{s,1}\times\vec{k'}_{i}\right]\times\hat{R'}_{s,1}\right\} \right]\times\hat{R'}_{s,1}\\
 & -\left[\hat{R'}_{s,1}\times\vec{k'}_{i}\right]\times\hat{R'}_{s,1}-2\left[\vec{k'}_{i}\cdot\hat{R'}_{s,1}\right]\hat{R'}_{s,1}\\
 & =-2\left[\vec{k'}_{i}\cdot\hat{R'}_{s,1}\right]\hat{R'}_{s,1},\end{align*}
 where $\left[\hat{R'}_{s,1}\times\left\{ \left[\hat{R'}_{s,1}\times\vec{k'}_{i}\right]\times\hat{R'}_{s,1}\right\} \right]\times\hat{R'}_{s,1}=\left[\hat{R'}_{s,1}\times\vec{k'}_{i}\right]\times\hat{R'}_{s,1}.$

For the incident wave traveling in the opposite direction, i.e., approaching
$\vec{R'}_{s,1}$ from $\vec{R'}_{s,2},$ one has \begin{eqnarray*}
\vec{k'}_{i,-}\equiv-\vec{k'}_{r}=-\vec{k'}_{r,+}, &  & \vec{k'}_{r,-}\equiv-\vec{k'}_{i}=-\vec{k'}_{i,+},\end{eqnarray*}
 where the subscript $\left(-\right)$ on $\vec{k'}_{i,-}$ denotes
the particular contribution where the incident wave $\vec{k'}_{i}$
is approaching $\vec{R'}_{s,1}$ from $\vec{R'}_{s,2}.$ In this case,
the changes in the resultant wave vector $\vec{k'}_{inner}$ at the
point $\vec{R'}_{s,1}$ due to $\vec{k'}_{i,-}$ at the location $\vec{R'}_{s,0}$
is given by \begin{align*}
\triangle\vec{k'}_{inner,-}\left(;\vec{R'}_{s,1},\vec{R'}_{s,0}\right) & =\vec{k'}_{r,-}-\vec{k'}_{i,-}=-\vec{k'}_{i,+}+\vec{k'}_{r,+}=\triangle\vec{k'}_{inner,+}\left(;\vec{R'}_{s,1},\vec{R'}_{s,0}\right).\end{align*}
 The resultant wave vector $\vec{k'}_{inner}$ acting at the point
$\vec{R'}_{s,1}$ due to incident wave approaching $\vec{R'}_{s,1}$
from $\vec{R'}_{s,0}$ and the other incident wave approaching $\vec{R'}_{s,1}$
from $\vec{R'}_{s,2}$ is therefore \begin{align*}
\triangle\vec{k'}_{inner}\left(;\vec{R'}_{s,1},\vec{R'}_{s,0}\right) & \equiv\triangle\vec{k'}_{inner,+}\left(;\vec{R'}_{s,1},\vec{R'}_{s,0}\right)+\triangle\vec{k'}_{inner,-}\left(;\vec{R'}_{s,1},\vec{R'}_{s,0}\right)\\
 & =-4\left[\vec{k'}_{i,b}\cdot\hat{R'}_{s,1}\right]\hat{R'}_{s,1},\end{align*}
 where the subscript $b$ of $\vec{k'}_{i,b}$ denotes the wave vector
for ambient fields inside cavity. 

The wave number $\left\Vert \vec{k'}_{i,b}\right\Vert $ that can
be fit in the bounded space of a resonator is restricted by the boundary
condition \begin{align*}
\left\Vert \vec{k'}_{i,b}\right\Vert  & =n\pi\left\Vert \vec{L}\right\Vert ^{-1}=n\pi\left\Vert \vec{R}_{s,2}\left(r'_{i},\vec{\Lambda}'_{s,2}\right)-\vec{R}_{s,1}\left(r'_{i},\vec{\Lambda}'_{s,1}\right)\right\Vert ^{-1}.\end{align*}
 The scalar product of $\vec{k'}_{i,b}$ and $\hat{R'}_{s,1}$ is
\begin{align*}
\vec{k'}_{i,b}\cdot\hat{R'}_{s,1} & =\left\Vert \vec{k'}_{i,b}\right\Vert \cos\theta_{inc},\end{align*}
 where the angle between the two vectors $\vec{k'}_{i,b}$ and $\vec{R}_{s,1}$
is equal to the angle of incidence $\theta_{inc},$ as shown in equation
(\ref{eq:angle-of-incidence-exp}). The momentum transfer is proportional
to \begin{eqnarray}
\triangle\vec{k'}_{inner}\left(;\vec{R'}_{s,1},\vec{R'}_{s,0}\right)=-\frac{4n\pi\cos\theta_{inc}}{\left\Vert \vec{R}_{s,2}\left(r'_{i},\vec{\Lambda}'_{s,2}\right)-\vec{R}_{s,1}\left(r'_{i},\vec{\Lambda}'_{s,1}\right)\right\Vert }\hat{R'}_{s,1}, &  & \left\{ \begin{array}{c}
0\leq\theta_{inc}<\pi/2,\\
\\n=1,2,\cdots.\end{array}\right.\label{eq:SPHERE-Delta-K-Inside}\end{eqnarray}

Similarly, the resultant wave vector $\vec{k'}_{outer}$ acting at
point $\vec{R'}_{s,1}+a\hat{R'}_{s,1}$ on the outer spherical surface,
where $a$ is the sphere thickness parameter, is given by \begin{align*}
\triangle\vec{k'}_{outer}\left(;\vec{R'}_{s,1}+a\hat{R'}_{s,1}\right) & =-4\left[\vec{k'}_{i,f}\cdot\hat{R'}_{s,1}\right]\hat{R'}_{s,1},\end{align*}
 where the subscript $f$ of $\vec{k'}_{i,f}$ denotes the wave vector
of the ambient fields in free space, and the factor 4 is due to the
fact there are two incidence wave vectors from opposite directions.
The free space wave number $\left\Vert \vec{k'}_{i,f}\right\Vert $
has no quantization restriction due to the boundary. And, the scalar
product of $\vec{k'}_{i,f}$ and $\hat{R'}_{s,1}$ is \begin{align*}
\vec{k'}_{i,f}\cdot\hat{R'}_{s,1} & =\left\Vert \vec{k'}_{i,f}\right\Vert \cos\left(\pi-\theta_{inc}\right)=-\left\Vert \vec{k'}_{i,f}\right\Vert \cos\theta_{inc}\hat{R'}_{s,1}.\end{align*}
 The momentum transfer is proportional to \begin{eqnarray}
\triangle\vec{k'}_{outer}\left(;\vec{R'}_{s,1}+a\hat{R'}_{s,1}\right)=4\left\Vert \vec{k'}_{i,f}\right\Vert \cos\theta_{inc}\hat{R'}_{s,1}, &  & \left\{ \begin{array}{c}
0\leq\theta_{inc}<\pi/2,\\
\\n=1,2,\cdots.\end{array}\right.\label{eq:SPHERE-Delta-K-Outside}\end{eqnarray}

\subsection{Hemisphere-Hemisphere}

For the hemisphere-hemisphere configuration, the preferred choice
for a system origin is that of $\vec{R}=0.$ The $N$th internal reflection
point is given by equation (\ref{eq:Points-on-Hemisphere-R-arbi}),
\begin{align}
\vec{R}_{h,N}\left(r'_{i},\grave{\vec{\Lambda}}_{h,N},\vec{\Lambda}'_{h,N},\vec{R}_{T,h}\right) & =\left\{ \sum_{i=1}^{3}\left[\nu_{T,h,i}+r'_{i}\Lambda'_{h,N,i}\right]^{2}\right\} ^{1/2}\sum_{i=1}^{3}\grave{\Lambda}_{h,N,i}\hat{e_{i}},\label{eq:Points-on-Hemisphere-R}\end{align}
 where the label $h$ denotes hemisphere; and \[
\left\{ \begin{array}{c}
\grave{\Lambda}_{h,N,1}\left(\grave{\theta}_{h,N},\grave{\phi}_{h,N}\right)=\sin\grave{\theta}_{h,N}\cos\grave{\phi}_{h,N},\\
\\\grave{\Lambda}_{h,N,2}\left(\grave{\theta}_{h,N},\grave{\phi}_{h,N}\right)=\sin\grave{\theta}_{h,N}\sin\grave{\phi}_{h,N},\\
\\\grave{\Lambda}_{h,N,3}\left(\grave{\theta}_{h,N}\right)=\cos\grave{\theta}_{h,N}.\;\;\;\;\end{array}\right.\]
 The definitions for $\Lambda'_{h,N,i},$ $i=1,2,3$ are identical
in form. The angular variables $\left(\grave{\theta}_{h,N},\grave{\phi}_{h,N}\right)$
are given in equations (\ref{eq:phi}) and (\ref{eq:theta}), where
the obvious notational changes are understood. The implicit angular
variable $\theta'_{s,N}$ is defined in equations (\ref{eq:thetaN-prime-0-to-90-degree-final})
and (\ref{eq:thetaN-prime-90-to-180-degree-final}); and $\phi'_{s,N},$
defined in equations (\ref{eq:phiN-prime-0-90-and-180-270-final})
and (\ref{eq:phiN-prime-90-180-and-270-360-final}). 

We have to determine the maximum number of internal reflections of
the wave in the hemisphere cavity before its escape. Three vectors,
$\vec{R'}_{0},$ $\xi_{i}\hat{k'}_{i}$ and $\vec{R'}_{h,i}\equiv\vec{R'}_{i},$
shown in Figure \ref{cap:plane-sphere-intersection-2} of appendix
A satisfy the relation \begin{align*}
\vec{R'}_{h,i=1}-\vec{R'}_{0} & =\xi_{i=1}\hat{k'}_{i=1},\end{align*}
 where the notation $h$ of $\vec{R'}_{h,i=1}$ denotes the hemisphere.
The path length squared is given by \begin{align*}
\left\Vert \vec{R'}_{h,i=1}-\vec{R'}_{0}\right\Vert ^{2} & =\left\Vert \vec{R'}_{h,1}\right\Vert ^{2}+\left\Vert \vec{R'}_{0}\right\Vert ^{2}-2\vec{R'}_{h,1}\cdot\vec{R'}_{0}=\left[r'_{i}\right]^{2}+\left\Vert \vec{R'}_{0}\right\Vert ^{2}-2r'_{i}\left\Vert \vec{R'}_{0}\right\Vert \cos\psi_{0,1}.\end{align*}
 Since $\left\Vert \vec{R'}_{h,i=1}-\vec{R'}_{0}\right\Vert ^{2}=\left\Vert \xi_{i=1}\hat{k'}_{i=1}\right\Vert ^{2}=\xi_{1}^{2},$
the angle $\psi_{0,1}$ is found from the last equation to be \begin{align*}
\psi_{0,1} & =\arccos\left(\frac{1}{2}\left\{ r'_{i}\left\Vert \vec{R'}_{0}\right\Vert ^{-1}+\left[r'_{i}\right]^{-1}\left\Vert \vec{R'}_{0}\right\Vert -\left[r'_{i}\left\Vert \vec{R'}_{0}\right\Vert \right]^{-1}\xi_{1}^{2}\right\} \right),\end{align*}
 where $\xi_{1}=\xi_{1,p}$ is given in equation (\ref{eq:positive-root-k}).
The angle $\psi_{1,2}$ measured between the two vectors $\vec{R'}_{h,1}$
and $\vec{R'}_{h,2}$ is \begin{align*}
\psi_{1,2} & =\arccos\left(\left[r'_{i}\right]^{-2}\vec{R'}_{h,1}\cdot\vec{R'}_{h,2}\right)=\arccos\left(\sum_{i=1}^{3}\Lambda'_{h,1,i}\Lambda'_{h,2,i}\right),\end{align*}
 where $\vec{R'}_{h,1}$ and $\vec{R'}_{h,2}$ have been explicitly
written for $N=1,2$ in equation (\ref{eq:Nth-Reflection-Point-sphere}),
or equivalently, \begin{align*}
\psi_{1,2} & =\pi-2\theta_{inc}\end{align*}
 from equation (\ref{eq:psi-i-i-plus-1-equal-pi-minus-2-theta-i}).
For a hemisphere, it is convenient to define a quantity \begin{align*}
\mathbb{Z}_{h,max} & =\frac{1}{\psi_{1,2}}\left[\pi-\psi_{0,1}\right],\end{align*}
 or explicitly, \begin{align}
\mathbb{Z}_{h,max} & =\frac{1}{\pi-2\theta_{inc}}\left[\pi-\arccos\left(\frac{1}{2}\left\{ r'_{i}\left\Vert \vec{R'}_{0}\right\Vert ^{-1}+\left[r'_{i}\right]^{-1}\left\Vert \vec{R'}_{0}\right\Vert -\left[r'_{i}\left\Vert \vec{R'}_{0}\right\Vert \right]^{-1}\xi_{1,p}^{2}\right\} \right)\right],\label{eq:N-max-Hemisphere-Greatest-Integer-Function}\end{align}
 where $\xi_{1,p}$ is given in equation (\ref{eq:positive-root-k})
and $\theta_{inc}$ is given in equation (\ref{eq:angle-of-incidence-exp}).
The maximum number of internal reflections is then simply \begin{align}
N_{h,max} & =\left[\mathbb{Z}_{h,max}\right]_{G},\label{eq:N-max-Hemisphere}\end{align}
 where the notation $\left[\mathbb{Z}_{h,max}\right]_{G}$ is the
greatest integer of $\mathbb{Z}_{h,max}$ and it is defined in equation
(\ref{eq:N-max-Hemisphere-Greatest-Integer-Function}). 

The distance $\left\Vert \vec{L}\right\Vert $ between the two immediate
neighboring reflection points of the hemisphere is \begin{align}
\left\Vert \vec{L}\right\Vert  & =\left\Vert \vec{R}_{h,2}\left(r'_{i},\grave{\vec{\Lambda}}_{h,2},\vec{\Lambda}'_{h,2},\vec{R}_{T,h}\right)-\vec{R}_{h,1}\left(r'_{i},\grave{\vec{\Lambda}}_{h,1},\vec{\Lambda}'_{h,1},\vec{R}_{T,h}\right)\right\Vert .\label{eq:hemisphere-distance-rela}\end{align}
 It should be noted that \begin{eqnarray}
\left\Vert \vec{R}_{h,2}-\vec{R}_{h,1}\right\Vert =\left\Vert \vec{R}_{h,j}-\vec{R}_{h,j-1}\right\Vert , &  & j=3,\cdots,N_{h,max}\label{eq:hemisphere-distance-rela-equiv}\end{eqnarray}
 and the only reason that $\vec{R}_{h,1}$ and $\vec{R}_{h,2}$ are
used is for the purpose of convenience. 

The change in wave vector direction upon reflection at the point $\hat{R'}_{h,1}$
inside the resonator, or at the location $\vec{R'}_{h,1}+a\hat{R'}_{h,1}$
outside of the hemisphere, is given by results found for the sphere
case, equations (\ref{eq:SPHERE-Delta-K-Inside}) and (\ref{eq:SPHERE-Delta-K-Outside}),
with obvious subscript changes, \begin{eqnarray}
\triangle\vec{k'}_{inner}\left(;\vec{R'}_{h,1},\vec{R'}_{h,0}\right)=-\frac{4n\pi\cos\theta_{inc}}{\left\Vert \vec{R}_{h,2}\left(r'_{i},\vec{\Lambda}'_{h,2}\right)-\vec{R}_{h,1}\left(r'_{i},\vec{\Lambda}'_{h,1}\right)\right\Vert }\hat{R'}_{h,1}, &  & \left\{ \begin{array}{c}
0\leq\theta_{inc}<\pi/2,\\
\\n=1,2,\cdots\end{array}\right.\label{eq:HEMISPHERE-Delta-K-Inside}\end{eqnarray}
 and \begin{eqnarray}
\triangle\vec{k'}_{outer}\left(;\vec{R'}_{h,1}+a\hat{R'}_{h,1}\right)=4\left\Vert \vec{k'}_{i,f}\right\Vert \cos\theta_{inc}\hat{R'}_{h,1}, &  & \left\{ \begin{array}{c}
0\leq\theta_{inc}<\pi/2,\\
\\n=1,2,\cdots.\end{array}\right.\label{eq:HEMISPHERE-Delta-K-Outside}\end{eqnarray}

The above results on $\triangle\vec{k'}_{inner}\left(;\vec{R'}_{h,1},\vec{R'}_{h,0}\right)$
and $\triangle\vec{k'}_{outer}\left(;\vec{R'}_{h,1}+a\hat{R'}_{h,1}\right)$
have been derived based upon the fact that there are multiple internal
reflections. For a sphere, the multiple internal reflections are inherent.
However, for a hemisphere, it is not necessarily true that all incoming
waves would result in multiple internal reflections. The criteria
for multiple internal reflections are to be established. For a given
initial incoming wave vector $\hat{k'}_{1},$ there can be multiple
or single internal reflections depending upon the location of point
of entry into the cavity, $\vec{R'}_{0}.$ Shown in Figure \ref{cap:plane-sphere-intersection-3-Critical-Angle}
of section (3.B.2) are two such reflections where the dashed vectors
represent the single reflection case and the non-dashed vectors represent
the multiple reflection case. Because all the processes occur in the
same plane of incidence, the relationship $\vec{R'}_{f}=-\lambda_{0}\vec{R'}_{0}$
with $\lambda_{0}\geq0$ has to be true. Therefore, we will have \begin{eqnarray}
\vec{R'}_{1}=\vec{R'}_{0}+\xi_{p}\hat{k'}_{1}, &  & \vec{R'}_{f}\equiv-\lambda_{0}\vec{R'}_{0}=\vec{R'}_{1}+\vec{k'}_{2}.\label{eq:HEMI-HEMI-Multi-Reflect-Set1}\end{eqnarray}
 After eliminating $\vec{R'}_{1}$ from the last two equations, we
find \begin{align}
\vec{R'}_{0} & =-\left[1+\lambda_{0}\right]^{-1}\left[\xi_{p}\hat{k'}_{1}+\vec{k'}_{2}\right].\label{eq:HEMI-HEMI-Multi-Reflect-Ro-equal-Set2}\end{align}
 The direction of the reflected wave vector $\vec{k'}_{2}$ cannot
be arbitrary because it has to obey the reflection law. The relationship
between the directions of an incident and the associated reflection
wave is shown in equation (\ref{eq:K-reflected-in-PARA-n-PERPE-N}).
Designating $\hat{n'}=-\vec{R'}_{1}/r'_{i},$ $\vec{k'}_{r}\rightarrow\vec{k'}_{2}$
and $\vec{k'}_{i}\rightarrow\vec{k'}_{1},$ the reflected wave vector
$\vec{k'}_{2}$ can be written in the form \begin{align*}
\vec{k'}_{2} & \propto\alpha_{r,\perp}\left[r'_{i}\right]^{-2}\left[\vec{R'}_{1}\times\vec{k'}_{1}\right]\times\vec{R'}_{1}-\alpha_{r,\parallel}\left[r'_{i}\right]^{-2}\vec{R'}_{1}\cdot\vec{k'}_{1}\vec{R'}_{1}.\end{align*}
 By introducing a proportionality factor $\lambda_{2},$ it becomes
\begin{align*}
\vec{k'}_{2} & =\lambda_{2}\alpha_{r,\perp}\left[r'_{i}\right]^{-2}\left[\vec{R'}_{1}\times\vec{k'}_{1}\right]\times\vec{R'}_{1}-\lambda_{2}\alpha_{r,\parallel}\left[r'_{i}\right]^{-2}\vec{R'}_{1}\cdot\vec{k'}_{1}\vec{R'}_{1}.\end{align*}
 The goal is to relate $\vec{R'}_{f},$ or $\lambda_{0},$ in terms
of $\vec{R'}_{0}.$ Substituting the expression for $\vec{R'}_{1}$
from equation (\ref{eq:HEMI-HEMI-Multi-Reflect-Set1}), we arrive
at \begin{align}
\vec{k'}_{2} & =-\xi_{p}^{2}\lambda_{2}\alpha_{r,\parallel}\left[r'_{i}\right]^{-2}\vec{k'}_{1}+\xi_{p}\lambda_{2}\left[r'_{i}\right]^{-2}\left\{ \alpha_{r,\perp}\left[\vec{R'}_{0}\times\vec{k'}_{1}\right]\times\hat{k'}_{1}-\alpha_{r,\parallel}\left[\vec{R'}_{0}\cdot\vec{k'}_{1}\hat{k'}_{1}+\left\Vert \vec{k'}_{1}\right\Vert \vec{R'}_{0}\right]\right\} \nonumber \\
 & +\lambda_{2}\left[r'_{i}\right]^{-2}\left\{ \alpha_{r,\perp}\left[\vec{R'}_{0}\times\vec{k'}_{1}\right]\times\vec{R'}_{0}-\alpha_{r,\parallel}\vec{R'}_{0}\cdot\vec{k'}_{1}\vec{R'}_{0}\right\} .\label{eq:HEMI-HEMI-Multi-Reflect-k2-equal-Set3}\end{align}
 Finally, equations (\ref{eq:HEMI-HEMI-Multi-Reflect-Ro-equal-Set2})
and (\ref{eq:HEMI-HEMI-Multi-Reflect-k2-equal-Set3}) are combined
to yield \begin{align}
\xi_{p}^{2}\alpha_{r,\parallel}\vec{k'}_{1}-\xi_{p}\left\{ \alpha_{r,\perp}\left[\vec{R'}_{0}\times\vec{k'}_{1}\right]\times\hat{k'}_{1}-\alpha_{r,\parallel}\left[\vec{R'}_{0}\cdot\vec{k'}_{1}\hat{k'}_{1}+\left\Vert \vec{k'}_{1}\right\Vert \vec{R'}_{0}\right]+\lambda_{2}^{-1}\left[r'_{i}\right]^{2}\hat{k'}_{1}\right\} \nonumber \\
+\alpha_{r,\parallel}\vec{R'}_{0}\cdot\vec{k'}_{1}\vec{R'}_{0}-\alpha_{r,\perp}\left[\vec{R'}_{0}\times\vec{k'}_{1}\right]\times\vec{R'}_{0}-\lambda_{2}^{-1}\left[r'_{i}\right]^{2}\left[1+\lambda_{0}\right]\vec{R'}_{0} & =0.\label{eq:HEMI-HEMI-Multi-Reflect-k2-equal-Set4}\end{align}
 Utilizing the formula $\left[\vec{A}\times\vec{B}\right]\times\vec{C}=\sum_{l=1}^{3}\left\{ \left[\vec{A}\cdot\vec{C}\right]B_{l}-\left[\vec{B}\cdot\vec{C}\right]A_{l}\right\} \hat{e_{l}},$
the cross products are evaluated as \[
\left\{ \begin{array}{c}
\left[\vec{R'}_{0}\times\vec{k'}_{1}\right]\times\hat{k'}_{1}\equiv\left\Vert \vec{k'}_{1}\right\Vert ^{-1}\left[\vec{R'}_{0}\times\vec{k'}_{1}\right]\times\vec{k'}_{1}=\sum_{l=1}^{3}\left\{ \left\Vert \vec{k'}_{1}\right\Vert ^{-1}\left[\vec{R'}_{0}\cdot\vec{k'}_{1}\right]k'_{1,l}-\left\Vert \vec{k'}_{1}\right\Vert r'_{0,l}\right\} \hat{e_{l}},\\
\\\left[\vec{R'}_{0}\times\vec{k'}_{1}\right]\times\vec{R'}_{0}=\sum_{l=1}^{3}\left\{ \left\Vert \vec{R'}_{0}\right\Vert ^{2}k'_{1,l}-\left[\vec{k'}_{1}\cdot\vec{R'}_{0}\right]r'_{0,l}\right\} \hat{e_{l}},\\
\\\begin{array}{ccc}
\vec{k'}_{1}=\sum_{l=1}^{3}k'_{1,l}\hat{e_{l}}, &  & \vec{R'}_{0}=\sum_{l=1}^{3}r'_{0,l}\hat{e_{l}}.\end{array}\end{array}\right.\]
 We can rewrite equation (\ref{eq:HEMI-HEMI-Multi-Reflect-k2-equal-Set4})
as \begin{align*}
\sum_{l=1}^{3}\left\{ \xi_{p}^{2}\alpha_{r,\parallel}k'_{1,l}+\xi_{p}\left(\alpha_{r,\parallel}\left[\vec{R'}_{0}\cdot\vec{k'}_{1}\left\Vert \vec{k'}_{1}\right\Vert ^{-1}k'_{1,l}+\left\Vert \vec{k'}_{1}\right\Vert r'_{0,l}\right]-\alpha_{r,\perp}\left[\vec{R'}_{0}\cdot\vec{k'}_{1}\left\Vert \vec{k'}_{1}\right\Vert ^{-1}k'_{1,l}\right.\right.\right.\\
\left.\left.-\left\Vert \vec{k'}_{1}\right\Vert r'_{0,l}\right]-\lambda_{2}^{-1}\left[r'_{i}\right]^{2}\left\Vert \vec{k'}_{1}\right\Vert ^{-1}k'_{1,l}\right)+\alpha_{r,\parallel}\vec{R'}_{0}\cdot\vec{k'}_{1}r'_{0,l}-\alpha_{r,\perp}\\
\left.\times\left[\left\Vert \vec{R'}_{0}\right\Vert ^{2}k'_{1,l}-\vec{k'}_{1}\cdot\vec{R'}_{0}r'_{0,l}\right]-\lambda_{2}^{-1}\left[r'_{i}\right]^{2}\left[1+\lambda_{0}\right]r'_{0,l}\right\} \hat{e_{l}} & =0,\end{align*}
 which leads to the component equations \begin{align*}
\xi_{p}^{2}\alpha_{r,\parallel}k'_{1,l}+\xi_{p}\left(\alpha_{r,\parallel}\left[\vec{R'}_{0}\cdot\vec{k'}_{1}\left\Vert \vec{k'}_{1}\right\Vert ^{-1}k'_{1,l}+\left\Vert \vec{k'}_{1}\right\Vert r'_{0,l}\right]-\alpha_{r,\perp}\left[\vec{R'}_{0}\cdot\vec{k'}_{1}\left\Vert \vec{k'}_{1}\right\Vert ^{-1}k'_{1,l}\right.\right.\\
\left.\left.-\left\Vert \vec{k'}_{1}\right\Vert r'_{0,l}\right]-\lambda_{2}^{-1}\left[r'_{i}\right]^{2}\left\Vert \vec{k'}_{1}\right\Vert ^{-1}k'_{1,l}\right)+\alpha_{r,\parallel}\vec{R'}_{0}\cdot\vec{k'}_{1}r'_{0,l}-\alpha_{r,\perp}\\
\times\left[\left\Vert \vec{R'}_{0}\right\Vert ^{2}k'_{1,l}-\vec{k'}_{1}\cdot\vec{R'}_{0}r'_{0,l}\right]-\lambda_{2}^{-1}\left[r'_{i}\right]^{2}\left[1+\lambda_{0}\right]r'_{0,l} & =0,\end{align*}
 where $l=1,2,3.$ For an isotropic system, $\alpha_{r,\perp}=\alpha_{r,\parallel}=\alpha_{r},$
the last equation reduces to \begin{align*}
\xi_{p}^{2}\alpha_{r}k'_{1,l}+\xi_{p}\left[2\alpha_{r}\left\Vert \vec{k'}_{1}\right\Vert r'_{0,l}-\lambda_{2}^{-1}\left[r'_{i}\right]^{2}\left\Vert \vec{k'}_{1}\right\Vert ^{-1}k'_{1,l}\right]\\
+2\alpha_{r}\vec{R'}_{0}\cdot\vec{k'}_{1}r'_{0,l}-\alpha_{r}\left\Vert \vec{R'}_{0}\right\Vert ^{2}k'_{1,l}-\lambda_{2}^{-1}\left[r'_{i}\right]^{2}\left[1+\lambda_{0}\right]r'_{0,l} & =0,\end{align*}
 where $l=1,2,3.$ Because there are three such relations of the above,
all three component equations are added to yield \begin{align}
\xi_{p}^{2}+\xi_{p}\left[\sum_{l=1}^{3}k'_{1,l}\right]^{-1}\sum_{l=1}^{3}\left[2\left\Vert \vec{k'}_{1}\right\Vert r'_{0,l}-\lambda_{2}^{-1}\left[r'_{i}\right]^{2}\left\Vert \vec{k'}_{1}\right\Vert ^{-1}k'_{1,l}\right]\nonumber \\
+\left[\sum_{l=1}^{3}k'_{1,l}\right]^{-1}\sum_{l=1}^{3}\left\{ 2\vec{R'}_{0}\cdot\vec{k'}_{1}r'_{0,l}-\left\Vert \vec{R'}_{0}\right\Vert ^{2}k'_{1,l}-\lambda_{2}^{-1}\left[r'_{i}\right]^{2}\left[1+\lambda_{0}\right]r'_{0,l}\right\}  & =0,\label{eq:Hem-Hemi-root-polynomial-scalar-poly}\end{align}
 where the both sides of the above equation have been multiplied by
$\left[\sum_{l=1}^{3}k'_{1,l}\right]^{-1}$ and $\alpha_{r}=1$ have
been chosen for simplicity. Since $\xi_{p}$ is just a positive root
of $\xi_{1}$ of equation (\ref{eq:root-polynomial-scalar-form}),
it satisfies the equation \begin{align*}
\xi_{1}^{2}\overbrace{\sum_{l=1}^{3}\left\Vert \vec{k'}_{1}\right\Vert ^{-2}\left[k'_{1,l}\right]^{2}}^{1}+2\xi_{1}\left\Vert \vec{k'}_{1}\right\Vert ^{-1}\sum_{l=1}^{3}r'_{0,l}k'_{1,l}+\sum_{l=1}^{3}\left[r'_{0,l}\right]^{2}-\left[r'_{i}\right]^{2} & =0,\end{align*}
 where the index $i$ in equation (\ref{eq:root-polynomial-scalar-form})
have been changed to $l.$ The reflection coefficient $\alpha_{r}$
have been set to a unity in equation (\ref{eq:Hem-Hemi-root-polynomial-scalar-poly})
for the very reason that $\alpha_{r}=1$ had already been imposed
in equation (\ref{eq:root-polynomial-scalar-form}). Because $\xi_{1}\equiv\xi_{p},$
and the fact that coefficient of $\xi_{p}^{2}=\xi_{1}^{2}=1$ in equations
(\ref{eq:root-polynomial-scalar-form}) and (\ref{eq:Hem-Hemi-root-polynomial-scalar-poly}),
the two polynomials must be identical. Therefore, subtracting equation
(\ref{eq:root-polynomial-scalar-form}) from equation (\ref{eq:Hem-Hemi-root-polynomial-scalar-poly}),
we obtain \begin{align*}
\xi_{p}\left\{ \left[\sum_{l=1}^{3}k'_{1,l}\right]^{-1}\sum_{l=1}^{3}\left[2\left\Vert \vec{k'}_{1}\right\Vert r'_{0,l}-\lambda_{2}^{-1}\left[r'_{i}\right]^{2}\left\Vert \vec{k'}_{1}\right\Vert ^{-1}k'_{1,l}\right]-2\left\Vert \vec{k'}_{1}\right\Vert ^{-1}\sum_{l=1}^{3}r'_{0,l}k'_{1,l}\right\} +\left[r'_{i}\right]^{2}\\
+\left[\sum_{l=1}^{3}k'_{1,l}\right]^{-1}\sum_{l=1}^{3}\left\{ 2\vec{R'}_{0}\cdot\vec{k'}_{1}r'_{0,l}-\left\Vert \vec{R'}_{0}\right\Vert ^{2}k'_{1,l}-\lambda_{2}^{-1}\left[r'_{i}\right]^{2}\left[1+\lambda_{0}\right]r'_{0,l}-\left[r'_{0,l}\right]^{2}\right\}  & =0.\end{align*}
 Because $\xi_{p}$ is a particular value for the root of $\xi_{1},$
for the case where $\xi_{p}\neq0,$ the above equation is satisfied
only when the coefficients of the different powers of $\xi_{p}$ vanish
independently. This is another way of stating that each coefficients
of the different powers of $\xi_{p}$ in equations (\ref{eq:root-polynomial-scalar-form})
and (\ref{eq:Hem-Hemi-root-polynomial-scalar-poly}) must be proportional
to each other. For the situation here, they must be identical due
to the fact that coefficients of $\xi_{p}^{2}=\xi_{1}^{2}=1.$ Hence,
we have the conditions: \begin{equation}
\left\{ \begin{array}{c}
\left[\sum_{l=1}^{3}k'_{1,l}\right]^{-1}\sum_{l=1}^{3}\left[2\left\Vert \vec{k'}_{1}\right\Vert r'_{0,l}-\lambda_{2}^{-1}\left[r'_{i}\right]^{2}\left\Vert \vec{k'}_{1}\right\Vert ^{-1}k'_{1,l}\right]-2\left\Vert \vec{k'}_{1}\right\Vert ^{-1}\sum_{l=1}^{3}r'_{0,l}k'_{1,l}=0,\\
\\\left[r'_{i}\right]^{2}+\left[\sum_{l=1}^{3}k'_{1,l}\right]^{-1}\sum_{l=1}^{3}\left\{ 2\vec{R'}_{0}\cdot\vec{k'}_{1}r'_{0,l}-\left\Vert \vec{R'}_{0}\right\Vert ^{2}k'_{1,l}-\lambda_{2}^{-1}\left[r'_{i}\right]^{2}\left[1+\lambda_{0}\right]r'_{0,l}-\left[r'_{0,l}\right]^{2}\right\} =0.\end{array}\right.\label{eq:Hem-Hemi-root-polynomial-scalar-coeffi}\end{equation}
 From the first expression of equation (\ref{eq:Hem-Hemi-root-polynomial-scalar-coeffi}),
we find \begin{align}
\lambda_{2}^{-1} & =2\left[r'_{i}\right]^{-2}\left[\sum_{l=1}^{3}k'_{1,l}\right]^{-1}\sum_{l=1}^{3}\left[\left\Vert \vec{k'}_{1}\right\Vert ^{2}-k'_{1,l}\sum_{m=1}^{3}k'_{1,m}\right]r'_{0,l}.\label{eq:Hem-Hemi-root-polynomial-scalar-Lambda2-inv}\end{align}
 Solving for $\lambda_{0}$ from the second expression of equation
(\ref{eq:Hem-Hemi-root-polynomial-scalar-coeffi}), we have \begin{align*}
\lambda_{0} & =\frac{\sum_{l=1}^{3}\left\{ k'_{1,l}+2\vec{R'}_{0}\cdot\vec{k'}_{1}r'_{0,l}\left[r'_{i}\right]^{-2}-\left\Vert \vec{R'}_{0}\right\Vert ^{2}k'_{1,l}\left[r'_{i}\right]^{-2}-\lambda_{2}^{-1}r'_{0,l}-\left[r'_{0,l}\right]^{2}\left[r'_{i}\right]^{-2}\right\} }{\lambda_{2}^{-1}\sum_{l=1}^{3}r'_{0,l}}\end{align*}
 or, substituting the expression of $\lambda_{2}^{-1}$ given in equation
(\ref{eq:Hem-Hemi-root-polynomial-scalar-Lambda2-inv}), we have the
result: \begin{align}
\lambda_{0} & =\frac{1}{2}\left[\sum_{n=1}^{3}k'_{1,n}\right]\left\{ \sum_{j=1}^{3}\sum_{l=1}^{3}\left[\left\Vert \vec{k'}_{1}\right\Vert ^{2}-k'_{1,l}\sum_{m=1}^{3}k'_{1,m}\right]r'_{0,l}r'_{0,j}\right\} ^{-1}\sum_{l=1}^{3}\left\{ k'_{1,l}\left[r'_{i}\right]^{2}-\left[r'_{0,l}\right]^{2}\right.\nonumber \\
 & \left.+2\vec{R'}_{0}\cdot\vec{k'}_{1}r'_{0,l}-\left\Vert \vec{R'}_{0}\right\Vert ^{2}k'_{1,l}-2r'_{0,l}\left[\sum_{l=1}^{3}k'_{1,l}\right]^{-1}\sum_{i=1}^{3}\left[\left\Vert \vec{k'}_{1}\right\Vert ^{2}-k'_{1,i}\sum_{m=1}^{3}k'_{1,m}\right]r'_{0,i}\right\} .\label{eq:Hem-Hemi-root-polynomial-scalar-Lambda0}\end{align}
 Referring back to Figure \ref{cap:plane-sphere-intersection-3-Critical-Angle},
the term $\lambda_{0}$ is connected to $\vec{R'}_{f}$ through the
relation $\vec{R'}_{f}\equiv-\lambda_{0}\vec{R'}_{0}.$ Therefore,
the criterion for waves to have multiple or single internal reflection
is contained in the controlled quantity $\lambda_{0}.$ The vector
$\vec{R'}_{0}$ is a quantity that must be specified initially. Because
$\lambda_{0}$ is a positive definite scalar, we can rewrite it as
\begin{align*}
\lambda_{0} & =\left\Vert \vec{R'}_{f}\right\Vert \left\Vert \vec{R'}_{0}\right\Vert ^{-1}=\left\Vert \vec{R'}_{f}\right\Vert \left\{ \sum_{i=1}^{3}\left[r'_{0,i}\right]^{2}\right\} ^{-1/2}.\end{align*}
 Substituting the above definition of $\lambda_{0}$ into equation
(\ref{eq:Hem-Hemi-root-polynomial-scalar-Lambda0}), the quantity
$\left\Vert \vec{R'}_{f}\right\Vert $ can be solved as \begin{align}
\left\Vert \vec{R'}_{f}\right\Vert  & =\frac{1}{2}\left\Vert \vec{R'}_{0}\right\Vert \left[\sum_{n=1}^{3}k'_{1,n}\right]\left\{ \sum_{j=1}^{3}\sum_{l=1}^{3}\left[\left\Vert \vec{k'}_{1}\right\Vert ^{2}-k'_{1,l}\sum_{m=1}^{3}k'_{1,m}\right]r'_{0,l}r'_{0,j}\right\} ^{-1}\sum_{l=1}^{3}\left\{ k'_{1,l}\left[r'_{i}\right]^{2}-\left[r'_{0,l}\right]^{2}\right.\nonumber \\
 & \left.+2\vec{R'}_{0}\cdot\vec{k'}_{1}r'_{0,l}-\left\Vert \vec{R'}_{0}\right\Vert ^{2}k'_{1,l}-2r'_{0,l}\left[\sum_{l=1}^{3}k'_{1,l}\right]^{-1}\sum_{i=1}^{3}\left[\left\Vert \vec{k'}_{1}\right\Vert ^{2}-k'_{1,i}\sum_{m=1}^{3}k'_{1,m}\right]r'_{0,i}\right\} .\label{eq:Hem-Hemi-Multi-Single-Ref-condi}\end{align}
 Because the hemisphere opening has a radius $r'_{i},$ the following
criteria are concluded: \begin{equation}
\left\{ \begin{array}{ccc}
\left\Vert \vec{R'}_{f}\right\Vert <r'_{i}, &  & single\; internal\; reflection,\\
\\\left\Vert \vec{R'}_{f}\right\Vert \geq r'_{i}, &  & multiple\; internal\; reflections,\end{array}\right.\label{eq:Hem-Hemi-Multi-Single-Ref-condi-Interpretation}\end{equation}
 where $\left\Vert \vec{R'}_{f}\right\Vert $ is given in equation
(\ref{eq:Hem-Hemi-Multi-Single-Ref-condi}) and $r'_{i}$ is the radius
of a hemisphere.

\subsection{Plate-Hemisphere}

The description of a surface is a study of the orientation of its
local normal $\hat{n'}_{p},$ which is shown in Figure \ref{cap:circular-plane}.
In spherical coordinates, the unit vectors $\hat{n'}_{p},$ $\hat{\theta'}_{p}$
and $\hat{\phi'}_{p}$ are expressed as \begin{equation}
\begin{array}{ccccc}
\hat{n'}_{p}=\sum_{i=1}^{3}\Lambda'_{p,i}\hat{e}_{i}, &  & \hat{\theta'}_{p}=\sum_{i=1}^{3}\frac{\partial\Lambda'_{p,i}}{\partial\theta'_{p}}\hat{e}_{i}, &  & \hat{\phi'}_{p}=\sum_{i=1}^{3}\frac{1}{\sin\theta'_{p}}\frac{\partial\Lambda'_{p,i}}{\partial\phi'_{p}}\hat{e}_{i},\end{array}\label{eq:unit-vectors-circular-plane}\end{equation}
 where \begin{equation}
\begin{array}{ccccc}
\Lambda'_{p,1}\left(\theta'_{p},\phi'_{p}\right)=\sin\theta'_{p}\cos\phi'_{p}, &  & \Lambda'_{p,2}\left(\theta'_{p},\phi'_{p}\right)=\sin\theta'_{p}\sin\phi'_{p}, &  & \Lambda'_{p,3}\left(\theta'_{p}\right)=\cos\theta'_{p}.\end{array}\label{eq:Capital-Lambda-prime-Plate-Def}\end{equation}
 It is easy to show that the set of unit vectors $\left(\hat{n'}_{p},\hat{\theta'}_{p},\hat{\phi'}_{p}\right)$
forms an orthonormal coordinates. Therefore, the points on plane can
be described by a \textbf{2D} coordinate system made of $\hat{\theta'}_{p}$
and $\hat{\phi'}_{p},$ \begin{align}
\vec{R'}_{p} & =\nu'_{p,\theta'_{p}}\hat{\theta'}_{p}+\nu'_{p,\phi'_{p}}\hat{\phi'}_{p}=\sum_{i=1}^{3}\left[\nu'_{p,\theta'_{p}}\frac{\partial\Lambda'_{p,i}}{\partial\theta'_{p}}+\frac{\nu'_{p,\phi'_{p}}}{\sin\theta'_{p}}\frac{\partial\Lambda'_{p,i}}{\partial\phi'_{p}}\right]\hat{e}_{i}.\label{eq:points-on-circular-plane}\end{align}
 If the plane's orientation constantly changes in time about its origin,
the points on the plane experience the velocity $d\vec{R'}_{p}/dt,$
\begin{align}
\dot{\vec{R'}}_{p}\equiv\frac{d\vec{R'}_{p}}{dt} & =\sum_{i=1}^{3}\left[\dot{\nu'}_{p,\theta'_{p}}\frac{\partial\Lambda'_{p,i}}{\partial\theta'_{p}}+\left\{ \nu'_{p,\theta'_{p}}\frac{\partial^{2}\Lambda'_{p,i}}{\partial\left[\theta'_{p}\right]^{2}}+\frac{\nu'_{p,\phi'_{p}}}{\sin\theta'_{p}}\left(\frac{\partial^{2}\Lambda'_{p,i}}{\partial\theta'_{p}\partial\phi'_{p}}-\cot\theta'_{p}\frac{\partial\Lambda'_{p,i}}{\partial\phi'_{p}}\right)\right\} \dot{\theta'}_{p}\right.\nonumber \\
 & \left.+\frac{\dot{\nu'}_{p,\phi'_{p}}}{\sin\theta'_{p}}\frac{\partial\Lambda'_{p,i}}{\partial\phi'_{p}}+\left\{ \nu'_{p,\theta'_{p}}\frac{\partial^{2}\Lambda'_{p,i}}{\partial\phi'_{p}\partial\theta'_{p}}+\frac{\nu'_{p,\phi'_{p}}}{\sin\theta'_{p}}\frac{\partial^{2}\Lambda'_{p,i}}{\partial\left[\phi'_{p}\right]^{2}}\right\} \dot{\phi'}_{p}\right]\hat{e}_{i},\label{eq:velocity-points-on-circular-plane}\end{align}
 where $\dot{\theta'}_{p},$ $\dot{\phi'}_{p}$ are the angular frequencies
and $\dot{\nu'}_{p,\theta'_{p}},$ $\dot{\nu'}_{p,\phi'_{p}}$ are
the lattice vibrations along the directions $\hat{\theta'}_{p}$ and
$\hat{\phi'}_{p},$ respectively. Here, it is understood that $\Lambda'_{p,3}$
is independent of $\phi'_{p}.$ Therefore, the differentiation of
$\Lambda'_{p,3}$ with respect to the $\phi'_{p}$ vanishes. For the
static plate in which there are no lattice vibrations, $\dot{\nu'}_{p,\theta'_{p}}$
and $\dot{\nu'}_{p,\phi'_{p}}$ vanishes. 

For the case of plate-hemisphere configuration shown in Figure \ref{cap:plate-and-hemisphere-complex},
the points on the plate are represented by the vector $\vec{R}_{p}$
relative to the system origin. Making the correspondence in equation
(\ref{eq:r-vector-cart-in-prime}), $\vec{R}\rightarrow\vec{R}_{n},$
$\vec{R}_{T}\rightarrow\vec{R}_{T,p}$ and $\vec{R'}\rightarrow\hat{n'}_{p},$
the two angular variable sets $\left(\theta'_{p},\phi'_{p}\right)$
and $\left(\theta,\phi\right)$ are connected through the relations
given in equations (\ref{eq:phi}) and (\ref{eq:theta}) with $r'_{i}\rightarrow1.$
Here $r'_{i}\rightarrow1$ because $\left\Vert \hat{n'}_{p}\right\Vert =1.$
Therefore, we obtain \begin{align}
\grave{\phi}_{p}\left(\theta'_{p},\phi'_{p},\nu_{T,p,1},\nu_{T,p,2}\right) & =\arctan\left(\frac{\nu_{T,p,2}+\sin\theta'_{p}\sin\phi'_{p}}{\nu_{T,p,1}+\sin\theta'_{p}\cos\phi'_{p}}\right),\label{eq:phi-Plate}\end{align}
 \begin{align}
\grave{\theta}_{p}\left(\theta'_{p},\phi'_{p},\vec{R}_{T,p}\right) & =\arctan\left(\frac{\left\{ \nu_{T,p,1}+\nu_{T,p,2}+\sin\theta'_{p}\left[\cos\phi'_{p}+\sin\phi'_{p}\right]\right\} \left[\nu_{T,p,3}+\cos\theta'_{p}\right]^{-1}}{\cos\left(\arctan\left(\frac{\nu_{T,p,2}+\sin\theta'_{p}\sin\phi'_{p}}{\nu_{T,p,1}+\sin\theta'_{p}\cos\phi'_{p}}\right)\right)+\sin\left(\arctan\left(\frac{\nu_{T,p,2}+\sin\theta'_{p}\sin\phi'_{p}}{\nu_{T,p,1}+\sin\theta'_{p}\cos\phi'_{p}}\right)\right)}\right),\label{eq:theta-Plate}\end{align}
 where the subscript $p$ of $\grave{\phi}_{p}$ and $\grave{\theta}_{p}$
indicates that these are the spherical variables for the points on
the plate shown in Figure \ref{cap:plate-and-hemisphere-complex},
and they are not that of the hemisphere. The vector $\vec{R}_{p}$
becomes \begin{align}
\vec{R}_{p} & =\vec{R}_{T,p}\left(\nu_{T,p,1},\nu_{T,p,2},\nu_{T,p,3}\right)+\vec{R'}_{p}=\sum_{i=1}^{3}\left[\nu_{T,p,i}+\nu'_{p,\theta'_{p}}\frac{\partial\Lambda'_{p,i}}{\partial\theta'_{p}}+\frac{\nu'_{p,\phi'_{p}}}{\sin\theta'_{p}}\frac{\partial\Lambda'_{p,i}}{\partial\phi'_{p}}\right]\hat{e}_{i}.\label{eq:Rp-vector-cart-in-prime-Plate}\end{align}
 The magnitude $\left\Vert \vec{R}_{p}\right\Vert $ is given by \begin{align*}
r_{p}\left(\vec{\Lambda}'_{p},\vec{R}_{T,p}\right) & \equiv\left\Vert \vec{R}_{p}\right\Vert =\left\{ \sum_{i=1}^{3}\left[\nu_{T,p,i}+\nu'_{p,\theta'_{p}}\frac{\partial\Lambda'_{p,i}}{\partial\theta'_{p}}+\frac{\nu'_{p,\phi'_{p}}}{\sin\theta'_{p}}\frac{\partial\Lambda'_{p,i}}{\partial\phi'_{p}}\right]^{2}\right\} ^{1/2}.\end{align*}
 In terms of the spherical coordinates, $\vec{R}_{p}$ is expressed
by \begin{align}
\vec{R}_{p}\left(\grave{\vec{\Lambda}}_{p},\vec{\Lambda}'_{p},\vec{R}_{T,p}\right) & =\left\{ \sum_{i=1}^{3}\left[\nu_{T,p,i}+\nu'_{p,\theta'_{p}}\frac{\partial\Lambda'_{p,i}}{\partial\theta'_{p}}+\frac{\nu'_{p,\phi'_{p}}}{\sin\theta'_{p}}\frac{\partial\Lambda'_{p,i}}{\partial\phi'_{p}}\right]^{2}\right\} ^{1/2}\sum_{i=1}^{3}\grave{\Lambda}_{p,i}\hat{e_{i}},\label{eq:Points-on-Plate-R}\end{align}
 where \begin{equation}
\begin{array}{ccccc}
\grave{\Lambda}_{p,1}\left(\grave{\theta}_{p},\grave{\phi}_{p}\right)=\sin\grave{\theta}_{p}\cos\grave{\phi}_{p}, &  & \grave{\Lambda}_{p,2}\left(\grave{\theta}_{p},\grave{\phi}_{p}\right)=\sin\grave{\theta}_{p}\sin\grave{\phi}_{p}, &  & \grave{\Lambda}_{p,3}\left(\grave{\theta}_{p}\right)=\cos\grave{\theta}_{p}.\end{array}\label{eq:Capital-Lambda-Plate-Def}\end{equation}
 Here, the subscript $p$ in $\vec{R}_{p}$ indicates that the vector
$\vec{R}_{p}$ describe the points on the plate. If the plane's orientation
constantly changes in time about its origin, then the same orientation
change observed relative to the system origin is given by the velocity
$d\vec{R}_{p}/dt,$ \begin{align}
\dot{\vec{R}}_{p}\equiv\frac{d\vec{R}_{p}}{dt} & =\left\{ \sum_{i=1}^{3}\left[\nu_{T,p,i}+\nu'_{p,\theta'_{p}}\frac{\partial\Lambda'_{p,i}}{\partial\theta'_{p}}+\frac{\nu'_{p,\phi'_{p}}}{\sin\theta'_{p}}\frac{\partial\Lambda'_{p,i}}{\partial\phi'_{p}}\right]^{2}\right\} ^{-1/2}\sum_{j=1}^{3}\sum_{k=1}^{3}\left(\left[\nu_{T,p,k}+\nu'_{p,\theta'_{p}}\frac{\partial\Lambda'_{p,k}}{\partial\theta'_{p}}\right.\right.\nonumber \\
 & \left.+\frac{\nu'_{p,\phi'_{p}}}{\sin\theta'_{p}}\frac{\partial\Lambda'_{p,k}}{\partial\phi'_{p}}\right]\left[\dot{\nu}_{T,p,k}+\left\{ \nu'_{p,\theta'_{p}}\frac{\partial^{2}\Lambda'_{p,k}}{\partial\left[\theta'_{p}\right]^{2}}+\frac{\nu'_{p,\phi'_{p}}}{\sin\theta'_{p}}\left(\frac{\partial^{2}\Lambda'_{p,k}}{\partial\theta'_{p}\partial\phi'_{p}}-\cot\theta'_{p}\frac{\partial\Lambda'_{p,k}}{\partial\phi'_{p}}\right)\right\} \dot{\theta'}_{p}\right.\nonumber \\
 & \left.+\left\{ \nu'_{p,\theta'_{p}}\frac{\partial^{2}\Lambda'_{p,k}}{\partial\phi'_{p}\partial\theta'_{p}}+\frac{\nu'_{p,\phi'_{p}}}{\sin\theta'_{p}}\frac{\partial^{2}\Lambda'_{p,k}}{\partial\left[\phi'_{p}\right]^{2}}\right\} \dot{\phi'}_{p}+\dot{\nu'}_{p,\theta'_{p}}\frac{\partial\Lambda'_{p,k}}{\partial\theta'_{p}}+\frac{\dot{\nu'}_{p,\phi'_{p}}}{\sin\theta'_{p}}\frac{\partial\Lambda'_{p,k}}{\partial\phi'_{p}}\right]\grave{\Lambda}_{p,j}\nonumber \\
 & \left.+\sum_{i=1}^{3}\left[\nu_{T,p,i}+\nu'_{p,\theta'_{p}}\frac{\partial\Lambda'_{p,i}}{\partial\theta'_{p}}+\frac{\nu'_{p,\phi'_{p}}}{\sin\theta'_{p}}\frac{\partial\Lambda'_{p,i}}{\partial\phi'_{p}}\right]^{2}\left[\frac{\partial\grave{\Lambda}_{p,j}}{\partial\grave{\theta}_{p}}\frac{\partial\grave{\theta}_{p}}{\partial\phi'_{p}}\dot{\theta}'_{p}+\frac{\partial\grave{\Lambda}_{p,j}}{\partial\grave{\phi}_{p}}\frac{\partial\grave{\phi}_{p}}{\partial\phi'_{p}}\dot{\phi}'_{p}\right]\right)\hat{e_{j}},\label{eq:velocity-points-on-circular-plane}\end{align}
 where it is understood that $\Lambda'_{p,3}$ and $\grave{\Lambda}_{p,3}$
are independent of $\phi'_{p}$ and $\grave{\phi}_{p},$ respectively;
and as a consequence, their differentiation with respect to $\phi'_{p}$
and $\grave{\phi}_{p}$ vanishes. Here $\dot{\theta'}_{p},$ $\dot{\phi'}_{p}$
are angular frequencies and $\dot{\nu}_{T,p,i}$ is the translation
speed of plate relative to system origin. Also, $\dot{\nu'}_{p,\theta'_{p}},$
$\dot{\nu'}_{p,\phi'_{p}}$ are lattice vibrations along directions
$\hat{\theta'}_{p}$ and $\hat{\phi'}_{p},$ respectively. For a static
plate in which there are no lattice vibrations, $\dot{\nu'}_{p,\theta'_{p}}$
and $\dot{\nu'}_{p,\phi'_{p}}$ vanishes. 

A cross-sectional view of the plate-hemisphere system is shown in
Figure \ref{cap:plate-hemisphere-plane-of-incidence-intersect-Complex}.
The initial wave vector $\vec{k'}_{i}$ traveling toward the hemisphere
would go through reflections according to the law of reflection and
finally exit. It then continues toward the plate and reflects from
it. Depending on the orientation of plate at the time of impact, the
wave would either escape to infinity or re-enter the hemisphere to
repeat the process all over again. 

The equation (\ref{eq:points-on-circular-plane}) defines points on
a plate, as shown in Figure \ref{cap:plate-and-hemisphere-complex}
of section (3.B.3), relative to the plate origin. If $S_{p}$ is a
set of points on a plate whose members are defined by $\vec{R'}_{p}$
of equation (\ref{eq:points-on-circular-plane}), the wave reflection
dynamics off the plate involve only those points of $S_{p}$ in the
intersection between the plate and the plane of incidence whose unit
normal is $\hat{n'}_{p,1}$ given in equation (\ref{eq:n-hat-p-i-eq-mag-times-ki-cross-Ro}).
In order to determine the intersection between the plate and the incidence
plane, the plate is first represented by a scalar field. From equation
(\ref{eq:unit-vectors-circular-plane}), the unit plate normal is
\begin{align*}
\hat{n'}_{p} & =\sum_{i=1}^{3}\Lambda'_{p,i}\hat{e}_{i}.\end{align*}
 The scalar field corresponding to the unit normal $\hat{n'}_{p}$
satisfies the relation \begin{eqnarray*}
\vec{\nabla'}f_{p}\left(\nu'_{1},\nu'_{2},\nu'_{3}\right)\equiv\sum_{i=1}^{3}\hat{e}_{i}\frac{\partial f_{p}}{\partial\nu'_{i}}=\sum_{i=1}^{3}\Lambda'_{p,i}\hat{e}_{i} & \rightarrow & \sum_{i=1}^{3}\left[\frac{\partial f_{p}}{\partial\nu'_{i}}-\Lambda'_{p,i}\right]\hat{e}_{i}=0.\end{eqnarray*}
 The individual component of the equation is given by \begin{eqnarray*}
\frac{\partial f_{p}}{\partial\nu'_{i}}-\Lambda'_{p,i}=0, &  & i=1,2,3,\end{eqnarray*}
 where $\Lambda'_{p,i}$ is independent of $\nu'_{i}.$ An integration
with respect to $\nu'_{i}$ yields the result \begin{align}
f_{p}\left(\nu'_{1},\nu'_{2},\nu'_{3}\right) & =\sum_{i=1}^{3}\Lambda'_{p,i}\nu'_{i},\label{eq:Scalar-Field-Plate}\end{align}
 where the integration constant is set to zero because the plate contains
its local origin. The intersection between the plane of incidence
and the plate, shown in Figure \ref{cap:circular-plane} of section
(3.B.3), satisfies the relation \begin{eqnarray}
f_{p}\left(\nu'_{1},\nu'_{2},\nu'_{3}\right)-f_{p,1}\left(\nu'_{1},\nu'_{2},\nu'_{3}\right)=0 & \rightarrow & \sum_{i=1}^{3}\left[\Lambda'_{p,i}+\left\Vert \vec{n'}_{p,1}\right\Vert ^{-1}\epsilon_{ijk}k'_{1,j}r'_{0,k}\right]\nu'_{i}=0,\label{eq:Scalar-Field-Plate-Plane-of-inci-intersect}\end{eqnarray}
 where $f_{p,1}\left(\nu'_{1},\nu'_{2},\nu'_{3}\right)$ is given
in equation (\ref{eq:f-p-i-plane-of-incidence}), and $\nu'_{i}$
is a scalar corresponding to the basis $\hat{e_{i}},$ of course.
We have, from equation (\ref{eq:points-on-circular-plane}), \begin{eqnarray}
\nu'_{i}=\nu'_{p,\theta'_{p}}\frac{\partial\Lambda'_{p,i}}{\partial\theta'_{p}}+\frac{\nu'_{p,\phi'_{p}}}{\sin\theta'_{p}}\frac{\partial\Lambda'_{p,i}}{\partial\phi'_{p}}, &  & i=1,2,3.\label{eq:Scalar-Field-Plate-Plane-vi-vtheta-vphi-rela}\end{eqnarray}
 Substituting $\nu'_{i}$ into equation (\ref{eq:Scalar-Field-Plate-Plane-of-inci-intersect}),
$\nu'_{p,\theta'_{p}}$ is solved as \begin{align}
\nu'_{p,\theta'_{p}} & =-\frac{\nu'_{p,\phi'_{p}}}{\sin\theta'_{p}}\frac{\sum_{i=1}^{3}\frac{\partial\Lambda'_{p,i}}{\partial\phi'_{p}}\left[\Lambda'_{p,i}+\left\Vert \vec{n'}_{p,1}\right\Vert ^{-1}\epsilon_{ijk}k'_{1,j}r'_{0,k}\right]}{\sum_{l=1}^{3}\frac{\partial\Lambda'_{p,l}}{\partial\theta'_{p}}\left[\Lambda'_{p,l}+\left\Vert \vec{n'}_{p,1}\right\Vert ^{-1}\epsilon_{lmn}k'_{1,m}r'_{0,n}\right]},\label{eq:Scalar-Field-Plate-Plane-Vp-theta}\end{align}
 where the summation over indices $j,$ $k,$ $m$ and $n$ is implicit,
also the quotient $\nu'_{p,\phi'_{p}}/\sin\theta'_{p}$ has been moved
out of the summation. The $\vec{R'}_{p}$ given in equation (\ref{eq:points-on-circular-plane})
is then rewritten as \begin{align}
\vec{R'}_{p} & =\frac{\nu'_{p,\phi'_{p}}}{\sin\theta'_{p}}\sum_{i=1}^{3}\left\{ \frac{\partial\Lambda'_{p,i}}{\partial\phi'_{p}}-\frac{\sum_{i=1}^{3}\frac{\partial\Lambda'_{p,i}}{\partial\phi'_{p}}\left[\Lambda'_{p,i}+\left\Vert \vec{n'}_{p,1}\right\Vert ^{-1}\epsilon_{ijk}k'_{1,j}r'_{0,k}\right]}{\sum_{l=1}^{3}\frac{\partial\Lambda'_{p,l}}{\partial\theta'_{p}}\left[\Lambda'_{p,l}+\left\Vert \vec{n'}_{p,1}\right\Vert ^{-1}\epsilon_{lmn}k'_{1,m}r'_{0,n}\right]}\frac{\partial\Lambda'_{p,i}}{\partial\theta'_{p}}\right\} \hat{e}_{i}.\label{eq:points-on-circular-plane-reduced}\end{align}
 Similarly, $\vec{R}_{p}$ given in equation (\ref{eq:Points-on-Plate-R})
is rewritten as \begin{align}
\vec{R}_{p} & =\left\{ \sum_{i=1}^{3}\left[\nu_{T,p,i}+\frac{\nu'_{p,\phi'_{p}}}{\sin\theta'_{p}}\left(\frac{\partial\Lambda'_{p,i}}{\partial\phi'_{p}}-\frac{\sum_{i=1}^{3}\frac{\partial\Lambda'_{p,i}}{\partial\phi'_{p}}\left[\Lambda'_{p,i}+\left\Vert \vec{n'}_{p,1}\right\Vert ^{-1}\epsilon_{ijk}k'_{1,j}r'_{0,k}\right]}{\sum_{l=1}^{3}\frac{\partial\Lambda'_{p,l}}{\partial\theta'_{p}}\left[\Lambda'_{p,l}+\left\Vert \vec{n'}_{p,1}\right\Vert ^{-1}\epsilon_{lmn}k'_{1,m}r'_{0,n}\right]}\frac{\partial\Lambda'_{p,i}}{\partial\theta'_{p}}\right)\right]^{2}\right\} ^{1/2}\nonumber \\
 & \times\sum_{i=1}^{3}\grave{\Lambda}_{p,i}\hat{e_{i}},\label{eq:Points-on-Plate-R-reduced}\end{align}
 where $\grave{\Lambda}_{p,i}$ is given in equation (\ref{eq:Capital-Lambda-Plate-Def}).
If $N_{h,max}$ is the maximum count of reflections within the hemisphere
before the wave escapes, the direction of the escaping wave, measured
with respect to the system origin $\vec{R}=0,$ is \begin{align}
\vec{k}_{N_{h,max}+1} & =\vec{R}_{h,N_{h,max}+1}-\vec{R}_{h,N_{h,max}},\label{eq:Plate-Hemi-exiting-K-direction-set1}\end{align}
 where $\vec{k}_{N_{h,max}+1}=\xi_{N_{h,max}+1}\vec{k'}_{N_{h,max}+1}.$
Similarly, by the correspondence $\vec{R}_{h,N_{h,max}+1}\rightarrow\vec{R}_{h,i+3}$
and $\vec{R}_{h,N_{h,max}}\rightarrow\vec{R}_{h,i+2}$ in Figure \ref{cap:plate-hemisphere-plane-of-incidence-intersect-Complex},
the direction of the escaping wave vector $\vec{k}$ is equivalently
described by the relation \begin{align}
\zeta\vec{k}_{N_{h,max}+1} & =\vec{R}_{p}-\vec{R}_{h,N_{h,max}}\label{eq:Plate-Hemi-exiting-K-direction-set2}\end{align}
 where $\zeta$ is an appropriate positive scale factor. Combining
equations (\ref{eq:Plate-Hemi-exiting-K-direction-set1}) and (\ref{eq:Plate-Hemi-exiting-K-direction-set2}),
$\vec{R}_{p}$ is solved as \begin{align}
\vec{R}_{p} & =\zeta\vec{R}_{h,N_{h,max}+1}+\left[1-\zeta\right]\vec{R}_{h,N_{h,max}}.\label{eq:Plate-Hemi-exiting-K-direction-set3}\end{align}
 Because both $\vec{R}_{p}$ and $\vec{k}_{N_{h,max}+1}$ belong to
a spanning set for the plane of incidence whose unit normal is $\hat{n'}_{p,1}$
given in equation (\ref{eq:n-hat-p-i-eq-mag-times-ki-cross-Ro}),
we observe that the following relationship \begin{align}
\vec{R}_{p}\times\vec{k}_{N_{h,max}+1} & =\left\{ \zeta\vec{R}_{h,N_{h,max}+1}+\left[1-\zeta\right]\vec{R}_{h,N_{h,max}}\right\} \times\vec{k}_{N_{h,max}+1}=\gamma\hat{n'}_{p,1}\label{eq:Plate-Hemi-exiting-K-direction-set4}\end{align}
 hold, where $\gamma$ is a proportional constant. Substituting the
explicit form for $\hat{n'}_{p,1}$ from equation (\ref{eq:n-hat-p-i-eq-mag-times-ki-cross-Ro})
into equation (\ref{eq:Plate-Hemi-exiting-K-direction-set4}), it
simplifies into the following equation \begin{align*}
\sum_{i=1}^{3}\left[\zeta\epsilon_{ijk}R_{h,N_{h,max}+1,j}k_{N_{h,max}+1,k}+\left[1-\zeta\right]\epsilon_{ijk}R_{h,N_{h,max},j}k_{N_{h,max}+1,k}+\gamma\left\Vert \vec{n'}_{p,1}\right\Vert ^{-1}\epsilon_{ijk}k'_{1,j}r'_{0,k}\right]\hat{e_{i}} & =0,\end{align*}
 and its component equations are given by \begin{align*}
\zeta\epsilon_{ijk}R_{h,N_{h,max}+1,j}k_{N_{h,max}+1,k}+\left[1-\zeta\right]\epsilon_{ijk}R_{h,N_{h,max},j}k_{N_{h,max}+1,k}+\gamma\left\Vert \vec{n'}_{p,1}\right\Vert ^{-1}\epsilon_{ijk}k'_{1,j}r'_{0,k} & =0,\end{align*}
 where $k_{N_{h,max}+1,k}=R_{h,N_{h,max}+1,k}-R_{h,N_{h,max},k}$
as described in equation (\ref{eq:Plate-Hemi-exiting-K-direction-set1}).
Finally, the scale factor $\zeta$ is solved as \begin{align}
\zeta\equiv\zeta_{i} & =\left[\epsilon_{ijk}R_{h,N_{h,max},j}R_{h,N_{h,max}+1,k}-\epsilon_{ijk}R_{h,N_{h,max},j}R_{h,N_{h,max},k}+\gamma\left\Vert \vec{n'}_{p,1}\right\Vert ^{-1}\epsilon_{ijk}k'_{1,j}r'_{0,k}\right]\nonumber \\
 & \times\left[\epsilon_{ijk}R_{h,N_{h,max},j}R_{h,N_{h,max}+1,k}-\epsilon_{ijk}R_{h,N_{h,max},j}R_{h,N_{h,max},k}-\epsilon_{ijk}R_{h,N_{h,max}+1,j}R_{h,N_{h,max}+1,k}\right.\nonumber \\
 & \left.+\epsilon_{ijk}R_{h,N_{h,max}+1,j}R_{h,N_{h,max},k}\right]^{-1},\label{eq:Plate-Hemi-exiting-K-scale}\end{align}
 where $i,j,k=1,2,3.$ Here, the notation $\zeta_{i}$ have been adopted
in place of $\zeta.$ It should be understood that for irrotational
\textbf{3D} vectors, $\zeta_{1}=\zeta_{2}=\zeta_{3}=\zeta.$ For vectors
in \textbf{2D} and \textbf{1D} space, it is understood then $\zeta_{3},$
$\zeta_{2}$ are absent, respectively. In current form, equation (\ref{eq:Plate-Hemi-exiting-K-scale})
is incomplete because $\gamma$ is still arbitrary. This happens because
$\nu'_{p,\theta'_{p}}$ and $\nu'_{p,\phi'_{p}}$ of $\vec{R}_{p},$
equation (\ref{eq:Points-on-Plate-R}), still needs to be related
to the scale parameter $\zeta_{i}.$ Substituting $\zeta_{i}$ for
$\zeta$ in equation (\ref{eq:Plate-Hemi-exiting-K-direction-set3}),
it is rewritten as \begin{align*}
\vec{R}_{p} & =\zeta_{i}\vec{R}_{h,N_{h,max}+1}+\left[1-\zeta_{i}\right]\vec{R}_{h,N_{h,max}}\end{align*}
 or using equation (\ref{eq:Points-on-Hemisphere-R}) to explicitly
substitute for $\vec{R}_{h,N_{h,max}+1}$ and $\vec{R}_{h,N_{h,max}}$
for $N=N_{h,max}+1,$ $N=N_{h,max},$ respectively; and, regrouping
the terms \begin{align}
\vec{R}_{p} & =\sum_{i=1}^{3}\left(\zeta_{i}\left\{ \sum_{j=1}^{3}\left[\nu_{T,h,j}+r'_{i}\Lambda'_{h,N_{h,max}+1,j}\right]^{2}\right\} ^{1/2}\grave{\Lambda}_{h,N_{h,max}+1,i}+\left[1-\zeta_{i}\right]\right.\nonumber \\
 & \left.\times\left\{ \sum_{j=1}^{3}\left[\nu_{T,h,j}+r'_{i}\Lambda'_{h,N_{h,max},j}\right]^{2}\right\} ^{1/2}\grave{\Lambda}_{h,N_{h,max},i}\right)\hat{e_{i}},\label{eq:Plate-Hemi-exiting-K-direction-Rp-in-Zetai}\end{align}
 where $\zeta_{1}=\zeta_{2}=\zeta_{3}=\zeta.$ The subscript $i$
of $r'_{i}$ is not a summation index. Equating the above result for
$\vec{R}_{p}$ with that of equation (\ref{eq:Points-on-Plate-R-reduced}),
we arrive at \begin{align*}
\sum_{i=1}^{3}\left(\left\{ \sum_{j=1}^{3}\left[\nu_{T,p,j}+\frac{\nu'_{p,\phi'_{p}}}{\sin\theta'_{p}}\left(\frac{\partial\Lambda'_{p,j}}{\partial\phi'_{p}}-\frac{\sum_{x=1}^{3}\frac{\partial\Lambda'_{p,x}}{\partial\phi'_{p}}\left[\Lambda'_{p,x}+\left\Vert \vec{n'}_{p,1}\right\Vert ^{-1}\epsilon_{xyz}k'_{1,y}r'_{0,z}\right]}{\sum_{l=1}^{3}\frac{\partial\Lambda'_{p,l}}{\partial\theta'_{p}}\left[\Lambda'_{p,l}+\left\Vert \vec{n'}_{p,1}\right\Vert ^{-1}\epsilon_{lmn}k'_{1,m}r'_{0,n}\right]}\frac{\partial\Lambda'_{p,j}}{\partial\theta'_{p}}\right)\right]^{2}\right\} ^{1/2}\right.\\
\times\grave{\Lambda}_{p,i}-\zeta_{i}\left\{ \sum_{j=1}^{3}\left[\nu_{T,h,j}+r'_{i}\Lambda'_{h,N_{h,max}+1,j}\right]^{2}\right\} ^{1/2}\grave{\Lambda}_{h,N_{h,max}+1,i}+\left[\zeta_{i}-1\right]\grave{\Lambda}_{h,N_{h,max},i}\\
\left.\times\left\{ \sum_{j=1}^{3}\left[\nu_{T,h,j}+r'_{i}\Lambda'_{h,N_{h,max},j}\right]^{2}\right\} ^{1/2}\right)\hat{e_{i}} & =0,\end{align*}
 and its component equations are \begin{align}
\left\{ \sum_{j=1}^{3}\left[\nu_{T,p,j}+\frac{\nu'_{p,\phi'_{p}}}{\sin\theta'_{p}}\left(\frac{\partial\Lambda'_{p,j}}{\partial\phi'_{p}}-\frac{\sum_{x=1}^{3}\frac{\partial\Lambda'_{p,x}}{\partial\phi'_{p}}\left[\Lambda'_{p,x}+\left\Vert \vec{n'}_{p,1}\right\Vert ^{-1}\epsilon_{xyz}k'_{1,y}r'_{0,z}\right]}{\sum_{l=1}^{3}\frac{\partial\Lambda'_{p,l}}{\partial\theta'_{p}}\left[\Lambda'_{p,l}+\left\Vert \vec{n'}_{p,1}\right\Vert ^{-1}\epsilon_{lmn}k'_{1,m}r'_{0,n}\right]}\frac{\partial\Lambda'_{p,j}}{\partial\theta'_{p}}\right)\right]^{2}\right\} ^{1/2}\nonumber \\
\times\grave{\Lambda}_{p,i}-\zeta_{i}\left\{ \sum_{j=1}^{3}\left[\nu_{T,h,j}+r'_{i}\Lambda'_{h,N_{h,max}+1,j}\right]^{2}\right\} ^{1/2}\grave{\Lambda}_{h,N_{h,max}+1,i}+\left[\zeta_{i}-1\right]\grave{\Lambda}_{h,N_{h,max},i}\nonumber \\
\times\left\{ \sum_{j=1}^{3}\left[\nu_{T,h,j}+r'_{i}\Lambda'_{h,N_{h,max},j}\right]^{2}\right\} ^{1/2} & =0,\label{eq:Plate-Hemi-exiting-K-directioni-Pre-ABC}\end{align}
where $i=1,2,3.$ Introducing the following definitions for convenience,
\begin{equation}
\left\{ \begin{array}{c}
\begin{array}{ccc}
A_{\zeta}=\left\{ \sum_{j=1}^{3}\left[\nu_{T,h,j}+r'_{i}\Lambda'_{h,N_{h,max}+1,j}\right]^{2}\right\} ^{1/2}, &  & B_{\zeta}=\left\{ \sum_{j=1}^{3}\left[\nu_{T,h,j}+r'_{i}\Lambda'_{h,N_{h,max},j}\right]^{2}\right\} ^{1/2},\end{array}\\
\\C_{\zeta}=-\left(\sum_{x=1}^{3}\frac{\partial\Lambda'_{p,x}}{\partial\phi'_{p}}\left[\Lambda'_{p,x}+\left\Vert \vec{n'}_{p,1}\right\Vert ^{-1}\epsilon_{xyz}k'_{1,y}r'_{0,z}\right]\right)\\
\times\left(\sum_{l=1}^{3}\frac{\partial\Lambda'_{p,l}}{\partial\theta'_{p}}\left[\Lambda'_{p,l}+\left\Vert \vec{n'}_{p,1}\right\Vert ^{-1}\epsilon_{lmn}k'_{1,m}r'_{0,n}\right]\right)^{-1},\end{array}\right.\label{eq:Plate-Hemi-exiting-K-scale-ABC-in-Zetai-Def}\end{equation}
 the relation shown in equation (\ref{eq:Plate-Hemi-exiting-K-directioni-Pre-ABC})
is rewritten as \begin{align*}
\left\{ \sum_{j=1}^{3}\left[\nu_{T,p,j}+\frac{\nu'_{p,\phi'_{p}}}{\sin\theta'_{p}}\left(\frac{\partial\Lambda'_{p,j}}{\partial\phi'_{p}}-C_{\zeta}\frac{\partial\Lambda'_{p,j}}{\partial\theta'_{p}}\right)\right]^{2}\right\} ^{1/2}\grave{\Lambda}_{p,i}-\zeta_{i}A_{\zeta}\grave{\Lambda}_{h,N_{h,max}+1,i}\\
+\left[\zeta_{i}-1\right]B_{\zeta}\grave{\Lambda}_{h,N_{h,max},i} & =0,\end{align*}
 where $i=1,2,3.$ There are three such relations, one for each value
of $i.$ It is convenient to combine additively all three relations
to form \begin{align*}
\left\{ \sum_{j=1}^{3}\left[\nu_{T,p,j}+\frac{\nu'_{p,\phi'_{p}}}{\sin\theta'_{p}}\left(\frac{\partial\Lambda'_{p,j}}{\partial\phi'_{p}}-C_{\zeta}\frac{\partial\Lambda'_{p,j}}{\partial\theta'_{p}}\right)\right]^{2}\right\} ^{1/2}\sum_{i=1}^{3}\grave{\Lambda}_{p,i}-\zeta_{i}A_{\zeta}\sum_{i=1}^{3}\grave{\Lambda}_{h,N_{h,max}+1,i}\\
+\left[\zeta_{i}-1\right]B_{\zeta}\sum_{i=1}^{3}\grave{\Lambda}_{h,N_{h,max},i} & =0.\end{align*}
 After regrouping the terms and squaring both sides, it becomes \begin{align*}
\underbrace{\sum_{j=1}^{3}\left[\nu_{T,p,j}+\frac{\nu'_{p,\phi'_{p}}}{\sin\theta'_{p}}\left(\frac{\partial\Lambda'_{p,j}}{\partial\phi'_{p}}-C_{\zeta}\frac{\partial\Lambda'_{p,j}}{\partial\theta'_{p}}\right)\right]^{2}}_{L_{\sum}} & =\left[\frac{\zeta_{i}A_{\zeta}\sum_{i=1}^{3}\grave{\Lambda}_{h,N_{h,max}+1,i}-\left[\zeta_{i}-1\right]B_{\zeta}\sum_{i=1}^{3}\grave{\Lambda}_{h,N_{h,max},i}}{\sum_{l=1}^{3}\grave{\Lambda}_{p,l}}\right]^{2}.\end{align*}
 The summation labeled $L_{\sum}$ is rewritten as \begin{align*}
L_{\sum} & =\left[\nu'_{p,\phi'_{p}}\right]^{2}\sum_{j=1}^{3}\frac{1}{\sin^{2}\theta'_{p}}\left[\frac{\partial\Lambda'_{p,j}}{\partial\phi'_{p}}-C_{\zeta}\frac{\partial\Lambda'_{p,j}}{\partial\theta'_{p}}\right]^{2}+\nu'_{p,\phi'_{p}}\sum_{j=1}^{3}\left[\frac{2\nu_{T,p,j}}{\sin\theta'_{p}}\left(\frac{\partial\Lambda'_{p,j}}{\partial\phi'_{p}}-C_{\zeta}\frac{\partial\Lambda'_{p,j}}{\partial\theta'_{p}}\right)+\nu_{T,p,j}^{2}\right].\end{align*}
 The above equation is simplified into a quadratic equation of $\nu'_{p,\phi'_{p}},$
\begin{align*}
\left[\nu'_{p,\phi'_{p}}\right]^{2}\sum_{j=1}^{3}\frac{1}{\sin^{2}\theta'_{p}}\left[\frac{\partial\Lambda'_{p,j}}{\partial\phi'_{p}}-C_{\zeta}\frac{\partial\Lambda'_{p,j}}{\partial\theta'_{p}}\right]^{2}+\nu'_{p,\phi'_{p}}\sum_{j=1}^{3}\left[\frac{2\nu_{T,p,j}}{\sin\theta'_{p}}\left(\frac{\partial\Lambda'_{p,j}}{\partial\phi'_{p}}-C_{\zeta}\frac{\partial\Lambda'_{p,j}}{\partial\theta'_{p}}\right)+\nu_{T,p,j}^{2}\right]\\
-\left[\frac{\zeta_{i}A_{\zeta}\sum_{i=1}^{3}\grave{\Lambda}_{h,N_{h,max}+1,i}-\left[\zeta_{i}-1\right]B_{\zeta}\sum_{i=1}^{3}\grave{\Lambda}_{h,N_{h,max},i}}{\sum_{l=1}^{3}\grave{\Lambda}_{p,l}}\right]^{2} & =0.\end{align*}
 The two roots $\nu'_{p,\phi'_{p}}$ are given by \begin{align}
\nu'_{p,\phi'_{p}} & =\left(-\sum_{j=1}^{3}\left[\frac{\nu_{T,p,j}}{\sin\theta'_{p}}\left(\frac{\partial\Lambda'_{p,j}}{\partial\phi'_{p}}-C_{\zeta}\frac{\partial\Lambda'_{p,j}}{\partial\theta'_{p}}\right)+\frac{1}{2}\nu_{T,p,j}^{2}\right]\pm\left\{ \frac{1}{4}\left(\sum_{j=1}^{3}\left[\frac{2\nu_{T,p,j}}{\sin\theta'_{p}}\left(\frac{\partial\Lambda'_{p,j}}{\partial\phi'_{p}}-C_{\zeta}\frac{\partial\Lambda'_{p,j}}{\partial\theta'_{p}}\right)+\nu_{T,p,j}^{2}\right]\right)^{2}\right.\right.\nonumber \\
 & \left.\left.+\sum_{j=1}^{3}\frac{1}{\sin^{2}\theta'_{p}}\left[\frac{\partial\Lambda'_{p,j}}{\partial\phi'_{p}}-C_{\zeta}\frac{\partial\Lambda'_{p,j}}{\partial\theta'_{p}}\right]^{2}\left[\frac{\zeta_{i}A_{\zeta}\sum_{i=1}^{3}\grave{\Lambda}_{h,N_{h,max}+1,i}-\left[\zeta_{i}-1\right]B_{\zeta}\sum_{i=1}^{3}\grave{\Lambda}_{h,N_{h,max},i}}{\sum_{l=1}^{3}\grave{\Lambda}_{p,l}}\right]^{2}\right\} ^{1/2}\right)\nonumber \\
 & \times\left(\sum_{j=1}^{3}\frac{1}{\sin^{2}\theta'_{p}}\left[\frac{\partial\Lambda'_{p,j}}{\partial\phi'_{p}}-C_{\zeta}\frac{\partial\Lambda'_{p,j}}{\partial\theta'_{p}}\right]^{2}\right)^{-1},\label{eq:Scalar-F-Plate-Pla-Vp-phi}\end{align}
 where $A_{\zeta},$ $B_{\zeta}$ and $C_{\zeta}$ are defined in
equation (\ref{eq:Plate-Hemi-exiting-K-scale-ABC-in-Zetai-Def}).
It is understood that one does not mix summation indices of $A_{\zeta},$
$B_{\zeta}$ and $C_{\zeta}$ with those already present above. The
result for $\nu'_{p,\phi'_{p}}$ is still incomplete because the factor
$\gamma$ in $\zeta_{i}$ needs to be fixed by normalization. Unfortunately,
the translation property of the plate, $\nu_{T,p,j},$ makes it difficult
to extract $\zeta_{i}$ out of the radical. Besides the stated difficulty
regarding $\zeta_{i},$ $\nu'_{p,\phi'_{p}}$ is still ambiguous in
deciding which of the two roots correspond to the actual reflection
point on the plate. Fortunately, for the plate-hemisphere system of
Figure \ref{cap:plate-and-hemisphere-complex}, the choice of system
origin is arbitrary. One can always choose the plate origin to be
the system origin and the translation of the plate can be equivalently
simulated by a translation of the hemisphere origin in the opposite
direction. Then, in the rest frame of the plate, the translational
motion of the plate is zero, i.e., $\nu_{T,p,j}=0.$ In this frame,
$\nu'_{p,\phi'_{p}}$ takes on much simplified form \begin{align*}
\nu'_{p,\phi'_{p}} & =\pm\sin\theta'_{p}\frac{\zeta_{i}A_{\zeta}\sum_{i=1}^{3}\grave{\Lambda}_{h,N_{h,max}+1,i}-\left[\zeta_{i}-1\right]B_{\zeta}\sum_{i=1}^{3}\grave{\Lambda}_{h,N_{h,max},i}}{\left\{ \sum_{j=1}^{3}\left[\frac{\partial\Lambda'_{p,j}}{\partial\phi'_{p}}-C_{\zeta}\frac{\partial\Lambda'_{p,j}}{\partial\theta'_{p}}\right]^{2}\right\} ^{1/2}\sum_{l=1}^{3}\grave{\Lambda}_{p,l}}.\end{align*}
 For the sign ambiguity in $\nu'_{p,\phi'_{p}},$ it can be quickly
fixed by noting that for $\nu_{T,p,j}=0,$ equation (\ref{eq:Plate-Hemi-exiting-K-directioni-Pre-ABC})
yields \begin{eqnarray}
\nu'_{p,\phi'_{p}}=\frac{\zeta_{i}\sum_{i=1}^{3}\left[A_{\zeta}\grave{\Lambda}_{h,N_{h,max}+1,i}-B_{\zeta}\grave{\Lambda}_{h,N_{h,max},i}\right]+B_{\zeta}\sum_{i=1}^{3}\grave{\Lambda}_{h,N_{h,max},i}}{\left[\sin\theta'_{p}\right]^{-1}\left\{ \sum_{j=1}^{3}\left[\frac{\partial\Lambda'_{p,j}}{\partial\phi'_{p}}-C_{\zeta}\frac{\partial\Lambda'_{p,j}}{\partial\theta'_{p}}\right]^{2}\right\} ^{1/2}\sum_{l=1}^{3}\grave{\Lambda}_{p,l}}, &  & \nu_{T,p,j}=0,\label{eq:Scalar-F-Plate-Pla-Vp-phi-No-Translation-pre}\end{eqnarray}
 where $A_{\zeta},$ $B_{\zeta}$ and $C_{\zeta}$ are defined in
equation (\ref{eq:Plate-Hemi-exiting-K-scale-ABC-in-Zetai-Def}) with
$\nu_{T,p,j}=0.$ It is to be noticed that for a situation where $\nu_{T,p,j}=0,$
$\grave{\Lambda}$ becomes identical to $\Lambda'$ in form. One can
obtain $\grave{\Lambda}$ simply by replacing the primed variables
with the unprimed ones in $\Lambda'.$ For convenience, $\zeta_{i}$
of equation (\ref{eq:Plate-Hemi-exiting-K-scale}) is rewritten as
\begin{align*}
\zeta_{i} & =C_{\gamma}^{-1}A_{\gamma}+\gamma C_{\gamma}^{-1}B_{\gamma},\end{align*}
where \begin{equation}
\left\{ \begin{array}{c}
\begin{array}{ccc}
A_{\gamma}=\epsilon_{ijk}R_{h,N_{h,max},j}\left[R_{h,N_{h,max}+1,k}-R_{h,N_{h,max},k}\right], &  & B_{\gamma}=\left\Vert \vec{n'}_{p,1}\right\Vert ^{-1}\epsilon_{ijk}k'_{1,j}r'_{0,k},\end{array}\\
\\C_{\gamma}=\epsilon_{ijk}\left[R_{h,N_{h,max},j}-R_{h,N_{h,max}+1,j}\right]\left[R_{h,N_{h,max}+1,k}-R_{h,N_{h,max},k}\right].\end{array}\right.\label{eq:Plate-Hemi-exiting-K-scale-ABC-in-GAMMA-Def}\end{equation}
 Furthermore introducing the definitions, \begin{equation}
\left\{ \begin{array}{c}
\begin{array}{ccc}
A_{\beta}=\sum_{i=1}^{3}\left[A_{\zeta}\grave{\Lambda}_{h,N_{h,max}+1,i}-B_{\zeta}\grave{\Lambda}_{h,N_{h,max},i}\right], &  & B_{\beta}=\sum_{i=1}^{3}\grave{\Lambda}_{h,N_{h,max},i},\end{array}\\
\\C_{\beta}=\left\{ \sum_{j=1}^{3}\left[\frac{\partial\Lambda'_{p,j}}{\partial\phi'_{p}}-C_{\zeta}\frac{\partial\Lambda'_{p,j}}{\partial\theta'_{p}}\right]^{2}\right\} ^{1/2}\sum_{l=1}^{3}\grave{\Lambda}_{p,l},\end{array}\right.\label{eq:Plate-Hemi-exiting-K-scale-ABC-in-BETA-Def}\end{equation}
 the $\nu'_{p,\phi'_{p}}$ of equation (\ref{eq:Scalar-F-Plate-Pla-Vp-phi-No-Translation-pre})
is rewritten as \begin{eqnarray*}
\nu'_{p,\phi'_{p}}=\left[C_{\beta}^{-1}C_{\gamma}^{-1}A_{\gamma}A_{\beta}+\gamma C_{\beta}^{-1}C_{\gamma}^{-1}B_{\gamma}A_{\beta}+C_{\beta}^{-1}B_{\zeta}B_{\beta}\right]\sin\theta'_{p}, &  & \nu_{T,p,j}=0.\end{eqnarray*}
 Substituting the last expression for $\nu'_{p,\phi'_{p}}$ into equation
(\ref{eq:Points-on-Plate-R-reduced}), we arrive at \begin{align}
\vec{R}_{p} & =\left[C_{\beta}^{-1}C_{\gamma}^{-1}A_{\gamma}A_{\beta}+\gamma C_{\beta}^{-1}C_{\gamma}^{-1}B_{\gamma}A_{\beta}+C_{\beta}^{-1}B_{\zeta}B_{\beta}\right]\left\{ \sum_{i=1}^{3}\left[\frac{\partial\Lambda'_{p,i}}{\partial\phi'_{p}}-C_{\zeta}\frac{\partial\Lambda'_{p,i}}{\partial\theta'_{p}}\right]^{2}\right\} ^{1/2}\sum_{i=1}^{3}\grave{\Lambda}_{p,i}\hat{e_{i}},\label{eq:Points-on-Plate-R-pre-final}\end{align}
 where $\nu_{T,p,i}=0.$ The vector cross product $\vec{R}_{p}\times\vec{k}_{N_{h,max}+1}$
is given by \begin{align*}
\vec{R}_{p}\times\vec{k}_{N_{h,max}+1} & =\sum_{i=1}^{3}\epsilon_{ijk}R_{p,j}k_{N_{h,max}+1,k}\hat{e_{i}}\end{align*}
 or \begin{align*}
\vec{R}_{p}\times\vec{k}_{N_{h,max}+1} & =\left[C_{\beta}^{-1}C_{\gamma}^{-1}A_{\gamma}A_{\beta}+\gamma C_{\beta}^{-1}C_{\gamma}^{-1}B_{\gamma}A_{\beta}+C_{\beta}^{-1}B_{\zeta}B_{\beta}\right]\left\{ \sum_{j=1}^{3}\left[\frac{\partial\Lambda'_{p,j}}{\partial\phi'_{p}}-C_{\zeta}\frac{\partial\Lambda'_{p,j}}{\partial\theta'_{p}}\right]^{2}\right\} ^{1/2}\\
 & \times\sum_{i=1}^{3}\epsilon_{ijk}\grave{\Lambda}_{p,j}k_{N_{h,max}+1,k}\hat{e_{i}}.\end{align*}
 Finally, substituting above vector cross product $\vec{R}_{p}\times\vec{k}_{N_{h,max}+1}$
into equation (\ref{eq:Plate-Hemi-exiting-K-direction-set4}) and
regrouping the terms, it becomes \begin{align*}
\sum_{i=1}^{3}\left(\left[C_{\beta}^{-1}C_{\gamma}^{-1}A_{\gamma}A_{\beta}+\gamma C_{\beta}^{-1}C_{\gamma}^{-1}B_{\gamma}A_{\beta}+C_{\beta}^{-1}B_{\zeta}B_{\beta}\right]\left\{ \sum_{j=1}^{3}\left[\frac{\partial\Lambda'_{p,j}}{\partial\phi'_{p}}-C_{\zeta}\frac{\partial\Lambda'_{p,j}}{\partial\theta'_{p}}\right]^{2}\right\} ^{1/2}\right.\\
\left.\times\epsilon_{ijk}\grave{\Lambda}_{p,j}k_{N_{h,max}+1,k}-\gamma\left\Vert \vec{n'}_{p,1}\right\Vert ^{-1}\epsilon_{ijk}k'_{1,j}r'_{0,k}\right)\hat{e_{i}} & =0,\end{align*}
 where $\hat{n'}_{p,1}=\left\Vert \vec{n'}_{p,1}\right\Vert ^{-1}\sum_{i=1}^{3}\epsilon_{ijk}k'_{1,j}r'_{0,k}\hat{e_{i}}$
have been used. And for the component equations \begin{align*}
\left[C_{\beta}^{-1}C_{\gamma}^{-1}A_{\gamma}A_{\beta}+\gamma C_{\beta}^{-1}C_{\gamma}^{-1}B_{\gamma}A_{\beta}+C_{\beta}^{-1}B_{\zeta}B_{\beta}\right]\left\{ \sum_{j=1}^{3}\left[\frac{\partial\Lambda'_{p,j}}{\partial\phi'_{p}}-C_{\zeta}\frac{\partial\Lambda'_{p,j}}{\partial\theta'_{p}}\right]^{2}\right\} ^{1/2}\\
\times\epsilon_{ijk}\grave{\Lambda}_{p,j}k_{N_{h,max}+1,k}-\gamma\left\Vert \vec{n'}_{p,1}\right\Vert ^{-1}\epsilon_{ijk}k'_{1,j}r'_{0,k} & =0,\end{align*}
 where $i=1,2,3.$ There are three such relations and they are additively
combined to yield \begin{align*}
\left[C_{\beta}^{-1}C_{\gamma}^{-1}A_{\gamma}A_{\beta}+\gamma C_{\beta}^{-1}C_{\gamma}^{-1}B_{\gamma}A_{\beta}+C_{\beta}^{-1}B_{\zeta}B_{\beta}\right]\left\{ \sum_{j=1}^{3}\left[\frac{\partial\Lambda'_{p,j}}{\partial\phi'_{p}}-C_{\zeta}\frac{\partial\Lambda'_{p,j}}{\partial\theta'_{p}}\right]^{2}\right\} ^{1/2}\\
\times\sum_{i=1}^{3}\epsilon_{ijk}\grave{\Lambda}_{p,j}k_{N_{h,max}+1,k}-\gamma\left\Vert \vec{n'}_{p,1}\right\Vert ^{-1}\sum_{i=1}^{3}\epsilon_{ijk}k'_{1,j}r'_{0,k} & =0.\end{align*}
 Finally, $\gamma$ is solved to give the result \begin{align}
\gamma\equiv\gamma_{o} & =\left(\left\Vert \vec{n'}_{p,1}\right\Vert ^{-1}\sum_{i=1}^{3}\epsilon_{ijk}k'_{1,j}r'_{0,k}-C_{\beta}^{-1}C_{\gamma}^{-1}B_{\gamma}A_{\beta}\left\{ \sum_{j=1}^{3}\left[\frac{\partial\Lambda'_{p,j}}{\partial\phi'_{p}}-C_{\zeta}\frac{\partial\Lambda'_{p,j}}{\partial\theta'_{p}}\right]^{2}\right\} ^{1/2}\right.\nonumber \\
 & \left.\times\sum_{i=1}^{3}\epsilon_{ijk}\grave{\Lambda}_{p,j}k_{N_{h,max}+1,k}\right)^{-1}\left(\left[C_{\beta}^{-1}C_{\gamma}^{-1}A_{\gamma}A_{\beta}+C_{\beta}^{-1}B_{\zeta}B_{\beta}\right]\right.\nonumber \\
 & \left.\times\left\{ \sum_{j=1}^{3}\left[\frac{\partial\Lambda'_{p,j}}{\partial\phi'_{p}}-C_{\zeta}\frac{\partial\Lambda'_{p,j}}{\partial\theta'_{p}}\right]^{2}\right\} ^{1/2}\sum_{i=1}^{3}\epsilon_{ijk}\grave{\Lambda}_{p,j}k_{N_{h,max}+1,k}\right).\label{eq:Plate-Hemi-exiting-K-scale-GAMMA}\end{align}
 The parameter $\nu'_{p,\phi'_{p}}$ is now completely defined, \begin{eqnarray}
\nu'_{p,\phi'_{p}}=\left[C_{\beta}^{-1}C_{\gamma}^{-1}A_{\gamma}A_{\beta}+\gamma_{o}C_{\beta}^{-1}C_{\gamma}^{-1}B_{\gamma}A_{\beta}+C_{\beta}^{-1}B_{\zeta}B_{\beta}\right]\sin\theta'_{p}, &  & \nu_{T,p,j}=0,\label{eq:Scalar-F-Plate-Pla-Vp-phi-No-Translation}\end{eqnarray}
 where $\left(A_{\zeta},B_{\zeta},C_{\zeta}\right),$ $\left(A_{\gamma},B_{\gamma},C_{\gamma}\right),$
$\left(A_{\beta},B_{\beta},C_{\beta}\right)$ and $\gamma_{o}$ are
given by equations (\ref{eq:Plate-Hemi-exiting-K-scale-ABC-in-Zetai-Def}),
(\ref{eq:Plate-Hemi-exiting-K-scale-ABC-in-GAMMA-Def}), (\ref{eq:Plate-Hemi-exiting-K-scale-ABC-in-BETA-Def})
and (\ref{eq:Plate-Hemi-exiting-K-scale-GAMMA}), respectively. With
$\nu'_{p,\phi'_{p}}$ defined in equation (\ref{eq:Scalar-F-Plate-Pla-Vp-phi-No-Translation}),
the reflection point on the plate is obtained from equation (\ref{eq:Points-on-Plate-R-reduced}),
\begin{align}
\vec{R}_{p} & =\left\{ \sum_{s=1}^{3}\left[\frac{\partial\Lambda'_{p,s}}{\partial\phi'_{p}}-\frac{\sum_{i=1}^{3}\frac{\partial\Lambda'_{p,i}}{\partial\phi'_{p}}\left[\Lambda'_{p,i}+\left\Vert \vec{n'}_{p,1}\right\Vert ^{-1}\epsilon_{ijk}k'_{1,j}r'_{0,k}\right]}{\sum_{l=1}^{3}\frac{\partial\Lambda'_{p,l}}{\partial\theta'_{p}}\left[\Lambda'_{p,l}+\left\Vert \vec{n'}_{p,1}\right\Vert ^{-1}\epsilon_{lmn}k'_{1,m}r'_{0,n}\right]}\frac{\partial\Lambda'_{p,s}}{\partial\theta'_{p}}\right]^{2}\right\} ^{1/2}\nonumber \\
 & \times\left[C_{\beta}^{-1}C_{\gamma}^{-1}A_{\gamma}A_{\beta}+\gamma_{o}C_{\beta}^{-1}C_{\gamma}^{-1}B_{\gamma}A_{\beta}+C_{\beta}^{-1}B_{\zeta}B_{\beta}\right]\sum_{i=1}^{3}\grave{\Lambda}_{p,i}\hat{e_{i}},\label{eq:Points-on-Plate-R-Final}\end{align}
 where $\nu_{T,p,j}=0.$ It should be noticed that for a situation
where $\nu_{T,p,j}=0,$ $\grave{\Lambda}$ becomes identical to $\Lambda'$
in form, and $\grave{\Lambda}$ can be obtained simply by replacing
the primed variables with the unprimed ones. 

To see if the wave reflected from the plate at location $\vec{R}_{p}$
re-enters the hemisphere cavity or escape to infinity, we consider
the reflected wave $\vec{k}_{r,N_{h,max}+1},$ \begin{align}
\vec{k}_{r,N_{h,max}+1} & =\alpha_{r,\perp}\left[\hat{n'}_{p}\times\vec{k}_{N_{h,max}+1}\right]\times\hat{n'}_{p}-\alpha_{r,\parallel}\hat{n'}_{p}\cdot\vec{k}_{N_{h,max}+1}\hat{n'}_{p},\label{eq:Reflected-Wave-Off-Plate-Vector-Form}\end{align}
 where the relation found in equation (\ref{eq:K-reflected-in-PARA-n-PERPE-N})
have been used. As always, it is convenient to express vectors in
component forms. Making the changes in variables $\left(l,m,n\right)\rightarrow\left(i,j,k\right),$
$\left(m,q,r\right)\rightarrow\left(j,l,m\right),$ $\hat{n'}\rightarrow\hat{n'}_{p}$
and $\vec{k'}_{i}\rightarrow\vec{k}_{N_{h,max}+1},$ the component
result of equation (\ref{eq:K-reflected-in-PARA-n-PERPE-N-COMPONENT})
is used to get \begin{align}
\vec{k}_{r,N_{h,max}+1} & =\sum_{i=1}^{3}\sum_{k=1}^{3}\left\{ \alpha_{r,\perp}\left[k_{N_{h,max}+1,i}n'_{p,k}n'_{p,k}-n'_{p,i}k_{N_{h,max}+1,k}n'_{p,k}\right]\right.\nonumber \\
 & \left.-\alpha_{r,\parallel}n'_{p,k}k_{N_{h,max}+1,k}n'_{p,i}\right\} \hat{e}_{i},\label{eq:Reflected-Wave-Off-Plate}\end{align}
 where $n'_{p,i}$ and $n'_{p,k}$ are coefficients of the normalized
$\hat{n'}_{p}.$ All wave vectors entering the hemisphere cavity satisfy
the relation \begin{eqnarray}
\vec{R}_{p}+\sum_{i=1}^{3}\left[\vec{\xi}_{\kappa}\cdot\hat{e_{i}}\right]\left[\vec{k}_{r,N_{h,max}+1}\cdot\hat{e_{i}}\right]\hat{e_{i}}-\vec{R}_{0}=0, &  & \vec{R}_{0}=\sum_{i=1}^{3}\left[\nu_{T,h,i}+r'_{0,i}\right]\hat{e_{i}},\label{eq:Plate-Hemi-K-Reenter-Criteria-Pre}\end{eqnarray}
 where $\vec{\xi}_{\kappa}$ is a real-valued positive scale vector
and $\vec{R}_{0}$ is the points on the opening face of hemisphere.
The scale vector $\vec{\xi}_{\kappa}$ has the form \begin{align*}
\vec{\xi}_{\kappa} & =\sum_{i=1}^{3}\xi_{\kappa,i}\hat{e_{i}}.\end{align*}
 With the scale vector $\vec{\xi}_{\kappa}$ defined above; and, $\vec{R}_{p}$
and $\vec{k}_{r,N_{h,max}+1}$ defined in equations (\ref{eq:Points-on-Plate-R-Final})
and (\ref{eq:Reflected-Wave-Off-Plate}), respectively, equation (\ref{eq:Plate-Hemi-K-Reenter-Criteria-Pre})
is rewritten in component form \begin{align*}
\sum_{i=1}^{3}\left(\left\{ \sum_{s=1}^{3}\left[\frac{\partial\Lambda'_{p,s}}{\partial\phi'_{p}}-\frac{\sum_{i=1}^{3}\frac{\partial\Lambda'_{p,i}}{\partial\phi'_{p}}\left[\Lambda'_{p,i}+\left\Vert \vec{n'}_{p,1}\right\Vert ^{-1}\epsilon_{ijk}k'_{1,j}r'_{0,k}\right]}{\sum_{l=1}^{3}\frac{\partial\Lambda'_{p,l}}{\partial\theta'_{p}}\left[\Lambda'_{p,l}+\left\Vert \vec{n'}_{p,1}\right\Vert ^{-1}\epsilon_{lmn}k'_{1,m}r'_{0,n}\right]}\frac{\partial\Lambda'_{p,s}}{\partial\theta'_{p}}\right]^{2}\right\} ^{1/2}\right.\\
\times\left[C_{\beta}^{-1}C_{\gamma}^{-1}A_{\gamma}A_{\beta}+\gamma_{o}C_{\beta}^{-1}C_{\gamma}^{-1}B_{\gamma}A_{\beta}+C_{\beta}^{-1}B_{\zeta}B_{\beta}\right]\grave{\Lambda}_{p,i}+\xi_{\kappa,i}\sum_{k=1}^{3}\left\{ \alpha_{r,\perp}\left[k_{N_{h,max}+1,i}\right.\right.\\
\left.\left.\left.\times n'_{p,k}n'_{p,k}-n'_{p,i}k_{N_{h,max}+1,k}n'_{p,k}\right]-\alpha_{r,\parallel}n'_{p,k}k_{N_{h,max}+1,k}n'_{p,i}\right\} -\nu_{T,h,i}-r'_{0,i}\right)\hat{e_{i}} & =0,\end{align*}
 which yields the component equations, \begin{align*}
\left\{ \sum_{s=1}^{3}\left[\frac{\partial\Lambda'_{p,s}}{\partial\phi'_{p}}-\frac{\sum_{i=1}^{3}\frac{\partial\Lambda'_{p,i}}{\partial\phi'_{p}}\left[\Lambda'_{p,i}+\left\Vert \vec{n'}_{p,1}\right\Vert ^{-1}\epsilon_{ijk}k'_{1,j}r'_{0,k}\right]}{\sum_{l=1}^{3}\frac{\partial\Lambda'_{p,l}}{\partial\theta'_{p}}\left[\Lambda'_{p,l}+\left\Vert \vec{n'}_{p,1}\right\Vert ^{-1}\epsilon_{lmn}k'_{1,m}r'_{0,n}\right]}\frac{\partial\Lambda'_{p,s}}{\partial\theta'_{p}}\right]^{2}\right\} ^{1/2}\left[C_{\beta}^{-1}C_{\gamma}^{-1}A_{\gamma}A_{\beta}\right.\\
\left.+\gamma_{o}C_{\beta}^{-1}C_{\gamma}^{-1}B_{\gamma}A_{\beta}+C_{\beta}^{-1}B_{\zeta}B_{\beta}\right]\grave{\Lambda}_{p,i}+\xi_{\kappa,i}\sum_{k=1}^{3}\left\{ \alpha_{r,\perp}\left[k_{N_{h,max}+1,i}n'_{p,k}n'_{p,k}\right.\right.\\
\left.\left.-n'_{p,i}k_{N_{h,max}+1,k}n'_{p,k}\right]-\alpha_{r,\parallel}n'_{p,k}k_{N_{h,max}+1,k}n'_{p,i}\right\} -\nu_{T,h,i}-r'_{0,i} & =0,\end{align*}
 where $i=1,2,3.$ Finally, $\xi_{\kappa,i}$ is solved as \begin{align}
\xi_{\kappa,i} & =\left(\nu_{T,h,i}+r'_{0,i}-\left\{ \sum_{s=1}^{3}\left[\frac{\partial\Lambda'_{p,s}}{\partial\phi'_{p}}-\frac{\sum_{i=1}^{3}\frac{\partial\Lambda'_{p,i}}{\partial\phi'_{p}}\left[\Lambda'_{p,i}+\left\Vert \vec{n'}_{p,1}\right\Vert ^{-1}\epsilon_{ijk}k'_{1,j}r'_{0,k}\right]}{\sum_{l=1}^{3}\frac{\partial\Lambda'_{p,l}}{\partial\theta'_{p}}\left[\Lambda'_{p,l}+\left\Vert \vec{n'}_{p,1}\right\Vert ^{-1}\epsilon_{lmn}k'_{1,m}r'_{0,n}\right]}\frac{\partial\Lambda'_{p,s}}{\partial\theta'_{p}}\right]^{2}\right\} ^{1/2}\right.\nonumber \\
 & \left.\times\left[C_{\beta}^{-1}C_{\gamma}^{-1}A_{\gamma}A_{\beta}+\gamma_{o}C_{\beta}^{-1}C_{\gamma}^{-1}B_{\gamma}A_{\beta}+C_{\beta}^{-1}B_{\zeta}B_{\beta}\right]\grave{\Lambda}_{p,i}\right)\left(\sum_{k=1}^{3}\left\{ \alpha_{r,\perp}\left[k_{N_{h,max}+1,i}n'_{p,k}n'_{p,k}\right.\right.\right.\nonumber \\
 & \left.\left.\left.-n'_{p,i}k_{N_{h,max}+1,k}n'_{p,k}\right]-\alpha_{r,\parallel}n'_{p,k}k_{N_{h,max}+1,k}n'_{p,i}\right\} \right)^{-1},\label{eq:Plate-Hemi-K-Reenter-Criteria}\end{align}
 where $i=1,2,3.$ 

The above result can be applied in setting the re-entry criteria.
Notice that $\vec{R}_{0}\leq r'_{i},$ which implies $r'_{0,i}\leq r'_{i},$
where $r'_{i}$ is the radius of the hemisphere. It can be concluded
then that all waves re-entering hemisphere cavity would satisfy the
condition $\xi_{\kappa,1}=\xi_{\kappa,2}=\xi_{\kappa,3}.$ On the
other hand, those waves that escapes to infinity cannot have all three
$\xi_{\kappa,i}$ equaling to a same constant. The re-entry condition
$\xi_{\kappa,1}=\xi_{\kappa,2}=\xi_{\kappa,3}$ is just another way
of stating the existence of parametric line along the vector $\vec{k}_{r,N_{h,max}+1}$
that happens to pierce through a hemisphere opening. When such a line
does not exist, the initial wave vector direction has to be rotated
accordingly to a new direction, such that in its rotated direction
there is a parametric line that pierces through the hemisphere opening;
it leads to the condition that all three $\xi_{\kappa,i}$ cannot
equal to a same constant. The re-entry criteria are now summarized
for bookkeeping purpose, \begin{equation}
\left\{ \begin{array}{c}
\xi_{\kappa,1}=\xi_{\kappa,2}=\xi_{\kappa,3}:\quad wave\; reenters\; hemisphere,\\
\\ELSE:\quad wave\; escapes\; to\; infinity,\end{array}\right.\label{eq:Plate-Hemi-K-Reenter-Criteria-Interpretation}\end{equation}
 where $ELSE$ is the case where $\xi_{\kappa,1}=\xi_{\kappa,2}=\xi_{\kappa,3}$
cannot be satisfied.

\section{Dynamical Casimir Force}

The original derivations and developments pertaining to the dynamical
Casimir force are included in this appendix. It is referenced by the
text of this paper to supply all the fine details. 
\vspace{0.25in}

\subsection{Formalism of Zero-Point Energy and its Force}

For massless fields, the energy-momentum relation is given by \begin{align}
\mathcal{H}'_{n_{s}}\equiv E_{Total} & =pc,\label{eq:Hamilton-equal-pc}\end{align}
 where $p$ is the momentum and $c$ the speed of light. The field
propagating in an arbitrary direction has a momentum  $\vec{p'}=\sum_{i=1}^{3}p'_{i}\hat{e_{i}}.$
The associated field energy-momentum relation is hence \begin{align*}
\mathcal{H}'_{n_{s}}-c\left\{ \sum_{i=1}^{3}\left[p'_{i}\right]^{2}\right\} ^{1/2} & =0.\end{align*}
 The differentiation of the above equation gives \begin{align}
d\left[\mathcal{H}'_{n_{s}}-c\left\{ \sum_{i=1}^{3}\left[p'_{i}\right]^{2}\right\} ^{1/2}\right]\equiv d\mathcal{H}'_{n_{s}}-cd\left\{ \sum_{i=1}^{3}\left[p'_{i}\right]^{2}\right\} ^{1/2} & =0.\label{eq:d-Hamilton-minus-d-momentum-mag}\end{align}
 The total differential energy $d\mathcal{H}'_{n_{s}}$ is \begin{align}
d\mathcal{H}'_{n_{s}} & =\sum_{i=1}^{3}\frac{\partial\mathcal{H}'_{n_{s}}}{\partial k'_{i}}\frac{\partial k'_{i}}{\partial p'_{i}}dp'_{i}=\sum_{i=1}^{3}\left(\left[n_{s}+\frac{1}{2}\right]\hbar\right)^{-1}\frac{\partial\mathcal{H}'_{n_{s}}}{\partial k'_{i}}dp'_{i},\label{eq:d-Hamilton-total-differential}\end{align}
 where the relation $p'_{i}=\left[n_{s}+\frac{1}{2}\right]\hbar k'_{i}$
has been used. The total differential momentum is \begin{align*}
d\left\{ \sum_{i=1}^{3}\left[p'_{i}\right]^{2}\right\} ^{1/2} & =\left\{ \sum_{i=1}^{3}\left[p'_{i}\right]^{2}\right\} ^{-1/2}\sum_{i=1}^{3}p'_{i}dp'_{i}.\end{align*}
 The combined result is \begin{align}
\sum_{i=1}^{3}\left[\left(\left[n_{s}+\frac{1}{2}\right]\hbar\right)^{-1}\frac{\partial\mathcal{H}'_{n_{s}}}{\partial k'_{i}}-\left\{ \sum_{i=1}^{3}\left[p'_{i}\right]^{2}\right\} ^{-1/2}cp'_{i}\right]dp'_{i} & =0.\label{eq:d-Hamilton-minus-d-momentum-mag-combined}\end{align}
 Because all the momentum differentials are linearly independent,
their coefficients are zero, \begin{eqnarray*}
\left(\left[n_{s}+\frac{1}{2}\right]\hbar\right)^{-1}\frac{\partial\mathcal{H}'_{n_{s}}}{\partial k'_{i}}-\left\{ \sum_{i=1}^{3}\left[p'_{i}\right]^{2}\right\} ^{-1/2}cp'_{i}=0, &  & i=1,2,3.\end{eqnarray*}
 There are three such equations. Then, additively combining the three
relations, and rearranging the terms, we have \begin{align*}
\left\{ \sum_{i=1}^{3}\left[p'_{i}\right]^{2}\right\} ^{1/2}\sum_{i=1}^{3}\left(\left[n_{s}+\frac{1}{2}\right]\hbar\right)^{-1}\frac{\partial\mathcal{H}'_{n_{s}}}{\partial k'_{i}} & =\sum_{i=1}^{3}cp'_{i}.\end{align*}
 Squaring both sides to get rid of the radical leads to \begin{align}
\left[\sum_{i=1}^{3}\left(\left[n_{s}+\frac{1}{2}\right]\hbar\right)^{-1}\frac{\partial\mathcal{H}'_{n_{s}}}{\partial k'_{i}}\right]^{2}\sum_{i=1}^{3}\left[p'_{i}\right]^{2} & =c^{2}\left[\sum_{i=1}^{3}p'_{i}\right]^{2}.\label{eq:dynamical-force-previous}\end{align}
 The summations $\sum_{i=1}^{3}\left[p'_{i}\right]^{2}$ and $\left[\sum_{i=1}^{3}p'_{i}\right]^{2}$
are rewritten as \begin{align*}
\sum_{i=1}^{3}\left[p'_{i}\right]^{2} & =\left[p'_{\alpha}\right]^{2}+\sum_{i=1}^{3}\left(1-\delta_{i\alpha}\right)\left[p'_{i}\right]^{2}=\left[p'_{\alpha}\right]^{2}+\sum_{i=1}^{3}\left(1-\delta_{i\alpha}\right)\left(\left[n_{s}+\frac{1}{2}\right]\hbar\right)^{2}\left[k'_{i}\right]^{2},\end{align*}
 \begin{align*}
\left[\sum_{i=1}^{3}p'_{i}\right]^{2} & =\left[p'_{\alpha}+\sum_{i=1}^{3}\left(1-\delta_{i\alpha}\right)p'_{i}\right]^{2}=\left[p'_{\alpha}\right]^{2}+2\sum_{i=1}^{3}\left(1-\delta_{i\alpha}\right)p'_{i}p'_{\alpha}+\left[\sum_{i=1}^{3}\left(1-\delta_{i\alpha}\right)p'_{i}\right]^{2}\\
 & =\left[p'_{\alpha}\right]^{2}+2\sum_{i=1}^{3}\left(1-\delta_{i\alpha}\right)\left[n_{s}+\frac{1}{2}\right]\hbar k'_{i}p'_{\alpha}+\left(\left[n_{s}+\frac{1}{2}\right]\hbar\right)^{2}\left[\sum_{i=1}^{3}\left(1-\delta_{i\alpha}\right)k'_{i}\right]^{2},\end{align*}
 where $p'_{i}$ has  been replaced by $\left[n_{s}+\frac{1}{2}\right]\hbar k'_{i}.$
Substituting the result into equation (\ref{eq:dynamical-force-previous})
and rearranging the terms in powers of $p'_{\alpha},$ we have \begin{align*}
\left(\left[\sum_{i=1}^{3}\frac{\partial\mathcal{H}'_{n_{s}}}{\partial k'_{i}}\right]^{2}-\left(\left[n_{s}+\frac{1}{2}\right]\hbar c\right)^{2}\right)\left[p'_{\alpha}\right]^{2}-2\left[n_{s}+\frac{1}{2}\right]\hbar\sum_{i=1}^{3}\left(1-\delta_{i\alpha}\right)\left(\left[n_{s}+\frac{1}{2}\right]\hbar c\right)^{2}k'_{i}p'_{\alpha}\\
-\left[\sum_{i=1}^{3}\left(1-\delta_{i\alpha}\right)\left(\left[n_{s}+\frac{1}{2}\right]\hbar c\right)^{2}k'_{i}\right]^{2}+\left[\sum_{i=1}^{3}\frac{\partial\mathcal{H}'_{n_{s}}}{\partial k'_{i}}\right]^{2}\sum_{i=1}^{3}\left(1-\delta_{i\alpha}\right)\left(\left[n_{s}+\frac{1}{2}\right]\hbar\right)^{2}\left[k'_{i}\right]^{2} & =0.\end{align*}
 Defining the following quantities, \begin{equation}
\left\{ \begin{array}{c}
\begin{array}{ccc}
C_{\alpha,1}=\sum_{i=1}^{3}\frac{\partial\mathcal{H}'_{n_{s}}}{\partial k'_{i}}, &  & C_{\alpha,2}=\sum_{i=1}^{3}\left(1-\delta_{i\alpha}\right)\left(\left[n_{s}+\frac{1}{2}\right]\hbar c\right)^{2}k'_{i},\end{array}\\
\\C_{\alpha,3}=\sum_{i=1}^{3}\left(1-\delta_{i\alpha}\right)\left(\left[n_{s}+\frac{1}{2}\right]\hbar\right)^{2}\left[k'_{i}\right]^{2},\end{array}\right.\label{eq:dynamical-force-C1-C2-C3-DEF}\end{equation}
 the above quadratic equation is rewritten as \begin{align*}
\left[C_{\alpha,1}^{2}-\left(\left[n_{s}+\frac{1}{2}\right]\hbar c\right)^{2}\right]\left[p'_{\alpha}\right]^{2}-2\left[n_{s}+\frac{1}{2}\right]\hbar C_{\alpha,2}p'_{\alpha}-C_{\alpha,2}^{2}+C_{\alpha,1}^{2}C_{\alpha,3} & =0.\end{align*}
 Finally, the root $p'_{\alpha}$ is found to be \begin{align}
p'_{\alpha} & =\frac{\left[n_{s}+\frac{1}{2}\right]\hbar C_{\alpha,2}}{C_{\alpha,1}^{2}-\left(\left[n_{s}+\frac{1}{2}\right]\hbar c\right)^{2}}+\left\{ \frac{\left[n_{s}+\frac{1}{2}\right]^{2}\hbar^{2}C_{\alpha,2}^{2}}{\left[C_{\alpha,1}^{2}-\left(\left[n_{s}+\frac{1}{2}\right]\hbar c\right)^{2}\right]^{2}}+\frac{C_{\alpha,2}^{2}-C_{\alpha,1}^{2}C_{\alpha,3}}{C_{\alpha,1}^{2}-\left(\left[n_{s}+\frac{1}{2}\right]\hbar c\right)^{2}}\right\} ^{1/2},\label{eq:dynamical-momentum-P-alpha}\end{align}
 where the positive root have been chosen since $p'_{\alpha}$ is
the magnitude of the $\alpha$th component of the total momentum $\vec{p'},$
therefore it is a positive scalar, $p'_{\alpha}\geq0.$ 

By definition, the force is equal to the rate of change of momentum,
\begin{align*}
\vec{\mathcal{F}'} & =\frac{d}{dt}\vec{p'}=\frac{d}{dt}\sum_{\alpha=1}^{3}p'_{\alpha}\hat{e_{\alpha}}=\sum_{\alpha=1}^{3}\left[\frac{dp'_{\alpha}}{dt}\hat{e_{\alpha}}+p'_{\alpha}\frac{d\hat{e_{\alpha}}}{dt}\right]=\sum_{\alpha=1}^{3}\frac{dp'_{\alpha}}{dt}\hat{e_{\alpha}}=\sum_{\alpha=1}^{3}\vec{\mathcal{F}'}_{\alpha}.\end{align*}
 The explicit expression for $\vec{\mathcal{F}'}_{\alpha}$ is found
to be \begin{align}
\vec{\mathcal{F}'}_{\alpha} & =\left\{ \left(\frac{C_{\alpha,1}C_{\alpha,4}\left[C_{\alpha,1}^{2}C_{\alpha,3}-C_{\alpha,2}^{2}\right]}{\left[C_{\alpha,1}^{2}-\left(\left[n_{s}+\frac{1}{2}\right]\hbar c\right)^{2}\right]^{2}}-\frac{2\left[n_{s}+\frac{1}{2}\right]^{2}\hbar^{2}C_{\alpha,1}C_{\alpha,2}^{2}C_{\alpha,4}}{\left[C_{\alpha,1}^{2}-\left(\left[n_{s}+\frac{1}{2}\right]\hbar c\right)^{2}\right]^{3}}-\frac{C_{\alpha,1}C_{\alpha,3}C_{\alpha,4}}{C_{\alpha,1}^{2}-\left(\left[n_{s}+\frac{1}{2}\right]\hbar c\right)^{2}}\right.\right.\nonumber \\
 & \left.-\frac{2\left[n_{s}+\frac{1}{2}\right]\hbar C_{\alpha,1}C_{\alpha,2}}{\left[C_{\alpha,1}^{2}-\left(\left[n_{s}+\frac{1}{2}\right]\hbar c\right)^{2}\right]^{2}}\right)\frac{dC_{\alpha,1}}{dt}+\left(\frac{\left[n_{s}+\frac{1}{2}\right]^{2}\hbar^{2}C_{\alpha,2}C_{\alpha,4}}{\left[C_{\alpha,1}^{2}-\left(\left[n_{s}+\frac{1}{2}\right]\hbar c\right)^{2}\right]^{2}}+\frac{C_{\alpha,2}C_{\alpha,4}}{C_{\alpha,1}^{2}-\left(\left[n_{s}+\frac{1}{2}\right]\hbar c\right)^{2}}\right.\nonumber \\
 & \left.\left.+\frac{\left[n_{s}+\frac{1}{2}\right]\hbar}{C_{\alpha,1}^{2}-\left(\left[n_{s}+\frac{1}{2}\right]\hbar c\right)^{2}}\right)\frac{dC_{\alpha,2}}{dt}-\frac{\frac{1}{2}C_{\alpha,1}^{2}C_{\alpha,4}}{C_{\alpha,1}^{2}-\left(\left[n_{s}+\frac{1}{2}\right]\hbar c\right)^{2}}\frac{dC_{\alpha,3}}{dt}\right\} \hat{e_{\alpha}},\label{eq:dynamical-force-general}\end{align}
 where \begin{align}
C_{\alpha,4} & =\left(\frac{\left[n_{s}+\frac{1}{2}\right]^{2}\hbar^{2}C_{\alpha,2}^{2}}{\left[C_{\alpha,1}^{2}-\left(\left[n_{s}+\frac{1}{2}\right]\hbar c\right)^{2}\right]^{2}}+\frac{C_{\alpha,2}^{2}-C_{\alpha,1}^{2}C_{\alpha,3}}{C_{\alpha,1}^{2}-\left(\left[n_{s}+\frac{1}{2}\right]\hbar c\right)^{2}}\right)^{-1/2}.\label{eq:dynamical-force-C4-DEF}\end{align}
 Before computing the three time derivatives $dC_{\alpha,1}/dt,$
$dC_{\alpha,2}/dt$ and $dC_{\alpha,3}/dt,$ we notice that $k'_{i}\left(n_{i}\right)=n_{i}f_{i}\left(L_{i}\right).$
Hence, the derivative $dk'_{i}/dt$ can be written as \begin{align}
\frac{dk'_{i}}{dt} & =\frac{\partial k'_{i}}{\partial n_{i}}\frac{dn_{i}}{dt}f_{i}\left(L_{i}\right)+n_{i}\frac{\partial f_{i}}{\partial L_{i}}\frac{dL_{i}}{dt}=f_{i}\left(L_{i}\right)\frac{\partial k'_{i}}{\partial n_{i}}\dot{n}_{i}+n_{i}\frac{\partial f_{i}}{\partial L_{i}}\dot{L}_{i}.\label{eq:dynamical-force-dk-dt}\end{align}
 The three derivatives $dC_{\alpha,1}/dt,$ $dC_{\alpha,2}/dt$ and
$dC_{\alpha,3}/dt$ are given by \begin{align}
\frac{dC_{\alpha,1}}{dt} & =\sum_{i=1}^{3}\sum_{j=1}^{3}\frac{\partial^{2}\mathcal{H}'_{n_{s}}}{\partial k'_{j}\partial k'_{i}}\frac{dk'_{j}}{dt}=\sum_{i=1}^{3}\sum_{j=1}^{3}\frac{\partial^{2}\mathcal{H}'_{n_{s}}}{\partial k'_{j}\partial k'_{i}}\left[f_{j}\left(L_{j}\right)\frac{\partial k'_{j}}{\partial n_{j}}\dot{n}_{j}+n_{j}\frac{\partial f_{j}}{\partial L_{j}}\dot{L}_{j}\right]\nonumber \\
 & =\sum_{i=1}^{3}\frac{\partial^{2}\mathcal{H}'_{n_{s}}}{\partial\left[k'_{i}\right]^{2}}\left[f_{i}\left(L_{i}\right)\frac{\partial k'_{i}}{\partial n_{i}}\dot{n}_{i}+n_{i}\frac{\partial f_{i}}{\partial L_{i}}\dot{L}_{i}\right]+\sum_{i=1}^{3}\sum_{j=1}^{3}\left\{ \left(1-\delta_{ij}\right)\frac{\partial^{2}\mathcal{H}'_{n_{s}}}{\partial k'_{j}\partial k'_{i}}\right.\nonumber \\
 & \left.\times\left[f_{j}\left(L_{j}\right)\frac{\partial k'_{j}}{\partial n_{j}}\dot{n}_{j}+n_{j}\frac{\partial f_{j}}{\partial L_{j}}\dot{L}_{j}\right]\right\} ,\label{eq:eq:dynamical-force-dC1-dt}\end{align}
 \begin{align}
\frac{dC_{\alpha,2}}{dt} & =\sum_{i=1}^{3}\left(1-\delta_{i\alpha}\right)\left[n_{s}+\frac{1}{2}\right]\left[f_{i}\left(L_{i}\right)\frac{\partial k'_{i}}{\partial n_{i}}\dot{n}_{i}+n_{i}\frac{\partial f_{i}}{\partial L_{i}}\dot{L}_{i}\right],\label{eq:eq:dynamical-force-dC2-dt}\end{align}
 \begin{align}
\frac{dC_{\alpha,3}}{dt} & =2\sum_{i=1}^{3}\left(1-\delta_{i\alpha}\right)\left[n_{s}+\frac{1}{2}\right]^{2}k'_{i}\left[f_{i}\left(L_{i}\right)\frac{\partial k'_{i}}{\partial n_{i}}\dot{n}_{i}+n_{i}\frac{\partial f_{i}}{\partial L_{i}}\dot{L}_{i}\right],\label{eq:eq:dynamical-force-dC3-dt}\end{align}
 where $C_{\alpha,1},$ $C_{\alpha,2}$ and $C_{\alpha,3}$ are defined
in equation (\ref{eq:dynamical-force-C1-C2-C3-DEF}). It is noted
that the derivative $dk'_{i}/dt,$ and also each of $dC_{\alpha,1}/dt,$
$dC_{\alpha,2}/dt$ and $dC_{\alpha,3}/dt,$ consists of two contributing
parts, one is proportional to $\dot{n}_{i}$ and the other involves
$\dot{L}_{i}.$ The force expression in equation (\ref{eq:dynamical-force-general})
has then two contributing parts. The force contribution involving
$\dot{n}_{i}$ has a physical meaning that the boundaries are being
driven to generate the extra wave modes that would otherwise be missing
when such drivers were not present. The force contribution involving
$\dot{L}_{i}$ is the effect of feedbacks from the moving boundaries.
This feedback effect due to the moving boundaries tends to either
cool or heat the conducting boundaries. For an isolated, non-driven
conducting boundaries, the force contribution proportional to $\dot{n}_{i}$
vanishes. The expression of force is then rewritten as \begin{align*}
\vec{\mathcal{F}'}_{\alpha} & =\left\{ \left(\frac{C_{\alpha,1}C_{\alpha,4}\left[C_{\alpha,1}^{2}C_{\alpha,3}-C_{\alpha,2}^{2}\right]}{\left[C_{\alpha,1}^{2}-\left(\left[n_{s}+\frac{1}{2}\right]\hbar c\right)^{2}\right]^{2}}-\frac{2\left[n_{s}+\frac{1}{2}\right]^{2}\hbar^{2}C_{\alpha,1}C_{\alpha,2}^{2}C_{\alpha,4}}{\left[C_{\alpha,1}^{2}-\left(\left[n_{s}+\frac{1}{2}\right]\hbar c\right)^{2}\right]^{3}}-\frac{C_{\alpha,1}C_{\alpha,3}C_{\alpha,4}}{C_{\alpha,1}^{2}-\left(\left[n_{s}+\frac{1}{2}\right]\hbar c\right)^{2}}\right.\right.\\
 & \left.-\frac{2\left[n_{s}+\frac{1}{2}\right]\hbar C_{\alpha,1}C_{\alpha,2}}{\left[C_{\alpha,1}^{2}-\left(\left[n_{s}+\frac{1}{2}\right]\hbar c\right)^{2}\right]^{2}}\right)\left(\sum_{i=1}^{3}\frac{\partial^{2}\mathcal{H}'_{n_{s}}}{\partial\left[k'_{i}\right]^{2}}n_{i}\frac{\partial f_{i}}{\partial L_{i}}\dot{L}_{i}+\sum_{i=1}^{3}\sum_{j=1}^{3}\left(1-\delta_{ij}\right)\frac{\partial^{2}\mathcal{H}'_{n_{s}}}{\partial k'_{j}\partial k'_{i}}n_{j}\frac{\partial f_{j}}{\partial L_{j}}\dot{L}_{j}\right)\\
 & +\left(\frac{\left[n_{s}+\frac{1}{2}\right]^{2}\hbar^{2}C_{\alpha,2}C_{\alpha,4}}{\left[C_{\alpha,1}^{2}-\left(\left[n_{s}+\frac{1}{2}\right]\hbar c\right)^{2}\right]^{2}}+\frac{C_{\alpha,2}C_{\alpha,4}}{C_{\alpha,1}^{2}-\left(\left[n_{s}+\frac{1}{2}\right]\hbar c\right)^{2}}+\frac{\left[n_{s}+\frac{1}{2}\right]\hbar}{C_{\alpha,1}^{2}-\left(\left[n_{s}+\frac{1}{2}\right]\hbar c\right)^{2}}\right)\sum_{i=1}^{3}\left(1-\delta_{i\alpha}\right)\\
 & \left.\times\left[n_{s}+\frac{1}{2}\right]n_{i}\frac{\partial f_{i}}{\partial L_{i}}\dot{L}_{i}-\frac{\frac{1}{2}C_{\alpha,1}^{2}C_{\alpha,4}}{C_{\alpha,1}^{2}-\left(\left[n_{s}+\frac{1}{2}\right]\hbar c\right)^{2}}2\sum_{i=1}^{3}\left(1-\delta_{i\alpha}\right)\left[n_{s}+\frac{1}{2}\right]^{2}k'_{i}n_{i}\frac{\partial f_{i}}{\partial L_{i}}\dot{L}_{i}\right\} \hat{e_{\alpha}}.\end{align*}
 It can be simplified with the following definitions, \begin{align}
C_{\alpha,5} & =\frac{C_{\alpha,1}C_{\alpha,4}\left[C_{\alpha,1}^{2}C_{\alpha,3}-C_{\alpha,2}^{2}\right]}{\left[C_{\alpha,1}^{2}-\left(\left[n_{s}+\frac{1}{2}\right]\hbar c\right)^{2}\right]^{2}}-\frac{2\left[n_{s}+\frac{1}{2}\right]^{2}\hbar^{2}C_{\alpha,1}C_{\alpha,2}^{2}C_{\alpha,4}}{\left[C_{\alpha,1}^{2}-\left(\left[n_{s}+\frac{1}{2}\right]\hbar c\right)^{2}\right]^{3}}\nonumber \\
 & -\frac{C_{\alpha,1}C_{\alpha,3}C_{\alpha,4}}{C_{\alpha,1}^{2}-\left(\left[n_{s}+\frac{1}{2}\right]\hbar c\right)^{2}}-\frac{2\left[n_{s}+\frac{1}{2}\right]\hbar C_{\alpha,1}C_{\alpha,2}}{\left[C_{\alpha,1}^{2}-\left(\left[n_{s}+\frac{1}{2}\right]\hbar c\right)^{2}\right]^{2}},\label{eq:dynamical-force-C5-DEF}\end{align}
 \begin{align}
C_{\alpha,6} & =\frac{\left[n_{s}+\frac{1}{2}\right]^{2}\hbar^{2}C_{\alpha,2}C_{\alpha,4}}{\left[C_{\alpha,1}^{2}-\left(\left[n_{s}+\frac{1}{2}\right]\hbar c\right)^{2}\right]^{2}}+\frac{C_{\alpha,2}C_{\alpha,4}}{C_{\alpha,1}^{2}-\left(\left[n_{s}+\frac{1}{2}\right]\hbar c\right)^{2}}+\frac{\left[n_{s}+\frac{1}{2}\right]\hbar}{C_{\alpha,1}^{2}-\left(\left[n_{s}+\frac{1}{2}\right]\hbar c\right)^{2}},\label{eq:dynamical-force-C6-DEF}\end{align}
 \begin{align}
C_{\alpha,7} & =\frac{C_{\alpha,1}^{2}C_{\alpha,4}}{C_{\alpha,1}^{2}-\left(\left[n_{s}+\frac{1}{2}\right]\hbar c\right)^{2}}.\label{eq:dynamical-force-C7-DEF}\end{align}
 The dynamical force can then be rewritten as \begin{align}
\vec{\mathcal{F}'}_{\alpha} & =\sum_{i=1}^{3}\left\{ n_{i}\frac{\partial f_{i}}{\partial L_{i}}\left[C_{\alpha,5}\frac{\partial^{2}\mathcal{H}'_{n_{s}}}{\partial\left[k'_{i}\right]^{2}}+\left(1-\delta_{i\alpha}\right)\left(C_{\alpha,6}-C_{\alpha,7}\left[n_{s}+\frac{1}{2}\right]k'_{i}\right)\left[n_{s}+\frac{1}{2}\right]\right]\dot{L}_{i}\right.\nonumber \\
 & \left.+\sum_{j=1}^{3}\left(1-\delta_{ij}\right)C_{\alpha,5}n_{j}\frac{\partial f_{j}}{\partial L_{j}}\frac{\partial^{2}\mathcal{H}'_{n_{s}}}{\partial k'_{j}\partial k'_{i}}\dot{L}_{j}\right\} \hat{e_{\alpha}},\label{eq:dynamical-force-L-dot-ONLY-3D}\end{align}
 where $C_{\alpha,5},$ $C_{\alpha,6}$ and $C_{\alpha,7}$ are defined
in equations (\ref{eq:dynamical-force-C5-DEF}), (\ref{eq:dynamical-force-C6-DEF})
and (\ref{eq:dynamical-force-C7-DEF}). The force equation (\ref{eq:dynamical-force-L-dot-ONLY-3D})
vanishes for the \textbf{1D} case, which is an expected result. The
reason is explained as follow: Recall that equation (\ref{eq:d-Hamilton-minus-d-momentum-mag-combined})
reads \begin{align*}
\sum_{i=1}^{3}\left[\left(\left[n_{s}+\frac{1}{2}\right]\hbar\right)^{-1}\frac{\partial\mathcal{H}'_{n_{s}}}{\partial k'_{i}}-\left\{ \sum_{i=1}^{3}\left[p'_{i}\right]^{2}\right\} ^{-1/2}cp'_{i}\right]dp'_{i} & =0.\end{align*}
 For the \textbf{1D} case, the summation runs only once and the above
expression simplifies to \begin{eqnarray*}
\left[\left(\left[n_{s}+\frac{1}{2}\right]\hbar\right)^{-1}\frac{\partial\mathcal{H}'_{n_{s}}}{\partial k'_{i}}-c\right]dp'_{i}=0 & \rightarrow & \left(\left[n_{s}+\frac{1}{2}\right]\hbar\right)^{-1}\frac{\partial\mathcal{H}'_{n_{s}}}{\partial k'_{i}}-c=0.\end{eqnarray*}
 This is a classic situation where the problem has been over specified.
For the \textbf{3D} case, equation (\ref{eq:d-Hamilton-minus-d-momentum-mag-combined})
is really a combination of two constraints, $\sum_{i=1}^{3}\left[p'_{i}\right]^{2}$
and $\mathcal{H}'_{n_{s}}.$ For the \textbf{1D} case, there is only
one constraint, $\mathcal{H}'_{n_{s}}.$ Hence, equation (\ref{eq:d-Hamilton-minus-d-momentum-mag-combined})
becomes an over specification. In order to avoid the problem caused
by over specifications in this formulation, the one dimensional force
expression can be obtained directly by differentiating equation (\ref{eq:Hamilton-equal-pc})
instead of using the above formulation for the three dimensional case.
We have then for the force expression in \textbf{1D} case: \textbf{}\begin{eqnarray*}
p'=\frac{1}{c}\mathcal{H}'_{n_{s}} & \rightarrow & \frac{dp'}{dt}=\frac{1}{c}\frac{\partial\mathcal{H}'_{n_{s}}}{\partial k'}\frac{dk'}{dt}=\frac{1}{c}\frac{\partial\mathcal{H}'_{n_{s}}}{\partial k'}\left[f\left(L\right)\frac{\partial k'}{\partial n}\dot{n}+n\frac{\partial f}{\partial L}\dot{L}\right].\end{eqnarray*}
 For an isolated, non-driven systems, \begin{align}
\vec{\mathcal{F}'} & =\frac{n}{c}\frac{\partial f}{\partial L}\frac{\partial\mathcal{H}'_{n_{s}}}{\partial k'}\dot{L}\hat{e},\label{eq:dynamical-force-L-dot-ONLY-1D}\end{align}
 where $\vec{\mathcal{F}'}$ is the force expression in \textbf{1D}
space. Here the subscript $\alpha$ of $\vec{\mathcal{F}'}_{\alpha}$
have been dropped for simplicity, since it is a one dimensional force.

\subsection{Equations of Motion for the Driven Parallel Plates}

Consider the one dimensional system of two parallel plates shown in
Figure \ref{cap:driven-parallel-plates-force}. Defining the boundary
length $L_{i,\Re}$ as the magnitude of a vector $\hat{e_{i}}\left[\vec{L}_{\Re}\cdot\hat{e_{i}}\right],$
where $\Re$ denotes the region, the following relation is found from
Figure \ref{cap:driven-parallel-plates-force}, \begin{align}
\vec{L}_{\Re} & =\vec{R}_{rp,\Re}-\vec{R}_{lp,\Re}=\sum_{i=1}^{3}\left[\vec{R}_{rp,\Re}\cdot\hat{e_{i}}-\vec{R}_{lp,\Re}\cdot\hat{e_{i}}\right]\hat{e_{i}}.\label{eq:L-general-vec}\end{align}
 Hence, the velocity $d\vec{L}_{\Re}/dt$ is \begin{align}
\frac{d\vec{L}_{\Re}}{dt} & =\frac{d\vec{R}_{rp,\Re}}{dt}-\frac{d\vec{R}_{lp,\Re}}{dt}=\sum_{i=1}^{3}\left[\frac{d\vec{R}_{rp,\Re}}{dt}\cdot\hat{e_{i}}-\frac{d\vec{R}_{lp,\Re}}{dt}\cdot\hat{e_{i}}\right]\hat{e_{i}}\label{eq:dL-dt-general-vec}\end{align}
 and the corresponding component magnitude is given by \begin{align}
\dot{L}_{i,\Re}\equiv\frac{d\vec{L}_{\Re}}{dt}\cdot\hat{e_{i}} & =\frac{d\vec{R}_{rp,\Re}}{dt}\cdot\hat{e_{i}}-\frac{d\vec{R}_{lp,\Re}}{dt}\cdot\hat{e_{i}}.\label{eq:dL-dt-general-mag}\end{align}
 Substituting the result $\dot{L}_{i,\Re}$ of equation (\ref{eq:dL-dt-general-mag})
for $\dot{L}_{\alpha}$ in the one dimensional dynamical force expression
of equation (\ref{eq:dynamical-force-L-dot-ONLY-1D}), \begin{align}
\vec{\mathcal{F}'}_{\alpha,\Re} & =\frac{n_{\alpha,\Re}}{c}\frac{\partial f_{\alpha,\Re}}{\partial L_{\alpha,\Re}}\frac{\partial\mathcal{H}'_{n_{s},\Re}}{\partial k'_{\alpha,\Re}}\left[\frac{d\vec{R}_{rp,\Re}}{dt}\cdot\hat{e_{\alpha}}-\frac{d\vec{R}_{lp,\Re}}{dt}\cdot\hat{e_{\alpha}}\right]\hat{e_{\alpha}},\label{eq:Field-F-Region-arbi}\end{align}
 where $\dot{L}_{i,\Re}\equiv\dot{L}_{\alpha}$ and $i\equiv\alpha.$
The subscript $\Re$ denotes the corresponding quantities associated
with the region $\Re=1,2,3,$ e.g., $\mathcal{H}'_{n_{s},\Re}$ denotes
the field energy in region $\Re.$ For simplicity, the following notational
convention is adopted \begin{eqnarray*}
\dot{R}_{a,b}=\frac{d\vec{R}_{a}}{dt}\cdot\hat{e_{b}}, &  & \ddot{R}_{a,b}=\frac{d^{2}\vec{R}_{a}}{dt^{2}}\cdot\hat{e_{b}},\end{eqnarray*}
 \begin{align}
g_{\alpha,\Re} & =\frac{n_{\alpha,\Re}}{c}\left(\frac{\partial f_{\alpha,\Re}}{\partial L_{\alpha,\Re}}\right)\left(\frac{\partial\mathcal{H}'_{n_{s},\Re}}{\partial k'_{\alpha,\Re}}\right).\label{eq:g-i-R-DEF}\end{align}
 The force expression of equation (\ref{eq:Field-F-Region-arbi})
is then rewritten as \begin{align}
\vec{\mathcal{F}'}_{\alpha,\Re} & =g_{\alpha,\Re}\left[\dot{R}_{rp,\Re,\alpha}-\dot{R}_{lp,\Re,\alpha}\right]\hat{e_{\alpha}},\label{eq:Field-F-Region-arbi-condensed-nota}\end{align}
 Before writing down equations of motion for each plates illustrated
in Figure \ref{cap:driven-parallel-plates-force}, the associated
center of mass point relative the the surface point vectors $\vec{R}_{rp,\Re,\alpha}$
for each plates needs to be determined. The center of mass point $\vec{R}_{rp,cm}$for
plate labeled {}``right plate'' in Figure \ref{cap:driven-parallel-plates-force}
is related to the surface point vector $\vec{R}_{rp,\Re}$ through
a relation \begin{align*}
\vec{R}_{rp,cm}\left(t\right) & =\vec{R}_{rp,\Re=2}\left(t\right)+\vec{R}_{rp,cm-\Re}\left(t\right),\end{align*}
 where $\vec{R}_{rp,cm-\Re}\left(t\right)\equiv\vec{R}_{rp,cm-2}\left(t\right)$
is a displacement between surface and the center of mass point. The
$\alpha$th component of the center of mass point $\vec{R}_{rp,cm}$
is then \begin{align}
R_{rp,cm,\alpha}\left(t\right)\equiv\hat{e_{\alpha}}\cdot\vec{R}_{rp,cm}\left(t\right) & =\hat{e_{\alpha}}\cdot\vec{R}_{rp,2}\left(t\right)+\hat{e_{\alpha}}\cdot\vec{R}_{rp,cm-2}\left(t\right)=R_{rp,2,\alpha}\left(t\right)+R_{rp,cm-2,\alpha}\left(t\right).\label{eq:Center-of-Mass-Rela}\end{align}
 The component of the center of mass point speed is given by \begin{align}
\dot{R}_{rp,cm,\alpha}\left(t\right) & =\dot{R}_{rp,2,\alpha}\left(t\right)+\dot{R}_{rp,cm-2,\alpha}\left(t\right).\label{eq:Center-of-Mass-Rela-Velocity}\end{align}
 Similarly, for the plate labeled {}``left plate,'' the center of
mass point is related to the surface vector point $\vec{R}_{lp,\Re=2}\left(t\right)$
by \begin{align*}
\vec{R}_{lp,cm}\left(t\right) & =\vec{R}_{lp,\Re=2}\left(t\right)-\vec{R}_{lp,cm-\Re}\left(t\right),\end{align*}
 and the component along the direction $\hat{e_{\alpha}}$ is \begin{eqnarray}
R_{lp,cm,\alpha}\left(t\right)=R_{lp,2,\alpha}\left(t\right)-R_{lp,cm-2,\alpha}\left(t\right), &  & \dot{R}_{lp,cm,\alpha}\left(t\right)=\dot{R}_{lp,2,\alpha}\left(t\right)-\dot{R}_{lp,cm-2,\alpha}\left(t\right).\label{eq:Center-of-Mass-Rela-Combo-Left-Plate-Region2}\end{eqnarray}
 Using the above center of mass relations, equations (\ref{eq:Center-of-Mass-Rela}),
(\ref{eq:Center-of-Mass-Rela-Velocity}) and (\ref{eq:Center-of-Mass-Rela-Combo-Left-Plate-Region2}),
along with the force equation (\ref{eq:Field-F-Region-arbi-condensed-nota}),
the net force acting on a plate labeled {}``right plate'' along
the direction of $\hat{e_{\alpha}}$ in the configuration shown in
Figure \ref{cap:driven-parallel-plates-force} is \begin{align*}
m_{rp}\ddot{R}_{rp,cm,\alpha} & =\left[\vec{\mathcal{F}'}_{\alpha,\Re=2}+\vec{\mathcal{F}'}_{\alpha,\Re=3}\right]\cdot\hat{e_{\alpha}}\end{align*}
 or \begin{align}
m_{rp}\ddot{R}_{rp,cm,\alpha} & =g_{\alpha,2}\left[\dot{R}_{rp,cm,\alpha}-\dot{R}_{lp,cm,\alpha}-\dot{R}_{rp,cm-2,\alpha}-\dot{R}_{lp,cm-2,\alpha}\right]\nonumber \\
 & +g_{\alpha,3}\left[\dot{R}_{dpr,cm,\alpha}-\dot{R}_{rp,cm,\alpha}-\dot{R}_{dpr,cm-2,\alpha}-\dot{R}_{rp,cm-2,\alpha}\right]\label{eq:Eq-of-Mot-Right-Plate-Gen}\end{align}
 where $m_{rp}$ is the mass of the {}``right plate.'' If the plate
surface is not vibrating longitudinally along the direction of $\hat{e_{\alpha}},$
the displacements $R_{rp,cm-2,\alpha}$ and $R_{dpr,cm-2,\alpha}$
are constants; hence, $\dot{R}_{rp,cm-2,\alpha}=\dot{R}_{dpr,cm-2,\alpha}=0.$
For static surfaces, the above net force relation simplifies to \begin{align}
m_{rp}\ddot{R}_{rp,cm,\alpha} & =s_{rp,2}g_{\alpha,2}\left[\dot{R}_{rp,cm,\alpha}-\dot{R}_{lp,cm,\alpha}\right]+s_{rp,3}g_{\alpha,3}\left[\dot{R}_{dpr,cm,\alpha}-\dot{R}_{rp,cm,\alpha}\right],\label{eq:Eq-of-Mot-Right-Plate-Static-pre}\end{align}
 where $s_{rp,2}$ and $s_{rp,3}$ have been inserted for convenience
due to the force sign convention to be set later. Similarly, for the
plate labeled {}``left plate'' in Figure \ref{cap:driven-parallel-plates-force},
the net force relation along the direction of $\hat{e_{\alpha}}$
is \begin{align*}
m_{lp}\ddot{R}_{lp,cm,\alpha}\left(t\right) & =\left[\vec{\mathcal{F}'}_{\alpha,\beta,\Re=1}+\vec{\mathcal{F}'}_{\alpha,\beta,\Re=2}\right]\cdot\hat{e_{\alpha}}\end{align*}
 or, for the case where plate surfaces do not have longitudinal vibrations,
\begin{align}
m_{lp}\ddot{R}_{lp,cm,\alpha} & =s_{lp,1}g_{\alpha,1}\left[\dot{R}_{lp,cm,\alpha}-\dot{R}_{dpl,cm,\alpha}\right]+s_{lp,2}g_{\alpha,2}\left[\dot{R}_{rp,cm,\alpha}-\dot{R}_{lp,cm,\alpha}\right],\label{eq:Eq-of-Mot-Left-Plate-Static-pre}\end{align}
 where $m_{lp}$ is a mass of {}``left plate'' and the terms $s_{lp,1}$
and $s_{lp,2}$ have been inserted for convenience due to the force
sign convention to be set later. We have now the two coupled differential
equations, \begin{align*}
m_{rp}\ddot{R}_{rp,cm,\alpha}+s_{rp,3}g_{\alpha,3}\dot{R}_{rp,cm,\alpha}-s_{rp,2}g_{\alpha,2}\dot{R}_{rp,cm,\alpha}+s_{rp,2}g_{\alpha,2}\dot{R}_{lp,cm,\alpha} & =s_{rp,3}g_{\alpha,3}\dot{R}_{dpr,cm,\alpha},\end{align*}
 \begin{align*}
m_{lp}\ddot{R}_{lp,cm,\alpha}+s_{lp,2}g_{\alpha,2}\dot{R}_{lp,cm,\alpha}-s_{lp,1}g_{\alpha,1}\dot{R}_{lp,cm,\alpha}-s_{lp,2}g_{\alpha,2}\dot{R}_{rp,cm,\alpha} & =-s_{lp,1}g_{\alpha,1}\dot{R}_{dpl,cm,\alpha}.\end{align*}
 Introducing the following definitions, \begin{equation}
\left\{ \begin{array}{ccc}
\eta_{1}=m_{rp}^{-1}\left[s_{rp,2}g_{\alpha,2}-s_{rp,3}g_{\alpha,3}\right], &  & \eta_{2}=-s_{rp,2}g_{\alpha,2}m_{rp}^{-1}\\
\\\eta_{3}=m_{lp}^{-1}\left[s_{lp,1}g_{\alpha,1}-s_{lp,2}g_{\alpha,2}\right], &  & \eta_{4}=s_{lp,2}g_{\alpha,2}m_{lp}^{-1},\\
\\\xi_{rp}=s_{rp,3}g_{\alpha,3}m_{rp}^{-1}\dot{R}_{dpr,cm,\alpha}, &  & \xi_{lp}=-s_{lp,1}g_{\alpha,1}m_{lp}^{-1},\\
\\R_{1}=R_{rp,cm,\alpha}, &  & R_{2}=R_{lp,cm,\alpha},\end{array}\right.\label{eq:eta-1-2-3-4-ci-lp-rp-DEF}\end{equation}
 the coupled differential equations are rewritten as \begin{eqnarray}
\ddot{R}_{1}-\eta_{1}\dot{R}_{1}-\eta_{2}\dot{R}_{2}=\xi_{rp}, &  & \ddot{R}_{2}-\eta_{3}\dot{R}_{2}-\eta_{4}\dot{R}_{1}=\xi_{lp}.\label{eq:Eq-of-Mot-Plate-System-Static}\end{eqnarray}
 The equations of motion shown in equation (\ref{eq:Eq-of-Mot-Plate-System-Static})
are a system of two linear second-order inhomogeneous differential
equations. In order to rewrite the coupled linear inhomogeneous differential
equation (\ref{eq:Eq-of-Mot-Plate-System-Static}) into a set of first-order
linear inhomogeneous equation, a set of new variables are defined
first, \begin{equation}
\left\{ \begin{array}{c}
\begin{array}{ccc}
\dot{R}_{1}=R_{3}, &  & \dot{R}_{2}=R_{4},\end{array}\\
\\\dot{R}_{3}=\ddot{R}_{1}=\xi_{rp}+\eta_{1}\dot{R}_{1}+\eta_{2}\dot{R}_{2}=\xi_{rp}+\eta_{1}R_{3}+\eta_{2}R_{4},\\
\\\dot{R}_{4}=\ddot{R}_{2}=\xi_{lp}+\eta_{3}\dot{R}_{2}+\eta_{4}\dot{R}_{1}=\xi_{lp}+\eta_{3}R_{4}+\eta_{4}R_{3}.\end{array}\right.\label{eq:Eq-of-Mot-Plate-Static-Coeffi-new-variables}\end{equation}
 Using these new variables defined in equation (\ref{eq:Eq-of-Mot-Plate-Static-Coeffi-new-variables}),
equation (\ref{eq:Eq-of-Mot-Plate-System-Static}) can be cast into
first-order inhomogeneous equation in matrix form, \begin{eqnarray*}
\dot{\vec{R}}=\widetilde{M}\cdot\vec{R}+\vec{\xi} & \rightarrow & \left[\begin{array}{c}
\dot{R}_{1}\\
\dot{R}_{2}\\
\dot{R}_{3}\\
\dot{R}_{4}\end{array}\right]=\left[\begin{array}{cccc}
0 & 0 & 1 & 0\\
0 & 0 & 0 & 1\\
0 & 0 & \eta_{1} & \eta_{2}\\
0 & 0 & \eta_{4} & \eta_{3}\end{array}\right]\left[\begin{array}{c}
R_{1}\\
R_{2}\\
R_{3}\\
R_{4}\end{array}\right]+\left[\begin{array}{c}
0\\
0\\
\xi_{rp}\\
\xi_{lp}\end{array}\right].\end{eqnarray*}
 The above first-order inhomogeneous equation is equivalent to \begin{eqnarray}
R_{1}=\int_{t_{0}}^{t}R_{3}dt', &  & R_{2}=\int_{t_{0}}^{t}R_{4}dt',\label{eq:Eq-of-Mot-Plate-System-Static-Matrix-R1-R2}\end{eqnarray}
 and \begin{align}
\underbrace{\left[\begin{array}{c}
\dot{R}_{3}\\
\dot{R}_{4}\end{array}\right]}_{\dot{\vec{R}}_{\eta}} & =\underbrace{\left[\begin{array}{cc}
\eta_{1} & \eta_{2}\\
\eta_{4} & \eta_{3}\end{array}\right]}_{\widetilde{M}_{\eta}}\underbrace{\left[\begin{array}{c}
R_{3}\\
R_{4}\end{array}\right]}_{\vec{R}_{\eta}}+\underbrace{\left[\begin{array}{c}
\xi_{rp}\\
\xi_{lp}\end{array}\right]}_{\vec{\xi}}.\label{eq:Eq-of-Mot-Plate-System-Static-Matrix}\end{align}
 For the homogeneous system \begin{align}
\underbrace{\left[\begin{array}{c}
\dot{R}_{3}\\
\dot{R}_{4}\end{array}\right]}_{\dot{\vec{R}}_{\eta}} & =\underbrace{\left[\begin{array}{cc}
\eta_{1} & \eta_{2}\\
\eta_{4} & \eta_{3}\end{array}\right]}_{\widetilde{M}_{\eta}}\underbrace{\left[\begin{array}{c}
R_{3}\\
R_{4}\end{array}\right]}_{\vec{R}_{\eta}},\label{eq:Eq-of-Mot-Plate-System-Static-Matrix-Homo}\end{align}
 the eigenvalues are found from the root of the characteristic equation
\begin{align*}
\det\left(\lambda\widetilde{I}-\widetilde{M}_{\eta}\right)\equiv\lambda^{2}-\left[\eta_{1}+\eta_{3}\right]\lambda+\eta_{1}\eta_{3}-\eta_{2}\eta_{4} & =0.\end{align*}
 The two eigenvalues are \begin{eqnarray}
\lambda_{3}=\frac{\eta_{1}+\eta_{3}}{2}+\left\{ \frac{1}{4}\left[\eta_{1}-\eta_{3}\right]^{2}+\eta_{2}\eta_{4}\right\} ^{1/2}, &  & \lambda_{4}=\frac{\eta_{1}+\eta_{3}}{2}-\left\{ \frac{1}{4}\left[\eta_{1}-\eta_{3}\right]^{2}+\eta_{2}\eta_{4}\right\} ^{1/2}.\label{eq:Eq-of-Mot-Plate-System-Static-Matrix-Eigenvalues}\end{eqnarray}
 And, the two corresponding eigenvectors are found to be \begin{eqnarray}
\vec{R}_{\lambda_{3}}=\grave{R}_{4}\left[\begin{array}{c}
\frac{\eta_{2}}{\lambda_{3}-\eta_{1}}\\
1\end{array}\right], &  & \grave{R}_{4}=\left\{ \left[\frac{\eta_{2}}{\lambda_{3}-\eta_{1}}\right]^{2}+1\right\} ^{-1/2},\label{eq:eigenvector-Lambda3}\end{eqnarray}
 and \begin{eqnarray}
\vec{R}_{\lambda_{4}}=\grave{R}_{3}\left[\begin{array}{c}
1\\
\frac{\lambda_{4}-\eta_{1}}{\eta_{2}}\end{array}\right], &  & \grave{R}_{3}=\left\{ 1+\left[\frac{\lambda_{4}-\eta_{1}}{\eta_{2}}\right]^{2}\right\} ^{-1/2},\label{eq:eigenvector-Lambda4}\end{eqnarray}
 where $\grave{R}_{3}$ and $\grave{R}_{4}$ are the normalization
constants. The solutions for the matrix equation (\ref{eq:Eq-of-Mot-Plate-System-Static-Matrix-Homo})
are then \begin{eqnarray*}
\vec{\phi}_{\lambda_{3}}=\vec{R}_{\lambda_{3}}\exp\left(\lambda_{3}t\right)=\grave{R}_{4}\left[\begin{array}{c}
\frac{\eta_{2}}{\lambda_{3}-\eta_{1}}\exp\left(\lambda_{3}t\right)\\
\exp\left(\lambda_{3}t\right)\end{array}\right], &  & \vec{\phi}_{\lambda_{4}}=\vec{R}_{\lambda_{4}}\exp\left(\lambda_{4}t\right)=\grave{R}_{3}\left[\begin{array}{c}
\exp\left(\lambda_{4}t\right)\\
\frac{\lambda_{4}-\eta_{1}}{\eta_{2}}\exp\left(\lambda_{4}t\right)\end{array}\right].\end{eqnarray*}
 The fundamental matrix solution $\widetilde{\Phi}\left(t\right)=\left[\vec{\phi}_{\lambda_{3}}\left(t\right),\vec{\phi}_{\lambda_{4}}\left(t\right)\right]$
is given by \begin{align}
\widetilde{\Phi}\left(t\right) & =\left[\begin{array}{cc}
\frac{\eta_{2}}{\lambda_{3}-\eta_{1}}\grave{R}_{4}\exp\left(\lambda_{3}t\right) & \grave{R}_{3}\exp\left(\lambda_{4}t\right)\\
\grave{R}_{4}\exp\left(\lambda_{3}t\right) & \frac{\lambda_{4}-\eta_{1}}{\eta_{2}}\grave{R}_{3}\exp\left(\lambda_{4}t\right)\end{array}\right].\label{eq:Fundamental-Matrix}\end{align}
 The fundamental matrix solution $\widetilde{\Phi}\left(t\right)$
has an inverse \begin{align*}
\widetilde{\Phi}^{-1}\left(t\right) & =\frac{1}{\det\left(\widetilde{\Phi}\left(t\right)\right)}\left[\begin{array}{cc}
\frac{\lambda_{4}-\eta_{1}}{\eta_{2}}\grave{R}_{3}\exp\left(\lambda_{4}t\right) & -\grave{R}_{3}\exp\left(\lambda_{4}t\right)\\
-\grave{R}_{4}\exp\left(\lambda_{3}t\right) & \frac{\eta_{2}}{\lambda_{3}-\eta_{1}}\grave{R}_{4}\exp\left(\lambda_{3}t\right)\end{array}\right],\end{align*}
 where \begin{align}
\det\left(\widetilde{\Phi}\left(t\right)\right) & =\left[\frac{\lambda_{4}-\eta_{1}}{\lambda_{3}-\eta_{1}}-1\right]\grave{R}_{3}\grave{R}_{4}\exp\left(\left[\lambda_{3}+\lambda_{4}\right]t\right).\label{eq:Fundamental-Matrix-Determinant}\end{align}
 The principal matrix solution $\widetilde{\Psi}\left(t,t_{0}\right)=\widetilde{\Phi}\left(t\right)\cdot\widetilde{\Phi}^{-1}\left(t_{0}\right)$
of equation (\ref{eq:Eq-of-Mot-Plate-System-Static-Matrix}) becomes
then \begin{align}
\widetilde{\Psi}\left(t,t_{0}\right) & =\frac{1}{\det\left(\widetilde{\Phi}\left(t_{0}\right)\right)}\left[\begin{array}{cc}
\psi_{11}\left(t,t_{0}\right) & \psi_{12}\left(t,t_{0}\right)\\
\psi_{21}\left(t,t_{0}\right) & \psi_{22}\left(t,t_{0}\right)\end{array}\right],\label{eq:Principal-Matrix}\end{align}
 where \begin{align}
\psi_{11}\left(t,t_{0}\right) & =\grave{R}_{3}\grave{R}_{4}\left[\frac{\lambda_{4}-\eta_{1}}{\lambda_{3}-\eta_{1}}\exp\left(\lambda_{3}t+\lambda_{4}t_{0}\right)-\exp\left(\lambda_{4}t+\lambda_{3}t_{0}\right)\right],\label{eq:Principal-Matrix-Psi-11}\end{align}
 \begin{align}
\psi_{12}\left(t,t_{0}\right) & =\grave{R}_{3}\grave{R}_{4}\left[\frac{\eta_{2}}{\lambda_{3}-\eta_{1}}\exp\left(\lambda_{4}t+\lambda_{3}t_{0}\right)-\frac{\eta_{2}}{\lambda_{3}-\eta_{1}}\exp\left(\lambda_{3}t+\lambda_{4}t_{0}\right)\right],\label{eq:Principal-Matrix-Psi-12}\end{align}
 \begin{align}
\psi_{21}\left(t,t_{0}\right) & =\grave{R}_{3}\grave{R}_{4}\left[\frac{\lambda_{4}-\eta_{1}}{\eta_{2}}\exp\left(\lambda_{3}t+\lambda_{4}t_{0}\right)-\frac{\lambda_{4}-\eta_{1}}{\eta_{2}}\exp\left(\lambda_{4}t+\lambda_{3}t_{0}\right)\right],\label{eq:Principal-Matrix-Psi-21}\end{align}
 \begin{align}
\psi_{22}\left(t,t_{0}\right) & =\grave{R}_{3}\grave{R}_{4}\left[\frac{\lambda_{4}-\eta_{1}}{\lambda_{3}-\eta_{1}}\exp\left(\lambda_{4}t+\lambda_{3}t_{0}\right)-\exp\left(\lambda_{3}t+\lambda_{4}t_{0}\right)\right].\label{eq:Principal-Matrix-Psi-22}\end{align}
 The inverse of principal matrix solution $\widetilde{\Psi}\left(t,t_{0}\right)$
is \begin{align}
\widetilde{\Psi}^{-1}\left(t,t_{0}\right) & =\frac{1}{\det\left(\widetilde{\Psi}\left(t,t_{0}\right)\right)\det\left(\widetilde{\Phi}\left(t_{0}\right)\right)}\left[\begin{array}{cc}
\psi_{22}\left(t,t_{0}\right) & -\psi_{12}\left(t,t_{0}\right)\\
-\psi_{21}\left(t,t_{0}\right) & \psi_{11}\left(t,t_{0}\right)\end{array}\right],\label{eq:Principal-Matrix-Inverse}\end{align}
 where \begin{align}
\det\left(\widetilde{\Psi}\left(t,t_{0}\right)\right) & =\left[\det\left(\widetilde{\Phi}\left(t_{0}\right)\right)\right]^{-2}\left[\psi_{11}\left(t,t_{0}\right)\psi_{22}\left(t,t_{0}\right)-\psi_{12}\left(t,t_{0}\right)\psi_{21}\left(t,t_{0}\right)\right].\label{eq:Principal-Matrix-Determinant}\end{align}
 Using a variation-of-parameters technique, the solution to the inhomogeneous
first-order differential equation (\ref{eq:Eq-of-Mot-Plate-System-Static-Matrix})
is \begin{align*}
\vec{R}_{\eta}\left(t\right) & =\widetilde{\Psi}\left(t,t_{0}\right)\cdot\vec{R}_{\eta}\left(t_{0}\right)+\widetilde{\Psi}\left(t,t_{0}\right)\cdot\int_{t_{0}}^{t}\widetilde{\Psi}^{-1}\left(t',t_{0}\right)\cdot\vec{\xi}\left(t'\right)dt',\end{align*}
 where it is understood the multiplications are that of the matrix
operations. Substituting into this integral equation the results for
$\vec{R}_{\eta}\left(t\right),$ $\vec{\xi}\left(t'\right),$ $\widetilde{\Psi}\left(t\right)$
and $\widetilde{\Psi}^{-1}\left(t'\right)$ given by equations (\ref{eq:Eq-of-Mot-Plate-System-Static-Matrix}),
(\ref{eq:Principal-Matrix}) and (\ref{eq:Principal-Matrix-Inverse}),
\begin{align*}
\left[\begin{array}{c}
R_{3}\left(t\right)\\
R_{4}\left(t\right)\end{array}\right] & =\frac{1}{\det\left(\widetilde{\Phi}\left(t_{0}\right)\right)}\left(\left[\begin{array}{cc}
\psi_{11}\left(t,t_{0}\right) & \psi_{12}\left(t,t_{0}\right)\\
\psi_{21}\left(t,t_{0}\right) & \psi_{22}\left(t,t_{0}\right)\end{array}\right]\left[\begin{array}{c}
R_{3}\left(t_{0}\right)\\
R_{4}\left(t_{0}\right)\end{array}\right]+\left[\begin{array}{cc}
\psi_{11}\left(t,t_{0}\right) & \psi_{12}\left(t,t_{0}\right)\\
\psi_{21}\left(t,t_{0}\right) & \psi_{22}\left(t,t_{0}\right)\end{array}\right]\right.\\
 & \left.\cdot\int_{t_{0}}^{t}\left\{ \frac{1}{\det\left(\widetilde{\Psi}\left(t',t_{0}\right)\right)}\left[\begin{array}{cc}
\psi_{22}\left(t',t_{0}\right) & -\psi_{12}\left(t',t_{0}\right)\\
-\psi_{21}\left(t',t_{0}\right) & \psi_{11}\left(t',t_{0}\right)\end{array}\right]\left[\begin{array}{c}
\xi_{rp}\left(t'\right)\\
\xi_{lp}\left(t'\right)\end{array}\right]\right\} dt'\right)\end{align*}
 or \begin{align}
R_{3}\left(t\right) & =\frac{1}{\det\left(\widetilde{\Phi}\left(t_{0}\right)\right)}\left\{ \psi_{11}\left(t,t_{0}\right)R_{3}\left(t_{0}\right)+\psi_{12}\left(t,t_{0}\right)R_{4}\left(t_{0}\right)+\frac{\psi_{11}\left(t,t_{0}\right)}{\det\left(\widetilde{\Phi}\left(t_{0}\right)\right)}\right.\nonumber \\
 & \times\left[\int_{t_{0}}^{t}\frac{\psi_{22}\left(t',t_{0}\right)\xi_{rp}\left(t'\right)}{\det\left(\widetilde{\Psi}\left(t',t_{0}\right)\right)}dt'-\int_{t_{0}}^{t}\frac{\psi_{12}\left(t',t_{0}\right)\xi_{lp}\left(t'\right)}{\det\left(\widetilde{\Psi}\left(t',t_{0}\right)\right)}dt'\right]+\frac{\psi_{12}\left(t,t_{0}\right)}{\det\left(\widetilde{\Phi}\left(t_{0}\right)\right)}\nonumber \\
 & \left.\times\left[\int_{t_{0}}^{t}\frac{\psi_{11}\left(t',t_{0}\right)\xi_{lp}\left(t'\right)}{\det\left(\widetilde{\Psi}\left(t',t_{0}\right)\right)}dt'-\int_{t_{0}}^{t}\frac{\psi_{21}\left(t',t_{0}\right)\xi_{rp}\left(t'\right)}{\det\left(\widetilde{\Psi}\left(t',t_{0}\right)\right)}dt'\right]\right\} ,\label{eq:Eq-of-Mot-R3}\end{align}
 \begin{align}
R_{4}\left(t\right) & =\frac{1}{\det\left(\widetilde{\Phi}\left(t_{0}\right)\right)}\left\{ \psi_{21}\left(t,t_{0}\right)R_{3}\left(t_{0}\right)+\psi_{22}\left(t,t_{0}\right)R_{4}\left(t_{0}\right)+\frac{\psi_{21}\left(t,t_{0}\right)}{\det\left(\widetilde{\Phi}\left(t_{0}\right)\right)}\right.\nonumber \\
 & \times\left[\int_{t_{0}}^{t}\frac{\psi_{22}\left(t',t_{0}\right)\xi_{rp}\left(t'\right)}{\det\left(\widetilde{\Psi}\left(t',t_{0}\right)\right)}dt'-\int_{t_{0}}^{t}\frac{\psi_{12}\left(t',t_{0}\right)\xi_{lp}\left(t'\right)}{\det\left(\widetilde{\Psi}\left(t',t_{0}\right)\right)}dt'\right]+\frac{\psi_{22}\left(t,t_{0}\right)}{\det\left(\widetilde{\Phi}\left(t_{0}\right)\right)}\nonumber \\
 & \left.\times\left[\int_{t_{0}}^{t}\frac{\psi_{11}\left(t',t_{0}\right)\xi_{lp}\left(t'\right)}{\det\left(\widetilde{\Psi}\left(t',t_{0}\right)\right)}dt'-\int_{t_{0}}^{t}\frac{\psi_{21}\left(t',t_{0}\right)\xi_{rp}\left(t'\right)}{\det\left(\widetilde{\Psi}\left(t',t_{0}\right)\right)}dt'\right]\right\} .\label{eq:Eq-of-Mot-R4}\end{align}
 It is noted from equation (\ref{eq:Eq-of-Mot-Plate-System-Static-Matrix-R1-R2}),
$R_{3}\left(t_{0}\right)$ and $R_{4}\left(t_{0}\right)$ are initial
speeds, \begin{eqnarray*}
\dot{R}_{rp,cm,\alpha}\left(t_{0}\right)\equiv\dot{R}_{1}\left(t_{0}\right)=R_{3}\left(t_{0}\right), &  & \dot{R}_{lp,cm,\alpha}\left(t_{0}\right)\equiv\dot{R}_{2}\left(t_{0}\right)=R_{4}\left(t_{0}\right).\end{eqnarray*}
 Hence, \begin{align}
\dot{R}_{rp,cm,\alpha}\left(t\right) & =\left[\frac{\lambda_{4}\left(;t_{0}\right)-\eta_{1}\left(;t_{0}\right)}{\lambda_{3}\left(;t_{0}\right)-\eta_{1}\left(;t_{0}\right)}-1\right]^{-1}\frac{\psi_{11}\left(t,t_{0}\right)\dot{R}_{rp,cm,\alpha}\left(t_{0}\right)+\psi_{12}\left(t,t_{0}\right)\dot{R}_{lp,cm,\alpha}\left(t_{0}\right)}{\exp\left(\left[\lambda_{3}\left(;t_{0}\right)+\lambda_{4}\left(;t_{0}\right)\right]t_{0}\right)}\nonumber \\
 & +\psi_{11}\left(t,t_{0}\right)\int_{t_{0}}^{t}\frac{\psi_{22}\left(t',t_{0}\right)\xi_{rp}\left(t'\right)-\psi_{12}\left(t',t_{0}\right)\xi_{lp}\left(t'\right)}{\psi_{11}\left(t',t_{0}\right)\psi_{22}\left(t',t_{0}\right)-\psi_{12}\left(t',t_{0}\right)\psi_{21}\left(t',t_{0}\right)}dt'+\psi_{12}\left(t,t_{0}\right)\nonumber \\
 & \times\int_{t_{0}}^{t}\frac{\psi_{11}\left(t',t_{0}\right)\xi_{lp}\left(t'\right)-\psi_{21}\left(t',t_{0}\right)\xi_{rp}\left(t'\right)}{\psi_{11}\left(t',t_{0}\right)\psi_{22}\left(t',t_{0}\right)-\psi_{12}\left(t',t_{0}\right)\psi_{21}\left(t',t_{0}\right)}dt',\label{eq:Eq-of-Mot-R1-Speed}\end{align}
 \begin{align}
\dot{R}_{lp,cm,\alpha}\left(t\right) & =\left[\frac{\lambda_{4}\left(;t_{0}\right)-\eta_{1}\left(;t_{0}\right)}{\lambda_{3}\left(;t_{0}\right)-\eta_{1}\left(;t_{0}\right)}-1\right]^{-1}\frac{\psi_{21}\left(t,t_{0}\right)\dot{R}_{rp,cm,\alpha}+\psi_{22}\left(t,t_{0}\right)\dot{R}_{lp,cm,\alpha}\left(t_{0}\right)}{\exp\left(\left[\lambda_{3}\left(;t_{0}\right)+\lambda_{4}\left(;t_{0}\right)\right]t_{0}\right)}\nonumber \\
 & +\psi_{21}\left(t,t_{0}\right)\int_{t_{0}}^{t}\frac{\psi_{22}\left(t',t_{0}\right)\xi_{rp}\left(t'\right)-\psi_{12}\left(t',t_{0}\right)\xi_{lp}\left(t'\right)}{\psi_{11}\left(t',t_{0}\right)\psi_{22}\left(t',t_{0}\right)-\psi_{12}\left(t',t_{0}\right)\psi_{21}\left(t',t_{0}\right)}dt'+\psi_{22}\left(t,t_{0}\right)\nonumber \\
 & \times\int_{t_{0}}^{t}\frac{\psi_{11}\left(t',t_{0}\right)\xi_{lp}\left(t'\right)-\psi_{21}\left(t',t_{0}\right)\xi_{rp}\left(t'\right)}{\psi_{11}\left(t',t_{0}\right)\psi_{22}\left(t',t_{0}\right)-\psi_{12}\left(t',t_{0}\right)\psi_{21}\left(t',t_{0}\right)}dt',\label{eq:Eq-of-Mot-R2-Speed}\end{align}
 where substitutions have been made for the determinants $\det\left(\widetilde{\Phi}\left(t_{0}\right)\right)$
and $\det\left(\widetilde{\Psi}\left(t',t_{0}\right)\right)$ from
equations (\ref{eq:Fundamental-Matrix-Determinant}) and (\ref{eq:Principal-Matrix-Determinant}).
It is to be understood that the notation $\left(;t_{0}\right)$ on
$\eta_{1},$ $\lambda_{3}$ and $\lambda_{4}$ implies implicit time
dependence for these terms. Finally, integration of both sides of
equations (\ref{eq:Eq-of-Mot-R1-Speed}) and (\ref{eq:Eq-of-Mot-R2-Speed})
with respect to time gives the results \begin{align}
R_{rp,cm,\alpha}\left(t\right) & =\left[\frac{\lambda_{4}\left(;t_{0}\right)-\eta_{1}\left(;t_{0}\right)}{\lambda_{3}\left(;t_{0}\right)-\eta_{1}\left(;t_{0}\right)}-1\right]^{-1}\int_{t_{0}}^{t}\left[\frac{\psi_{11}\left(\tau,t_{0}\right)\dot{R}_{rp,cm,\alpha}\left(t_{0}\right)+\psi_{12}\left(\tau,t_{0}\right)\dot{R}_{lp,cm,\alpha}\left(t_{0}\right)}{\exp\left(\left[\lambda_{3}\left(;t_{0}\right)+\lambda_{4}\left(;t_{0}\right)\right]t_{0}\right)}\right.\nonumber \\
 & +\psi_{11}\left(\tau,t_{0}\right)\int_{t_{0}}^{\tau}\frac{\psi_{22}\left(t',t_{0}\right)\xi_{rp}\left(t'\right)-\psi_{12}\left(t',t_{0}\right)\xi_{lp}\left(t'\right)}{\psi_{11}\left(t',t_{0}\right)\psi_{22}\left(t',t_{0}\right)-\psi_{12}\left(t',t_{0}\right)\psi_{21}\left(t',t_{0}\right)}dt'+\psi_{12}\left(\tau,t_{0}\right)\nonumber \\
 & \left.\times\int_{t_{0}}^{\tau}\frac{\psi_{11}\left(t',t_{0}\right)\xi_{lp}\left(t'\right)-\psi_{21}\left(t',t_{0}\right)\xi_{rp}\left(t'\right)}{\psi_{11}\left(t',t_{0}\right)\psi_{22}\left(t',t_{0}\right)-\psi_{12}\left(t',t_{0}\right)\psi_{21}\left(t',t_{0}\right)}dt'\right]d\tau+R_{rp,cm,\alpha}\left(t_{0}\right),\label{eq:Eq-of-Mot-R1-Position}\end{align}
 \begin{align}
R_{lp,cm,\alpha}\left(t\right) & =\left[\frac{\lambda_{4}\left(;t_{0}\right)-\eta_{1}\left(;t_{0}\right)}{\lambda_{3}\left(;t_{0}\right)-\eta_{1}\left(;t_{0}\right)}-1\right]^{-1}\int_{t_{0}}^{t}\left[\frac{\psi_{21}\left(\tau,t_{0}\right)\dot{R}_{rp,cm,\alpha}\left(t_{0}\right)+\psi_{22}\left(\tau,t_{0}\right)\dot{R}_{lp,cm,\alpha}\left(t_{0}\right)}{\exp\left(\left[\lambda_{3}\left(;t_{0}\right)+\lambda_{4}\left(;t_{0}\right)\right]t_{0}\right)}\right.\nonumber \\
 & +\psi_{21}\left(\tau,t_{0}\right)\int_{t_{0}}^{\tau}\frac{\psi_{22}\left(t',t_{0}\right)\xi_{rp}\left(t'\right)-\psi_{12}\left(t',t_{0}\right)\xi_{lp}\left(t'\right)}{\psi_{11}\left(t',t_{0}\right)\psi_{22}\left(t',t_{0}\right)-\psi_{12}\left(t',t_{0}\right)\psi_{21}\left(t',t_{0}\right)}dt'+\psi_{22}\left(\tau,t_{0}\right)\nonumber \\
 & \left.\times\int_{t_{0}}^{\tau}\frac{\psi_{11}\left(t',t_{0}\right)\xi_{lp}\left(t'\right)-\psi_{21}\left(t',t_{0}\right)\xi_{rp}\left(t'\right)}{\psi_{11}\left(t',t_{0}\right)\psi_{22}\left(t',t_{0}\right)-\psi_{12}\left(t',t_{0}\right)\psi_{21}\left(t',t_{0}\right)}dt'\right]d\tau+R_{lp,cm,\alpha}\left(t_{0}\right),\label{eq:Eq-of-Mot-R2-Position}\end{align}
 where the terms $\psi_{11},$ $\psi_{12},$ $\psi_{21}$ and $\psi_{22}$
are defined in equations (\ref{eq:Principal-Matrix-Psi-11}), (\ref{eq:Principal-Matrix-Psi-12}),
(\ref{eq:Principal-Matrix-Psi-21}) and (\ref{eq:Principal-Matrix-Psi-22}).
The remaining integrations are straightforward; hence, their explicit
forms are not shown. 

As a closing remark of this section, one may argue that for the static
case, $\dot{R}_{rp,cm,\alpha}\left(t_{0}\right)$ and $\dot{R}_{lp,cm,\alpha}\left(t_{0}\right)$
must be zero because the conductors seem to be fixed in position.
This argument is flawed for any wall totally fixed in position upon
impact would require an infinite amount of energy. One has to consider
the conservation of momentum simultaneously. The wall has to have
moved by the amount $\bigtriangleup R_{wall}=\dot{R}_{wall}\triangle t,$
where $\triangle t$ is the total duration of impact, and $\dot{R}_{wall}$
is calculated from the momentum conservation and it is non-zero. The
same argument can be applied to the apparatus shown in Figure \ref{cap:driven-parallel-plates-force}.
For that system \begin{eqnarray*}
\left\Vert \vec{p}_{virtual-photon}\right\Vert =\frac{1}{c}\mathcal{H}'_{n_{s},\Re}\left(t_{0}\right), &  & \left\{ \begin{array}{c}
\dot{R}_{rp,cm,\alpha}\left(t_{0}\right)=\left\Vert \dot{\vec{R}}_{lp,3}\left(t_{0}\right)+\dot{\vec{R}}_{rp,2}\left(t_{0}\right)\right\Vert ,\\
\\\dot{R}_{lp,cm,\alpha}\left(t_{0}\right)=\left\Vert \dot{\vec{R}}_{rp,1}\left(t_{0}\right)+\dot{\vec{R}}_{lp,2}\left(t_{0}\right)\right\Vert \end{array}\right.\end{eqnarray*}
 or, for simplicity, assuming an impact along the normal direction,
\begin{eqnarray*}
\dot{R}_{rp,cm,\alpha}\left(t_{0}\right)=\frac{2}{m_{rp}c}\left\Vert \mathcal{H}'_{n_{s},3}\left(t_{0}\right)-\mathcal{H}'_{n_{s},2}\left(t_{0}\right)\right\Vert , &  & \dot{R}_{lp,cm,\alpha}\left(t_{0}\right)=\frac{2}{m_{lp}c}\left\Vert \mathcal{H}'_{n_{s},1}\left(t_{0}\right)-\mathcal{H}'_{n_{s},2}\left(t_{0}\right)\right\Vert ,\end{eqnarray*}
 where the difference under the magnitude symbol implies that the
energies from different regions act to counteract each other.

\end{document}